\documentclass[sigconf, nonacm, pdfa]{acmart}

\newcommand\vldbdoi{10.14778/3718057.3718059}
\newcommand\vldbpages{1278 - 1290}
\newcommand\vldbvolume{18}
\newcommand\vldbissue{5}
\newcommand\vldbyear{2025}
\newcommand\vldbauthors{\authors}
\newcommand\vldbtitle{\shorttitle} 
\newcommand\vldbavailabilityurl{URL_TO_YOUR_ARTIFACTS}
\newcommand\vldbpagestyle{empty} 

\usepackage{float}
\usepackage{graphicx}
\usepackage{enumerate}
\usepackage{enumitem}
\usepackage{subfigure}
\usepackage{color}
\usepackage{mathtools}
\usepackage{epstopdf}
\usepackage{balance}
\usepackage{tabularx}

\usepackage[linesnumbered,vlined,ruled,noend]{algorithm2e}

\newcommand{\cut}[1]{}

\newcommand{\todo}[1]{\textcolor{red}{TODO: #1}}

\newcommand{\xy}[1]{\textcolor{cyan}{#1}}
\newcommand{\revision}[1]{\textcolor{black}{#1}}

\newtheorem{theorem}{Theorem}
\newtheorem{definition}{Definition}
\newtheorem{property}{Property}

\newtheorem{example}{Example}
\newtheorem{lemma}{Lemma}

\newtheorem{assumption}{Assumption}

\newtheorem{procedure}{Procedure}

\DeclareMathOperator{\cost}{cost}

\usepackage{xspace}
\newcommand{\sysname}{\emph{Esc}\xspace}

\begin{document}
\title{Esc: An Early-Stopping Checker for Budget-aware Index Tuning}


\author{Xiaoying Wang}
\affiliation{%
  \institution{Simon Fraser Universtiy}
  \city{Burnaby}
  \country{Canada}
}
\email{xiaoying\_wang@sfu.ca}

\author{Wentao Wu}
\affiliation{%
  \institution{Microsoft Research}
  \city{Redmond}
  \country{USA}
}
\email{wentao.wu@microsoft.com}

\author{Vivek Narasayya}
\affiliation{%
  \institution{Microsoft Research}
  \city{Redmond}
  \country{USA}
}
\email{viveknar@microsoft.com}

\author{Surajit Chaudhuri}
\affiliation{%
  \institution{Microsoft Research}
  \city{Redmond}
  \country{USA}
}
\email{surajitc@microsoft.com}




\begin{abstract}

Index tuning is a time-consuming process. 
One major performance bottleneck in existing index tuning systems is the large amount of ``what-if'' query optimizer calls that estimate the cost of a given pair of query and index configuration without materializing the indexes.
There has been recent work on \emph{budget-aware} index tuning that limits the amount of what-if calls allowed in index tuning.
Existing budget-aware index tuning algorithms, however, typically make fast progress early on in terms of the best configuration found but slow down when more and more what-if calls are allocated.
This observation of ``diminishing return'' on index quality leads us to introduce \emph{early stopping} for budget-aware index tuning, where user specifies a threshold on the tolerable loss of index quality and we stop index tuning if the projected loss with the remaining budget is below the threshold.
We further propose \sysname, a low-overhead early-stopping checker that realizes this new functionality.
Experimental evaluation on top of both industrial benchmarks and real customer workloads demonstrates that \sysname can significantly reduce the number of what-if calls made during budget-aware index tuning while incurring little or zero improvement loss and little extra computational overhead compared to the overall index tuning time.
\end{abstract}

\maketitle

\pagestyle{\vldbpagestyle}
\begingroup\small\noindent\raggedright\textbf{PVLDB Reference Format:}\\
\vldbauthors. \vldbtitle. PVLDB, \vldbvolume(\vldbissue): \vldbpages, \vldbyear.\\
\href{https://doi.org/\vldbdoi}{doi:\vldbdoi}
\endgroup
\begingroup
\renewcommand\thefootnote{}\footnote{\noindent
This work is licensed under the Creative Commons BY-NC-ND 4.0 International License. Visit \url{https://creativecommons.org/licenses/by-nc-nd/4.0/} to view a copy of this license. For any use beyond those covered by this license, obtain permission by emailing \href{mailto:info@vldb.org}{info@vldb.org}. Copyright is held by the owner/author(s). Publication rights licensed to the VLDB Endowment. \\
\raggedright Proceedings of the VLDB Endowment, Vol. \vldbvolume, No. \vldbissue\ %
ISSN 2150-8097. \\
\href{https://doi.org/\vldbdoi}{doi:\vldbdoi} \\
}\addtocounter{footnote}{-1}\endgroup


\vspace{-0.5em}
\section{Introduction}

\begin{figure}[t]
\centering
    \includegraphics[clip, trim=4cm 4cm 4.5cm 2.5cm, width=0.95\columnwidth]{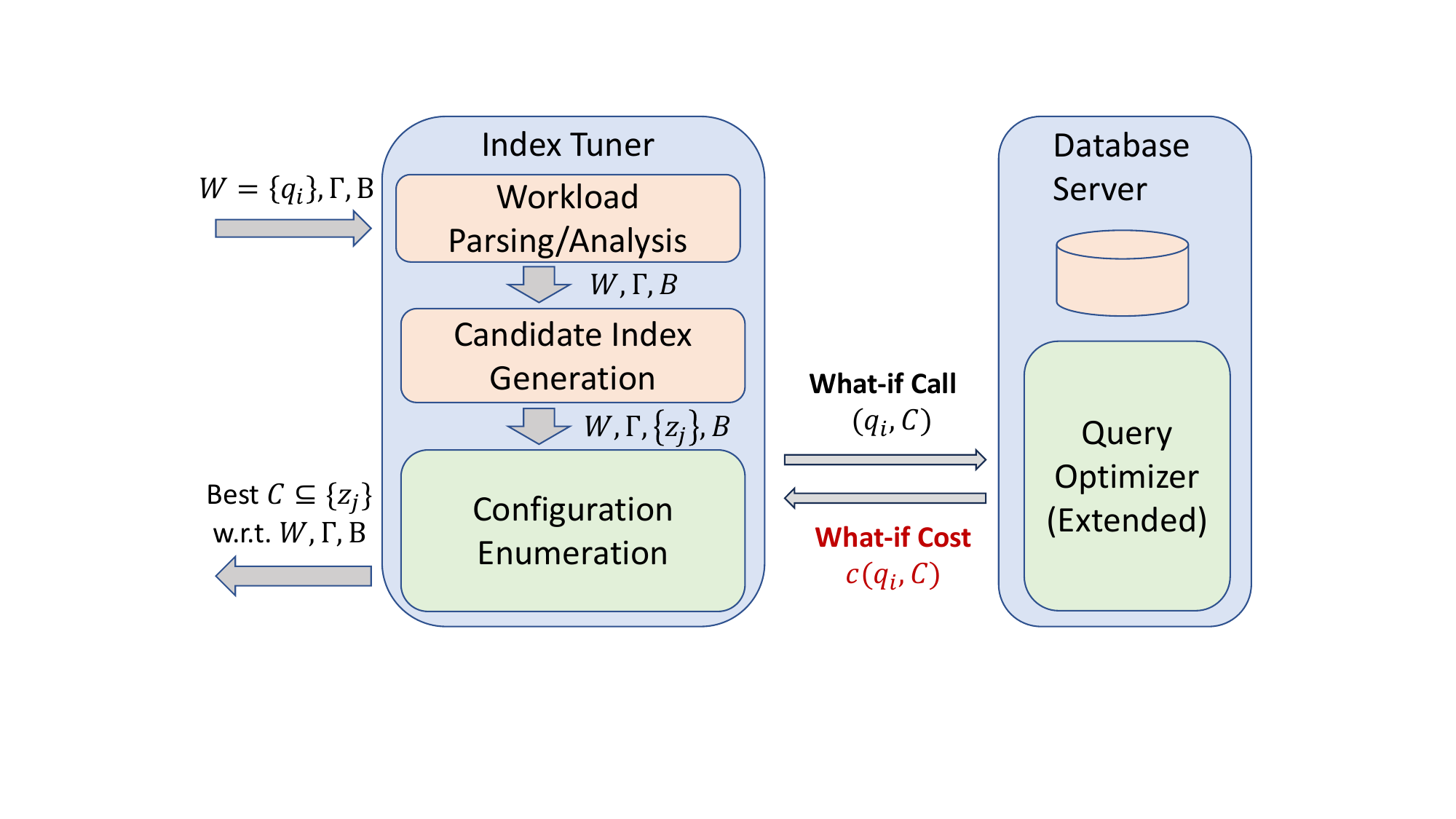}
\vspace{-1em}
\caption{The architecture of cost-based index tuning using what-if optimizer calls, where $W$ 
is the input workload and $q_i\in W$ is a single query, $\Gamma$ is a set of tuning constraints, $\{z_j\}$ is the set of candidate indexes generated for $W$, and $C\subseteq\{z_j\}$ represents an index configuration during enumeration.}
\label{fig:what-if-architecture}
\end{figure}

Index tuning is a time-consuming process that may take hours to finish for large and complex workloads.
Existing index tuners typically adopt a cost-based tuning architecture~\cite{ChaudhuriN97,ValentinZZLS00}, as illustrated in Figure~\ref{fig:what-if-architecture}.
It consists of three main components: (1) \emph{workload parsing and analysis}, which parses each query in the workload and extracts \emph{indexable columns}, e.g., columns that appear in selection and join predicates; (2) \emph{candidate index generation}, which puts together the extracted indexable columns to generate a set of indexes that can potentially reduce the execution cost of the input workload; and (3) \emph{configuration enumeration}, which looks for a subset (a.k.a., configuration) from the candidate indexes that meets the input constraints (e.g., maximum configuration size or amount of storage to be taken by the indexes) while minimizing the input workload cost.
To evaluate the cost of a given query and configuration pair, index tuners rely on the so-called ``what-if'' utility~\cite{ChaudhuriN98}.
It is an extended API of the query optimizer that can estimate the cost by viewing the indexes contained by the configuration as ``hypothetical indexes'' instead of materializing them in a storage system, which would be much more costly. Nevertheless, what-if optimizer calls are not free---they are at least as expensive as a regular query optimizer call.
As a result, they become the major bottleneck when tuning large and/or complex workloads~\cite{ml-index-tuning-overview}.

To address this challenge, some technologies have been developed, such as cost derivation~\cite{ChaudhuriN97}, caching/reusing what-if calls~\cite{PapadomanolakisDA07} that requires code changes to the query optimizer beyond the what-if API, or ML-based cost approximation~\cite{SiddiquiWNC22}.
Recent research has proposed \emph{budget-aware} index tuning, which
constrains the number of what-if calls allowed during configuration enumeration~\cite{WuWSWNCB22}.
Here, the main challenge shifts from reducing the number of what-if calls in classic index tuning to prioritizing what-if calls w.r.t. the importance of query-configuration pairs in budget-aware index tuning.
This problem is termed as \emph{budget allocation}, and there has been recent work on optimizing budget allocation in a \emph{dynamic} manner that skips \emph{inessential} what-if calls at index tuning runtime by utilizing lower and upper bounds of what-if costs~\cite{Wii}.

\begin{figure}[t]
\centering
\subfigure[\textbf{TPC-H}, \emph{two-phase greedy search}]{ \label{fig:lc-example:twophase:tpch:origin}
    \includegraphics[width=0.48\columnwidth]{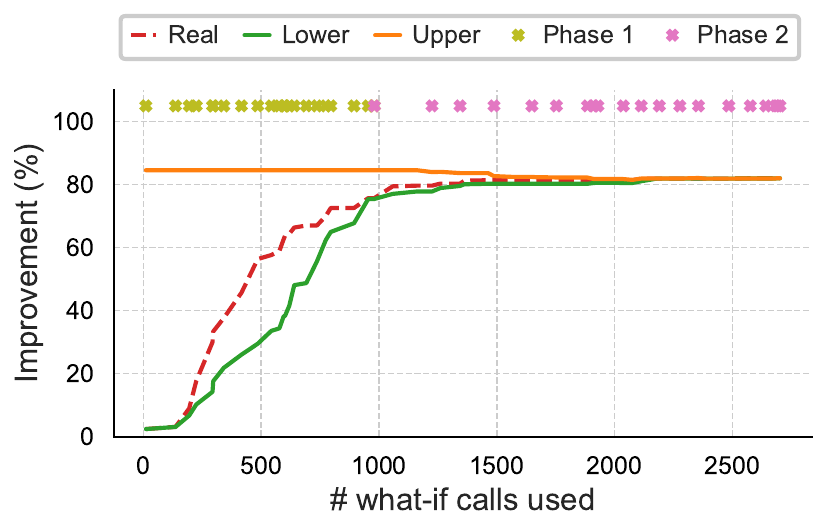}}
\subfigure[\textbf{Real-D}, \emph{MCTS}]{ \label{fig:lc-example:mcts:real-d:origin}
    \includegraphics[width=0.48\columnwidth]{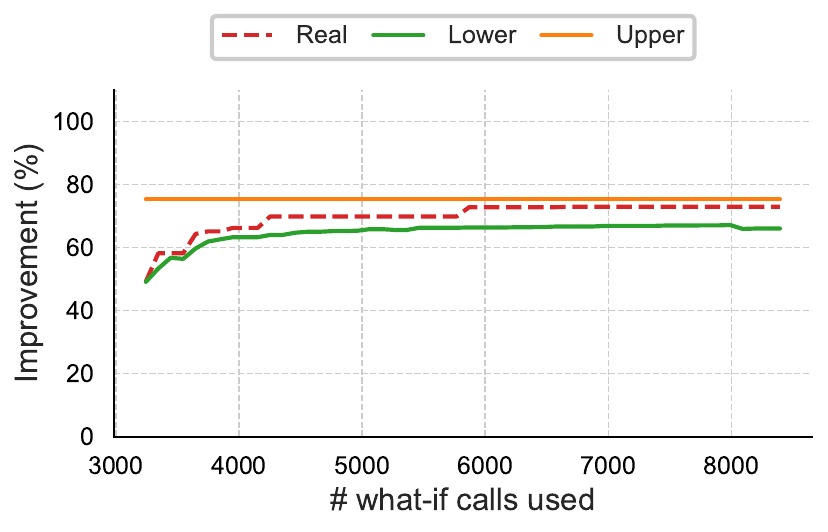}}
\vspace{-1.5em}
\caption{\revision{Examples of index tuning curves of \emph{two-phase greedy search} and \emph{MCTS}, where we set the number of indexes allowed $K=20$ and the budget on what-if calls $B=20,000$.}}
\label{fig:tuning-curve-example}
\end{figure}

In practice, we have observed the following ``diminishing return'' behavior of existing budget-aware index tuning algorithms: they typically make fast progress at the beginning in terms of the best index configuration found, but their progress slows down as more budget on what-if calls is allocated.
To put our discussion in context, Figure~\ref{fig:tuning-curve-example} presents examples of the \emph{index tuning curve} (ITC) when using two state-of-the-art budget-aware index tuning algorithms (see Section~\ref{sec:preliminaries}), namely, \emph{two-phase greedy search} and \emph{Monte Carlo tree search} (\emph{MCTS} for short), to tune the \textbf{TPC-H} benchmark workload and a real customer workload \textbf{Real-D} (see Section~\ref{sec:eval:settings:workloads}).
We defer a formal discussion of ITC to Section~\ref{sec:esc:index-tuning-curve}.
Roughly speaking, the ITC represents a \emph{function} that maps from the number of what-if calls made to the \emph{percentage improvement} of the best configuration found, where the percentage improvement is defined as
\vspace{-0.5em}
\begin{equation}\label{eq:percentage-impr}
  \eta(W, C)=\frac{c(W, \emptyset)-c(W, C)}{c(W, \emptyset)}=1-\frac{c(W, C)}{c(W, \emptyset)}.  
\vspace{-0.5em}
\end{equation}
Here, $W$ represents the input workload, $C$ represents a configuration, and $\emptyset$ represents the \emph{existing configuration} that index tuning starts from.
$c(W, C)=\sum_{q\in W} c(q, C)$ represents the what-if cost of the workload $W$ on top of the configuration $C$, which is the sum of the what-if costs of individual queries contained by $W$.
In each plot of Figure~\ref{fig:tuning-curve-example}, we use the red dashed line to represent the corresponding ITC.
Intuitively, the ITC is a \emph{profile} of the index tuner that characterizes its progress made so far with respect to the amount of budget on what-if calls being allocated.

This ``diminishing return'' behavior of existing budget-aware index tuning algorithms motivates us to introduce \emph{early stopping}.
Specifically, let $\epsilon$ (e.g., $\epsilon=5\%$) be a user-given threshold that controls the \emph{loss} on the percentage improvement, i.e., the gap between the percentage improvement of the best configuration found so far and the percentage improvement of the final best configuration with all budget allocated.
If the projected improvement loss is below $\epsilon$ after certain amount of what-if calls are made, then we can safely terminate index tuning.
Early stopping enables further savings on the number of what-if calls made in index tuning, and the savings can often be considerable.
For example, as shown in Figure~\ref{fig:lc-example:twophase:tpch:origin}, \emph{two-phase greedy search} requires making around 2,700 what-if calls to tune the \textbf{TPC-H} workload without early stopping.
However, it actually makes no further progress (i.e., the best index configuration found does not change) after 1,000 what-if calls are made.
Therefore, we would have saved 1,700 what-if calls, i.e., a reduction of 63\%. 
While early stopping has been a well-known technique in the machine learning (ML) literature for preventing ``overfitting'' when training an ML model with an iterative method such as \emph{gradient descent}~\cite{yao2007early,Prechelt2012,RaskuttiWY14}, to the best of our knowledge we are the first to introduce it for index tuning with a very different goal of saving the amount of what-if calls.


Enabling early stopping for budget-aware index tuning, however, raises new challenges.
First, to project the further improvement loss that is required by triggering early stopping, we need to know (1) the percentage improvement of the best configuration found so far and (2) the percentage improvement of the final best configuration \emph{assuming that all budget were allocated}.
Unfortunately, both are not available at the time point where the projection needs to be made.
While it is clear that (2) is not available, one may wonder why (1) is also not available.
Note that the best configuration found so far in budget-aware index tuning is based on \emph{derived cost} (see Section~\ref{sec:preliminaries:budget-aware-index-tuning}) rather than the true what-if cost~\cite{WuWSWNCB22}.
Technically, we can obtain (1) by making an extra what-if call for each query in the workload with the best configuration found.
However, this is too expensive to be affordable in practice when tuning a large workload.
Second, even if we know (1) and (2) so that we can compute the gap between (1) and (2) to verify whether the projected further improvement loss is below the threshold $\epsilon$, it is unclear \emph{when} this verification should be performed.
Conducting this verification at the beginning of index tuning seems unnecessary, as the index tuner is expected to make fast progress; however, if this verification happens too late, then most of the savings given by early stopping will vanish.

To address these challenges, in this paper we propose \sysname, a low-overhead early-stopping checker for budget-aware index tuning. It is based on the following main ideas:
\begin{itemize}[leftmargin=*]
    \item Instead of measuring the gap between (1) and (2),
    which cannot be obtained in practice, we develop a \emph{lower-bound} for (1) and an \emph{upper-bound} for (2) and then measure the gap between the lower and upper bounds. Clearly, if this gap is below the threshold $\epsilon$, then the gap between (1) and (2) is also below $\epsilon$.
    Figure~\ref{fig:tuning-curve-example} also presents the lower and upper bounds of each index tuning curve.
    \item To avoid verifying early-stopping either too early or too late, we develop a general approach that performs early-stopping verification by monitoring \emph{improvement rate} of the ITC. Specifically, we measure the degree of \emph{convexity}/\emph{concavity} of the ITC based on the variation observed in its improvement rate, and we only verify early stopping when the ITC becomes \emph{concave}.
\end{itemize}
In more detail, we develop the lower and upper bounds of percentage improvement by piggybacking on the previous work~\cite{Wii}.
\revision{While~\cite{Wii} lays the foundation of deriving lower and upper bounds for what-if cost, the bounds work only for individual what-if calls but not the entire workload.
The extension to workload-level bounds is nontrivial---a straightforward approach that simply sums up call-level bounds would lead to workload-level bounds that are too conservative to be useful (Section~\ref{sec:workload-lower-upper-bounds:general}).
Following this observation, we develop new mechanisms to improve over the naive workload-level bounds: (i) a \emph{simulated greedy search} procedure that is designed for optimizing the bounds in the context of greedy search, which has been leveraged by both \emph{two-phase greedy search} and \emph{MCTS} as a basic building block (Section~\ref{sec:workload-lower-upper-bounds:optimizations}) and (ii) a generic approach to refining the bounds by modeling \emph{index interactions}~\cite{SchnaitterPG09} at workload-level (Section~\ref{sec:lower-refinement:index-interaction}).}
On the other hand, 
there can be multiple \emph{concave} stages of an ITC, and only the \emph{final} concave stage is worth early-stopping verification. 
For instance, this final stage of the ITC shown by Figure~\ref{fig:lc-example:mcts:real-d:origin} begins after 6,000 what-if calls are made.
It is challenging to identify whether a \emph{concave} stage is the final one, and we further propose techniques to address this challenge and therefore reduce the chance of unnecessary early-stopping verification.

To summarize, this paper makes the following contributions:
\begin{itemize}[leftmargin=*]
    \item We introduce early stopping for budget-aware index tuning as a new mechanism that can result in significant savings on the number of what-if calls made (Section~\ref{sec:early-stopping}).
    \item We propose \sysname, a novel framework that enables early-stopping in budget-aware index tuning by developing lower/upper bounds of workload-level what-if cost (Section~\ref{sec:workload-lower-upper-bounds}) with refinement by exploiting index interactions (Section~\ref{sec:lower-refinement:index-interaction}) and lightweight verification schemes that leverage improvement rates and convexity/concavity properties of the index tuning curve (Section~\ref{sec:early-stopping-verification}).
    \item We conduct extensive experimental evaluation using both industrial benchmarks and real workloads, and empirical results demonstrate that \sysname can significantly reduce the number of what-if calls for state-of-the-art budget-aware tuning algorithms with little extra computational overhead and little or no improvement loss on the final configuration returned (Section~\ref{sec:eval}). 
\end{itemize}


Last but not least, while we focus on budget-aware index tuning algorithms in this work, \revision{early stopping can be applied to other index tuning algorithms such as (i) classic index tuning algorithms with \emph{unlimited} budget of what-if calls~\cite{KossmannHJS20, Wii}, which can be viewed as a special case of budget-aware index tuning and (ii) \emph{anytime} index tuning algorithms~\cite{dta}, which are more sophisticated than budget-aware index tuning by constraining the overall index tuning time.
Some of the technologies developed in this work, such as (a) the lower/upper bounds of workload-level what-if cost and (b) the general early-stopping verification scheme based on monitoring improvement rates of the index tuning curve, remain applicable, though their efficacy requires further investigation and evaluation.}
We leave this as an interesting direction for future work.

\vspace{-0.5em}
\section{Preliminaries}\label{sec:preliminaries}

We present an overview of the problem of budget allocation and existing budget-aware index tuning algorithms.

\vspace{-0.5em}
\subsection{Budget-aware Index Tuning}
\label{sec:preliminaries:budget-aware-index-tuning}

Budget-aware index tuning aims to constrain the amount of what-if calls that can be made during index tuning, in particular, during index configuration enumeration.
An essential problem of budget-aware index tuning is \emph{budget allocation}, i.e., determining on which query-configuration pairs to make what-if calls.
For any query-configuration pair without making what-if call, we use the \emph{derived cost} from \emph{cost derivation}~\cite{ChaudhuriN97}, defined by
\vspace{-0.5em}
\begin{equation}\label{eq:derived-cost}
    d(q,C)=\min\nolimits_{S\subseteq C}c(q, S),
\vspace{-0.5em}
\end{equation}
as an approximation of its true what-if cost.
There are two existing algorithms that address this budget allocation problem: (1) \emph{two-phase greedy search} and (2) \emph{Monte Carlo tree search} (MCTS).
\revision{Based on the empirical study in~\cite{Wii}, the gap between derived cost and the true what-if cost is below 5\% for 80\% to 90\% of the what-if calls made by these two budget-aware index tuning algorithms.}

\begin{figure}[t]
\centering
    \includegraphics[clip, trim=5cm 6.5cm 5.5cm 4.5cm, width=\columnwidth]{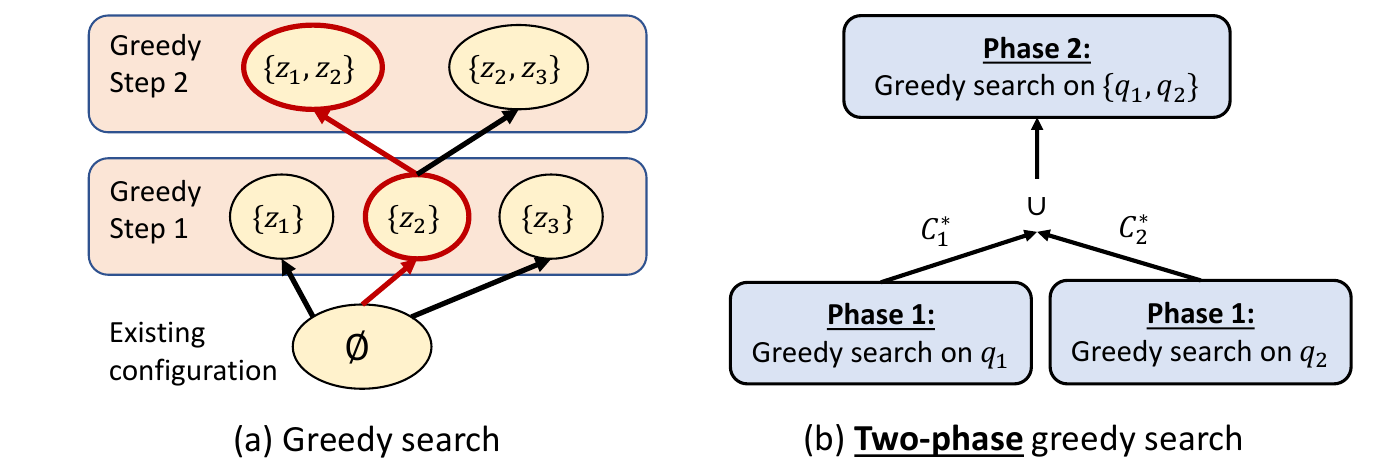}  
\vspace{-2.5em}
\caption{Example of budget-aware greedy search.}
\label{fig:greedy}
\end{figure}

\vspace{-0.5em}
\subsubsection{Two-phase Greedy Search}

A classic configuration enumeration algorithm is \emph{greedy search}~\cite{ChaudhuriN97}, as illustrated in Figure~\ref{fig:greedy}(a).
It is a step-by-step procedure where it selects the next best candidate index in each \emph{greedy step} that minimizes the workload cost, until the selected index configuration meets the given constraints.
An improved version is the so-called \emph{two-phase greedy search}~\cite{ChaudhuriN97}, which first runs greedy search on top of each query to find its best candidate indexes and then runs greedy search again for the entire workload by taking the union of the best candidate indexes found for the individual queries. Figure~\ref{fig:greedy}(b) presents an example of \emph{two-phase greedy search} with two queries in the workload.
What-if calls are allocated in a ``first come first serve'' manner.
\emph{Two-phase greedy search} can achieve state-of-the-art performance~\cite{ChaudhuriN97,KossmannHJS20,WuWSWNCB22,Wii} in terms of the final index configuration found and has also been integrated into commercial database tuning software such as the Database Tuning Advisor (DTA) developed for Microsoft SQL Server~\cite{dta}, 


\vspace{-0.5em}
\subsubsection{Monte Carlo Tree Search}

To better tackle the trade-off between exploration and exploitation in budget allocation, previous work~\cite{WuWSWNCB22} proposed a budget-aware index tuning algorithm based on Monte Carlo tree search (\emph{MCTS}). It models budget allocation as a Markov decision process (MDP) and allocates what-if calls with the goal of maximizing the ``reward'' that is defined by the percentage improvement (ref. Equation~\ref{eq:percentage-impr}).
After budget allocation is done, it runs greedy search again to find the best index configuration with the lowest derived cost (ref. Equation~\ref{eq:derived-cost}).
It has been shown that \emph{MCTS} outperforms \emph{two-phase greedy search} under limited budget on the number of what-if calls~\cite{WuWSWNCB22}.


\vspace{-0.5em}
\subsection{What-if Call Interception}

The two budget-aware index tuning algorithms discussed above allocate what-if calls at a \emph{macro} level by treating each what-if call as a black box.
That is, they use the what-if cost (or its approximation, e.g., derived cost) as the only signal to decide the next what-if call to be made.
This results in wasted budget on \emph{inessential} what-if calls that can be accurately approximated by their derived costs without affecting the result of index tuning.
To skip these inessential what-if calls, previous work developed Wii~\cite{Wii}, a what-if call \emph{interception} mechanism that enables \emph{dynamic} budget allocation in index tuning.
The main idea there is to use lower/upper bounds of what-if cost: a what-if call can be skipped if the gap between the lower and upper bounds is sufficiently small.
We present more details in Section~\ref{sec:overview:framework}.
In this paper, we will build on top of these call-level lower/upper bounds to develop \sysname that enables \emph{early stopping} at workload-level index tuning.
Moreover, in budget-constrained index tuning, skipping these inessential what-if calls can sharpen the efficacy of budget allocation by \emph{reallocating} the budget to what-if calls that cannot be skipped. This results in improved versions of \emph{two-phase greedy search} and \emph{MCTS} algorithms with Wii integrated.

\begin{table}[t]
\footnotesize
\centering
\begin{tabular}{|l|l|}
\hline
\textbf{Notation} & \textbf{Description} \\
\hline
\hline
$c(q, C)$ & The what-if cost of a QCP $(q, C)$ \\
$c(W, C)$ & The what-if cost of a WCP $(W, C)$ \\
$\eta(W, C)$ & The percentage improvement of a WCP $(W, C)$ \\
\hline\hline
$d(q, C)$ & The derived cost of a QCP $(q, C)$ \\
$d(W, C)$ & The derived cost of a WCP $(W, C)$ \\
$L(q, C)$ & The lower bound of $c(q, C)$ \\
$L(W, C)$ & The lower bound of $c(W, C)$ \\
$U(q, C)$ & The upper bound of $c(q, C)$ \\
$U(W, C)$ & The upper bound of $c(W, C)$ \\
\hline\hline
$\Delta(q, C)$ & The CI of $q$ given $C$ \\
$\delta(q, z, C)$ & The MCI of an index $z$ w.r.t. $C$ and $q$ \\
$u(q, z)$ & The MCI upper bound of an index $z$ w.r.t. $q$ \\
\hline
\end{tabular}
\caption{\revision{Notation and terminology (QCP: query-configuration pair; WCP: workload-configuration pair; CI: cost improvement; MCI: marginal cost improvement; $q$: a query; $W$: a workload; $z$: an index; $C$: an index configuration).}}
\label{tab:notation}
\vspace{-2em}
\end{table}

\vspace{-0.5em}
\section{Early Stopping in Index Tuning}
\label{sec:early-stopping}


We start with the problem formulation of early stopping in budget-aware index tuning and then present an overview of the solution that is based on lower/upper bounds of what-if cost. 
\revision{Table~\ref{tab:notation} summarizes the notation and terminology that will be used.}

\vspace{-0.5em}
\subsection{Problem Formulation}
\label{sec:framework:problem-formulation}
Let $B$ be the budget on the number of what-if calls.
At time $t$, i.e., when $t$ what-if calls have been allocated, we want to decide if it is safe to skip allocating the remaining $B-t$ what-if calls without much loss on the improvement of the final index configuration returned.
Formally, let $C_t^*$ be the configuration found with $t\leq B$ what-if calls allocated.
\emph{That is, after $t$ what-if calls we can only use derived cost when running the remaining part of configuration search.}
Under this notation, $C_B^*$ is the configuration found with all $B$ what-if calls allocated.
We stop index tuning if 
\vspace{-0.5em}
\begin{equation}\label{eq:early-exit}
\eta(W, C_B^*)-\eta(W, C_t^*)\leq\epsilon,    
\vspace{-0.5em}
\end{equation}
where $0<\epsilon<1$ is a user-defined threshold.
By Equation~\ref{eq:percentage-impr}, 
\vspace{-0.5em}
\begin{equation}\label{eq:early-exit:direct-computation}
c(W, C_t^*)-c(W, C_B^*)\leq\epsilon\cdot c(W, \emptyset).
\vspace{-0.5em}
\end{equation}
Unfortunately, computing the left side of Equation~\ref{eq:early-exit:direct-computation} is impossible since $c(W, C_B^*)$ would only be known when all the $B$ what-if calls were allocated, which negates the very purpose of \emph{early stopping}.
Moreover, the computation of $c(W, C_t^*)$ would require making $|W|$ extra what-if calls for each time point $t$, which would be prohibitively expensive for large workloads.
As a result, we need a different approach instead of utilizing Equation~\ref{eq:early-exit:direct-computation} directly.

\vspace{-0.5em}
\subsection{A Framework by Lower/Upper Bounds}
\label{sec:overview:framework}

We develop a lower bound $\eta_L(W, C_t^*)$ for $\eta(W, C_t^*)$ and an upper bound $\eta_U(W, C_B^*)$ for $\eta(W, C_B^*)$. That is, $\eta_L(W, C_t^*)\leq \eta(W, C_t^*)$ and $\eta(W, C_B^*)\leq \eta_U(W, C_B^*)$. As a result, if $\eta_U(W, C_B^*)-\eta_L(W, C_t^*)\leq\epsilon$, it then implies $\eta(W, C_B^*)-\eta(W, C_t^*)\leq\epsilon$ (i.e., Equation~\ref{eq:early-exit}). 

Figure~\ref{fig:framework} illustrates this framework in detail.
The $x$-axis represents the number of what-if calls allocated, whereas the $y$-axis represents the percentage improvement of the corresponding best configuration found.
Ideally, we should compare the \emph{true} percentage improvements $\eta(W, C_t^*)$ and $\eta(W, C_B^*)$; however, since the true improvements are not observable, we instead compare the lower and upper bounds $\eta_L(W, C_t^*)$ and $\eta_U(W, C_B^*)$.

\subsubsection{Conversion to Lower/Upper Bounds on What-if Costs}
Our problem is equivalent to developing an upper bound $U(W, C_t^*)\geq c(W, C_t^*)$ and a lower bound $L(W, C_B^*)\leq c(W, C_B^*)$.
As a result, $\eta_L(W, C_t^*)\leq \eta(W, C_t^*)$ and $\eta_U(W, C_B^*)\geq \eta(W, C_B^*)$.


To derive $L(W, C_B^*)$ and $U(W, C_t^*)$, we consider a more fundamental problem: Given an arbitrary configuration $C$, derive a lower bound $L(W, C)$ and an upper bound $U(W, C)$ such that $L(W, C)\leq c(W, C)\leq U(W, C)$.
Since $c(W, C)=\sum_{q\in W}c(q, C)$, it is natural to first consider \emph{call-level} lower and upper bounds $L(q, C)$ and $U(q, C)$ for a given query $q$ such that $L(q, C)\leq c(q, C)\leq U(q, C)$. For this purpose, we reuse the results developed in previous work~\cite{Wii}. Below we provide a summary of the call-level lower/upper bounds. We will discuss extensions to workload-level bounds in Section~\ref{sec:workload-lower-upper-bounds}.

\subsubsection{Call-level Upper Bound}
\label{sec:bounds:call-level:upper-bound}

We assume the following \emph{monotonicity} property of the what-if cost:
\begin{assumption}[Monotonicity]\label{assumption:monotone}
Let $C_1$ and $C_2$ be two index configurations where $C_1\subseteq C_2$. Then $c(q, C_2)\leq c(q, C_1)$.
\end{assumption}
\revision{That is, including more indexes into a configuration does not increase its what-if cost.}
We then have the derived cost $d(q, C)\geq c(q, C)$, which is a valid upper bound, i.e.,
$U(q, C) = d(q, C).$


\begin{figure}[t]
\centering
\includegraphics[clip, trim=1cm 3cm 2cm 1.5cm, width=0.9\columnwidth]{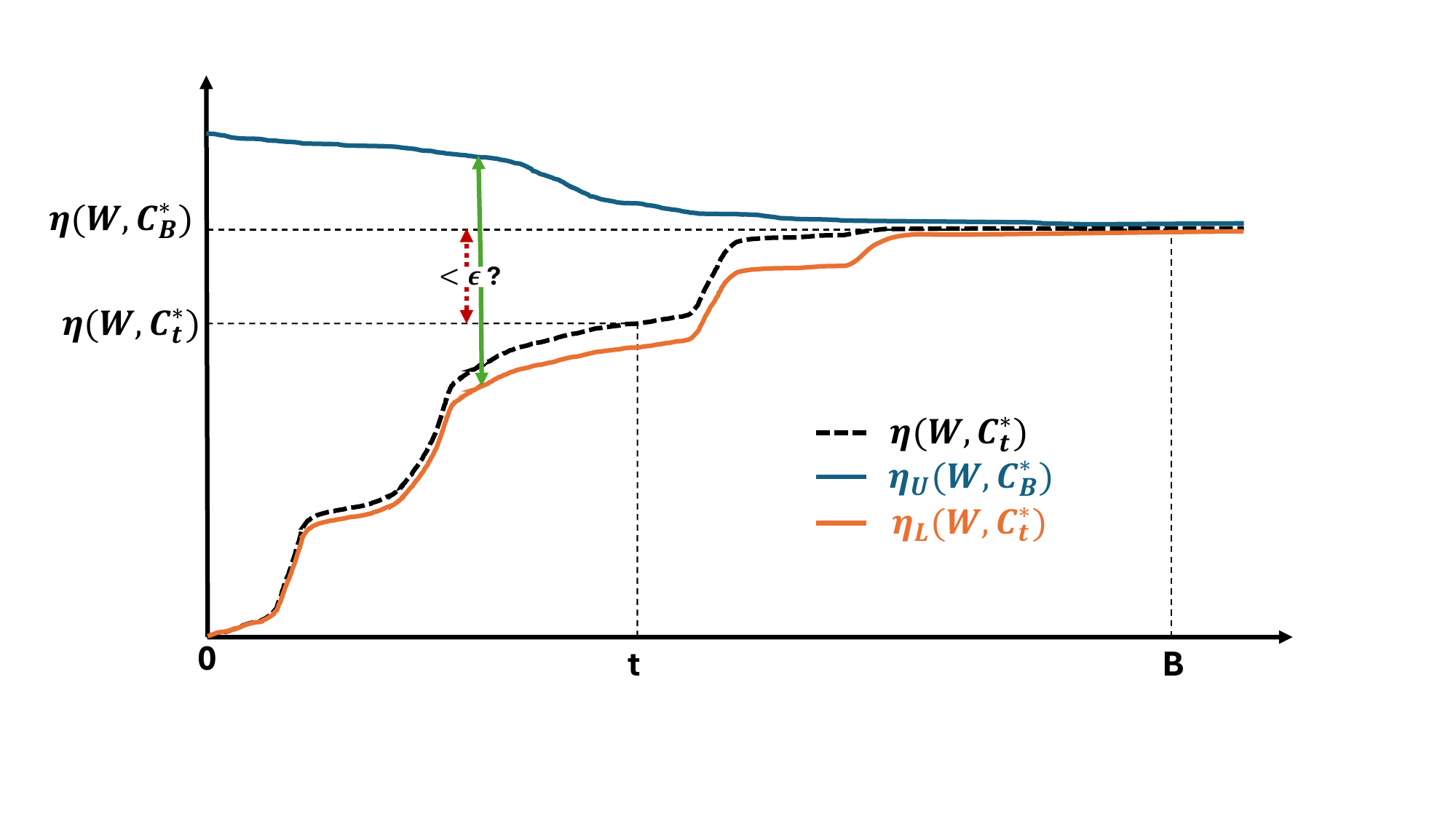}  
\vspace{-1em}
\caption{A framework for early-stopping in budget-aware index tuning based on workload-level bounds of what-if cost.}
\label{fig:framework}
\end{figure}

\subsubsection{\revision{Call-level Lower Bound}}

\revision{
We define the \emph{cost improvement} of the query $q$ given the configuration $C$ as
$\Delta(q, C)=c(q, \emptyset)-c(q, C).$
Moreover, we define the \emph{marginal cost improvement} (MCI) of an index $z$ with respect to a configuration $C$ as
$\delta(q, z, C)=c(q, C)-c(q, C\cup\{z\}).$
Let $C=\{z_1, ..., z_m\}$. We can rewrite CI in terms of the MCI's, i.e.,
$\Delta(q, C)=\sum\nolimits_{j=1}^m\delta(q, z_j, C_{j-1})\leq \sum\nolimits_{j=1}^m u(q, z_j),$
where $C_0=\emptyset$, $C_j=C_{j-1}\cup\{z_j\}$, and $u(q, z_j)$ is an \emph{upper bound} of the MCI $\delta(q, z_j, C_{j-1})$, for $1\leq j\leq m$.}
Hence,
we can set the lower bound
\vspace{-0.5em}
\begin{equation}\label{eq:lower-bound:call-level:v1}
  L(q, C)=c(q,\emptyset)-\sum\nolimits_{j=1}^m u(q, z_j)\leq c(q,C).  
\vspace{-0.5em}
\end{equation}

\subsubsection{MCI Upper Bounds}
\label{sec:bounds:call-level:mci}

We further assume the following \emph{submodularity} property of the what-if cost:
\begin{assumption}[Submodularity]\label{assumption:submodular}
Given two configurations $X$ and $Y$ s.t. $X\subseteq Y$ and an index $z\not\in Y$, we have
$c(q, Y)-c(q, Y\cup\{z\})\leq c(q, X)-c(q, X\cup\{z\}).$
Or equivalently, $\delta(q, z, Y)\leq \delta(q, z, X)$.
\end{assumption}
\revision{That is, the MCI of an index $z$ diminishes when $z$ is included into a \emph{larger} configuration with more indexes.}

Assume monotonicity and submodularity of the cost function $c(q, X)$. Let $\Omega_q$ be the \emph{best} possible configuration for $q$ assuming that all candidate indexes have been created. 
We can set
\begin{equation}\label{eq:upper-bound:mci:call-level}
    u(q,z)=\min\{c(q,\emptyset), \Delta(q,\Omega_q), \Delta(q, \{z\})\}.
\end{equation}
In practice, there are situations where we do not know $c(q, \{z\})$ and thus $\Delta(q, \{z\})$.
In previous work~\cite{Wii}, the authors proposed a lightweight approach to estimate $c(q, \{z\})$ based on the \emph{coverage} of $\{z\}$ with respect to $\Omega_q$, assuming that $c(q,\Omega_q)$ is known.

\vspace{-0.5em}
\section{Workload-level Bounds}
\label{sec:workload-lower-upper-bounds}

We now discuss how to leverage the call-level lower and upper bounds on what-if cost to establish lower/upper bounds that can be used at workload-level.
We discuss both general-purpose bounds as well as optimized bounds for greedy search, which has been an essential step in state-of-the-art budget-aware index tuning algorithms such as \emph{two-phase greedy search} and \emph{MCTS}.



\vspace{-0.5em}
\subsection{General-Purpose Bounds}
\label{sec:workload-lower-upper-bounds:general}

\subsubsection{Upper Bound of Workload Cost}
\label{sec:workload-level-bounds:upper-bound}

The upper bound $U(W, C_t^*)$ can just be set to the derived cost $d(W, C_t^*)$, since we can show
$$d(W, C)=\sum\nolimits_{q\in W} d(q, C)\geq\sum\nolimits_{q\in W} c(q, C) = c(W, C)$$
for an arbitrary index configuration $C$.
To obtain $C_t^*$, however, we need to continue with the index tuning algorithm on top of the current best configuration $C_t$ found \emph{without making more what-if calls}.
As an example, we will illustrate this \emph{simulation} process for greedy search in Section~\ref{sec:workload-level-bounds:greedy-search:simulation}.


\vspace{-0.5em}
\subsubsection{Lower Bound of Workload Cost}



Let $C_B^*=\{z_1, ..., z_k\}$ for some $k\leq K$. By Equation~\ref{eq:lower-bound:call-level:v1}, we could have set 
$$L(W, C_B^*) =\sum\nolimits_{q\in W}L(q, C_B^*)=\sum\nolimits_{q\in W}\Big(c(q, \emptyset)-\sum\nolimits_{i=1}^k u(q,z_i)\Big).$$
Unfortunately, this lower bound cannot be computed, because we do not know $C_B^*$ and therefore the $\{z_1, ..., z_k\}$ at time $t<B$.
However, for each query $q\in W$, if we order all candidate indexes $z$ decreasingly with respect to their $u(q, z)$ and then take the top $K$ candidate indexes in this ranking, it is easy to show that 
$$\sum\nolimits_{i=1}^k u(q,z_i)\leq \sum\nolimits_{z\in \mathcal{U}(q, K)}u(q, z),$$
where $\mathcal{U}(q, K)$ represents the set of candidate indexes of $q$ with the top-$K$ largest MCI upper bounds.
Therefore, we can instead set 
\begin{equation}\label{eq:workload-level:lower-bound}
L(W, C_B^*)= \sum\nolimits_{q\in W}\Big(c(q, \emptyset)-\sum\nolimits_{z\in \mathcal{U}(q, K)}u(q, z)\Big).
\end{equation}
However, while this lower bound can be used for any budget-aware tuning algorithm, it may be too conservative.
We next present optimizations of this lower bound for greedy search.

\vspace{-0.5em}
\subsection{Optimizations for Greedy Search}
\label{sec:workload-lower-upper-bounds:optimizations}





Now let $C_B^*=\{z_1, ..., z_k\}$ for some $k\leq K$ where $z_i$ represents the index selected by greedy search at the $i$-th step ($1\leq i\leq k$) with $B$ what-if calls allocated.
The lower bound by applying Equation~\ref{eq:lower-bound:call-level:v1},
\begin{equation}\label{eq:query-level:lower-bound:uncomputable}
    L(W, C_B^*) = \sum\nolimits_{q\in W}\Big(c(q, \emptyset)-\sum\nolimits_{i=1}^k u^{(i)}(q,z_i)\Big),
\end{equation}
cannot be computed. Here, $u^{(i)}(q, z)$ is the $u(q, z)$ after the greedy step $i$ and we use Procedure~\ref{proc:maintain-mci:greedy-search} to update the MCI upper bounds~\cite{Wii}:
\begin{procedure}\label{proc:maintain-mci:greedy-search}
For each index $z$ that has not been selected by greedy search, we update $u(q, z)$ as follows:
\begin{enumerate}[leftmargin=*,label=(\alph*)]
    \item Initialize $u(q, z)=\min\{c(q,\emptyset), \Delta(q, \Omega_q)\}$ for each index $z$.
    \item During each greedy step $1\leq k\leq K$, update 
    $$u(q, z)=c(q, C_{k-1})-c(q, C_{k-1}\cup\{z\})=\delta(q,z, C_{k-1})$$
    if both $c(q, C_{k-1})$ and $c(q, C_{k-1}\cup\{z\})$ are available, where $C_k$ is the configuration selected by the greedy step $k$ and $C_0=\emptyset$.
\end{enumerate}
\end{procedure}
Our idea is to further develop an upper bound for $\sum\nolimits_{i=1}^k u^{(i)}(q,z_i)$ by running a \emph{simulated greedy search} procedure described below.


\vspace{-0.5em}
\subsubsection{Simulated Greedy Search}
\label{sec:workload-level-bounds:greedy-search:simulation}

For ease of exposition, consider tuning a workload with a single query $q$ using greedy search.

\begin{procedure}\label{proc:simulation-greedy}
At time $t$ (i.e., when $t<B$ what-if calls have been made), run greedy search to get up to $K$ indexes in total, where each greedy step $j$ selects the index $z'_j$ with the maximum $u^{(j)}(q, z'_j)>0$. 
\end{procedure}
Let the configuration found by Procedure~\ref{proc:simulation-greedy} be $C_t^u=\{z'_1, z'_2, ..., z'_l\}$ where $l\leq K$.
If $l<K$, then it means that any remaining index $z$ satisfies $u(q, z)=0$.
As a result, we can assume $l=K$.


\begin{theorem}\label{theorem:query-level-lower-bound}
$\sum\nolimits_{j=1}^K u^{(j)}(q, z'_j)\geq\sum\nolimits_{i=1}^k u^{(i)}(q,z_i)$.
As a result,
\begin{equation}\label{eq:c_prim_q_C_t_u_K}
  L(q, C_B^*)=c(q,\emptyset)-\sum\nolimits_{j=1}^K u^{(j)}(q, z'_j) 
\end{equation}
is a lower bound of the what-if cost $c(q, C_B^*)$ for greedy search. 
\end{theorem}
Due to space constraints, all proofs are deferred to the full version of this paper~\cite{full-version}.
We next generalize this result to multi-query workload with the understanding that the index $z'_j$ is selected for the entire workload $W$ with the maximum $u^{(j)}(W, z'_j)>0$, i.e.,
\vspace{-0.5em}
\begin{equation}\label{eq:lower-bound-workload-level}
L(W, C_B^*)=c(W,\emptyset)-\sum\nolimits_{j=1}^K u^{(j)}(W, z'_j),
\vspace{-0.5em}
\end{equation}
where $u(W, z)=\sum_{q\in W} u(q, z)$.

Moreover, as we mentioned in Section~\ref{sec:workload-level-bounds:upper-bound}, the simulated greedy search outlined in Procedure~\ref{proc:simulation-greedy} can be reused for computing the upper bound $U(W, C_t^*)$ with slight modification.
Details of this revised simulated greedy search are included in the full version~\cite{full-version}.



\vspace{-0.5em}
\subsubsection{Lower Bound for Two-phase Greedy Search}
We update the MCI upper-bounds for \emph{two-phase greedy search} as follows:
\begin{procedure}\label{proc:mci-update:two-phase}
For index $z$ and query $q$, update $u(q, z)$ as follows:
\begin{enumerate}[leftmargin=*,label=(\alph*)]
    \item Initialize $u(q, z)=\min\{c(q,\emptyset), \Delta(q, \Omega_q)\}$ for each index $z$.
    \item In Phase 1, update $u(q, z)$ based on Equation~\ref{eq:upper-bound:mci:call-level}.
    \item In Phase 2, during each greedy step $1\leq k\leq K$, update 
    $$u(q, z)=c(q, C_{k-1})-c(q, C_{k-1}\cup\{z\})=\delta(q,z, C_{k-1})$$
    if both $c(W, C_{k-1})$ and $c(q, C_{k-1}\cup\{z\})$ are available, where $C_k$ is the configuration selected by greedy search in step $k$ ($C_0=\emptyset$) and $z$ has not been included in $C_k$.
\end{enumerate}
\end{procedure}
The update step (c) excludes pathological cases where $c(W, C_k)$ is unknown but both $c(q, C_k)$ and $c(q, C_k\cup\{z\})$ are known for a particular query $q$ (due to Phase 1).

\begin{theorem}\label{theorem:lower-bound-workload-level}
The $L(W, C_B^*)$ defined in Equation~\ref{eq:lower-bound-workload-level}
remains a lower bound of $c(W, C_B^*)$ for \emph{two-phase greedy search} if we maintain the MCI upper-bounds by following Procedure~\ref{proc:mci-update:two-phase}.
\end{theorem}

\vspace{-0.5em}
\subsubsection{Lower Bound for Monte Carlo Tree Search}
We can use the same simulated greedy search to obtain $L(W, C_B^*)$, given that there is a final greedy search stage in \emph{MCTS} after all budget allocation is done.
However, we are only able to use Equation~\ref{eq:upper-bound:mci:call-level} for maintaining the MCI upper bounds---we can prove that it is safe to do so using the same argument as in \emph{two-phase greedy search} when $t$ is in Phase 1 (see the full version~\cite{full-version}).
It remains future work to investigate further improvement over Equation~\ref{eq:upper-bound:mci:call-level} for \emph{MCTS}.


\vspace{-0.5em}
\section{Refinement with Index Interaction}
\label{sec:lower-refinement:index-interaction}

Our approach of computing the lower bounds $L(q, C_B^*)$ and $L(W, C_B^*)$ in Equations~\ref{eq:c_prim_q_C_t_u_K} and~\ref{eq:lower-bound-workload-level} basically sums up the MCI Upper-bounds of individual indexes.
This ignores potential \emph{index interactions}, as illustrated by the following example.

\begin{figure}
\centering
\includegraphics[clip, trim=6cm 10cm 8cm 5cm, width=\columnwidth]{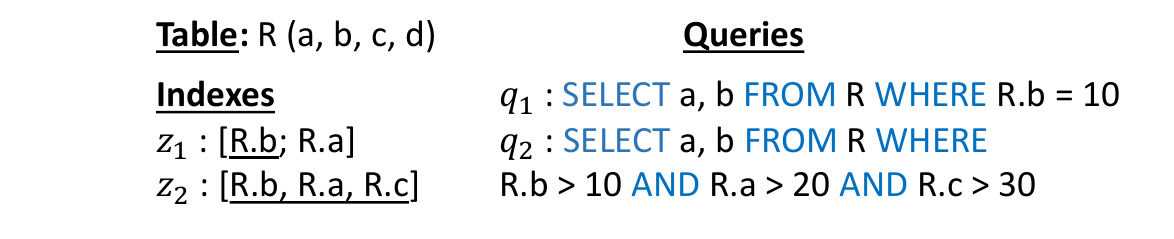}  
\vspace{-3em}
\caption{An example of index interaction} 
\label{fig:index-interaction-example}
\end{figure}

\begin{example}[Index Interaction]\label{example:interaction}
As shown in Figure~\ref{fig:index-interaction-example}, let $R$ be a table with four columns $a$, $b$, $c$, and $d$.
Let $z_1$ and $z_2$ be two indexes on $R$, where $z_1$ has a single key column $b$ with $a$ as an included column, and $z_2$ has a compound key with three columns $b$, $a$, and $c$ in order.
Consider the SQL query $q_1$ in Figure~\ref{fig:index-interaction-example}.
Both $z_1$ and $z_2$ have very similar, if not the same, cost improvement for $q_1$, as one can use an index scan on top of either $z_1$ and $z_2$ to evaluate $q_1$ without consulting the table $R$.
As a result, if $z_1$ (resp. $z_2$) has been included in some configuration, including $z_2$ (resp. $z_1$) cannot further improve the cost of $q_1$.
In other words, we have roughly the same cost improvements for $z_1$, $z_2$, and $\{z_1, z_2\}$, i.e., $\Delta(q_1, \{z_1\})\approx\Delta(q_1, \{z_2\})\approx\Delta(q_1, \{z_1,z_2\})$.
\end{example}

Note that index interaction is \emph{query-dependent}. 
To see this, consider the same $z_1$ and $z_2$ in Example~\ref{example:interaction} but a different SQL query $q_2$ in Figure~\ref{fig:index-interaction-example}. 
Since $z_1$ can hardly be used for evaluating $q_2$, we have $\Delta(q_2, \{z_1\})\approx 0$ (see~\cite{full-version} for details).
As a result, in the presence of both $z_1$ and $z_2$, the query optimizer will pick $z_2$ over $z_1$; hence, we have $\Delta(q_2, \{z_1, z_2\})=\Delta(q_2, \{z_2\})\approx \Delta(q_2, \{z_1\})+\Delta(q_2, \{z_2\})$.
Therefore, $z_1$ and $z_2$ \emph{do not interact} in the case of $q_2$.

\vspace{-0.5em}
\subsection{Index Interaction}
\label{sec:early-stopping:workload:lower-refinement:index-interaction}

Motivated by Example~\ref{example:interaction}, given two indexes $z_1$, $z_2$ and a query $q$, we define the \emph{index interaction} between $z_1$ and $z_2$ w.r.t. $q$ as 
$$\mathcal{I}(z_1, z_2 | q)=\frac{\Delta_U(q, \{z_1, z_2\}) - \Delta(q, \{z_1, z_2\})}{\Delta_U(q, \{z_1, z_2\}) - \Delta_L(q, \{z_1, z_2\})}.$$
Here, $\Delta_L(q, \{z_1, z_2\})=\max\{\Delta(q, \{z_1\}), \Delta(q, \{z_2\})\}$ is a lower bound of $\Delta(q, \{z_1, z_2\})$ based on Assumption~\ref{assumption:monotone} (i.e., monotonicity), and $\Delta_U(q, \{z_1, z_2\})=\Delta(q, \{z_1\}) + \Delta(q, \{z_2\})$ is an upper bound of $\Delta(q, \{z_1, z_2\})$ based on Assumption~\ref{assumption:submodular} (i.e., submodularity).


We now extend the above definition to define the interaction between an index $z$ and an index configuration $C$ w.r.t. a query $q$:
$$\mathcal{I}(z, C | q)=\frac{\Delta_U(q, C\cup \{z\}) - \Delta(q, C\cup \{z\})}{\Delta_U(q, C\cup \{z\}) - \Delta_L(q, C\cup \{z\})}.$$
Similarly, $\Delta_L(q, C\cup \{z\})=\max\{\Delta(q, C), \Delta(q, \{z\})\}$ is a lower bound of $\Delta(q, C\cup \{z\})$ by Assumption~\ref{assumption:monotone},  
and $\Delta_U(q, C\cup \{z\})=\Delta(q, C)+\Delta(q, \{z\})$ is an upper bound of $\Delta(q, C\cup \{z\})$ by Assumption~\ref{assumption:submodular}.

\vspace{-0.5em}
\subsection{A Similarity-based Approach}
\label{sec:early-stopping:workload:lower-refinement:similarity}

Note that the interaction $\mathcal{I}(z, C | q)$ defined above cannot be directly computed 
if we do not have knowledge about $\Delta(q, C)$ and $\Delta(q, C\cup\{z\}).$
Therefore, we propose an \emph{implicit} approach to measure index interaction based on the \emph{similarity} between indexes.
Intuitively, if two indexes are similar, e.g., they share similar key columns where one is a prefix of the other, then it is likely that one of them cannot improve the workload cost given the presence of the other.
As a result, there is strong interaction between the two indexes.

Specifically, given a query $q$ and two indexes $z_1, z_2$, we compute the similarity $\mathcal{S}(z_1, z_2 | q)$ between $z_1$ and $z_2$ w.r.t. $q$ as follows:
\begin{enumerate}[leftmargin=*]
    \item Convert the query and indexes into feature vectors $\vec{\mathbf{q}}$, $\vec{\mathbf{z}}_1$, and $\vec{\mathbf{z}}_2$. We reuse the feature representation in previous work~\cite{SiddiquiJ00NC22,Wii} for this purpose. In more detail, we collect all indexable columns from the workload. Let $D$ be the number of indexable columns collected. We then represent $\vec{\mathbf{q}}$, $\vec{\mathbf{z}}_1$, and $\vec{\mathbf{z}}_2$ as $D$-dimensional vectors. We assign weights to each indexable column in the query representation $\vec{\mathbf{q}}$ by using the approach proposed in ISUM~\cite{SiddiquiJ00NC22}.
    \revision{Specifically, the weight of a column is computed based on its corresponding table size and the number of candidate indexes that contain it.}
    We further assign weights to each indexable column in the index representation $\vec{\mathbf{z}}$ by using the approach proposed in Wii~\cite{Wii}. \revision{Specifically, the weight of a column is determined by its position in the index $z$, e.g., whether it is a \emph{key column} or an \emph{included column} of $z$.}
    \item Project the index vectors onto the query vector using dot product, i.e., $\vec{\mathbf{z}}_i^{\mathbf{q}}=\vec{\mathbf{z}}_i\cdot \vec{\mathbf{q}}$ for $i\in\{1,2\}$. Note that the resulting vectors $\vec{\mathbf{z}}_i^{\mathbf{q}}$ for $i\in\{1,2\}$ remain $D$-dimensional vectors. This projection filters out columns in $\vec{\mathbf{z}}_i$ that do not appear in $\vec{\mathbf{q}}$ and therefore do not have impact on the query performance of $q$.
    \item Calculate the cosine similarity 
    $\mathcal{S}(z_1, z_2 | q)=\frac{\vec{\mathbf{z}}_1^{\mathbf{q}} \cdot \vec{\mathbf{z}}_2^{\mathbf{q}}}{\|\vec{\mathbf{z}}_1^{\mathbf{q}} \|\cdot\|\vec{\mathbf{z}}_2^{\mathbf{q}}\|}.$
\end{enumerate}
We can further extend $\mathcal{S}(z_1, z_2 | q)$ to represent the similarity between an index $z$ and an index configuration $C$ w.r.t. a query $q$:
$\mathcal{S}(z, C | q)=\frac{\vec{\mathbf{z}}^{\mathbf{q}} \cdot \vec{\mathbf{C}}^{\mathbf{q}}}{\|\vec{\mathbf{z}}^{\mathbf{q}} \|\cdot\|\vec{\mathbf{C}}^{\mathbf{q}}\|}.$
All we need is a feature representation $\vec{\mathbf{C}}$ of the configuration $C$.
For this purpose, we use the same approach as in Wii~\cite{Wii}, where we featurize an index configuration as a $D$-dimensional vector as follows. For each dimension $d$ ($1\leq d\leq D$), we take the maximum of the feature values from the corresponding dimensions $d$ of the feature representations of the indexes contained by the configuration.
\revision{The intuition is that, if an indexable column appears in multiple indexes of the configuration, we take the largest weight that represents its most significant role (e.g., a leading key column in some index).}




Ideally, we would wish the $\mathcal{S}(z, C | q)$ to be equal to $\mathcal{I}(z, C | q)$.
Unfortunately, this is not the case.
\revision{To shed some light on this, we conduct an empirical study to measure the \emph{correlation} between pairwise index interaction $\mathcal{I}(z_1, z_2 | q)$ and pairwise index similarity $\mathcal{S}(z_1, z_2 | q)$, using the workloads summarized in Table~\ref{tab:databases}.
Specifically, we pick the most costly queries for each workload and evaluate the what-if costs of all single indexes (i.e., singleton configurations) for each query. We then select the top 50 indexes w.r.t. their cost improvement (CI) in decreasing order and evaluate the what-if costs of all $50\times 49=2,450$ configurations that contain a pair of the top-50 indexes.
Finally, we compute the pairwise index interaction and the pairwise index similarity of these index pairs.
Figure~\ref{fig:index-interaction-example} presents their correlation for the two most costly queries of \textbf{TPC-H}, and similar results over the other queries and workloads are included in the full version~\cite{full-version}.}
We observe that there is no strong correlation between the two.
Instead, for most of the queries, there is a sudden jump on the pairwise index interaction when the pairwise index similarity increases.
That is, when the pairwise index similarity exceeds a certain threshold (e.g., 0.2), the pairwise index interaction will increase to a high value (e.g., close to 1).
This motivates us to propose a threshold-based mechanism to utilize the index similarity to characterize the impact of index interaction. 

\begin{figure}
\centering
\subfigure{\includegraphics[width=0.23\textwidth]{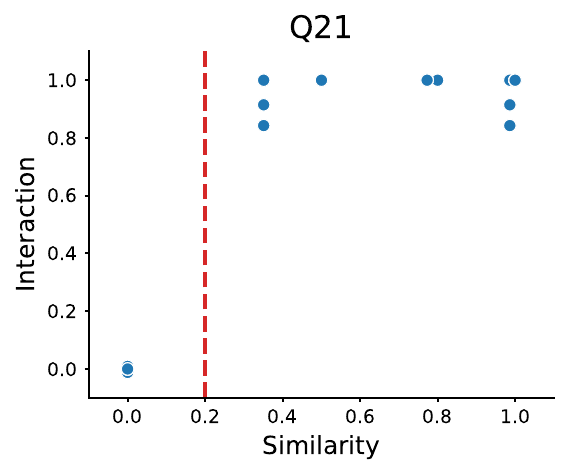}}
\subfigure{\includegraphics[width=0.23\textwidth]{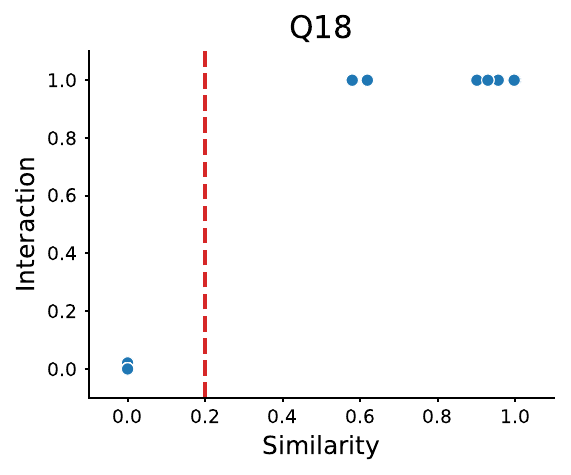}}
\vspace{-1.5em}
\caption{\revision{Relationship between pairwise index interaction and pairwise index similarity (TPC-H).}}
\label{fig:interaction-tpch:example}
\end{figure}

\vspace{-0.5em}
\subsection{Refined Workload-Level Lower Bound}
\label{sec:refinement:index-interactions:lower-bound}

Our basic idea is the following.
During each step of the simulated greedy search (SGS) when selecting the next index to be included, we consider not only the benefit of the index, but also its interaction with the indexes \emph{that have been selected in previous steps} of SGS.
Specifically, we quantify the \emph{conditional benefit} $\mu^{(j)}(q, z'_j)$ of the candidate index $z'_j$ based on its interaction with the SGS-selected configuration $C_{j-1}=\{z'_1,...,z'_{j-1}\}$ and use it to replace the MCI upper bound $u^{(j)}(q, z'_j)$ in Procedure~\ref{proc:simulation-greedy} as follows:
\begin{equation}\label{eq:conditional-benefit}
\mu^{(j)}(q, z'_j) = 
    \begin{cases}
    0, & \text{if } \mathcal{S}(z'_j, C_{j-1} | q) > \tau; \\
    u^{(j)}(q, z'_j), & \text{otherwise}.
    \end{cases}  
\end{equation}
Here, $0\leq\tau\leq 1$ is a threshold.
In our experimental evaluation (see Section~\ref{sec:eval}), we found that 
this threshold-based mechanism can significantly improve the lower bound for \emph{two-phase greedy search} but remains ineffective for \emph{MCTS}, due to the presence of many query-index pairs with unknown what-if costs.
We therefore further propose an optimization for \emph{MCTS}.
Specifically, for a query-index pair $(q, z)$ with unknown what-if cost, we initialize its MCI upper bound by averaging the MCI upper bounds of indexes with known what-if costs that are similar to $z$ w.r.t. $q$ (see~\cite{full-version} for details).
\vspace{-0.5em}
\section{Early-Stopping Verification}
\label{sec:early-stopping-verification}


Based on the workload-level lower/upper bounds in Sections~\ref{sec:workload-lower-upper-bounds} and~\ref{sec:lower-refinement:index-interaction}, we develop \sysname, an early-stopping checker for budget-aware index tuning.
One main technical challenge faced by \sysname is to understand \emph{when} to invoke early-stopping verification.
While one can employ simple strategies such as a fixed-step verification scheme where a verification is invoked every $s$ what-if calls, as we will see in our experimental evaluation (Section~\ref{sec:eval}) such strategies may incur high computation overhead since obtaining the lower and upper bounds (e.g., by using the simulated greedy search procedure in Section~\ref{sec:workload-level-bounds:greedy-search:simulation}) comes with a cost.
In this section, we present our solutions to this problem.
We start by giving a heuristic solution to \emph{two-phase greedy search} that exploits special structural properties of this algorithm (Section~\ref{sec:early-stopping-verification:heuristic}).
We then propose a generic solution (Section~\ref{sec:integration:early-stopping:check}) by only leveraging improvement rates and convexity properties of the index tuning curve (Section~\ref{sec:esc:index-tuning-curve}) without requiring any algorithm-specific knowledge.

\vspace{-0.5em}
\subsection{Heuristic Verification Scheme}
\label{sec:early-stopping-verification:heuristic}


There is some trade-off in terms of \emph{when} to invoke early-stopping verification (ESV):
if we invoke ESV too frequently, then the computation overhead may become considerable; on the other hand, if we invoke ESV insufficiently, then we may miss opportunities for stopping index tuning earlier and allocate more what-if calls than necessary.
Clearly, in the early stages of index tuning, there is no need to check for early-stopping, as the index tuning algorithm is still making rapid progress. Ideally, one needs to \emph{detect} when the progress of the index tuning algorithm starts to slow down. 

For \emph{two-phase greedy search}, this \emph{inflection point} is not difficult to tell.
As an example, consider Figure~\ref{fig:lc-example:twophase:tpch:origin} where we run \emph{two-phase greedy search} to tune the \textbf{TPC-H} workload.
In Figure~\ref{fig:lc-example:twophase:tpch:origin} we have marked each greedy step within both Phase 1 and Phase 2.
We observe that the progress starts to slow down significantly after the search enters Phase 2, especially during or after the first greedy step of Phase 2. 
As a result, we can simply skip Phase 1 and start checking early-stopping at the beginning of each greedy step of Phase 2.
Our experiments in Section~\ref{sec:eval} confirm that this simple scheme can result in effective early-stopping 
while keeping the computation overhead negligible. 

This heuristic early-stopping verification scheme clearly cannot work for other algorithms such as \emph{MCTS}.
However, the above discussion hinted us to focus on looking for similar \emph{inflection points} of index tuning curves.
It leads to a generic early-stopping verification scheme that only relies on improvement rates and convexity properties of index tuning curves, as we will present next.


\vspace{-0.5em}
\subsection{Index Tuning Curve Properties}
\label{sec:esc:index-tuning-curve}

We define the \emph{index tuning curve} (ITC) as a function that maps from the number of what-if calls allocated at time $t$ to the percentage improvement $\eta(W, C_t^*)$ of the corresponding best index configuration found. By definition, the ITC is \emph{monotonically non-decreasing}.
The dash line in Figure~\ref{fig:framework} presents an example of ITC.

\begin{figure}
\centering
\includegraphics[clip, trim=1cm 4cm 2cm 3.5cm, width=0.9\columnwidth]{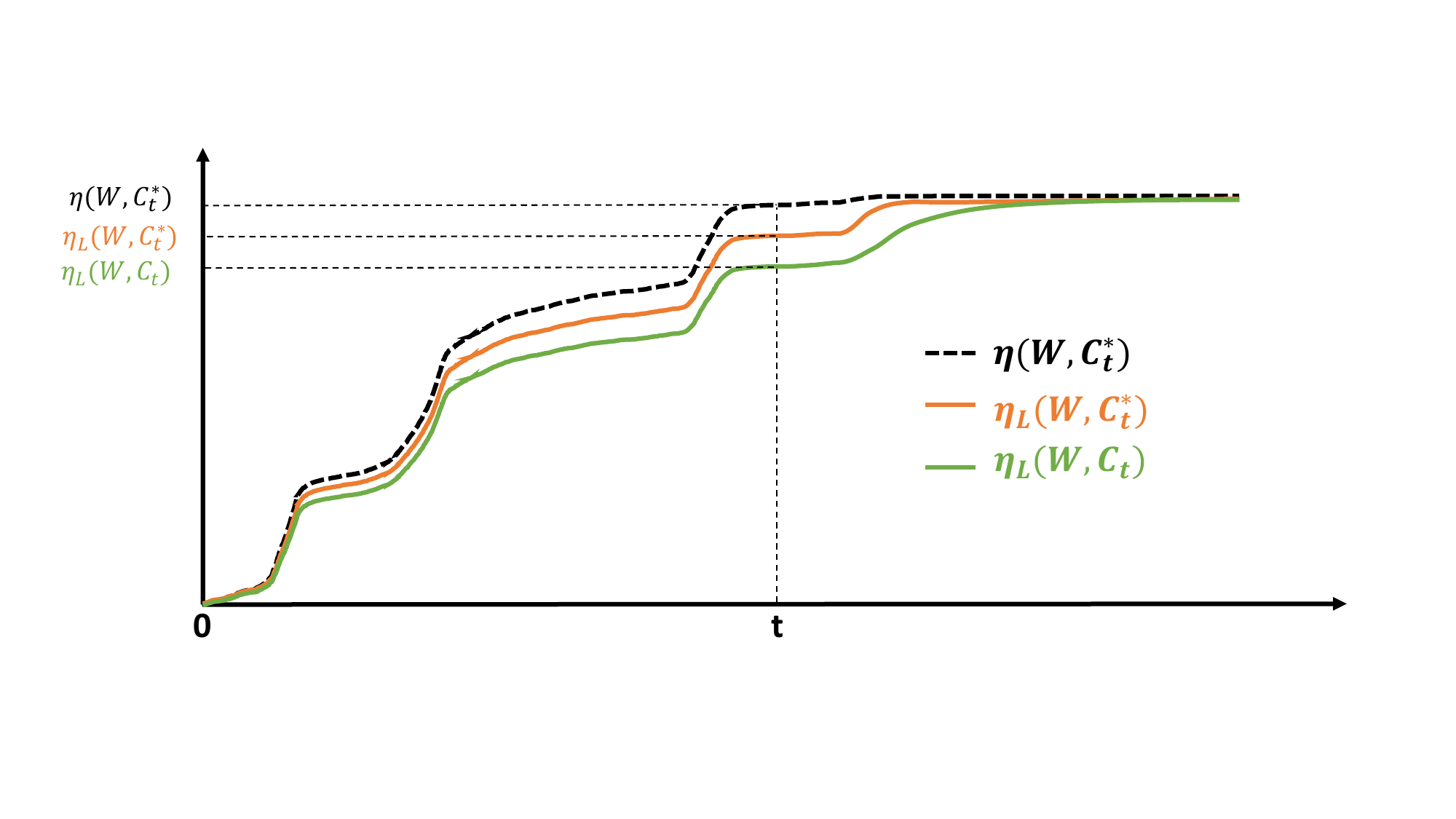}  
\vspace{-1em}
\caption{Characterization of the relationship between different definitions of index tuning curve.}
\label{fig:index-tuning-curve}
\end{figure}

Unfortunately, as we have discussed in Section~\ref{sec:framework:problem-formulation}, the ITC defined above cannot be directly observed without making 
extra what-if calls. 
One option is to replace $\eta(W, C_t^*)$ with its lower bound $\eta_L(W, C_t^*)$. However, the computation of $\eta_L(W, C_t^*)=1-\frac{d(W, C_t^*)}{c(W, \emptyset)}$ is not free (e.g., requiring running the simulated greedy search) and we therefore choose to use 
$\eta_L(W, C_t)=1-\frac{d(W, C_t)}{c(W, \emptyset)}$, where $C_t$ is the \emph{observed} best configuration at time $t$ without continuing tuning,
in lieu of $\eta_L(W, C_t^*)$.
$\eta_L(W, C_t)$ is directly available at time $t$ without extra computation.
Assuming monotonicity of what-if cost (i.e., Assumption~\ref{assumption:monotone}), we have $\eta(W, C_t)\leq \eta_L(W, C_t^*)$, because $d(W, C_t)\geq d(W, C_t^*)$ given that $C_t$ is a subset of $C_t^*$.
Figure~\ref{fig:index-tuning-curve} characterizes the relationship between different definitions of ITC.



\vspace{-0.5em}
\subsubsection{Improvement Rate}

Suppose that we check \emph{early stopping} at $n$ time points with $B_j$ what-if calls allocated at time point $j$, where $1\leq j\leq n$.
We call this sequence $\{B_j\}_{j=1}^n$ an \emph{early-stopping verification scheme} (ESVS).
Let the observed percentage improvement at time point $j$ be $I_j$, i.e., $I_j=\eta_L(W, C_{B_j})$.
We further define a starting point $(B_0, I_0)$ where we have known both $B_0$ and $I_0$.
By default, we choose $B_0=0$ and $I_0=0$.

\begin{definition}[Improvement Rate]
We define the \emph{improvement rate} $r_j$ at time point $j$ as
$r_j = \frac{I_j - I_0}{B_j - B_0}.$
\end{definition}
The \emph{projected improvement} at time point $j$ for budget $b$ of what-if calls (i.e., by making $b-B_j$ more what-if calls) is then defined as
\begin{equation}\label{eq:proj-impr}
p_j(b)=I_j + r_j\cdot (b-B_j). 
\end{equation}
For the default case where $B_0=0$ and $I_0=0$, we have
$p_j(b)=I_j\cdot\frac{b}{B_j}.$
For ease of exposition, we will use this default setup in the rest of our discussion throughout this section.

\begin{definition}[Latest Improvement Rate]
We define the \emph{latest improvement rate} $l_j$ at time point $j$ as
$l_j = \frac{I_j - I_{j-1}}{B_j - B_{j-1}}.$
\end{definition}

\subsubsection{Convexity and Concavity}

Let $I=f(b)$ be the function that represents the index tuning curve.
That is, $f(b)=\eta_L(W, C_b)$ where $C_b$ is the observed best configuration with $b$ what-if calls allocated.

\begin{lemma}\label{lemma:concave}
If $f$ is strictly concave and twice-differentiable, then $f'(b)<\frac{f(b)}{b}$ for any $0<b\leq B$.
\end{lemma}

We have the following immediate result based on Lemma~\ref{lemma:concave}:
\begin{theorem}\label{theorem:concave}
If $f$ is strictly concave and twice-differentiable, then $l_j<r_j$ for a given \emph{early-stopping verification scheme} $\{B_j\}_{j=1}^n$.
\end{theorem}

We have a similar result for a \emph{convex} index tuning curve:
\begin{theorem}\label{theorem:convex}
If $f$ is strictly convex and twice-differentiable, then $l_j>r_j$ for a given \emph{early-stopping verification scheme} $\{B_j\}_{j=1}^n$.
\end{theorem}

\subsubsection{Summary and Discussion}
\begin{figure}
\centering
\includegraphics[clip, trim=3cm 4cm 2cm 3cm, width=0.9\columnwidth]{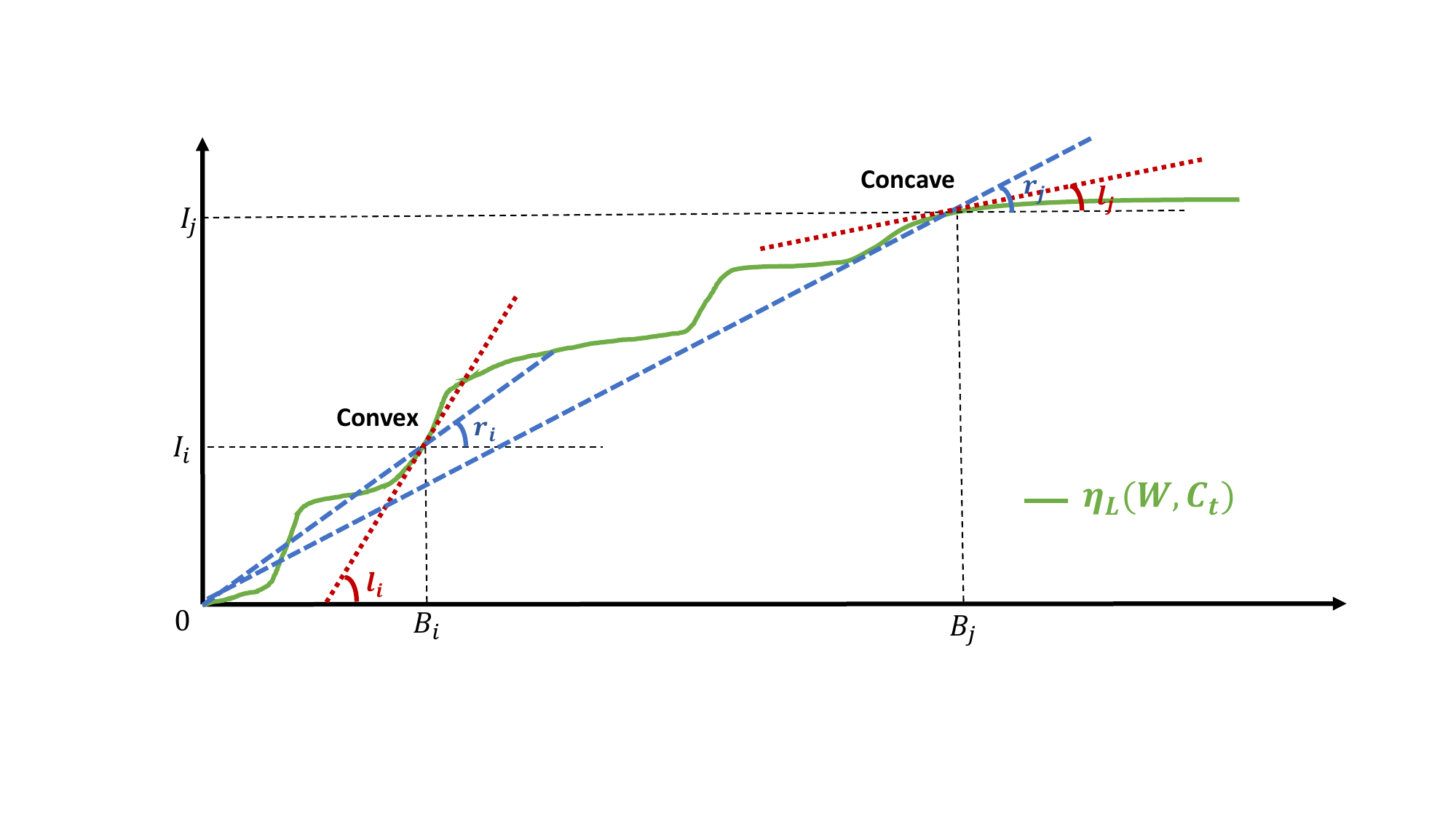}  
\vspace{-1em}
\caption{Relationship between improvement rates and convexity/concavity of index tuning curve. The latest improvement rate $l_j$ approximates the \emph{tangent} of the index tuning curve at the point $(B_j, I_j)$.}
\label{fig:improvement-rate}
\end{figure}

\revision{The previous analysis implies some potential relationship between the improvement rates that we defined and the \emph{convexity}/\emph{concavity} properties of an index tuning curve: (1) if the index tuning curve in $(B_{j-1}, B_j)$ is convex, i.e., it is making accelerating progress, then we will observe $l_j > r_j$;  (2) on the other hand, if the index tuning curve in $(B_{j-1}, B_j)$ is concave, then we will observe $l_j < r_j$.
Figure~\ref{fig:improvement-rate} illustrates this relationship.}


In practice, an index tuning curve can be partitioned into ranges where in each range the curve can fall into one of the three categories: (1) \emph{convex}, (2) \emph{concave}, and (3) \emph{flat} (i.e., $l_j=0$).
In general, we would expect that the curve is more likely to be convex in early stages of index tuning and is more likely to be concave or flat towards the end of tuning.
This observation leads us to develop a generic ESVS that will be detailed next, where we leverage the convexity of the ITC to skip unnecessary invocations of early-stopping verification and put the overall verification overhead under control.

\vspace{-0.5em}
\subsection{Generic Verification Scheme}
\label{sec:integration:early-stopping:check}

We start from the aforementioned simple ESVS with fixed step size $s$, i.e., $B_j=B_{j-1}+s$, where $s$ can be a small number of what-if calls.
We then compute $l_j$ and $r_j$ at each $B_j$ accordingly.

Now consider a specific time point $j$. If we observe that $l_j>r_j$, then it is likely that the index tuning curve in $(B_{j-1}, B_j)$ is convex. Note that the condition in Theorem~\ref{theorem:convex} is not necessary, so the convexity is not guaranteed when observing $l_j>r_j$.
In this case we can \emph{skip} the early-stopping verification, 
because the index tuner is still making \emph{accelerating} progress.
On the other hand, if we observe that $l_j<r_j$, then it is likely that the index tuning curve in $(B_{j-1}, B_j)$ is concave, i.e., the progress is \emph{decelerating}, which implies that we perhaps can perform a verification.

There are some subtleties in the above proposal.
First, although it is reasonable to assume that the index tuning curve will \emph{eventually} become concave/flat, it is not guaranteed that the index tuner has entered this final stage of tuning when $l_j<r_j$ is observed.
Second, even if the index tuner has entered the final stage, the deceleration process may be slow before we can conclude that the improvement loss will be lower than the user-given threshold $\epsilon$, which voids the necessity of the (expensive) early-stopping verification.

\vspace{-0.5em}
\subsubsection{Significance of Concavity}
\label{sec::generic-esvs:significance}
\revision{To address these challenges, we measure the \emph{significance} of the potential concavity of the index tuning curve.
For this purpose, we \emph{project} the percentage improvement at $B_{j+1}$ using the improvement rates $l_j$ and $r_j$ and compare it with $I_{j+1}$ to decide whether we want to invoke early-stopping verification (ESV) at the time point $j+1$.}
Specifically, we define the projected \emph{improvement gap} between the projected improvements $p_{j+1}^r$ and $p_{j+1}^l$ (using Equation~\ref{eq:proj-impr}) as $\Delta_{j+1}=p_{j+1}^r-p_{j+1}^l.$
Clearly, $\Delta_{j+1}>0$ since $l_j<r_j$. \revision{Moreover, the larger $\Delta_{j+1}$ is, the more significant the corresponding \emph{concavity} is. Therefore, intuitively, we should have a higher probability of invoking ESV.}

Now consider the relationship between $I_{j+1}$ and $p_{j+1}^{l,r}$. We have the following three possible cases:
\begin{itemize}[leftmargin=*]
    \item $p_{j+1}^l<p_{j+1}^r<I_{j+1}$: This suggests that $f$ grows even faster than $r_j$ when moving from $B_j$ to $B_{j+1}$, which implies that a verification at $j+1$ is unnecessary.
    \item $p_{j+1}^l< I_{j+1} < p_{j+1}^r$: This suggests that $f$ grows more slowly than $r_j$ but faster than $l_j$. We further define $\delta_{j+1} = p_{j+1}^r - I_{j+1}$ and define the \emph{significance of concavity} $\sigma_{j+1}$ as
        $\sigma_{j+1}=\frac{\delta_{j+1}}{\Delta_{j+1}}.$
    Clearly, $0<\delta_{j+1}<\Delta_{j+1}$. We then set a threshold $0<\sigma<1$ and perform an early-stopping verification if $\sigma_{j+1} \geq \sigma$.
    \item $I_{j+1}<p_{j+1}^l$: This suggests that $f$ grows even more slowly than $l_j$, which implies that a verification at $j+1$ is perhaps helpful.
\end{itemize}

\vspace{-0.5em}
\subsubsection{A Probabilistic Mechanism for Invoking ESV}
One problem is that, if the observed improvement is flat (i.e., $l_i=0$) but the lower and upper bounds are not converging yet, then it may result in unnecessary ESV invocations. 
We therefore need to further consider the convergence of the bounds.
Specifically, we use the following probabilistic mechanism for invoking ESV.
We define $\rho_j=\frac{U_j(W, C_B^*) - L_j(W, C_t^*)}{\epsilon}$ as the \emph{relative gap} w.r.t. the threshold $\epsilon$ of improvement loss.
Instead of always invoking ESV as was outlined in Section~\ref{sec::generic-esvs:significance}, we invoke it with probability $\lambda_j=\frac{1}{\rho_j}$.

\vspace{-0.5em}
\subsubsection{Refinement of Improvement Rates}

If early-stopping verification is invoked at $B_{j+1}$, there will be two possible outcomes:
\begin{itemize}[leftmargin=*]
    \item The early-stopping verification returns true, then we terminate index tuning accordingly.
    \item The early-stopping verification returns false. In this case, we let $L_{j+1}(W, C_t^*)$ and $U_{j+1}(W, C_B^*)$ be the lower and upper bounds returned. We can use $L_{j+1}$ and $U_{j+1}$ to further refine the improvement rates $l_{j+1}$ and $r_{j+1}$. Specifically, we have 
    $p_{j+2}^r=I_{j+1}+r_{j+1}\cdot s<U_{j+1}$ and
    $p_{j+2}^l=I_{j+1}+l_{j+1}\cdot s<U_{j+1},$
    which gives $r_{j+1}<\frac{U_{j+1}-I_{j+1}}{s}$ and $l_{j+1}<\frac{U_{j+1}-I_{j+1}}{s}$.
    Therefore,
    $r_{j+1}=\min\{\frac{I_{j+1}}{B_{j+1}}, \frac{U_{j+1}-I_{j+1}}{s}\},$ and
    $l_{j+1}=\min\{\frac{I_{j+1}-I_j}{s}, \frac{U_{j+1}-I_{j+1}}{s}\}.$
    This refinement can be applied to all later steps $j+3, j+4, \cdots$ as well.
\end{itemize}

\begin{figure*}
\centering
\subfigure[Time Overhead]{ \label{fig:twophase:tpch:k20:extra-time-overhead}
    \includegraphics[width=0.49\columnwidth]{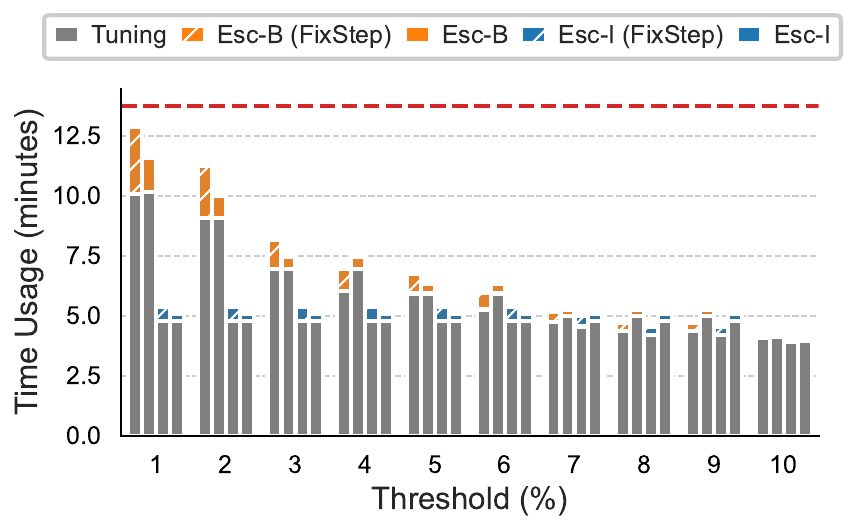}}
\subfigure[Improvement Loss]{ \label{fig:twophase:tpch:k20:impr-loss}
    \includegraphics[width=0.49\columnwidth]{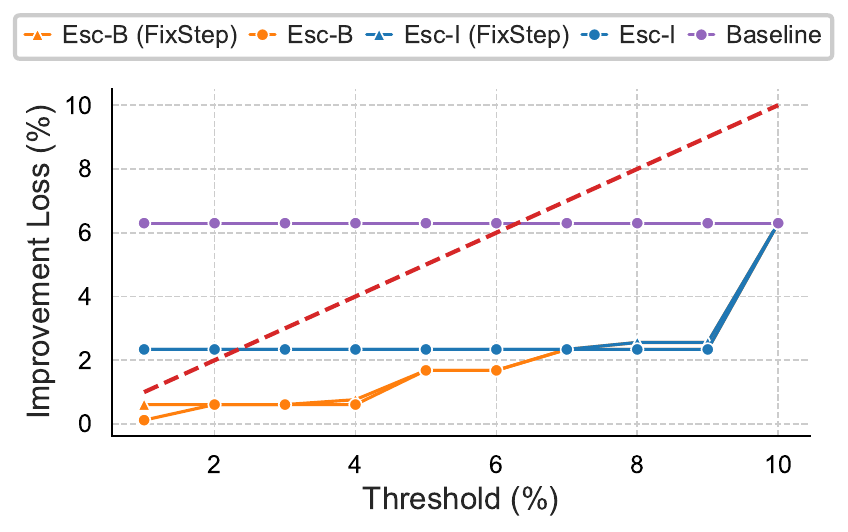}}
\subfigure[What-If Call Savings]{ \label{fig:twophase:tpch:k20:call-save}
    \includegraphics[width=0.49\columnwidth]{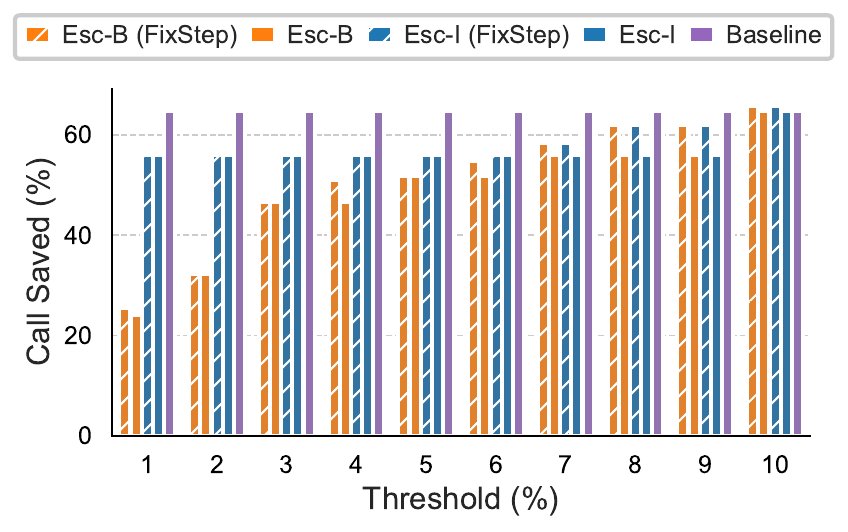}}
\subfigure[Learning Curve]{ \label{fig:twophase:tpch:k20:lc}
    \includegraphics[width=0.49\columnwidth]{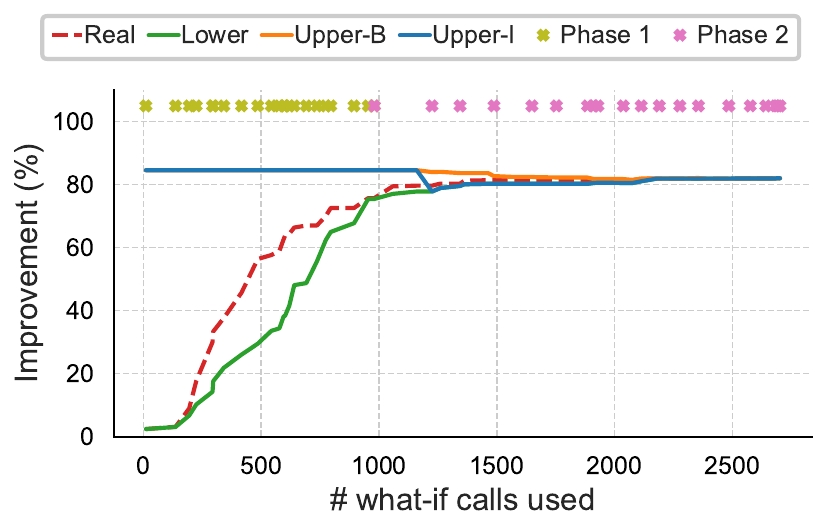}}
\vspace{-1.5em}
\caption{\revision{Two-phase greedy search, TPC-H, $K=20$, $B=20k$.}}
\label{fig:twophase:tpch:k20}
\vspace{-1.5em}
\end{figure*}


\begin{figure*}
\centering
\subfigure[Time Overhead]{ \label{fig:twophase:tpcds:k20:extra-time-overhead}
    \includegraphics[width=0.49\columnwidth]{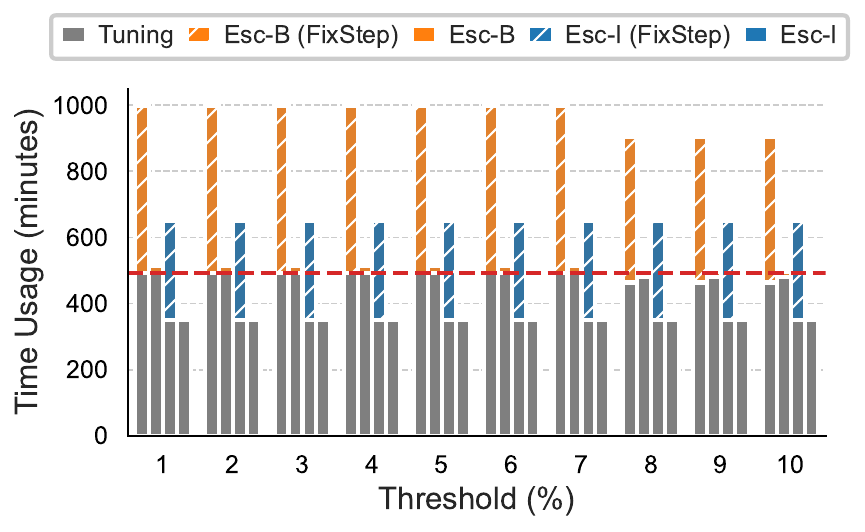}}
\subfigure[Improvement Loss]{ \label{fig:twophase:tpcds:k20:impr-loss}
    \includegraphics[width=0.49\columnwidth]{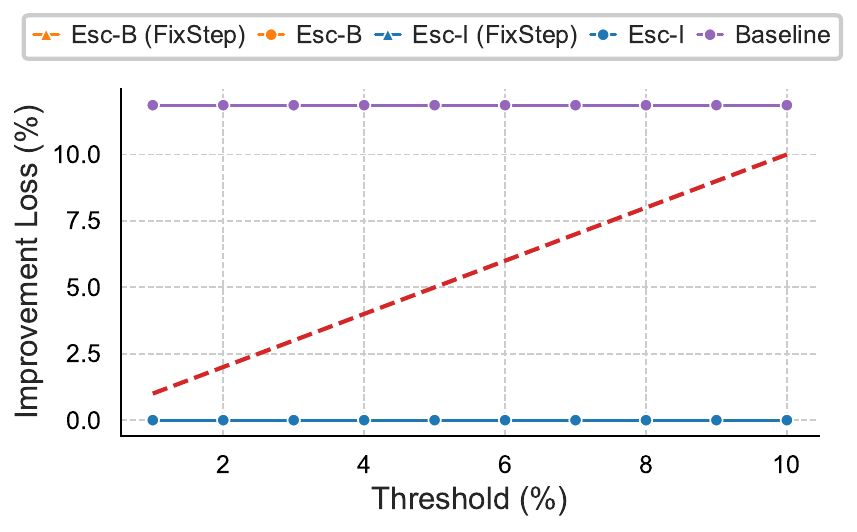}}
\subfigure[What-If Call Savings]{ \label{fig:twophase:tpcds:k20:call-save}
    \includegraphics[width=0.49\columnwidth]{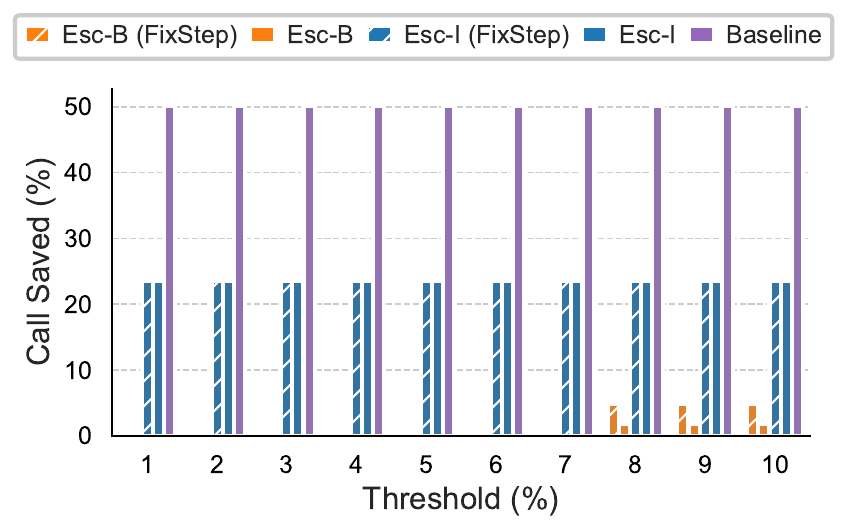}}
\subfigure[Learning Curve]{ \label{fig:twophase:tpcds:k20:lc}
    \includegraphics[width=0.49\columnwidth]{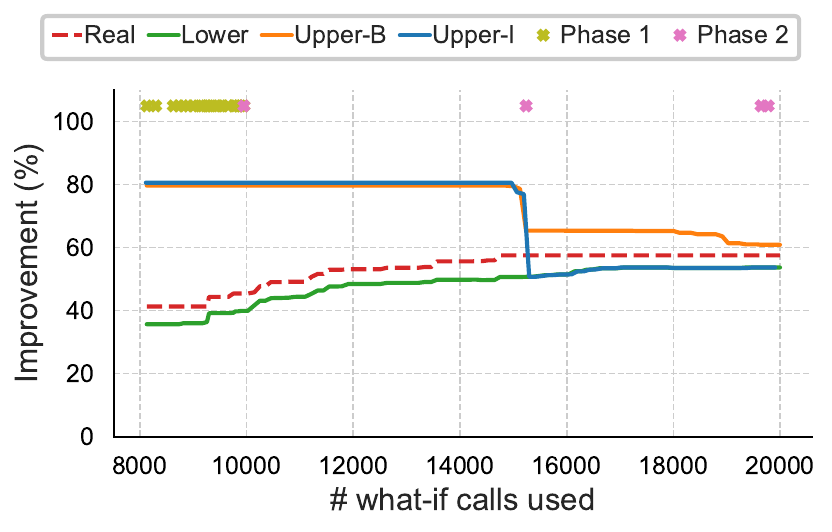}}
\vspace{-1.5em}
\caption{\revision{Two-phase greedy search, TPC-DS, $K=20$, $B=20k$.}}
\label{fig:twophase:tpcds:k20}
\vspace{-1.5em}
\end{figure*}

\begin{figure*}
\centering
\subfigure[Time Overhead]{ \label{fig:twophase:real-d:k20:extra-time-overhead}
    \includegraphics[width=0.49\columnwidth]{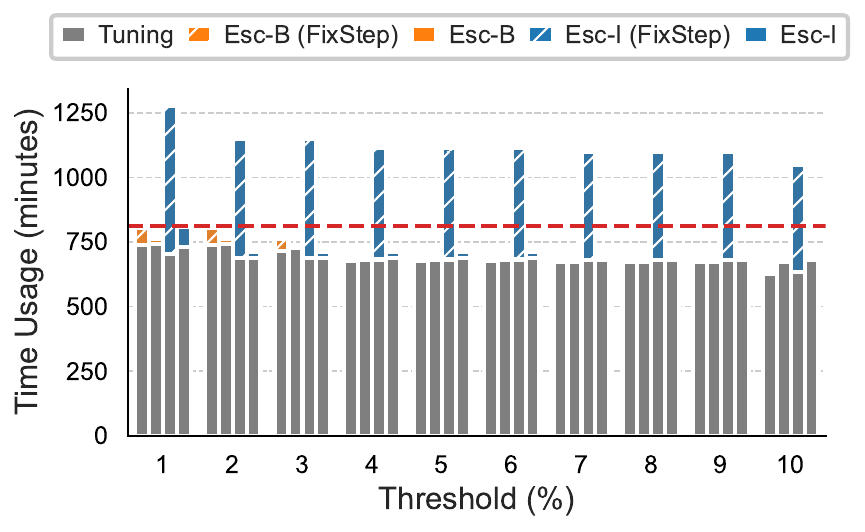}}
\subfigure[Improvement Loss]{ \label{fig:twophase:real-d:k20:impr-loss}
    \includegraphics[width=0.49\columnwidth]{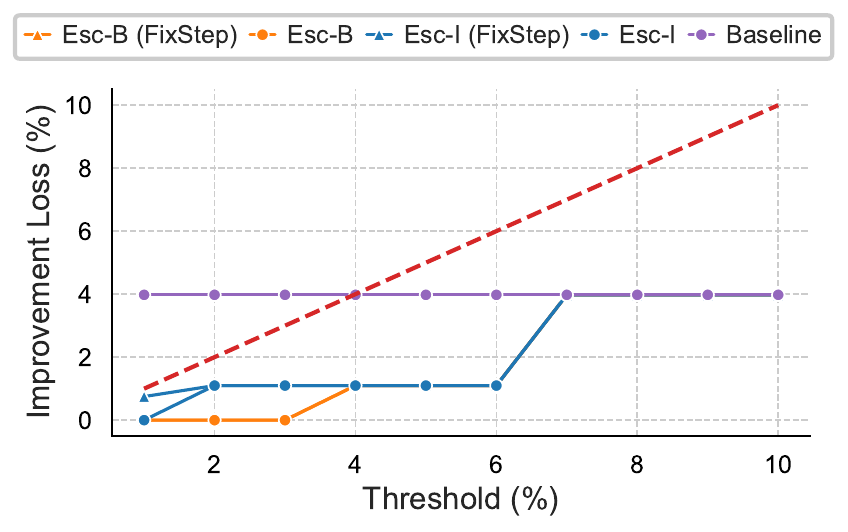}}
\subfigure[What-If Call Savings]{ \label{fig:twophase:real-d:k20:call-save}
    \includegraphics[width=0.49\columnwidth]{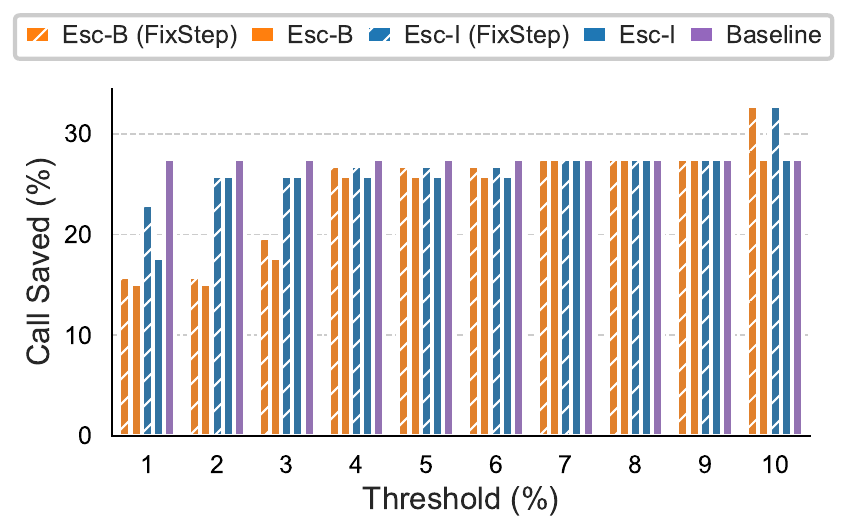}}
\subfigure[Learning Curve]{ \label{fig:twophase:real-d:k20:lc}
    \includegraphics[width=0.49\columnwidth]{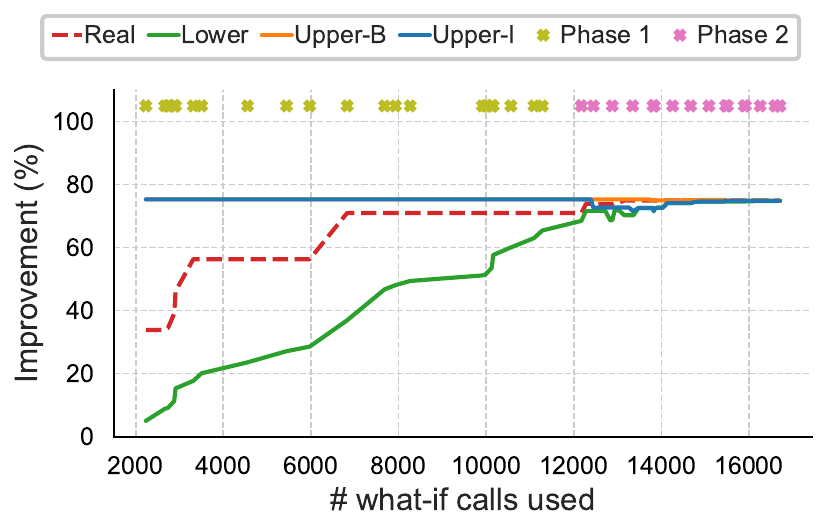}}
\vspace{-1.5em}
\caption{\revision{Two-phase greedy search, Real-D, $K=20$, $B=20k$.}}
\label{fig:twophase:real-d:k20}
\vspace{-1.5em}
\end{figure*}



\section{Evaluation}
\label{sec:eval}


We conduct extensive experimental evaluation of \sysname and report the evaluation results in this section.

\vspace{-0.5em}
\subsection{Experiment Settings}

\subsubsection{Databases and Workloads}
\label{sec:eval:settings:workloads}

\begin{table}[t] 
\footnotesize
\centering
\begin{tabular}{|l|r|r|r|r|r|r|}
\hline
\textbf{Name} & \textbf{DB Size} & \textbf{\#Queries} & \textbf{\#Tables} & \textbf{\#Joins} 
& \textbf{\#Scans}
& \revision{\textbf{\#Indexes}}\\
\hline
\hline
\textbf{TPC-H} & \emph{sf}=10 & 22 & 8 & 2.8 
& 3.7 & \revision{168} \\
\textbf{TPC-DS} & \emph{sf}=10 & 99 & 24 & 7.7 
& 8.8 & \revision{848} \\
\textbf{JOB} & 9.2GB & 33 & 21 & 7.9 & 2.5 
& \revision{66} \\
\hline
\hline
\textbf{Real-D} & 587GB & 32 & 7,912 & 15.6 
& 17 & \revision{417}\\
\textbf{Real-M} & 26GB & 31 & 474 & 13.3 
& 14.3 & \revision{642} \\
\hline
\end{tabular}
\caption{Summary of database and workload statistics.}
\vspace{-1em}
\label{tab:databases}
\end{table}

We use standard benchmarks as well as real customer workloads in our experiments.
For benchmark workloads, we use (1) \textbf{TPC-H}, (2) \textbf{TPC-DS}, and (3) the ``Join Order Benchmark'' (\textbf{JOB})~\cite{job-queries}.
We also use two real workloads, denoted by \textbf{Real-D} and \textbf{Real-M}.
\revision{Table~\ref{tab:databases} summarizes some basic properties of the workloads, in terms of schema complexity (e.g., the number of tables), query complexity (e.g., the average number of joins and table scans contained by a query), database/workload size, and the number of candidate indexes found for index tuning.}

\vspace{-0.5em}
\subsubsection{Budget-aware Index Tuning Algorithms}
We focus on evaluating two state-of-the-art budget-aware index tuning algorithms, (1) \emph{two-phase greedy search} and (2) \emph{MCTS}, as well as their enhanced versions with Wii, i.e., \emph{what-if call interception}~\cite{Wii}.



\vspace{-0.5em}
\subsubsection{Variants of Early-Stopping Verification Schemes}

We use the heuristic ESVS in Section~\ref{sec:early-stopping-verification:heuristic} for \emph{two-phase greedy search} and use the generic ESVS in Section~\ref{sec:integration:early-stopping:check} for \emph{MCTS}.
We compare four variants: (1) \textbf{\sysname-B}, where we use the corresponding ESVS with lower/upper bounds that do not consider index interaction; (2) \textbf{\sysname-I}, which further uses index interaction to refine the lower bound, as discussed in Section~\ref{sec:refinement:index-interactions:lower-bound}; (3) \textbf{\sysname-B (FixStep)}, which is a baseline of \textbf{\sysname-B} that instead adopts the fixed-step ESVS; and similarly, (4) \textbf{\sysname-I (FixStep)}, a baseline of \textbf{\sysname-I} with the fixed-step ESVS.

\vspace{-0.5em}
\subsubsection{Evaluation Metrics}

We vary the improvement-loss threshold $\epsilon$ from 1\% to 10\% in our evaluation.
For each $\epsilon$, let $b_{\epsilon}$ be the number of what-if calls allocated when early-stopping is triggered, and let $\tilde{B}$ be the number of what-if calls allocated without early-stopping.
Note that $\tilde{B}$ can be smaller than the budget $B$ on the number of what-if calls, because algorithms such as greedy search can terminate if no better configuration can be found (regardless of whether there is remaining budget on the number of what-if calls).
We then measure the following performance metrics of early-stopping:
(a) \emph{extra time overhead of early-stopping verification}, which is measured as the total time spent on invoking early-stopping verification;
(b) \emph{improvement loss}, defined as $\Delta(b_{\epsilon})=\eta(W, C_B^*)-\eta(W, C_{b_{\epsilon}}^*)$;
and (c) \emph{savings on the number of what-if calls}, defined as $(1-\frac{b_{\epsilon}}{\tilde{B}})\times 100\%$.

\vspace{-0.5em}
\subsubsection{Other Experimental Settings}
We vary the number of indexes allowed $K\in\{10, 20\}$.
We set the budget on what-if calls $B=20,000$ to make sure that index tuning can finish without early stopping; otherwise, early stopping would have never been triggered, which is correct but a tedious situation.
Moreover, we set the threshold of index interaction for refinement of the lower-bound in Section~\ref{sec:refinement:index-interactions:lower-bound} to be $\tau=0.2$, based on our empirical study in~\cite{full-version}.
For the generic ESVS in Section~\ref{sec:integration:early-stopping:check} and the baseline fixed-step ESVS, we set the step size $s=100$ (see~\cite{full-version} for results with $s=500$); furthermore, we set the threshold $\sigma = 0.5$ for the significance of concavity.






\subsubsection{\revision{Baselines}}

\revision{We also compare \sysname with baseline approaches that are based on simple heuristics.
Specifically, for \emph{two-phase greedy search}, we compare \sysname with a baseline that simply stops tuning after the first phase of greedy search; for \emph{MCTS}, we compare \sysname with a baseline that simply stops tuning if the observed percentage improvement $I_j$ over the existing configuration is greater than some fixed threshold (we set the threshold to be 30\% in our evaluation).}





\begin{figure*}
\centering
\subfigure[Time Overhead]{ \label{fig:mcts:tpch:k20:extra-time-overhead}
    \includegraphics[width=0.49\columnwidth]{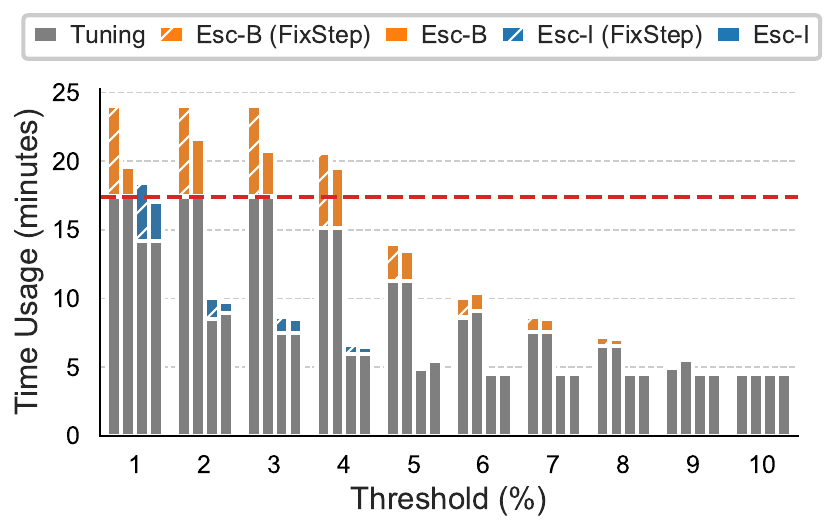}}
\subfigure[Improvement Loss]{ \label{fig:mcts:tpch:k20:impr-loss}
    \includegraphics[width=0.49\columnwidth]{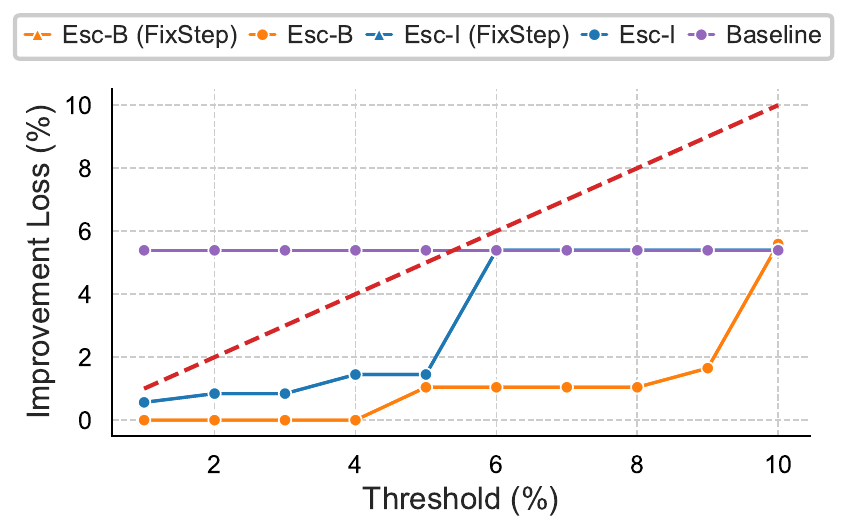}}
\subfigure[What-If Call Savings]{ \label{fig:mcts:tpch:k20:call-save}
    \includegraphics[width=0.49\columnwidth]{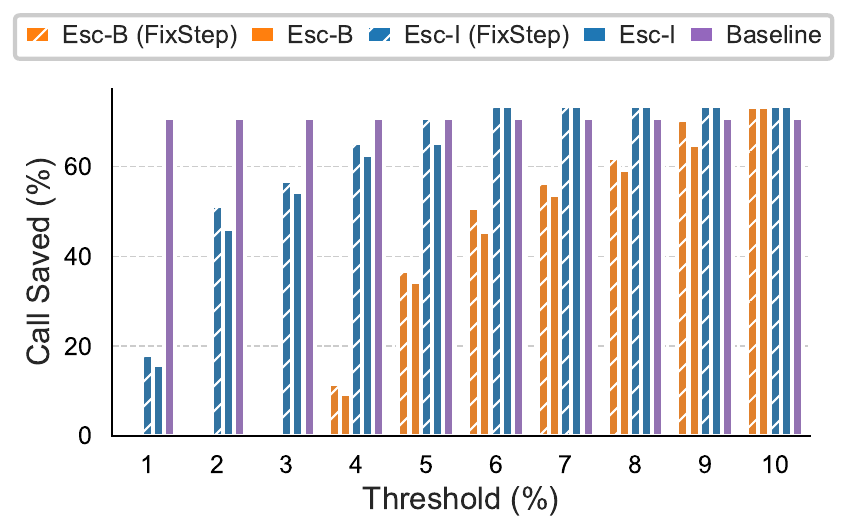}}
\subfigure[Learning Curve]{ \label{fig:mcts:tpch:k20:lc}
    \includegraphics[width=0.49\columnwidth]{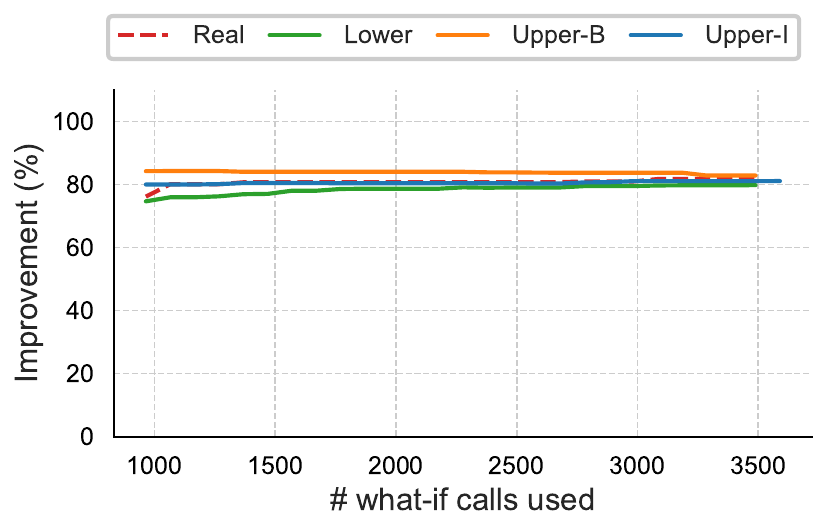}}
\vspace{-1.5em}
\caption{\revision{MCTS, TPC-H, $K=20$, $B=20k$.}}
\label{fig:mcts:tpch:k20}
\vspace{-2em}
\end{figure*}

\begin{figure*}
\centering
\subfigure[Time Overhead]{ \label{fig:mcts:real-d:k20:extra-time-overhead}
    \includegraphics[width=0.49\columnwidth]{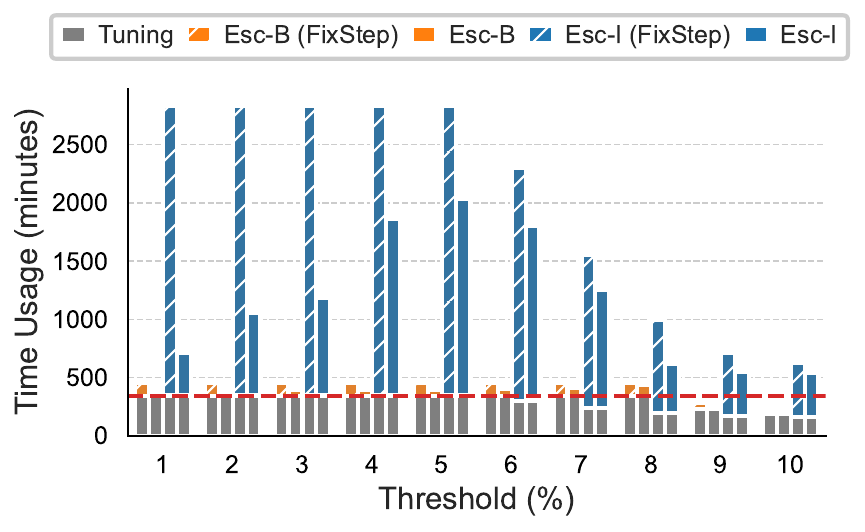}}
\subfigure[Improvement Loss]{ \label{fig:mcts:real-d:k20:impr-loss}
    \includegraphics[width=0.49\columnwidth]{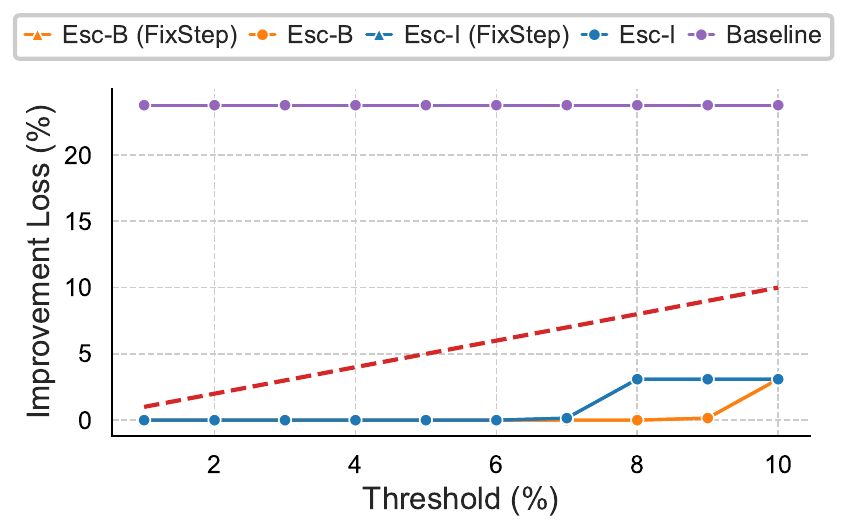}}
\subfigure[What-If Call Savings]{ \label{fig:mcts:real-d:k20:call-save}
    \includegraphics[width=0.49\columnwidth]{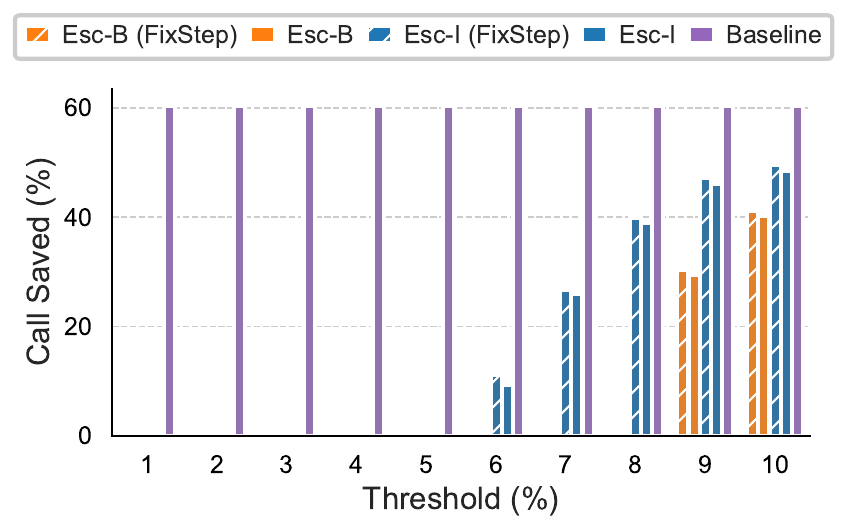}}
\subfigure[Learning Curve]{ \label{fig:mcts:real-d:k20:lc}
    \includegraphics[width=0.49\columnwidth]{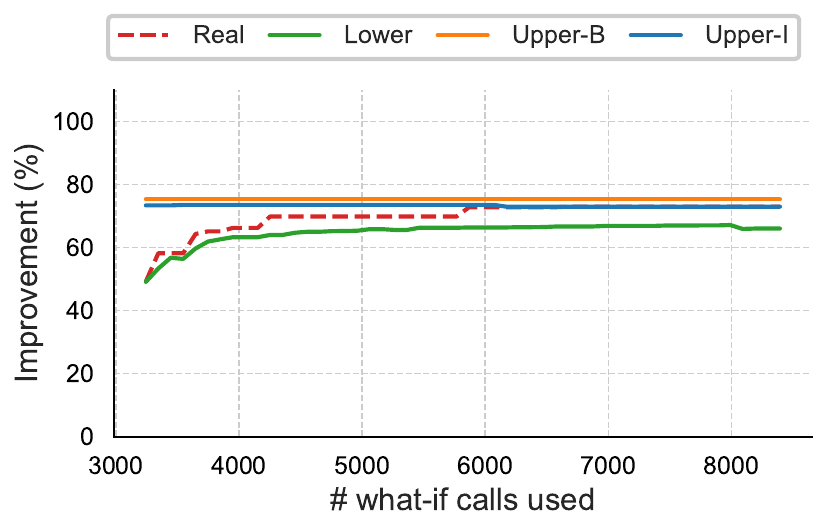}}
\vspace{-1.5em}
\caption{\revision{MCTS, Real-D, $K=20$, $B=20k$.}}
\label{fig:mcts:real-d:k20}
\vspace{-2em}
\end{figure*}

\vspace{-1em}
\subsection{Two-phase Greedy Search}

Figures~\ref{fig:twophase:tpch:k20} 
to~\ref{fig:twophase:real-d:k20} 
present the results when running \emph{two-phase greedy search} on top of \textbf{TPC-H}, \textbf{TPC-DS}, and \textbf{Real-D}. 
The results on \textbf{JOB} and \textbf{Real-M} are included in~\cite{full-version}.
\revision{In each figure, we present (a) the extra time overhead (in minutes) of early-stopping verification, (b) the improvement loss when early-stopping is triggered, (c) the savings on the number of what-if calls, and (d) the index tuning curve as well as the corresponding lower and upper bounds.}

\vspace{-0.5em}
\subsubsection{\revision{Extra Time Overhead of Early-Stopping Verification}}

\revision{As a reference point, in each plot (a) the red dashed line represents the corresponding index tuning time \emph{without} early-stopping verification, whereas the gray bars represent the net index tuning time \emph{with} early-stopping verification.}
We observe that the extra time overhead of both \textbf{\sysname-B} and \textbf{\sysname-I} is negligible compared to the index tuning time, across all workloads tested.
On the other hand, \textbf{\sysname-B (FixStep)} and \textbf{\sysname-I (FixStep)} sometimes result in considerable extra time overhead. 
For example, as shown in Figure~\ref{fig:twophase:tpcds:k20:extra-time-overhead}, on \textbf{TPC-DS} the extra time overhead of \textbf{\sysname-B (FixStep)} is comparable to the index tuning time when varying the threshold $\epsilon$ from 1\% to 7\%.
\revision{Overall, the savings in terms of end-to-end index tuning time by applying \sysname resonate with the corresponding savings on what-if calls shown in each plot (c).
}


\vspace{-0.5em}
\subsubsection{Improvement Loss}

The red dashed line in each plot (b) delineates the acceptable improvement loss.
That is, any improvement loss above that line violates the threshold $\epsilon$ set by the user.
We observe that violation occurs rarely, 
e.g., when setting $\epsilon=1\%$ on \textbf{TPC-H} and using \textbf{\sysname-I} for early stopping.
Moreover, the actual improvement loss is often much smaller than the threshold $\epsilon$ when early-stopping is triggered.
One reason for this is that our lower bound $\eta_L(W, C_t^*)$ and upper bound $\eta_U(W, C_B^*)$ are more conservative than the actual improvements $\eta(W, C_t^*)$ and $\eta(W, C_B^*)$ needed for triggering early-stopping (ref. Section~\ref{sec:overview:framework}).

\vspace{-0.5em}
\subsubsection{Savings on What-If Calls}

The plot (c) in each figure represents the (percentage) savings on the number of what-if calls.
We have the following observations.
First, the savings typically increase as the threshold $\epsilon$ increases.
Intuitively, a less stringent $\epsilon$ can trigger early-stopping sooner.
Second, the savings vary on different workloads.
For example, with $\epsilon=5\%$, the savings are around 60\% on \textbf{TPC-H}; 
however, the savings drop to 25\% on \textbf{TPC-DS} and 
\textbf{Real-D}. 
We can understand this better by looking at the corresponding index tuning curve in the plot (d).
Third, considering index interaction typically leads to an improved upper bound, which results in more savings on what-if calls.

\vspace{-0.5em}
\subsubsection{\revision{Comparison with Baseline}}

\revision{We now compare \sysname with the baseline approach that simply stops tuning after the first phase of greedy search, in terms of the improvement loss and the savings on what-if calls.
As shown by the plots (b) and (c) of each figure, the baseline can achieve higher savings on what-if calls but can suffer from significantly higher improvement loss.
For example, as Figure~\ref{fig:twophase:tpcds:k20:impr-loss} shows, on \textbf{TPC-DS} the improvement loss of the baseline is around 12\% while \sysname has zero improvement loss.}





\begin{figure*}
\centering
\subfigure[Time Overhead]{ \label{fig:twophase_covskip:real-m:k20:extra-time-overhead}
    \includegraphics[width=0.49\columnwidth]{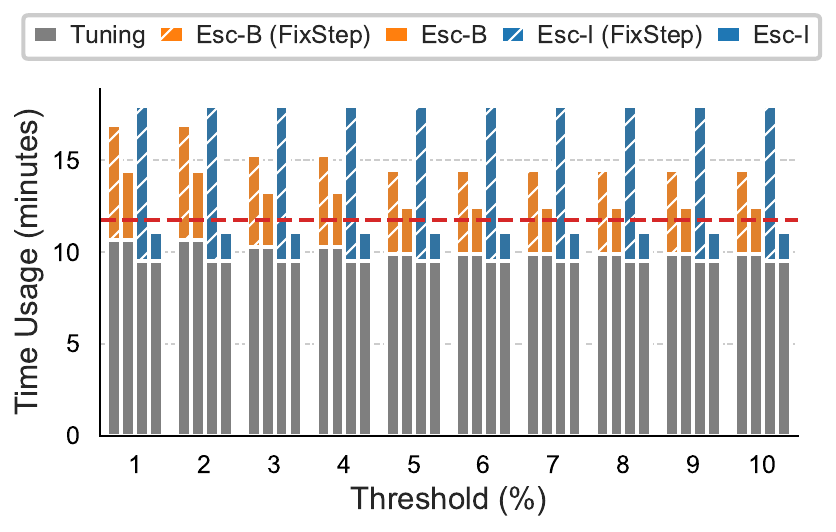}}
\subfigure[Improvement Loss]{ \label{fig:twophase_covskip:real-m:k20:impr-loss}
    \includegraphics[width=0.49\columnwidth]{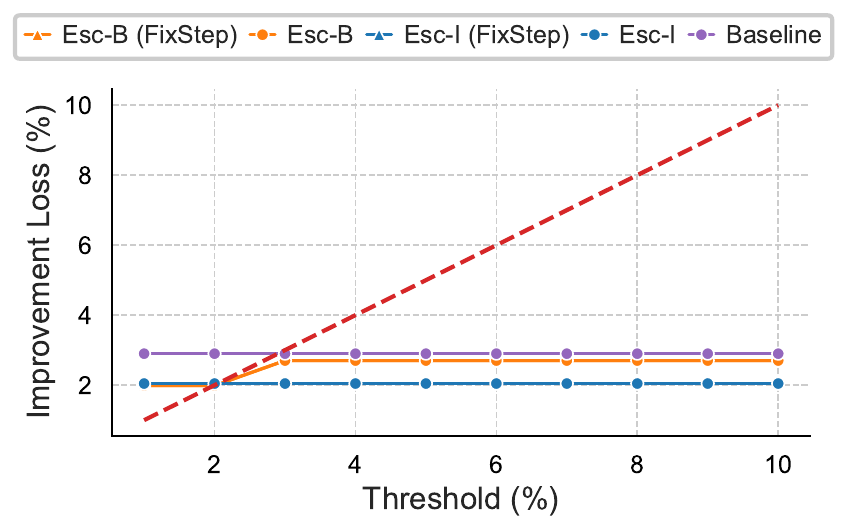}}
\subfigure[What-If Call Savings]{ \label{fig:twophase_covskip:real-m:k20:call-save}
    \includegraphics[width=0.49\columnwidth]{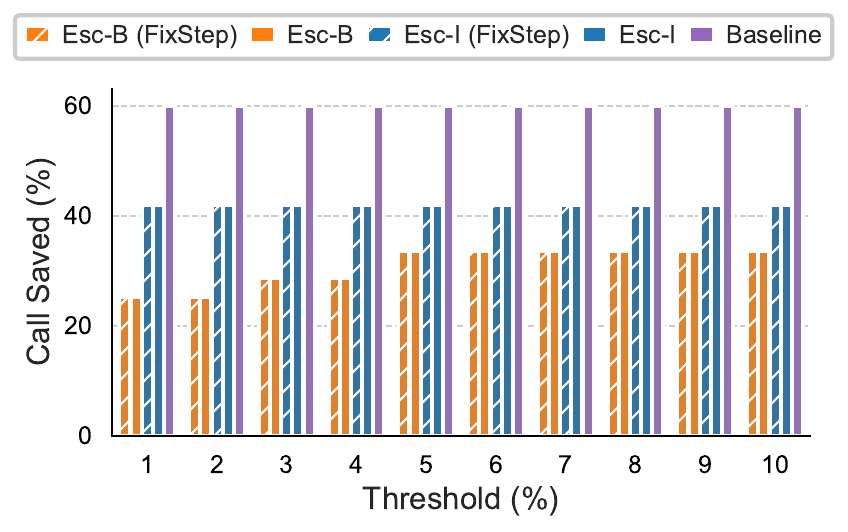}}
\subfigure[Learning Curve]{ \label{fig:twophase_covskip:real-m:k20:lc}
    \includegraphics[width=0.49\columnwidth]{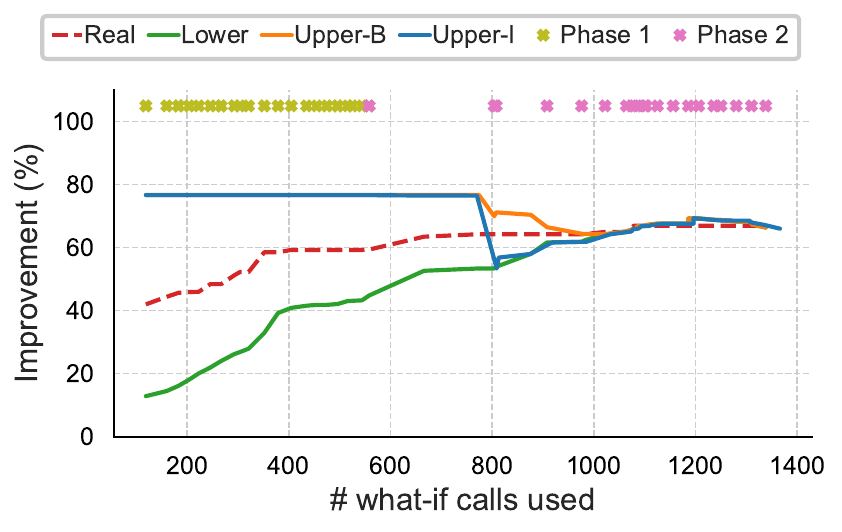}}
\vspace{-1.5em}
\caption{\revision{Two-phase greedy search (with Wii-Coverage), Real-M, $K=20$, $B=20k$.}}
\label{fig:twophase_covskip:real-m:k20}
\vspace{-1.5em}
\end{figure*}

\begin{figure*}
\centering
\subfigure[\textbf{JOB}]{ \label{fig:tpg-esc-vs-dta:job}
    \includegraphics[width=0.23\textwidth]{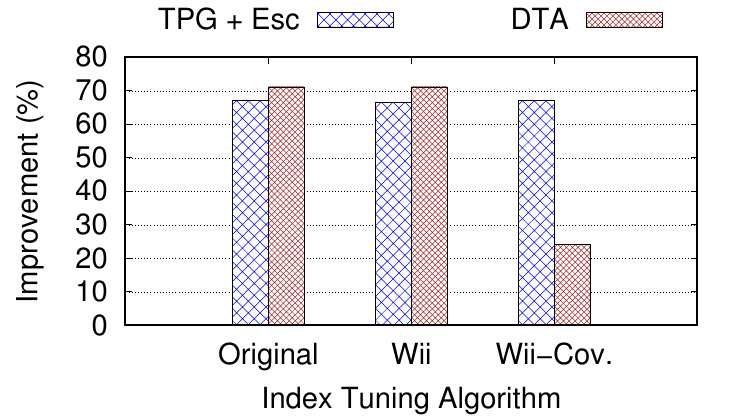}}
\subfigure[\textbf{TPC-DS}]{ \label{fig:tpg-esc-vs-dta:tpcds}
    \includegraphics[width=0.23\textwidth]{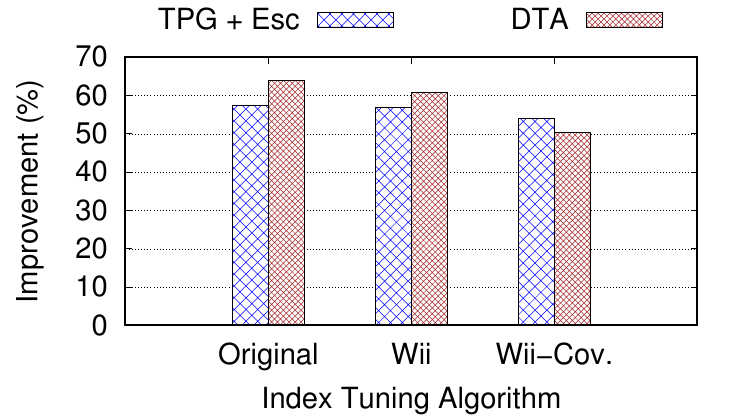}}
\subfigure[\textbf{Real-D}]{ \label{fig:tpg-esc-vs-dta:real-d}
    \includegraphics[width=0.23\textwidth]{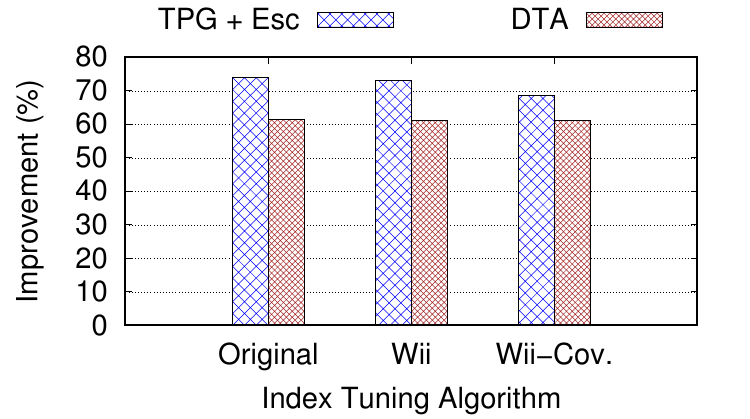}}
\subfigure[\textbf{Real-M}]{ \label{fig:tpg-esc-vs-dta:real-m}
    \includegraphics[width=0.23\textwidth]{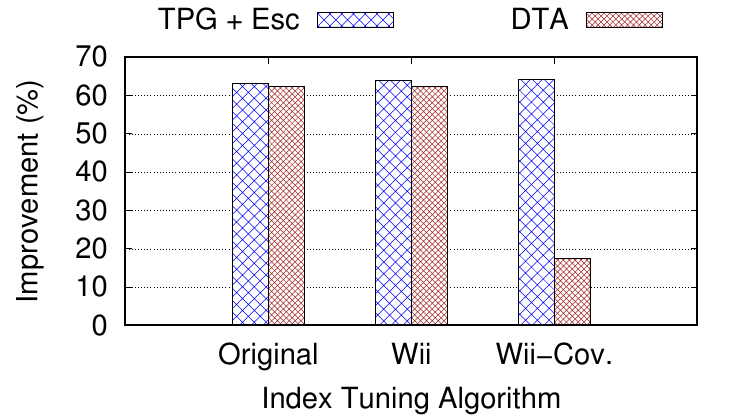}}
\vspace{-1.5em}
\caption{\revision{Comparison of two-phase greedy (TPG) search with \sysname (without or with what-if call interception) against DTA.}}
\label{fig:tpg-esc-vs-dta}
\vspace{-1.5em}
\end{figure*}

\vspace{-1em}
\subsection{Monte Carlo Tree Search}

Figures~\ref{fig:mcts:tpch:k20} and~\ref{fig:mcts:real-d:k20} present the results for \emph{MCTS} on \textbf{TPC-H} and \textbf{Real-D}. 
The results on the other workloads can be found in~\cite{full-version}.

\vspace{-0.5em}
\subsubsection{\revision{Extra Time Overhead of Early-Stopping Verification}}

Again, we observe that the extra time overhead of early-stopping verification is negligible compared to the index tuning time in most of the cases tested.
However, we also notice a few cases where the extra time overhead of early-stopping verification is considerable.
This typically happens when it is difficult to trigger early-stopping using the lower and upper bounds.
As a result, all the ESV invocations are unnecessary, which indicates opportunities for further improvement of the generic ESVS proposed in Section~\ref{sec:integration:early-stopping:check}.

Meanwhile, the generic ESVS again significantly reduces the extra time overhead compared to the fixed-step ESVS, by comparing \textbf{\sysname-B} and \textbf{\sysname-I} with \textbf{\sysname-B (FixStep)} and \textbf{\sysname-I (FixStep)}, respectively.
Moreover, like in \emph{two-phase greedy search}, the relationship between the extra time overhead of \textbf{\sysname-B} and \textbf{\sysname-I} is inconclusive.
In general, each invocation of early-stopping verification using \textbf{\sysname-B} is less expensive than using \textbf{\sysname-I}, because considering index interactions requires more computation.
However, since \textbf{\sysname-I} improves the upper bound $\eta_U(W, C_B^*)$, it can trigger early-stopping sooner, which leads to fewer invocations of early-stopping verification.
Therefore, the overall extra time overhead of \textbf{\sysname-I} can be smaller than that of \textbf{\sysname-B}, as showcased in Figure~\ref{fig:mcts:tpch:k20:extra-time-overhead} for \textbf{TPC-H}.
On the other hand, the overall extra time overhead of \textbf{\sysname-I} is considerably larger than that of \textbf{\sysname-B} for the workload \textbf{Real-D}, as evidenced by Figure~\ref{fig:mcts:real-d:k20:extra-time-overhead}.
\revision{Regarding the savings on end-to-end tuning time, for \textbf{TPC-H} the savings are similar to the corresponding savings on what-if calls, as Figure~\ref{fig:mcts:tpch:k20:call-save} shows; for \textbf{Real-D} the savings are similar when \textbf{\sysname-B} is used but are vanished when \textbf{\sysname-I} is used due to its much higher computation overhead.}


\vspace{-0.5em}
\subsubsection{Improvement Loss}

Like in \emph{two-phase greedy search}, we see almost no violation of the improvement-loss threshold $\epsilon$ when early-stopping is triggered for \emph{MCTS}.
Moreover, the actual improvement loss is typically much lower than the threshold $\epsilon$.

\vspace{-0.5em}
\subsubsection{Savings on What-If Calls}

The (percentage) savings on the number of what-if calls again vary across the workloads tested.
For example, on \textbf{TPC-H} we can save 60\% what-if calls by using \textbf{\sysname-I} when the improvement-loss threshold $\epsilon$ is set to 5\%, as shown in Figure~\ref{fig:mcts:tpch:k20:call-save}.
The actual improvement loss when early-stopping is triggered, however, is less than 2\% instead of the 5\% threshold, based on Figure~\ref{fig:mcts:tpch:k20:impr-loss}.
For \textbf{Real-D} we can only start saving on what-if calls with $\epsilon>5\%$, though we can save up to 40\% what-if calls when setting $\epsilon=10\%$ and using \textbf{\sysname-B}, as Figure~\ref{fig:mcts:real-d:k20:call-save} indicates.
Note that, although we can save up to 50\% what-if calls by using \textbf{\sysname-I}, its extra time overhead is prohibitively high based on Figure~\ref{fig:mcts:real-d:k20:extra-time-overhead}, while the extra time overhead of using \textbf{\sysname-B} is significantly lower than the overall index tuning time.
Moreover, a larger threshold $\epsilon$ typically leads to larger savings on the what-if calls, as it is easier for the gap between the lower and upper bounds to meet the threshold.

\vspace{-0.5em}
\subsubsection{\revision{Comparison with Baseline}}

\revision{Compared to \sysname, the baseline approach that simply stops tuning after observing 30\% improvement again can suffer from significant improvement loss. For example, as Figure~\ref{fig:mcts:real-d:k20:impr-loss} shows, the improvement loss of the baseline on \textbf{Real-D} is around 25\%, whereas \sysname has almost no loss.
One could argue that having a threshold different than the 30\% used may make a difference; however, choosing an appropriate threshold \emph{upfront} for the baseline approach is itself a challenging problem.}

\vspace{-1em}
\subsection{What-If Call Interception}

We have observed several cases where early-stopping offers little or no benefit, e.g., when running \emph{two-phase greedy search} on top of \textbf{Real-M}, 
or when running \emph{MCTS} on top of \textbf{TPC-DS} and \textbf{Real-M}, as shown in the full version~\cite{full-version}.
The main reason for this inefficacy is the slow convergence of the gap between the lower and upper bounds used for triggering early-stopping.
This phenomenon can be alleviated by using Wii, the what-if call interception mechanism developed in~\cite{Wii}, which skips \emph{inessential} what-if calls whose what-if costs are close to their derived costs.

For example, the heuristic ESVS in Section~\ref{sec:early-stopping-verification:heuristic} only invokes early-stopping verification when \emph{two-phase greedy search} enters Phase 2, when the upper bound is expected to drop sharply.
With Wii integrated into \emph{two-phase greedy search}, it can enter Phase 2 faster by skipping inessential what-if calls in Phase 1.
As a result, we can expect \sysname to be more effective for Wii-enhanced \emph{two-phase greedy search}.
To demonstrate this, we present the corresponding results for \textbf{Real-M} in Figure~\ref{fig:twophase_covskip:real-m:k20} using the Wii-enhanced \emph{two-phase greedy search} with the coverage-based refinement.
We observe that the savings on the number of what-if calls can further increase to 30\% (using \textbf{\sysname-B}) and 40\% (using \textbf{\sysname-I}), as Figure~\ref{fig:twophase_covskip:real-m:k20:call-save} presents.

\vspace{-0.5em}
\paragraph*{\revision{Remarks}}

\revision{While Wii can often significantly bring down the number of what-if calls, this is a side effect that is not by design.
Indeed, the goal of Wii is only to skip inessential what-if calls.
Nevertheless, it does reduce the number of what-if calls that need to be made---if this number is smaller than the given budget we will see a (sometimes significant) drop on the total number of what-if calls made.
Therefore, the contributions of early stopping and Wii in terms of reducing what-if calls are orthogonal and should not be directly compared.
That is, there are cases where Wii can and cannot reduce the number of what-if calls while early stopping can make similar (e.g., 20\% to 40\%) reductions.}




\vspace{-1em}
\subsection{\revision{Comparison with DTA}}

\revision{
To understand the overall benefit of budget-aware index tuning with \sysname enabled, when compared to \emph{other} index tuning algorithms, we further compare \emph{two-phase greedy search} with \sysname (TPG-\sysname) against DTA, which employs \emph{anytime} index tuning techniques~\cite{dta} that can achieve state-of-the-art tuning performance~\cite{KossmannHJS20}.
In our evaluation, we set the threshold of improvement loss $\epsilon=5\%$.
We measure the corresponding time spent by TPG-\sysname and use that as the tuning time allowed for DTA~\cite{dta-utility}, for a fair comparison.
}

\revision{
Figure~\ref{fig:tpg-esc-vs-dta} presents the results. We omit the results on \textbf{TPC-H} as TPG-\sysname and DTA achieve the same 79\% improvement.
We have the following observations on the other workloads.
On \textbf{JOB}, TPG-\sysname significantly outperforms DTA when Wii-coverage 
is enabled (67\% by TPG-\sysname vs. 24\% by DTA).
On \textbf{TPC-DS}, TPG-\sysname and DTA perform similarly.
On \textbf{Real-D}, TPG-\sysname outperforms DTA by around 10\%.
On \textbf{Real-M}, TPG-\sysname significantly outperforms DTA, again when Wii-coverage is enabled (64\% by TPG-\sysname vs. 17\% by DTA).
Overall, we observe that TPG-\sysname either performs similarly to DTA or outperforms DTA by a noticeable margin in terms of percentage improvement, within the same amount of tuning time.
Note that DTA leverages additional optimizations (e.g., ``table subset'' selection~\cite{AgrawalCN00,dta}, index merging~\cite{ChaudhuriN99}, prioritized index selection~\cite{dta}, etc.) that we did not implement for TPG-\sysname.
On the other hand, it remains interesting to see the further improvement on DTA by integrating \sysname, which is beyond the scope of this paper.}

\vspace{-0.5em}
\subsection{\revision{Discussion and Future Work}}

\paragraph*{\revision{Violation of Improvement Loss}}

\revision{
Violation is very rare based on our evaluation results, but it can happen if the assumptions about the what-if cost function, i.e., monotonicitiy and submodularity, are invalid.
In such situations, the lower and upper bounds derived for the workload-level what-if cost are also invalid and therefore can mislead the early-stopping checker.
One possible solution is then to validate the assumptions of monotonicity and submodularity while checking for early stopping.
If validation fails frequently, then we will have lower confidence on the validity of the bounds and thus we can stop running the early-stopping checker to avoid potential violation on the promised improvement loss.
}

\vspace{-0.5em}
\paragraph*{\revision{Hard Cases}}

\revision{
As an example, the \textbf{TPC-DS} results in Figure~\ref{fig:twophase:tpcds:k20} represent a difficult case for \sysname when applied to \emph{two-phase greedy search}.
From Table~\ref{tab:databases}, we observe a large search space for \emph{two-phase greedy search} over \textbf{TPC-DS} with 848 candidate indexes.
Moreover, the workload size of \textbf{TPC-DS} with 99 queries is also considerably larger than the other workloads in Table~\ref{tab:databases}.
As a result, the heuristic early-stopping verification scheme designed for two-phase greedy search (Section~\ref{sec:early-stopping-verification:heuristic}) works less effectively, because verification will not be invoked until entering the second phase of greedy search.
Lots of what-if calls have been made in the first phase as well as the first step of the second phase, before the bounds start converging sharply.
To improve on this case, we have to make the bounds converge earlier, which is challenging given the conservative nature of the bounds.
We therefore leave this for future work.}

\vspace{-0.5em}
\section{Related Work}

\paragraph*{Cost-based Index Tuning}

Offline index tuning has been extensively studied in the literature (e.g.,~\cite{Whang85,ChaudhuriN97,dta,ValentinZZLS00,BrunoC05,DashPA11,dexter-2,SchlosserK019,KossmannHJS20,WuWSWNCB22}).
\revision{Early work focused on index configuration enumeration algorithms, including, e.g., \emph{Drop}~\cite{Whang85}, \emph{AutoAdmin}~\cite{ChaudhuriN97}, \emph{DTA}~\cite{dta}, \emph{DB2Advisor}~\cite{ValentinZZLS00}, \emph{Relaxation}~\cite{BrunoC05}, \emph{CoPhy}~\cite{DashPA11}, \emph{Dexter}~\cite{dexter-2}, and \emph{Extend}~\cite{SchlosserK019}. We refer the readers to the recent benchmark studies~\cite{KossmannHJS20,ZhouLZL24} for more details and performance comparisons of these solutions.}
More recent work has been focusing on addressing scalability issues of index tuning when dealing with large and complex workloads (e.g.,~\cite{SiddiquiJ00NC22,SiddiquiWNC22,WuWSWNCB22,Wred,Wii,YuZSY24})
and query performance regressions when the recommended indexes are actually deployed (e.g.,~\cite{DingDWCN18,DingDM0CN19,ShiCL22,ZhaoCSM22,Wu25}).
The latter essentially addresses the problem of modeling query execution cost in the context of index tuning, and there has been lots of work devoted to this problem (e.g.,~\cite{Ganapathi-berkeley09,AkdereCRUZ12-brown-icde,LiKNC12,WuCZTHN13,WuCHN13,WuWHN14,WuNS16,MarcusP19,MarcusNMZAKPT19,SunL19,SiddiquiJQPL20,PaulCLS21,hilprecht2022zero,WuW24}).
There has also been recent work on \emph{online} index tuning with a focus of applying deep learning and reinforcement learning technologies (e.g.~\cite{PereraORB21,abs-1801-05643,LanBP20,PereraORB22}).
Online index tuning assumes a continuous workload model where queries are observed in a streaming manner, which is different from offline index tuning that assumes all queries have been observed before index tuning starts.

\vspace{-0.5em}
\paragraph*{Learning Curve and Early Stopping}

Our notion of index tuning curve is akin to the term ``learning curve'' in the machine learning (ML) literature, which is used to characterize the performance of an iterative ML algorithm as a function of its training time or number of iterations~\cite{learnign-curve-prediction-for-HPO-IJCAI-15,learnign-curve-prediction-for-HPO-ICLR-17}.
It is a popular tool for visualizing the concept of \emph{overfitting}: although the performance of the ML model on the training dataset improves over time, its performance on the test dataset often degrades eventually. 
The study of learning curve has led to \emph{early stopping} as a form of regularization used to avoid overfitting when training an ML model with an iterative method such as \emph{gradient descent}~\cite{yao2007early,Prechelt2012,RaskuttiWY14}.
Early-stopping in budget-aware index tuning, however, is different, with the goal of saving what-if calls instead of improving index quality, though the generic early-stopping verification scheme developed in Section~\ref{sec:integration:early-stopping:check} relies on the convexity/concavity properties of the index tuning curve.





\vspace{-0.5em}
\paragraph*{Index Interaction}

Some early work (e.g.~\cite{ChoenniBC93,FinkelsteinST88,WhangWS81}) has noted down the importance of modeling index interactions.
A more systematic study of index interaction was performed by Schnaitter et al.~\cite{SchnaitterPG09}, and our definition of index interaction presented in Section~\ref{sec:early-stopping:workload:lower-refinement:index-interaction} can be viewed as a simplified case of the definition proposed in that work.
Here, we are only concerned with the interaction between the next index to be selected and the indexes that have been selected in the simulated greedy search outlined by Procedure~\ref{proc:simulation-greedy}.
In contrast, the previous work~\cite{SchnaitterPG09} aims to quantify any pairwise index interaction within a given configuration, with respect to the presence of all other indexes within the same configuration.
To compute the index interaction so defined, one then needs to enumerate \emph{all possible subsets} of the configuration, which is computationally much more expensive.
Since we need a rough but efficient way of quantifying index interaction, we do not pursue the definition proposed by~\cite{SchnaitterPG09} due to its computational complexity.

\vspace{-0.5em}
\section{Conclusion}

We have presented \sysname, an early-stopping checker for budget-aware index tuning.
It extends call-level lower and upper bounds of what-if cost to develop workload-level improvement bounds that converge efficiently as index tuning proceeds.
It further adopts a generic early-stopping verification scheme that exploits the convexity/concavity properties of the index tuning curve to skip unnecessary invocations of early-stopping verification.
Evaluation on top of both industrial benchmarks and real customer workloads demonstrates that \sysname can effectively terminate index tuning with improvement loss below the user-specified threshold while at the same time significantly reduce the amount of what-if calls made for index tuning.
Moreover, the extra computation time introduced by early-stopping verification is negligible compared to the overall index tuning time.

\clearpage


\balance
\bibliographystyle{ACM-Reference-Format}
\bibliography{./paper}

\clearpage

\appendix




\begin{figure*}
\centering
\subfigure[Time Overhead]{ \label{fig:twophase:job:k20:extra-time-overhead}
    \includegraphics[width=0.49\columnwidth]{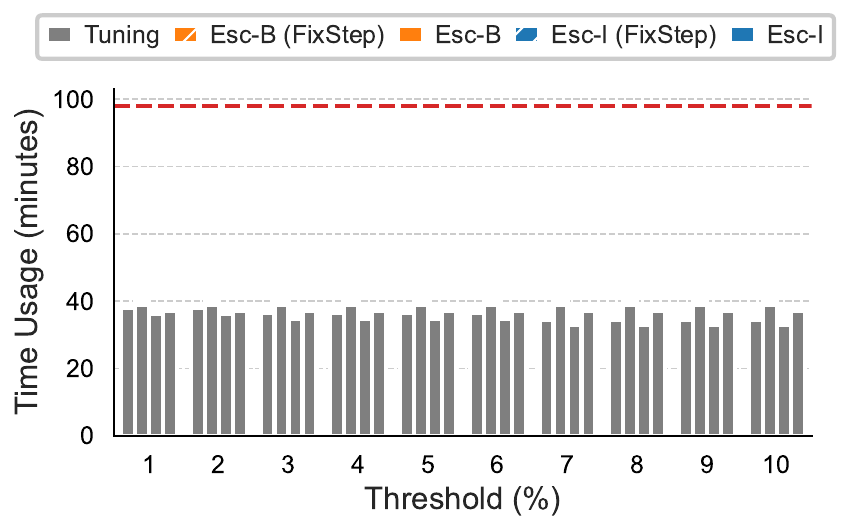}}
\subfigure[Improvement Loss]{ \label{fig:twophase:job:k20:impr-loss}
    \includegraphics[width=0.49\columnwidth]{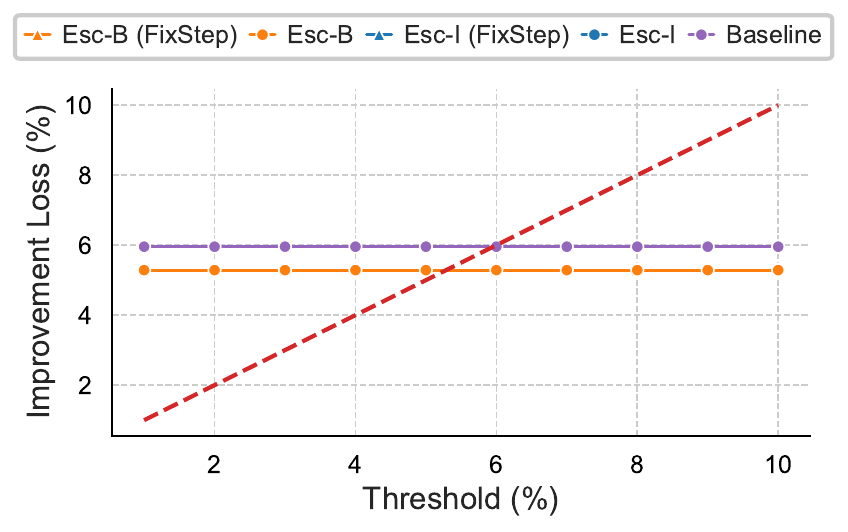}}
\subfigure[What-If Call Savings]{ \label{fig:twophase:job:k20:call-save}
    \includegraphics[width=0.49\columnwidth]{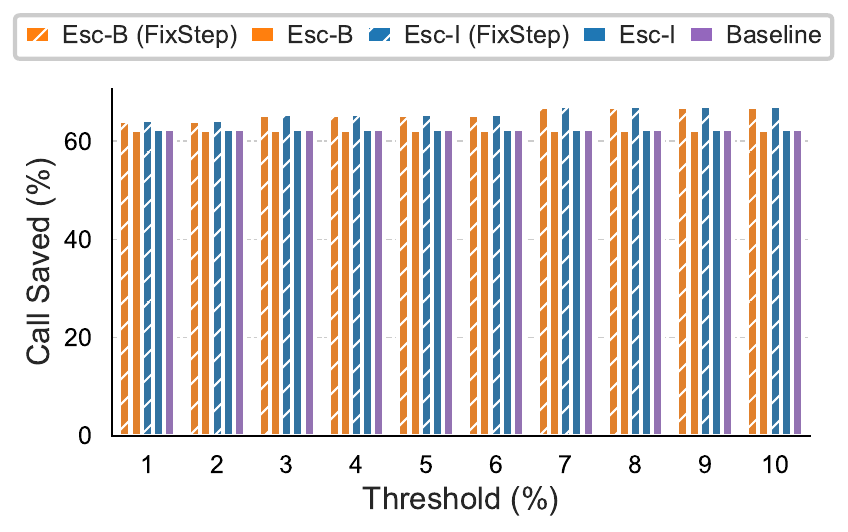}}
\subfigure[Learning Curve]{ \label{fig:twophase:job:k20:lc}
    \includegraphics[width=0.49\columnwidth]{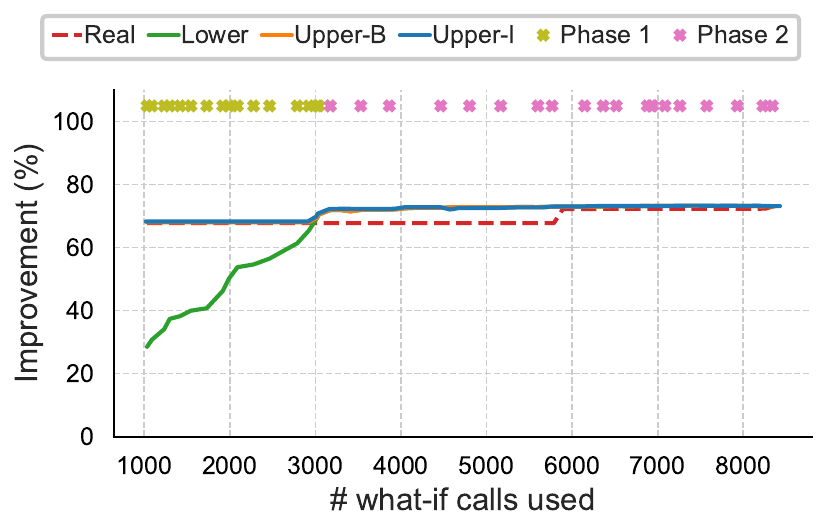}}
\vspace{-1.5em}
\caption{Two-phase greedy search, JOB, $K=20$, $B=20k$.}
\label{fig:twophase:job:k20}
\vspace{-1em}
\end{figure*}


\begin{figure*}
\centering
\subfigure[Time Overhead]{ \label{fig:twophase:real-m:k20:extra-time-overhead}
    \includegraphics[width=0.49\columnwidth]{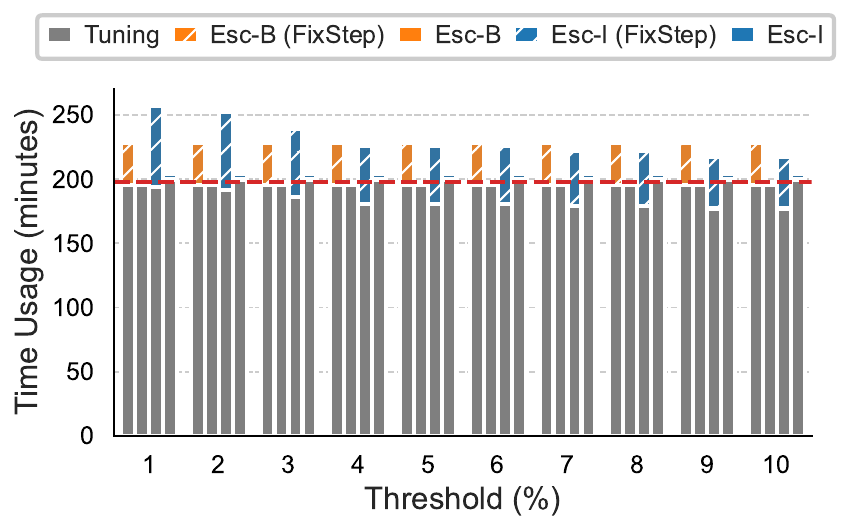}}
\subfigure[Improvement Loss]{ \label{fig:twophase:real-m:k20:impr-loss}
    \includegraphics[width=0.49\columnwidth]{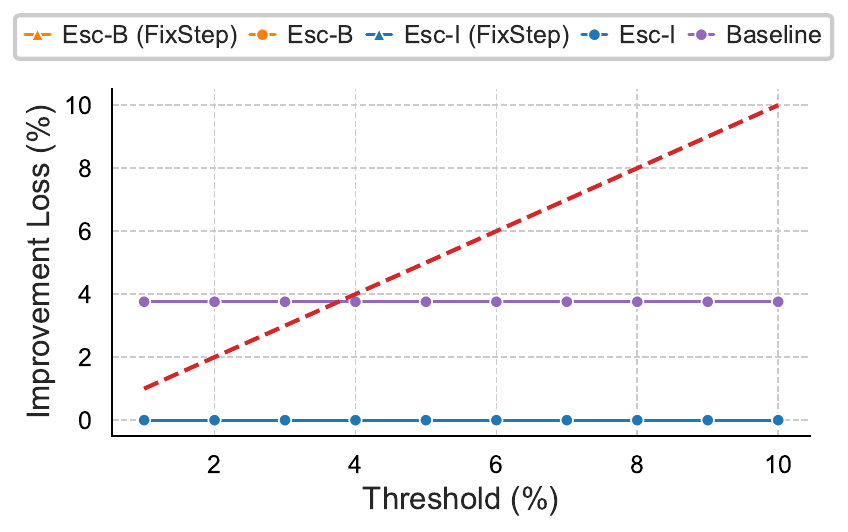}}
\subfigure[What-If Call Savings]{ \label{fig:twophase:real-m:k20:call-save}
    \includegraphics[width=0.49\columnwidth]{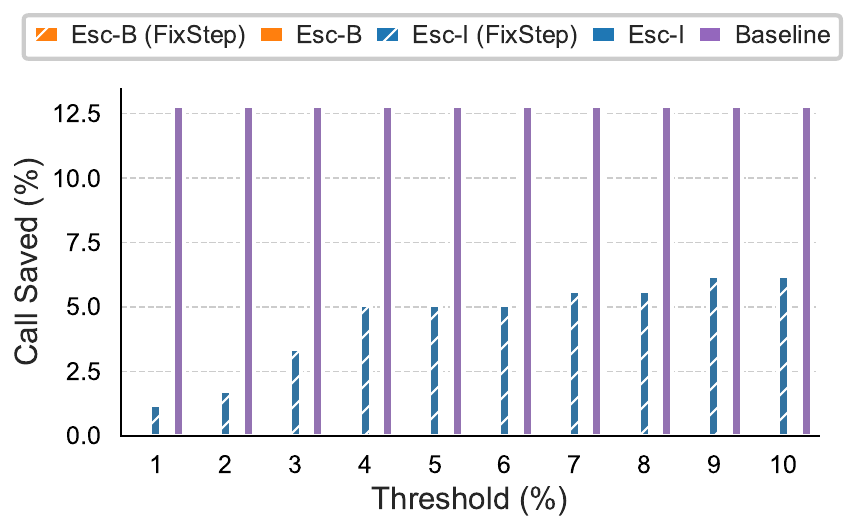}}
\subfigure[Learning Curve]{ \label{fig:twophase:real-m:k20:lc}
    \includegraphics[width=0.49\columnwidth]{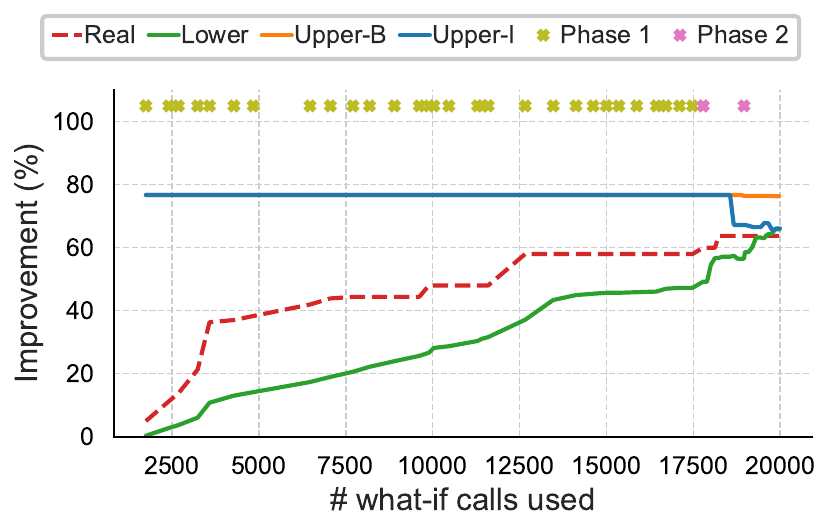}}
\vspace{-1.5em}
\caption{Two-phase greedy search, Real-M, $K=20$, $B=20k$.}
\label{fig:twophase:real-m:k20}
\vspace{-1em}
\end{figure*}



\begin{figure*}
\centering
\subfigure[Time Overhead]{ \label{fig:mcts:tpcds:k20:Time Usage}
    \includegraphics[width=0.49\columnwidth]{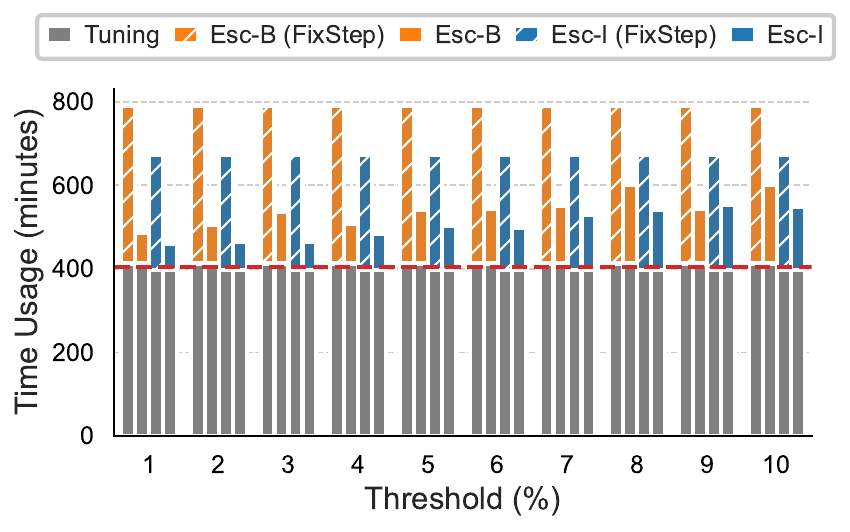}}
\subfigure[Improvement Loss]{ \label{fig:mcts:tpcds:k20:impr-loss}
    \includegraphics[width=0.49\columnwidth]{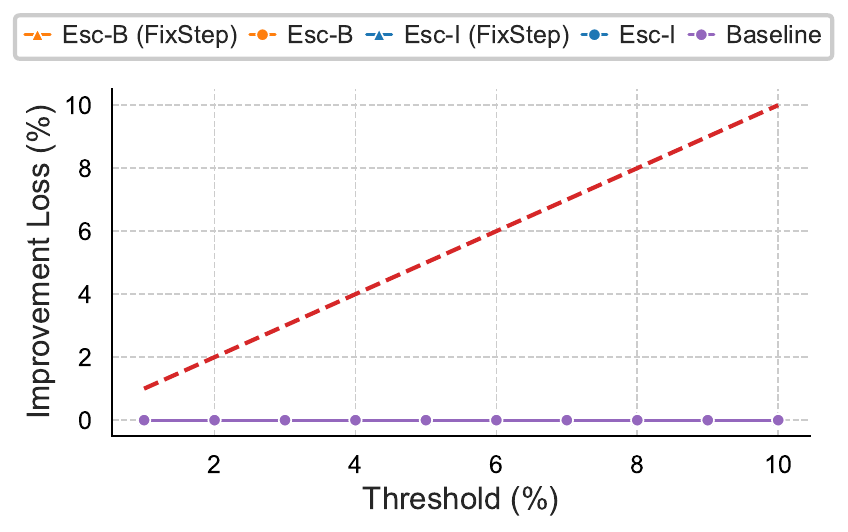}}
\subfigure[What-If Call Savings]{ \label{fig:mcts:tpcds:k20:call-save}
    \includegraphics[width=0.49\columnwidth]{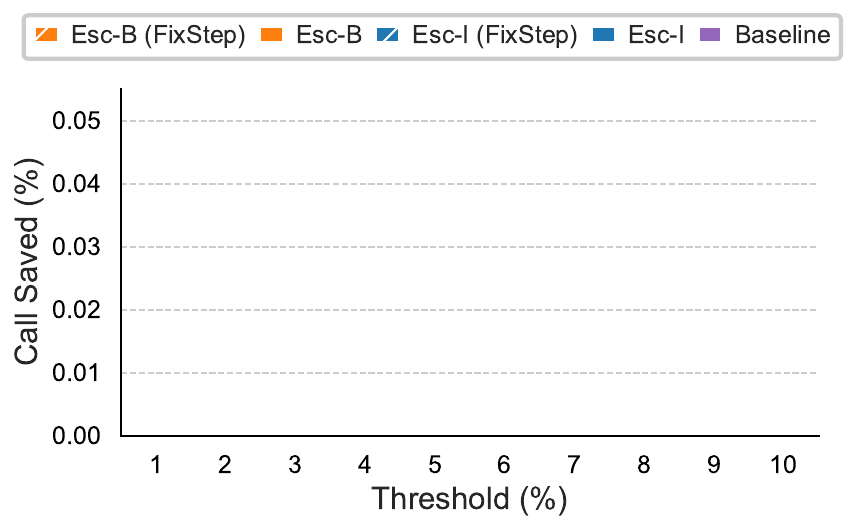}}
\subfigure[Learning Curve]{ \label{fig:mcts:tpcds:k20:lc}
    \includegraphics[width=0.49\columnwidth]{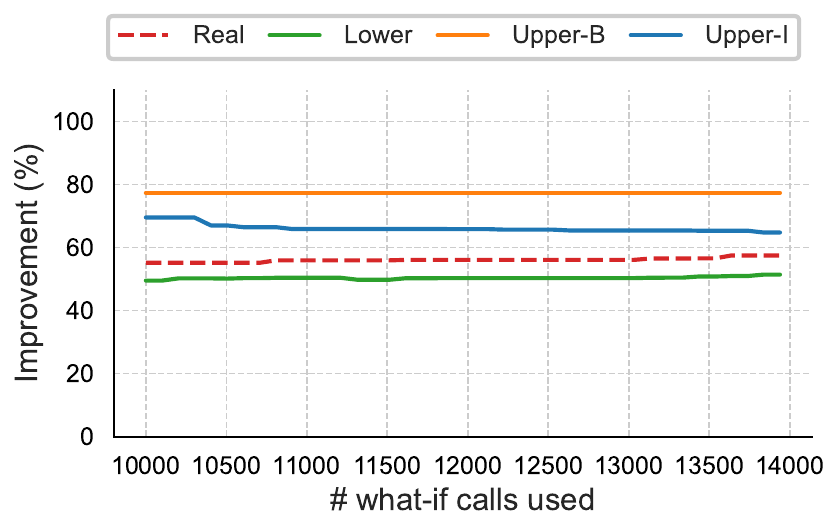}}
\vspace{-1.5em}
\caption{MCTS, TPC-DS, $K=20$, $B=20k$}
\label{fig:mcts:tpcds:k20}
\vspace{-1em}
\end{figure*}



\begin{figure*}
\centering
\subfigure[Time Overhead]{ \label{fig:mcts:job:k20:extra-time-overhead}
    \includegraphics[width=0.49\columnwidth]{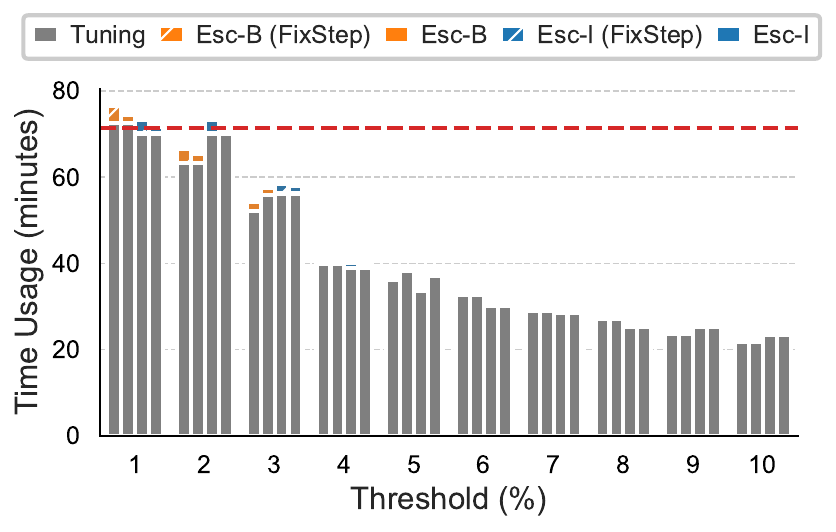}}
\subfigure[Improvement Loss]{ \label{fig:mcts:job:k20:impr-loss}
    \includegraphics[width=0.49\columnwidth]{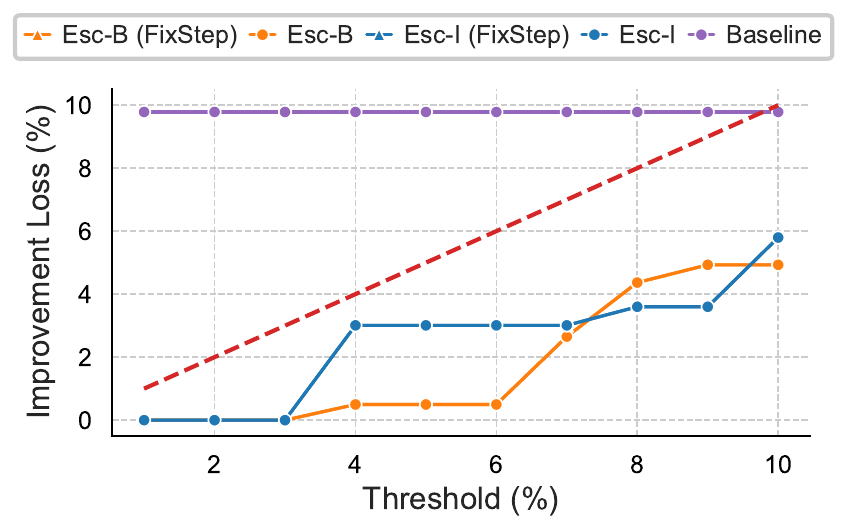}}
\subfigure[What-If Call Savings]{ \label{fig:mcts:job:k20:call-save}
    \includegraphics[width=0.49\columnwidth]{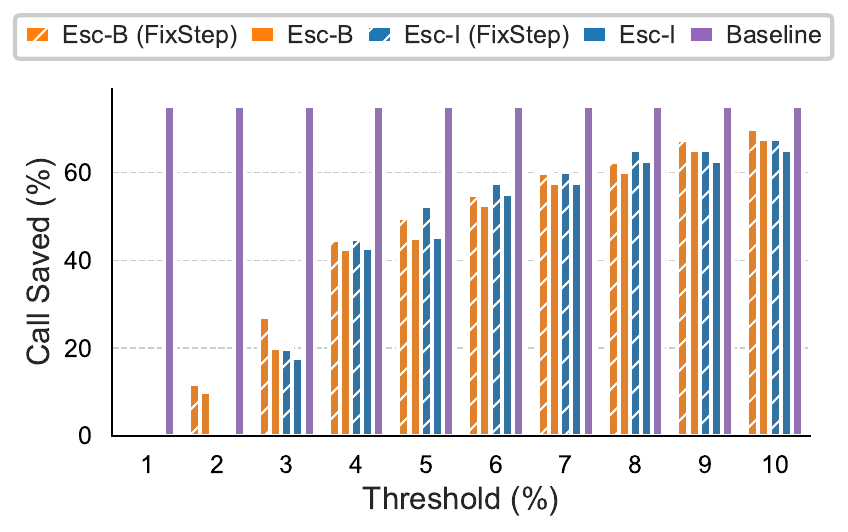}}
\subfigure[Learning Curve]{ \label{fig:mcts:job:k20:lc}
    \includegraphics[width=0.49\columnwidth]{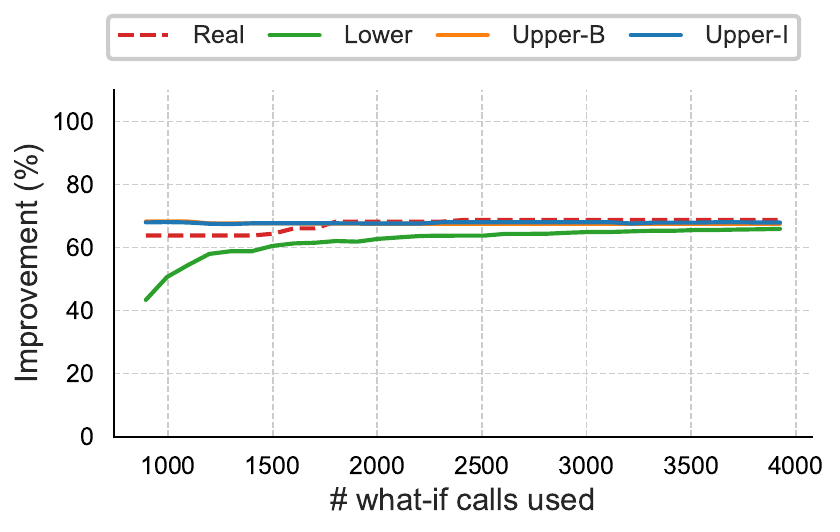}}
\vspace{-1.5em}
\caption{MCTS, JOB, $K=20$, $B=20k$.}
\label{fig:mcts:job:k20}
\vspace{-1em}
\end{figure*}



\begin{figure*}
\centering
\subfigure[Time Overhead]{ \label{fig:mcts:real-m:k20:Time Usage}
    \includegraphics[width=0.49\columnwidth]{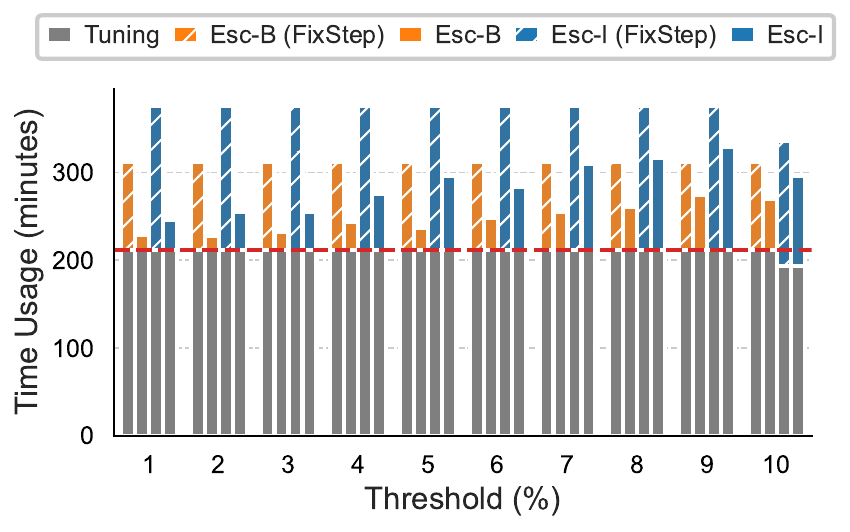}}
\subfigure[Improvement Loss]{ \label{fig:mcts:real-m:k20:impr-loss}
    \includegraphics[width=0.49\columnwidth]{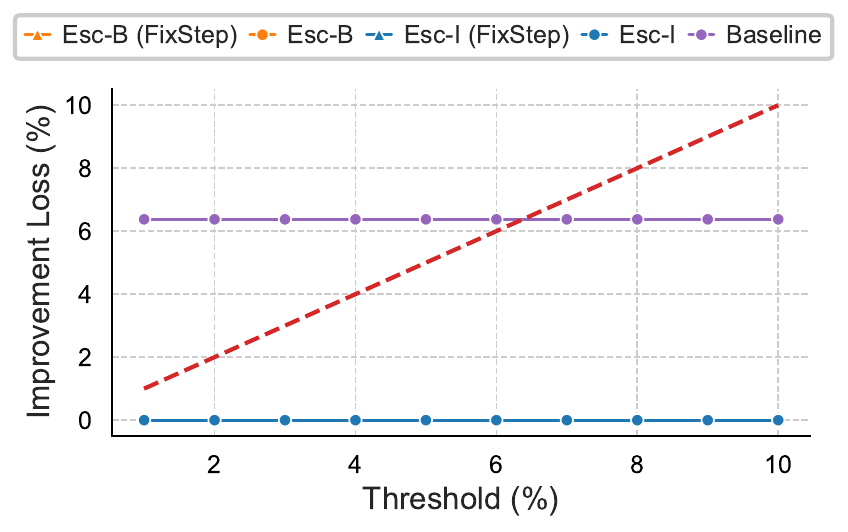}}
\subfigure[What-If Call Savings]{ \label{fig:mcts:real-m:k20:call-save}
    \includegraphics[width=0.49\columnwidth]{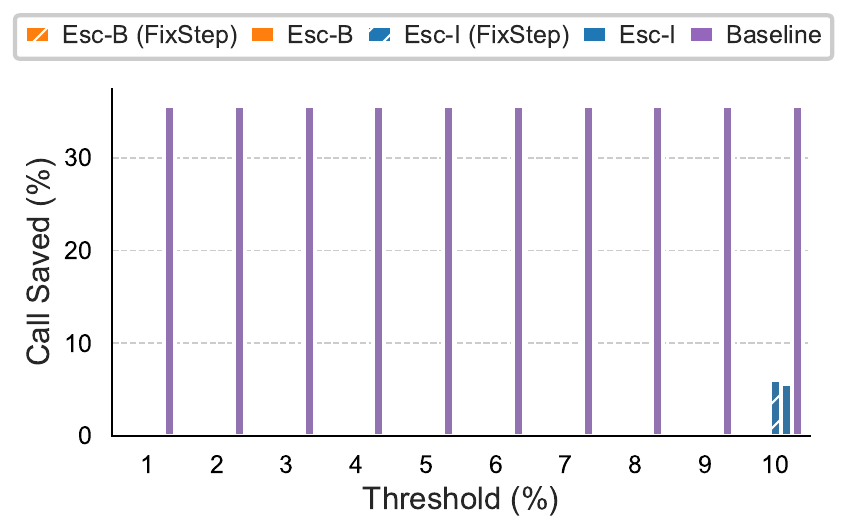}}
\subfigure[Learning Curve]{ \label{fig:mcts:real-m:k20:lc}
    \includegraphics[width=0.49\columnwidth]{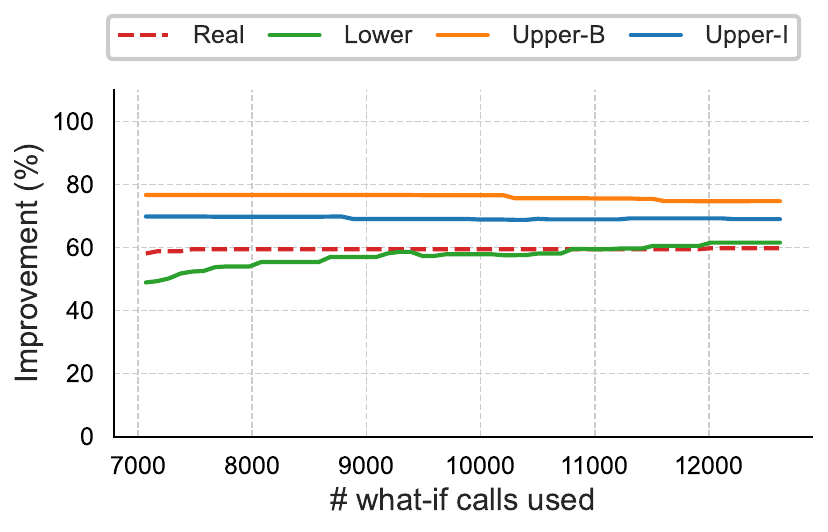}}
\vspace{-1.5em}
\caption{MCTS, Real-M, $K=20$, $B=20k$}
\label{fig:mcts:real-m:k20}
\vspace{-1em}
\end{figure*}

\section{More Evaluation Results}

\subsection{More Results with $K=20$}

Figures~\ref{fig:twophase:job:k20} and~\ref{fig:twophase:real-m:k20} presents the evaluation results of \sysname when running \emph{two-phase greedy search} on top of \textbf{JOB} and \textbf{Real-M} with $K=20$.
Figures~\ref{fig:mcts:tpcds:k20},~\ref{fig:mcts:job:k20}, and~\ref{fig:mcts:real-m:k20} present the evaluation results of \sysname when running \emph{MCTS} on top of \textbf{TPC-DS}, \textbf{JOB}, and \textbf{Real-M} with $K=20$.
In both cases of \textbf{TPC-DS} and \textbf{Real-M}, early-stopping was almost never triggered by \sysname when varying the improvement-loss threshold $\epsilon$, though it should have been triggered by observing the index tuning curves.
This indicates opportunities for further improvement of \sysname.

\vspace{-1em}
\subsection{Results with $K=20$ and Wii}

\begin{figure*}
\centering
\subfigure[Time Overhead]{ \label{fig:twophase_skip:tpch:k20:overhead}
    \includegraphics[width=0.49\columnwidth]{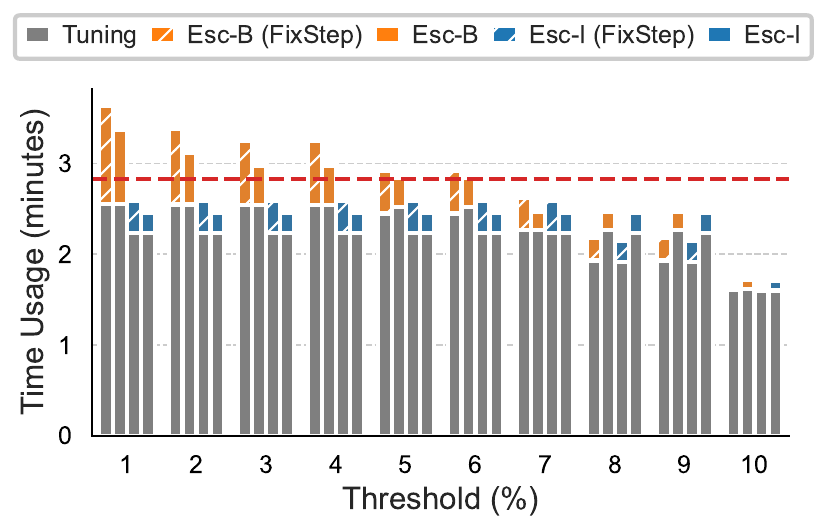}}
\subfigure[Improvement Loss]{ \label{fig:twophase_skip:tpch:k20:impr-loss}
    \includegraphics[width=0.49\columnwidth]{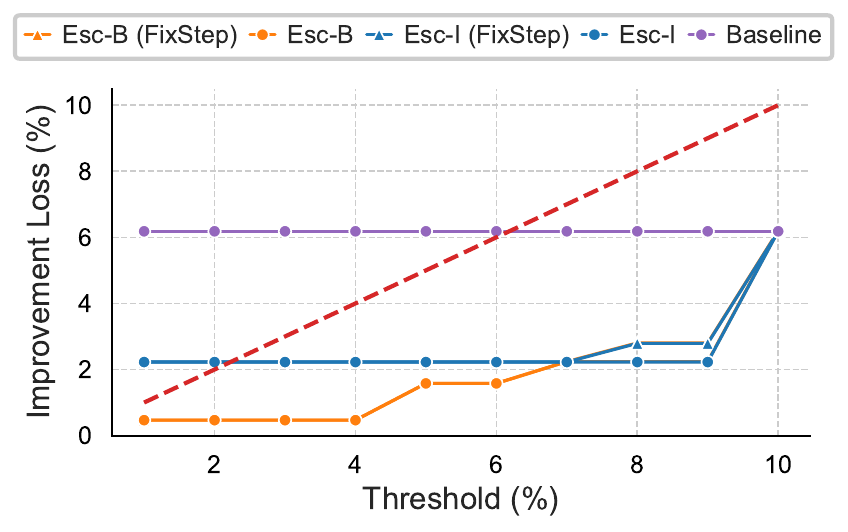}}
\subfigure[What-If Call Savings]{ \label{fig:twophase_skip:tpch:k20:call-save}
    \includegraphics[width=0.49\columnwidth]{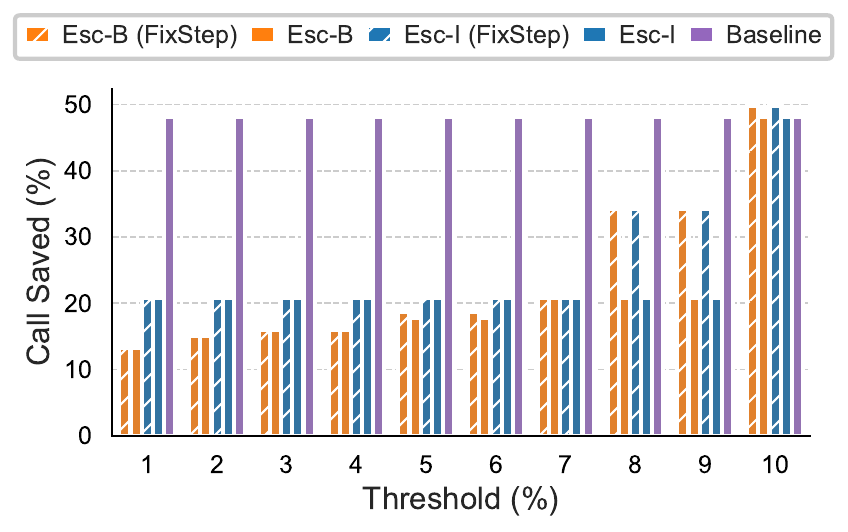}}
\subfigure[Learning Curve]{ \label{fig:twophase_skip:tpch:k20:lc}
    \includegraphics[width=0.49\columnwidth]{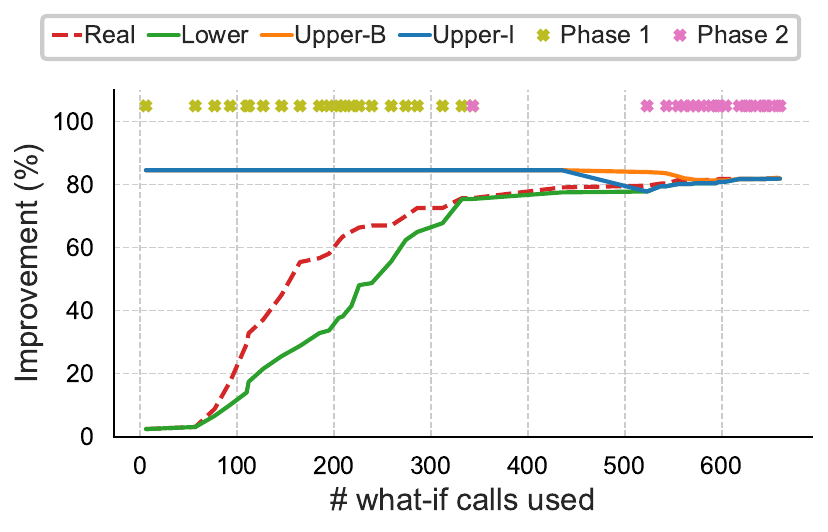}}
\vspace{-1.5em}
\caption{Two-phase greedy search (with Wii), TPC-H, $K=20$, $B=20k$}
\label{fig:twophase_skip:tpch:k20}
\vspace{-1em}
\end{figure*}

\begin{figure*}
\centering
\subfigure[Time Overhead]{ \label{fig:twophase_skip:tpcds:k20:overhead}
    \includegraphics[width=0.49\columnwidth]{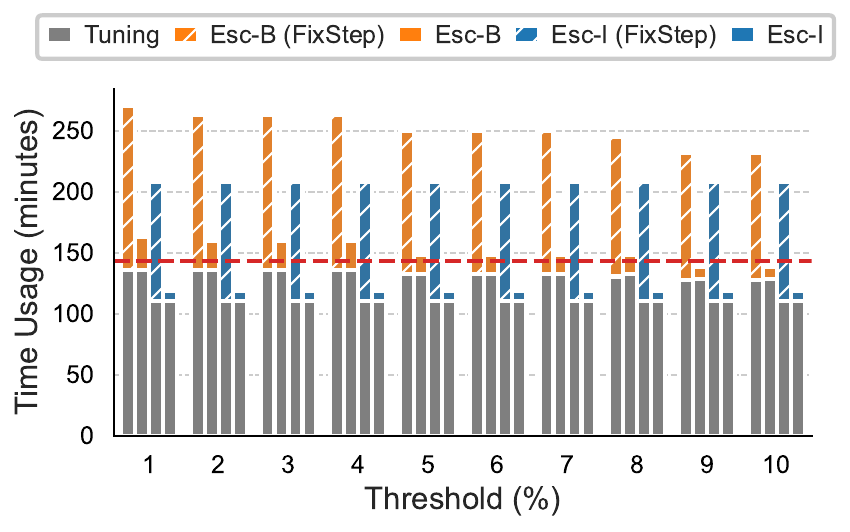}}
\subfigure[Improvement Loss]{ \label{fig:twophase_skip:tpcds:k20:impr-loss}
    \includegraphics[width=0.49\columnwidth]{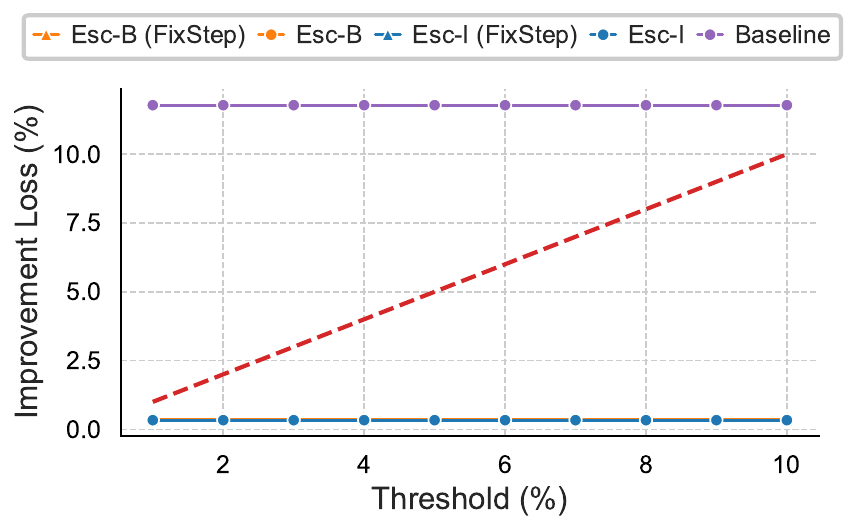}}
\subfigure[What-If Call Savings]{ \label{fig:twophase_skip:tpcds:k20:call-save}
    \includegraphics[width=0.49\columnwidth]{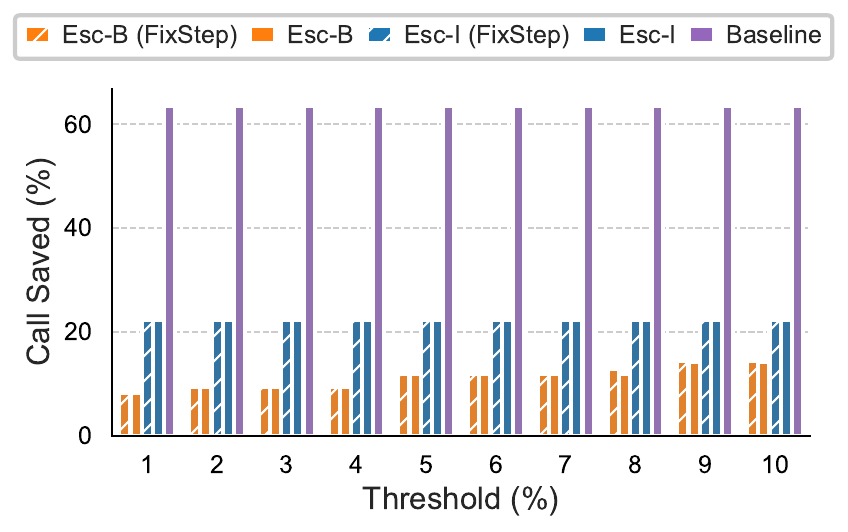}}
\subfigure[Learning Curve]{ \label{fig:twophase_skip:tpcds:k20:lc}
    \includegraphics[width=0.49\columnwidth]{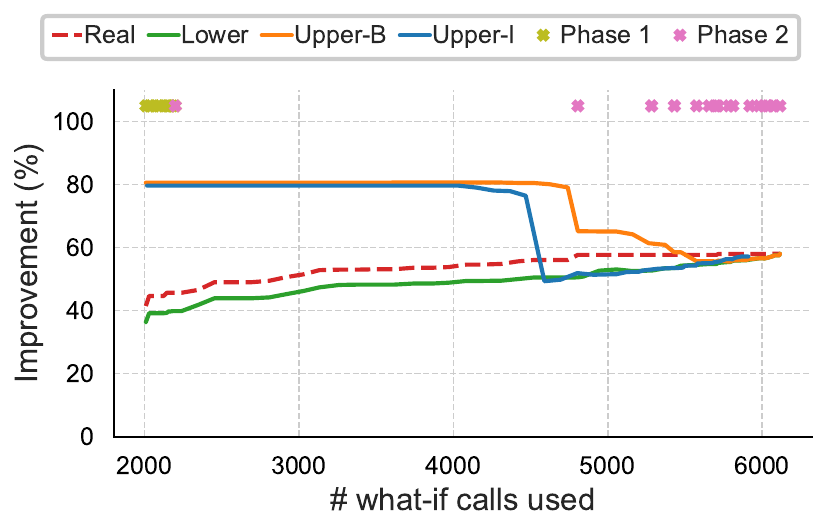}}
\vspace{-1.5em}
\caption{Two-phase greedy search (with Wii), TPC-DS, $K=20$, $B=20k$}
\label{fig:twophase_skip:tpcds:k20}
\vspace{-1em}
\end{figure*}

\begin{figure*}
\centering
\subfigure[Time Overhead]{ \label{fig:twophase_skip:job:k20:overhead}
    \includegraphics[width=0.49\columnwidth]{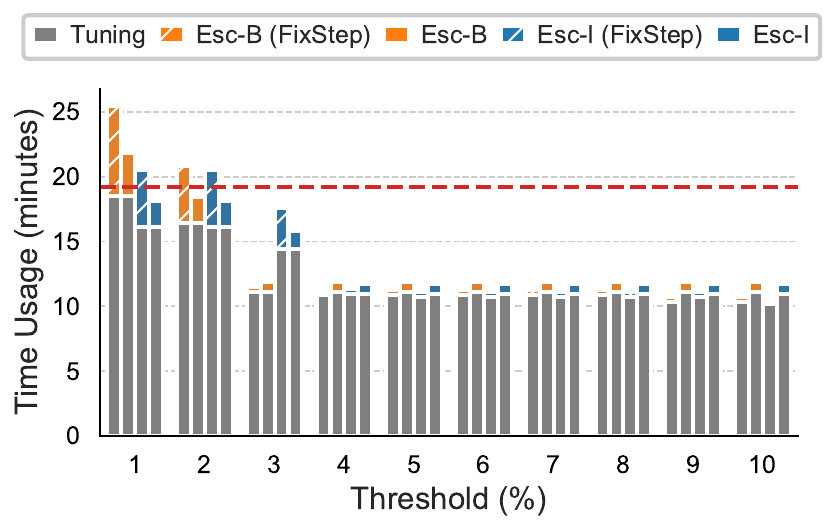}}
\subfigure[Improvement Loss]{ \label{fig:twophase_skip:job:k20:impr-loss}
    \includegraphics[width=0.49\columnwidth]{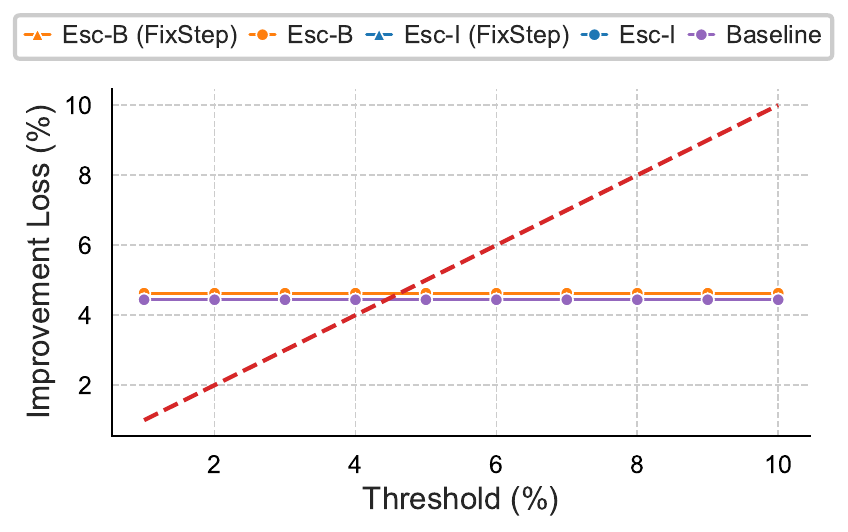}}
\subfigure[What-If Call Savings]{ \label{fig:twophase_skip:job:k20:call-save}
    \includegraphics[width=0.49\columnwidth]{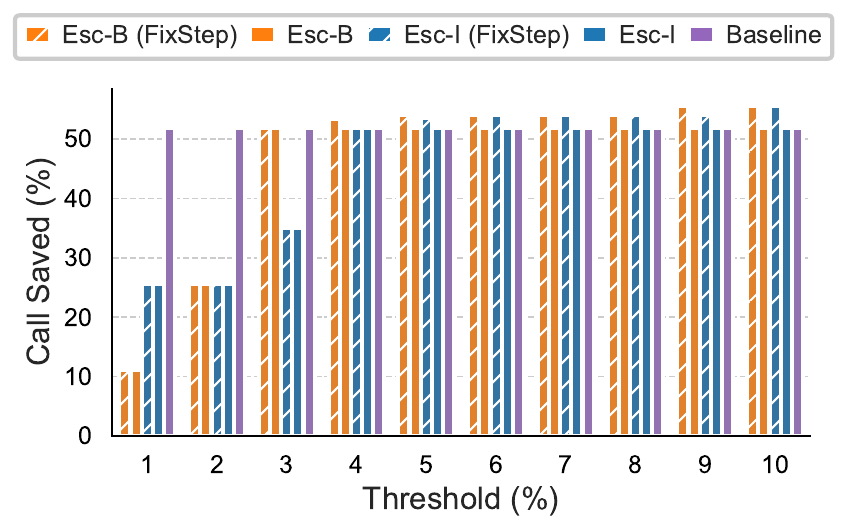}}
\subfigure[Learning Curve]{ \label{fig:twophase_skip:job:k20:lc}
    \includegraphics[width=0.49\columnwidth]{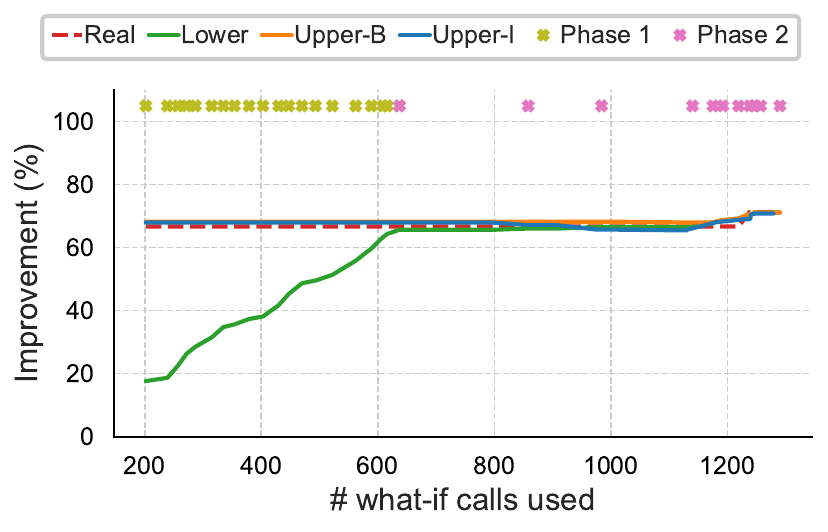}}
\vspace{-1.5em}
\caption{Two-phase greedy search (with Wii), JOB, $K=20$, $B=20k$}
\label{fig:twophase_skip:job:k20}
\vspace{-1em}
\end{figure*}

\begin{figure*}
\centering
\subfigure[Time Overhead]{ \label{fig:twophase_skip:real-d:k20:overhead}
    \includegraphics[width=0.49\columnwidth]{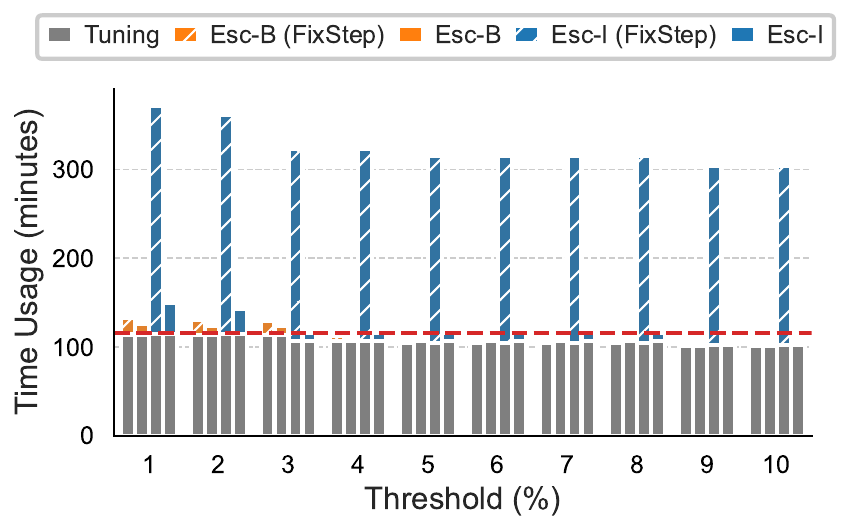}}
\subfigure[Improvement Loss]{ \label{fig:twophase_skip:real-d:k20:impr-loss}
    \includegraphics[width=0.49\columnwidth]{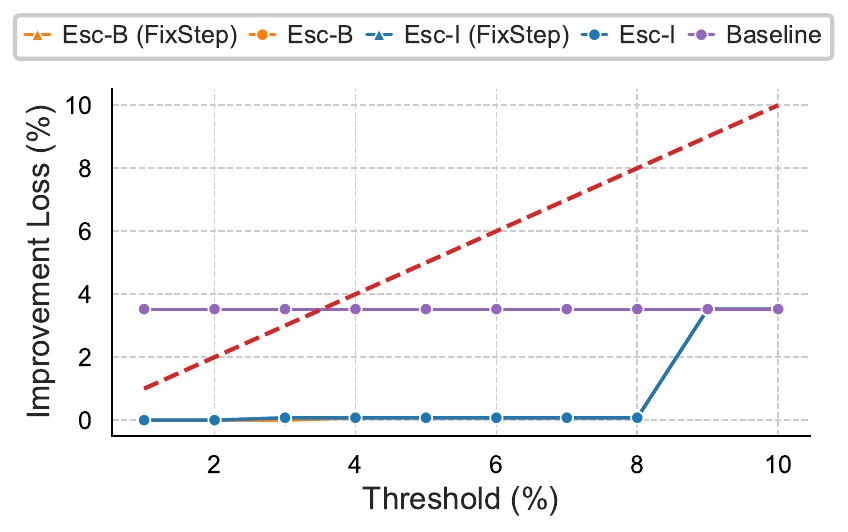}}
\subfigure[What-If Call Savings]{ \label{fig:twophase_skip:real-d:k20:call-save}
    \includegraphics[width=0.49\columnwidth]{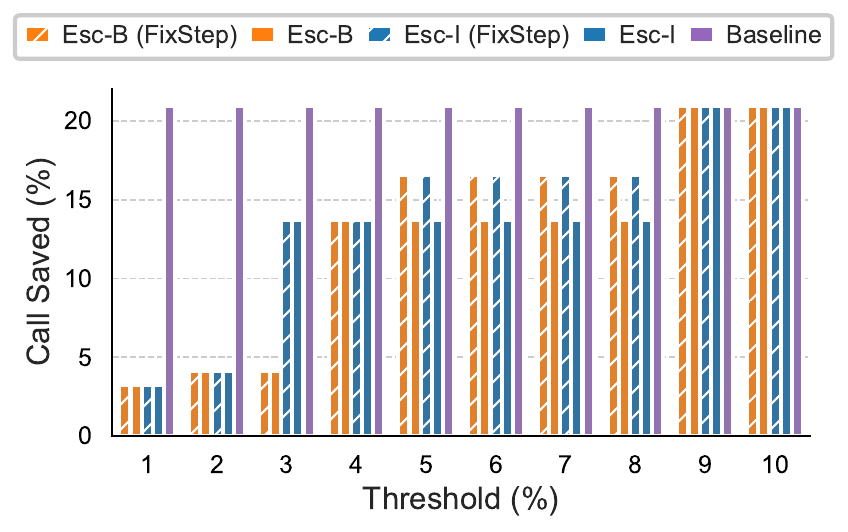}}
\subfigure[Learning Curve]{ \label{fig:twophase_skip:real-d:k20:lc}
    \includegraphics[width=0.49\columnwidth]{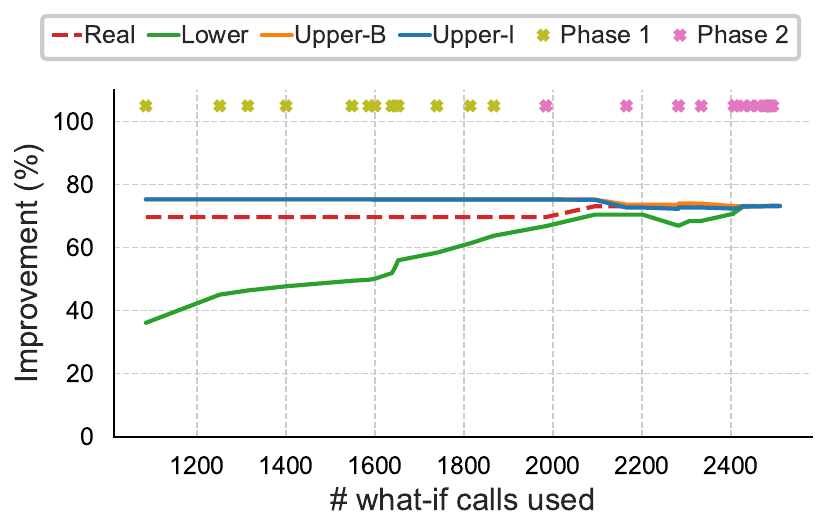}}
\vspace{-1.5em}
\caption{Two-phase greedy search (with Wii), Real-D, $K=20$, $B=20k$}
\label{fig:twophase_skip:real-d:k20}
\vspace{-1em}
\end{figure*}

\begin{figure*}
\centering
\subfigure[Time Overhead]{ \label{fig:twophase_skip:real-m:k20:extra-time-overhead}
    \includegraphics[width=0.49\columnwidth]{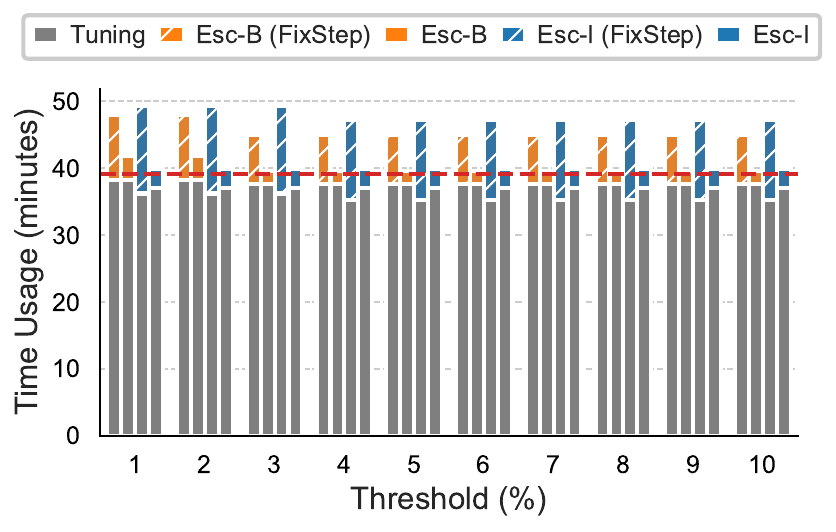}}
\subfigure[Improvement Loss]{ \label{fig:twophase_skip:real-m:k20:impr-loss}
    \includegraphics[width=0.49\columnwidth]{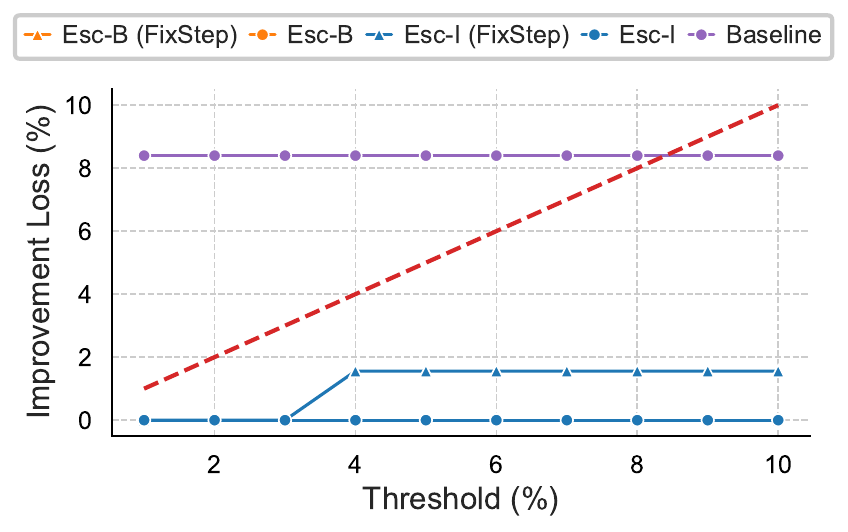}}
\subfigure[What-If Call Savings]{ \label{fig:twophase_skip:real-m:k20:call-save}
    \includegraphics[width=0.49\columnwidth]{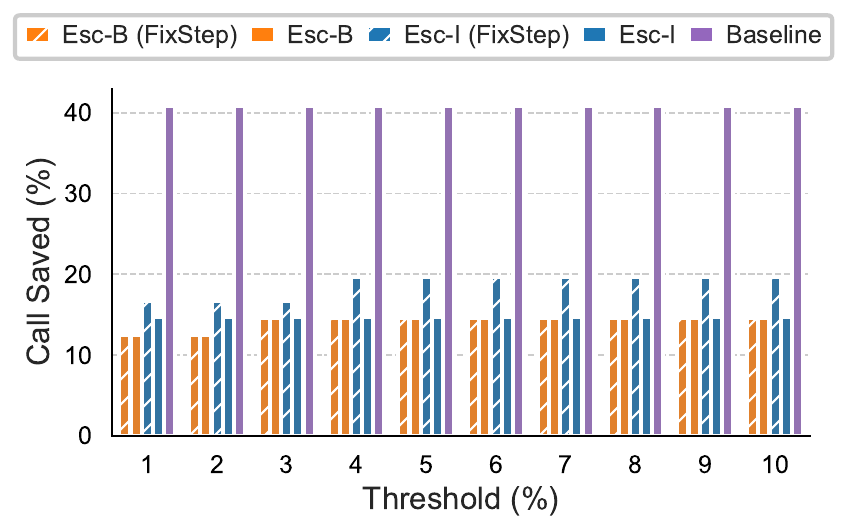}}
\subfigure[Learning Curve]{ \label{fig:twophase_skip:real-m:k20:lc}
    \includegraphics[width=0.49\columnwidth]{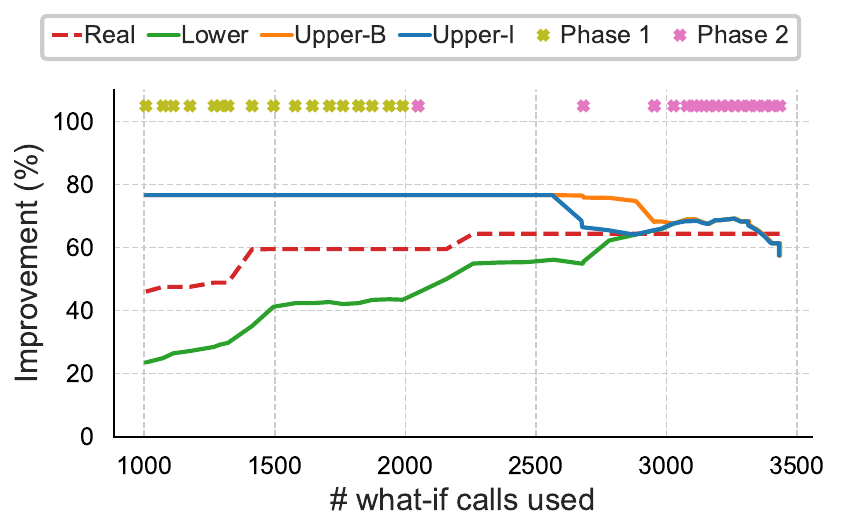}}
\vspace{-1.5em}
\caption{Two-phase greedy search (with Wii), Real-M, $K=20$, $B=20k$.}
\label{fig:twophase_skip:real-m:k20}
\vspace{-1em}
\end{figure*}

\begin{figure*}
\centering
\subfigure[Time Overhead]{ \label{fig:mcts_skip:tpch:k20:overhead}
    \includegraphics[width=0.49\columnwidth]{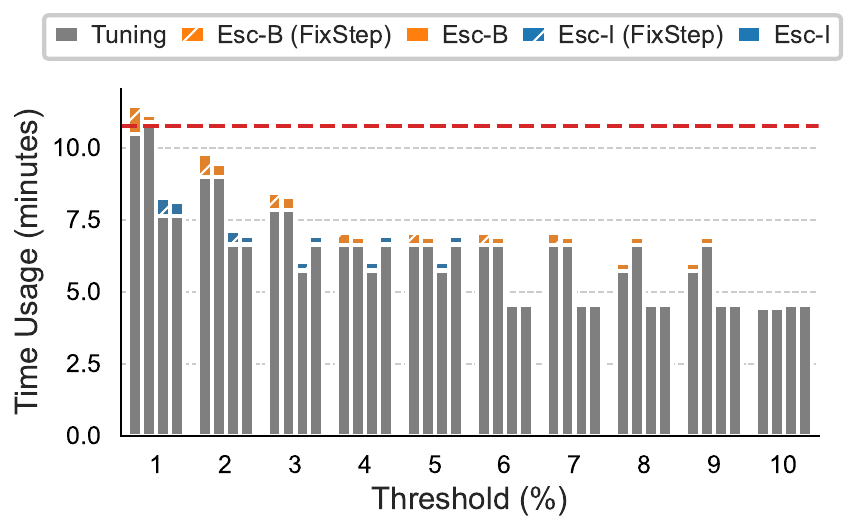}}
\subfigure[Improvement Loss]{ \label{fig:mcts_skip:tpch:k20:impr-loss}
    \includegraphics[width=0.49\columnwidth]{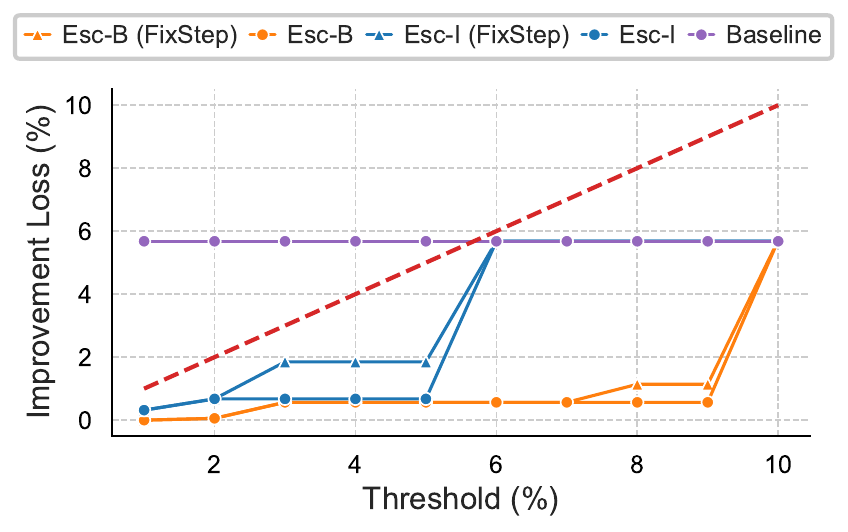}}
\subfigure[What-If Call Savings]{ \label{fig:mcts_skip:tpch:k20:call-save}
    \includegraphics[width=0.49\columnwidth]{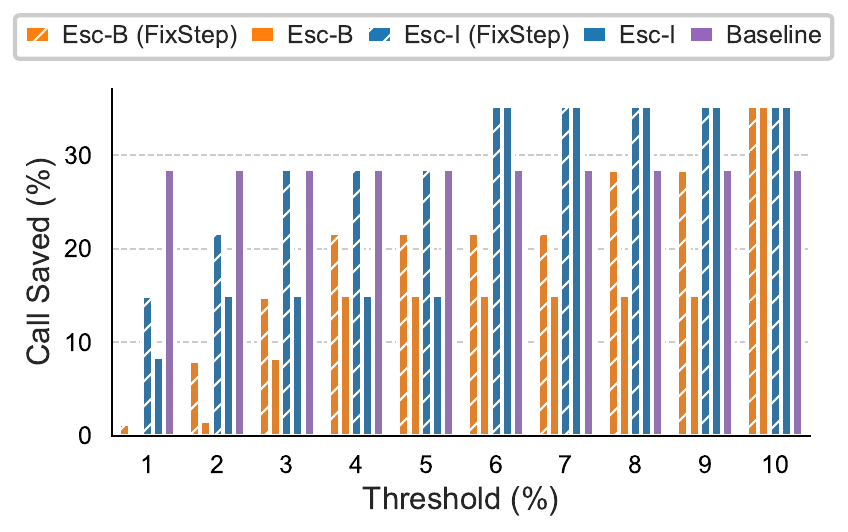}}
\subfigure[Learning Curve]{ \label{fig:mcts_skip:tpch:k20:lc}
    \includegraphics[width=0.49\columnwidth]{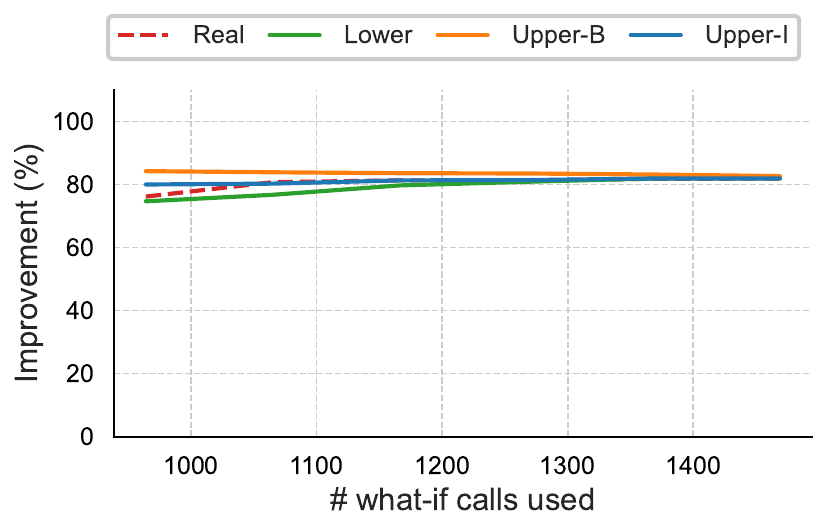}}
\vspace{-1.5em}
\caption{MCTS (with Wii), TPC-H, $K=20$, $B=20k$}
\label{fig:mcts_skip:tpch:k20}
\vspace{-1em}
\end{figure*}

\begin{figure*}
\centering
\subfigure[Time Overhead]{ \label{fig:mcts_skip:tpcds:k20:overhead}
    \includegraphics[width=0.49\columnwidth]{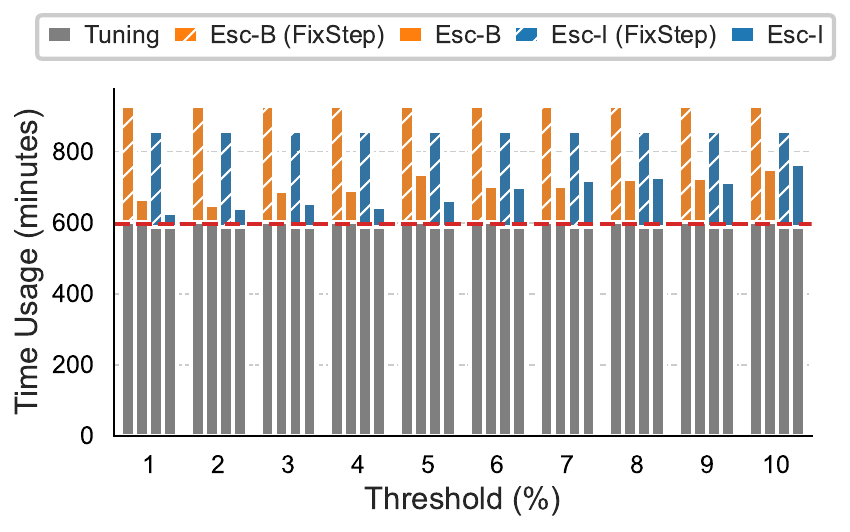}}
\subfigure[Improvement Loss]{ \label{fig:mcts_skip:tpcds:k20:impr-loss}
    \includegraphics[width=0.49\columnwidth]{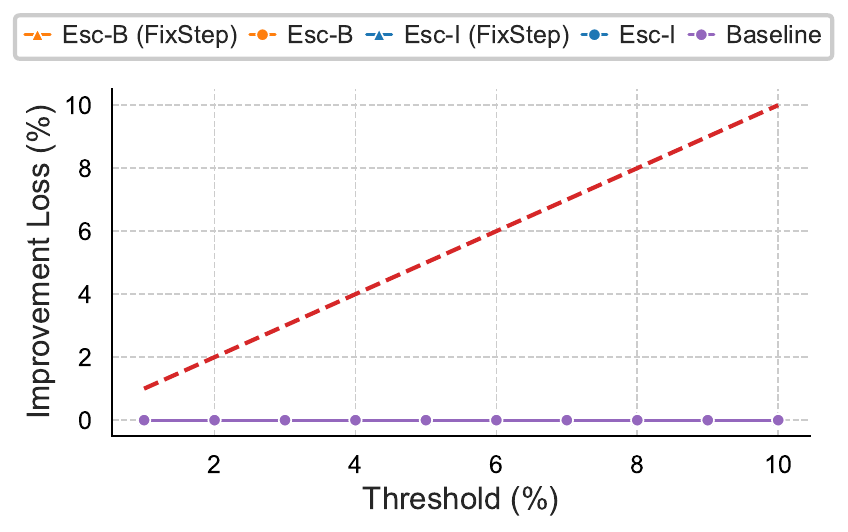}}
\subfigure[What-If Call Savings]{ \label{fig:mcts_skip:tpcds:k20:call-save}
    \includegraphics[width=0.49\columnwidth]{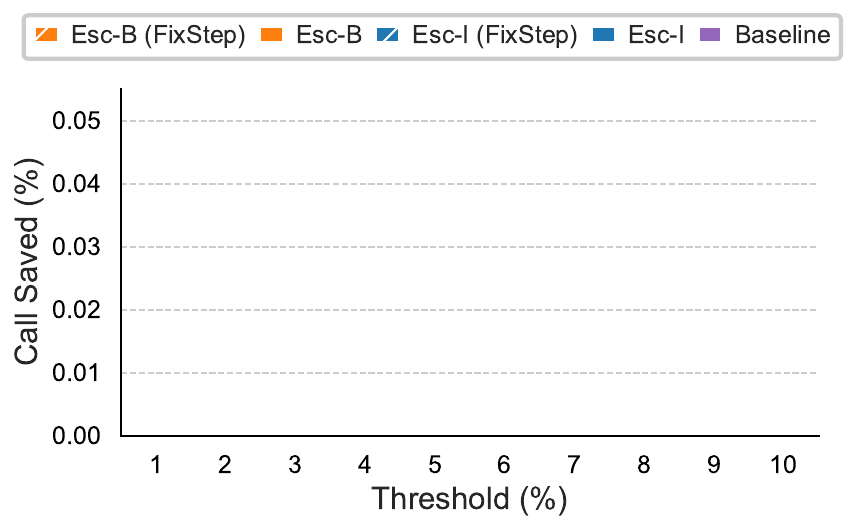}}
\subfigure[Learning Curve]{ \label{fig:mcts_skip:tpcds:k20:lc}
    \includegraphics[width=0.49\columnwidth]{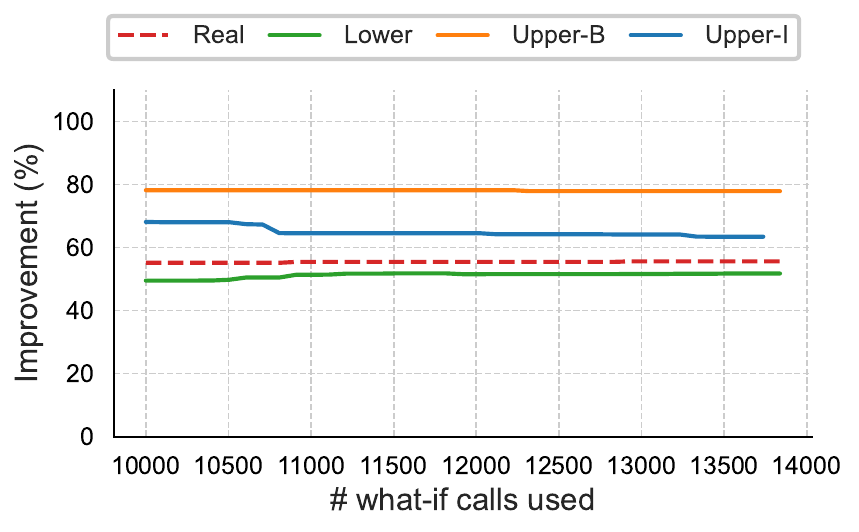}}
\vspace{-1.5em}
\caption{MCTS (with Wii), TPC-DS, $K=20$, $B=20k$}
\label{fig:mcts_skip:tpcds:k20}
\vspace{-1em}
\end{figure*}

\begin{figure*}
\centering
\subfigure[Time Overhead]{ \label{fig:mcts_skip:job:k20:overhead}
    \includegraphics[width=0.49\columnwidth]{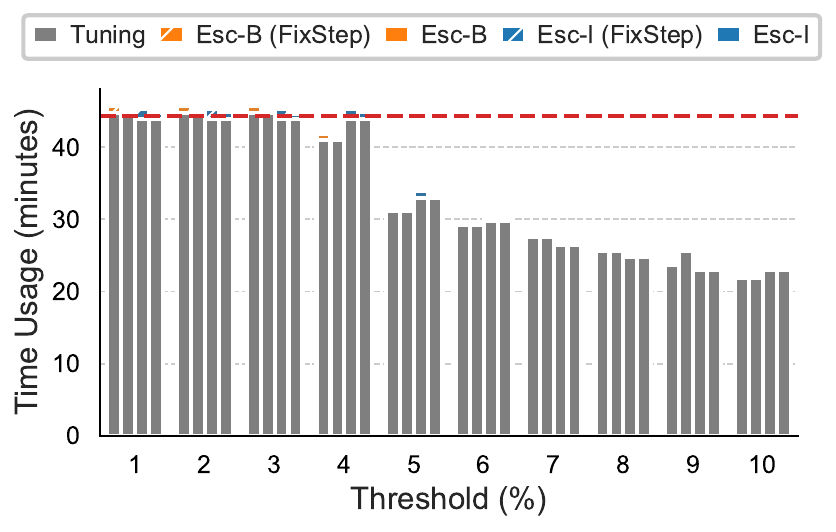}}
\subfigure[Improvement Loss]{ \label{fig:mcts_skip:job:k20:impr-loss}
    \includegraphics[width=0.49\columnwidth]{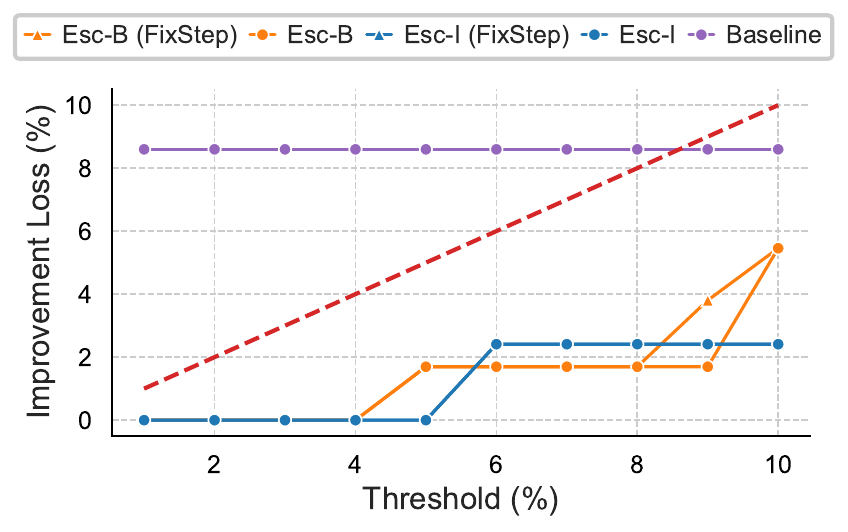}}
\subfigure[What-If Call Savings]{ \label{fig:mcts_skip:job:k20:call-save}
    \includegraphics[width=0.49\columnwidth]{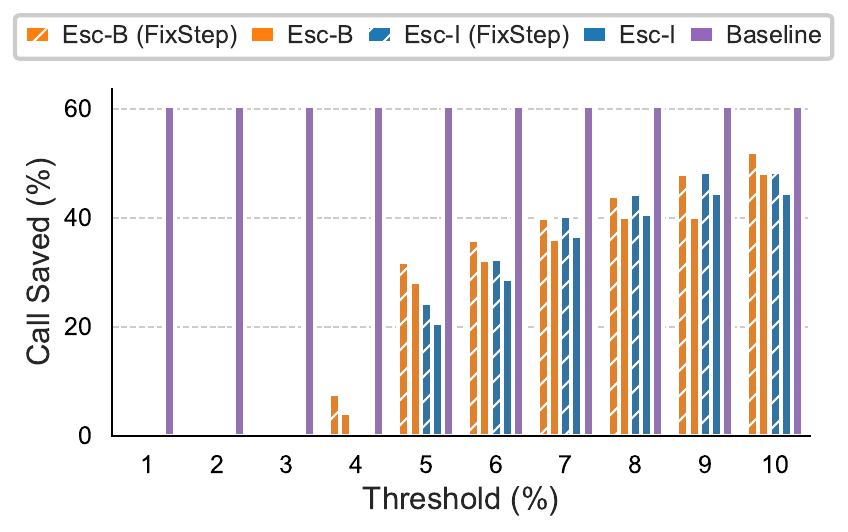}}
\subfigure[Learning Curve]{ \label{fig:mcts_skip:job:k20:lc}
    \includegraphics[width=0.49\columnwidth]{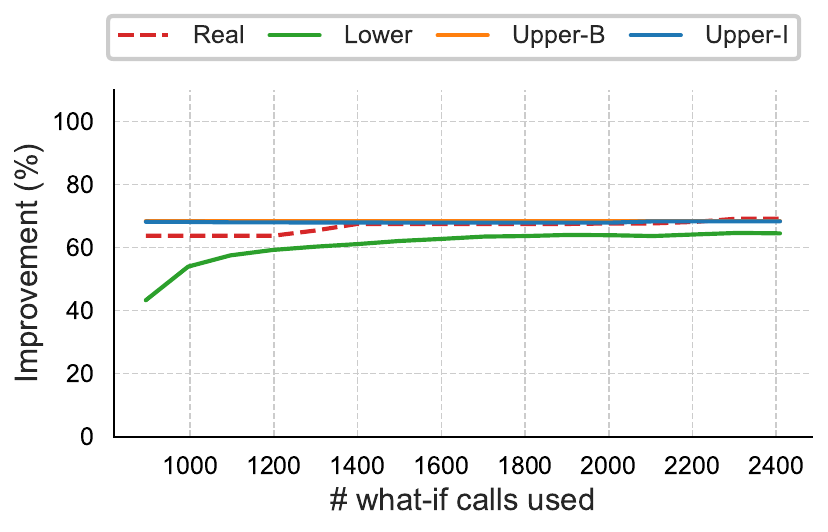}}
\vspace{-1.5em}
\caption{MCTS (with Wii), JOB, $K=20$, $B=20k$}
\label{fig:mcts_skip:job:k20}
\vspace{-1em}
\end{figure*}

\begin{figure*}
\centering
\subfigure[Time Overhead]{ \label{fig:mcts_skip:real-d:k20:overhead}
    \includegraphics[width=0.49\columnwidth]{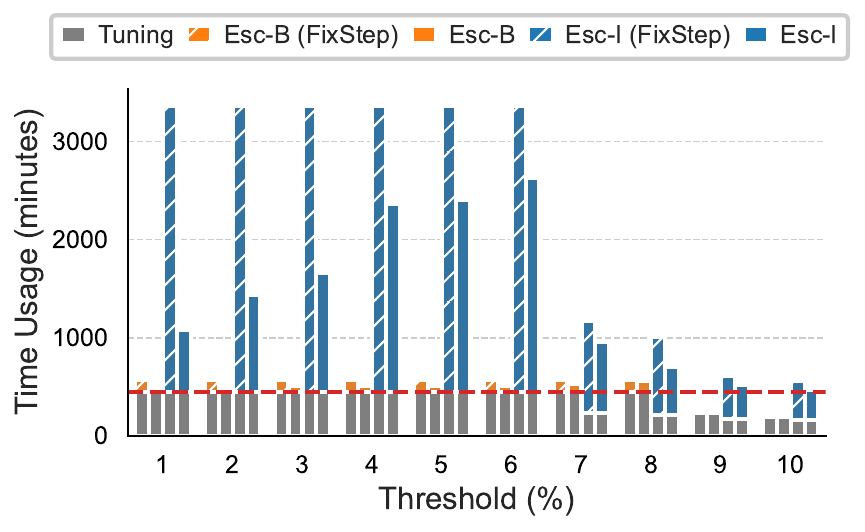}}
\subfigure[Improvement Loss]{ \label{fig:mcts_skip:real-d:k20:impr-loss}
    \includegraphics[width=0.49\columnwidth]{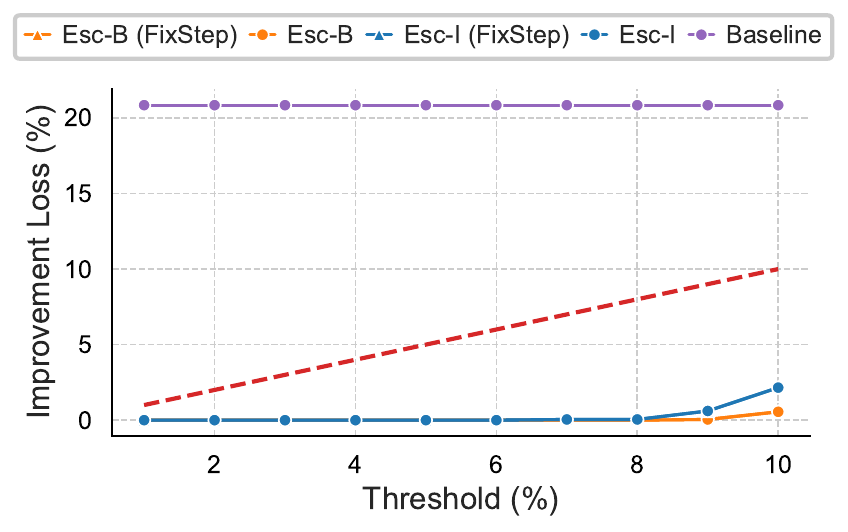}}
\subfigure[What-If Call Savings]{ \label{fig:mcts_skip:real-d:k20:call-save}
    \includegraphics[width=0.49\columnwidth]{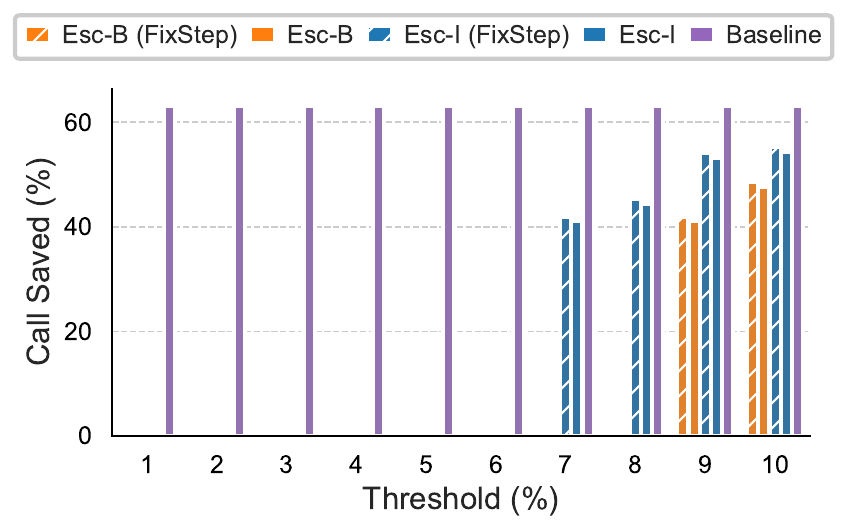}}
\subfigure[Learning Curve]{ \label{fig:mcts_skip:real-d:k20:lc}
    \includegraphics[width=0.49\columnwidth]{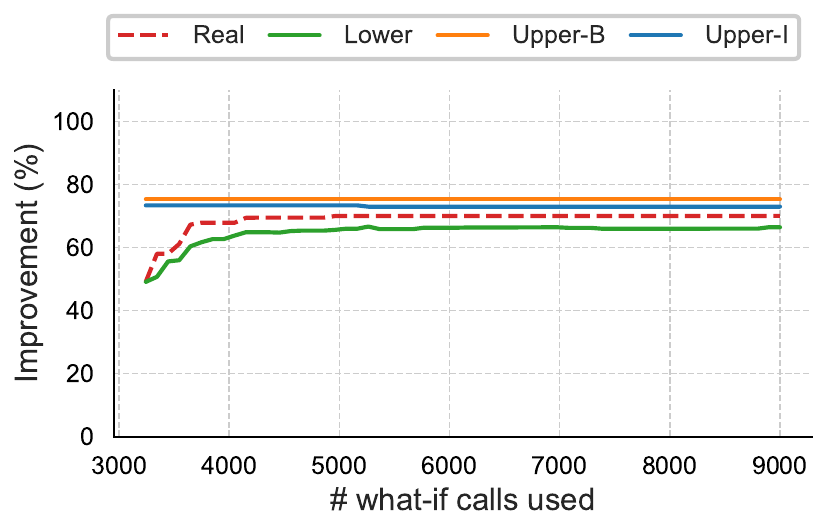}}
\vspace{-1.5em}
\caption{MCTS (with Wii), Real-D, $K=20$, $B=20k$}
\label{fig:mcts_skip:real-d:k20}
\vspace{-1em}
\end{figure*}

Figures~\ref{fig:twophase_skip:tpch:k20} to~\ref{fig:twophase_skip:real-m:k20} present evaluation results of running Wii-enhanced version of \emph{two-phase greedy search} on the workloads \textbf{TPC-H}, \textbf{TPC-DS}, \textbf{JOB}, \textbf{Real-D}, and \textbf{Real-M}.
Correspondingly, Figures~\ref{fig:mcts_skip:tpch:k20} to~\ref{fig:mcts_skip:real-m:k20} present evaluation results of running Wii-enhanced version of \emph{MCTS} on these workloads. 
We set $K=20$ in both evaluations.

Compared to the corresponding results from running the original version of \emph{two-phase greedy search}, 
Wii significantly improves the ``convergence'' of \emph{two-phase greedy search}, i.e., the number of what-if calls used when \emph{two-phase greedy search} terminates.
For example, while \emph{two-phase greedy search} used up all 20,000 what-if calls on \textbf{TPC-DS} (as Figure~\ref{fig:twophase:tpcds:k20:lc} shows), Wii brings the number of what-if calls down to around 6,000 (as Figure~\ref{fig:twophase_skip:tpcds:k20:lc} shows).
Moreover, with negligible Time Usage of using \textbf{\sysname-I} (ref. Figure~\ref{fig:twophase_skip:tpcds:k20:overhead}) and zero improvement loss (ref. Figure~\ref{fig:twophase_skip:tpcds:k20:impr-loss}), \sysname can further save around 20\% more what-if calls (ref. Figure~\ref{fig:twophase_skip:tpcds:k20:call-save}).

For \emph{MCTS}, compared to the corresponding results from running the original version, the impact of Wii on the number of what-if calls used is not significant.
However, we still observe some remarkable savings on what-if calls for certain workloads.
For instance, on \textbf{TPC-H}, Wii reduces the number of what-if calls from 3,500 (ref. Figure~\ref{fig:mcts:tpch:k20:lc}) to 1,500 (ref. Figure~\ref{fig:mcts_skip:tpch:k20:lc}), whereas using \textbf{\sysname-I} further brings in 30\% savings on what-if calls (ref. Figure~\ref{fig:mcts_skip:tpch:k20:call-save}) with almost no extra time overhead (ref. Figure~\ref{fig:mcts_skip:tpch:k20:overhead}) when setting the improvement-loss threshold $\epsilon=6\%$.

\vspace{-1em}
\subsection{Results with $K=20$ and Wii-Coverage}

\begin{figure*}
\centering
\subfigure[Time Overhead]{ \label{fig:twophase_covskip:tpch:k20:overhead}
    \includegraphics[width=0.49\columnwidth]{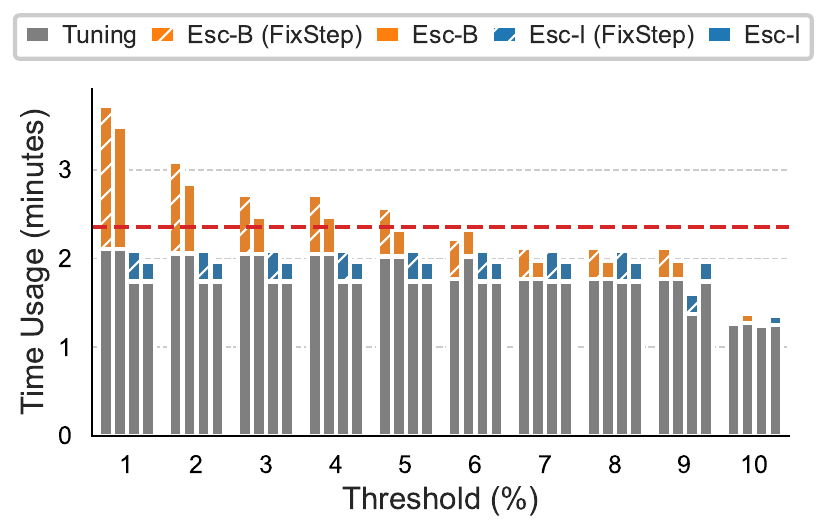}}
\subfigure[Improvement Loss]{ \label{fig:twophase_covskip:tpch:k20:impr-loss}
    \includegraphics[width=0.49\columnwidth]{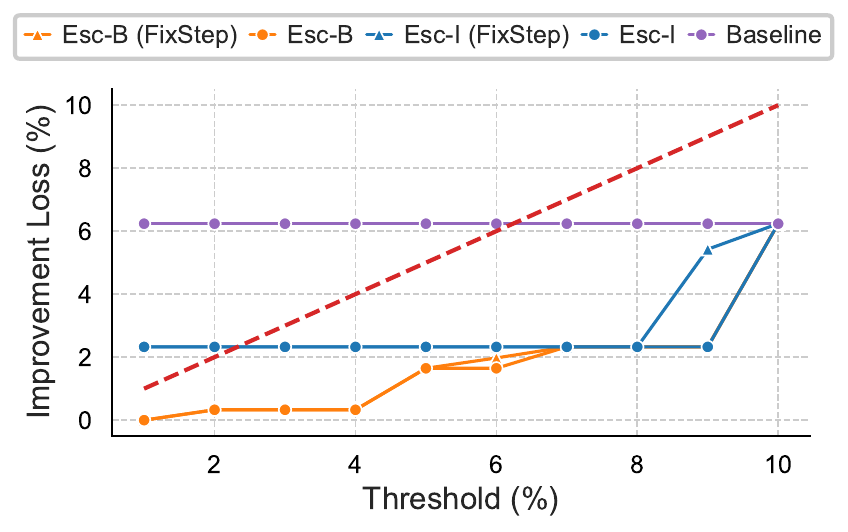}}
\subfigure[What-If Call Savings]{ \label{fig:twophase_covskip:tpch:k20:call-save}
    \includegraphics[width=0.49\columnwidth]{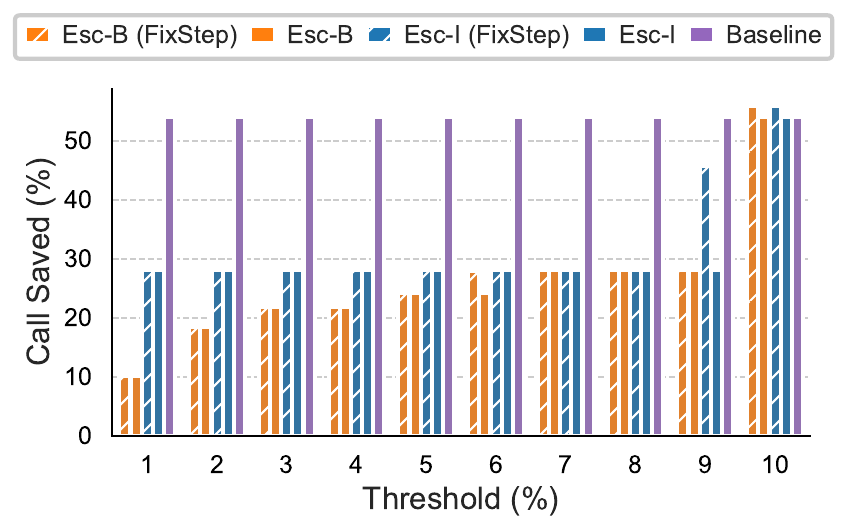}}
\subfigure[Learning Curve]{ \label{fig:twophase_covskip:tpch:k20:lc}
    \includegraphics[width=0.49\columnwidth]{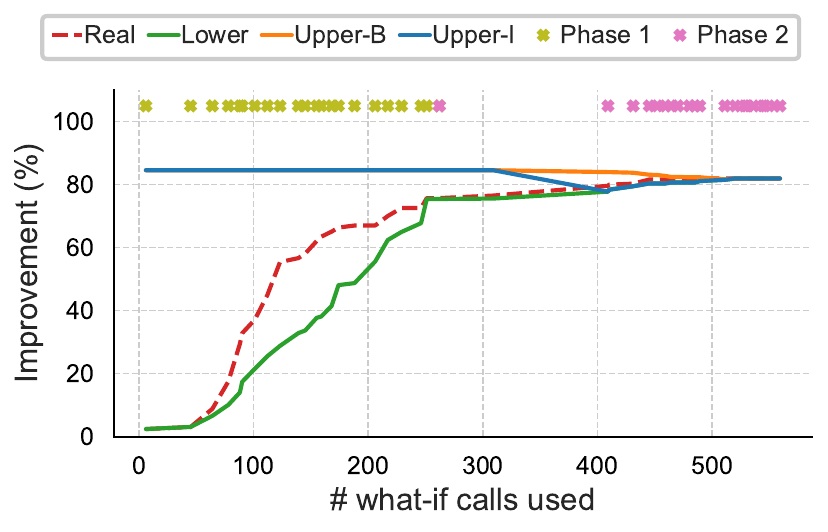}}
\vspace{-1.5em}
\caption{Two-phase greedy search (with Wii-Coverage), TPC-H, $K=20$, $B=20k$}
\label{fig:twophase_covskip:tpch:k20}
\vspace{-1em}
\end{figure*}

\begin{figure*}
\centering
\subfigure[Time Overhead]{ \label{fig:twophase_covskip:tpcds:k20:overhead}
    \includegraphics[width=0.49\columnwidth]{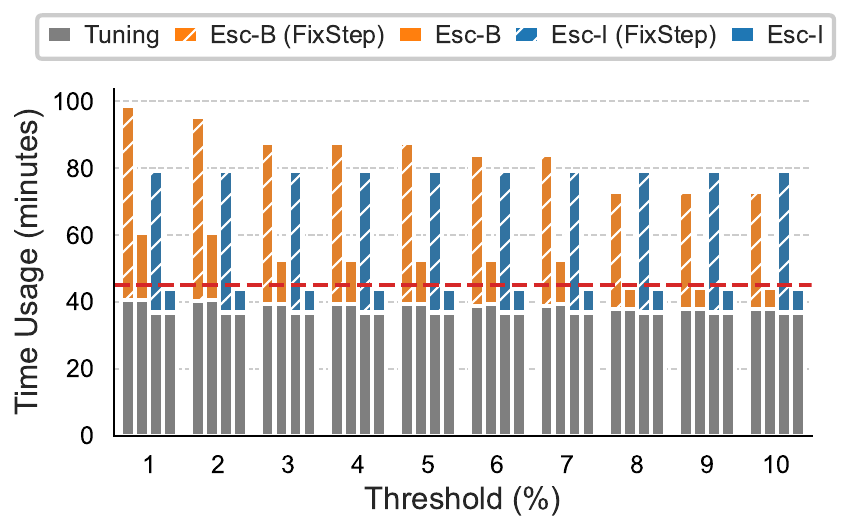}}
\subfigure[Improvement Loss]{ \label{fig:twophase_covskip:tpcds:k20:impr-loss}
    \includegraphics[width=0.49\columnwidth]{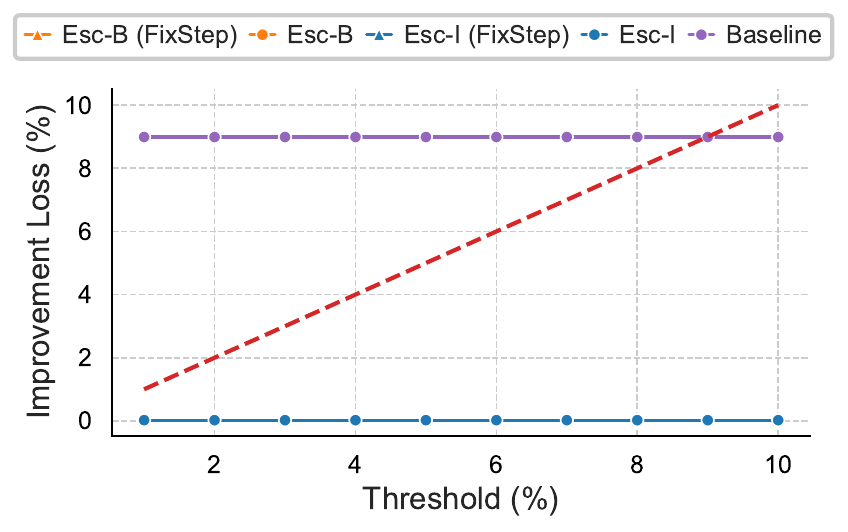}}
\subfigure[What-If Call Savings]{ \label{fig:twophase_covskip:tpcds:k20:call-save}
    \includegraphics[width=0.49\columnwidth]{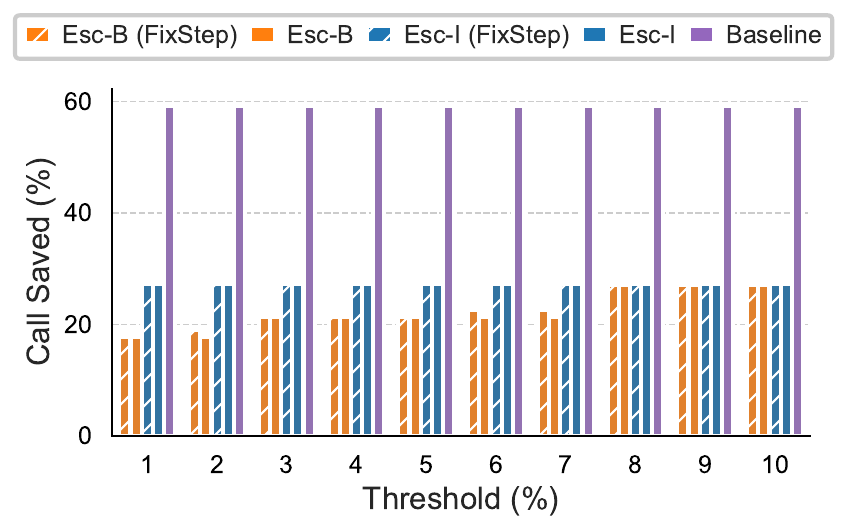}}
\subfigure[Learning Curve]{ \label{fig:twophase_covskip:tpcds:k20:lc}
    \includegraphics[width=0.49\columnwidth]{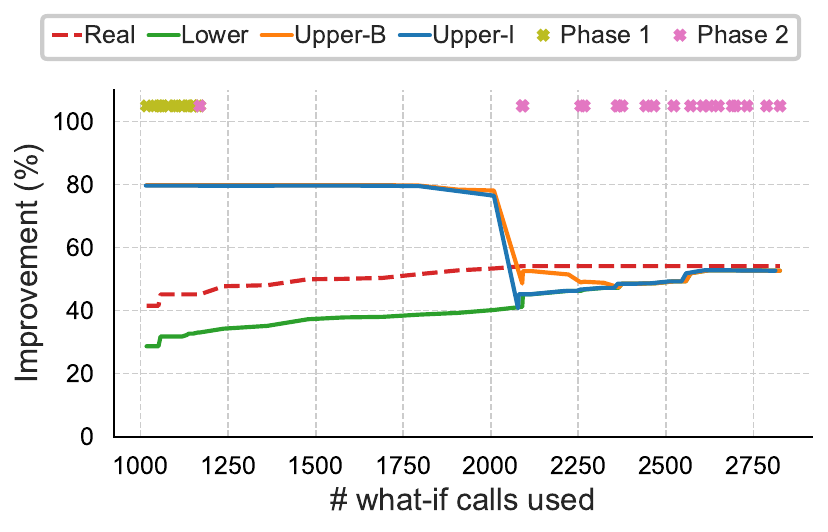}}
\vspace{-1.5em}
\caption{Two-phase greedy search (with Wii-Coverage), TPC-DS, $K=20$, $B=20k$}
\label{fig:twophase_covskip:tpcds:k20}
\vspace{-1em}
\end{figure*}

\begin{figure*}
\centering
\subfigure[Time Overhead]{ \label{fig:twophase_covskip:job:k20:overhead}
    \includegraphics[width=0.49\columnwidth]{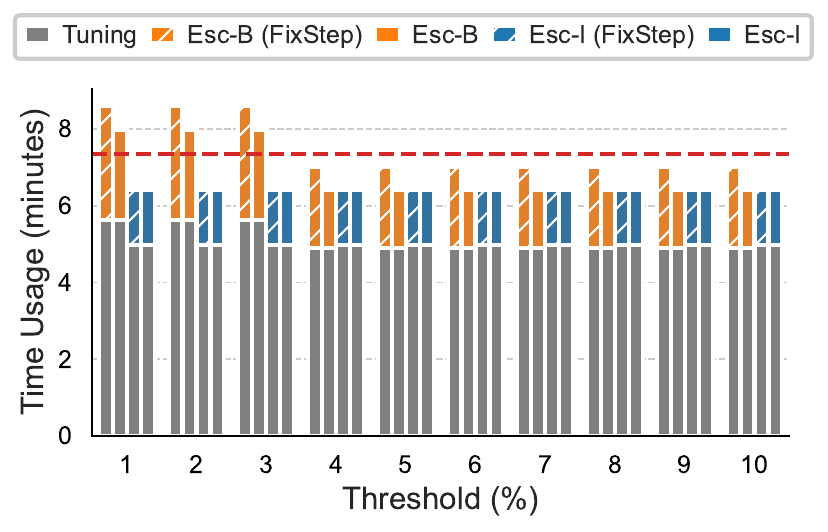}}
\subfigure[Improvement Loss]{ \label{fig:twophase_covskip:job:k20:impr-loss}
    \includegraphics[width=0.49\columnwidth]{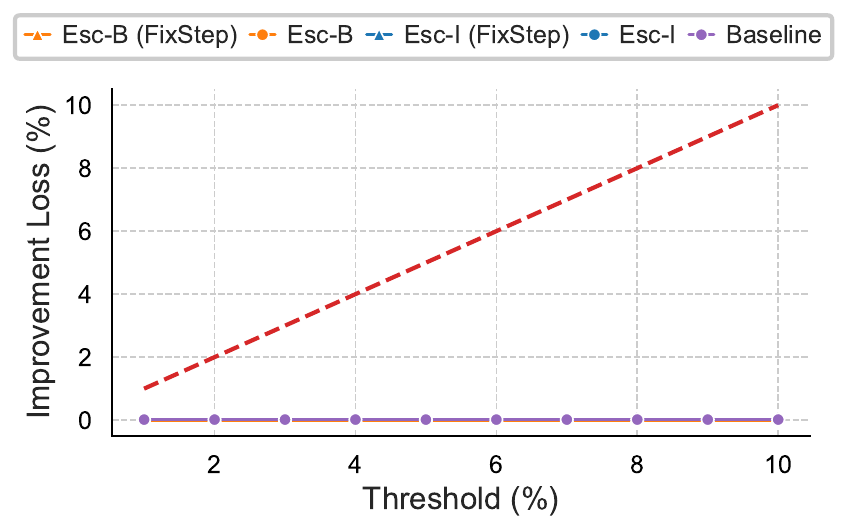}}
\subfigure[What-If Call Savings]{ \label{fig:twophase_covskip:job:k20:call-save}
    \includegraphics[width=0.49\columnwidth]{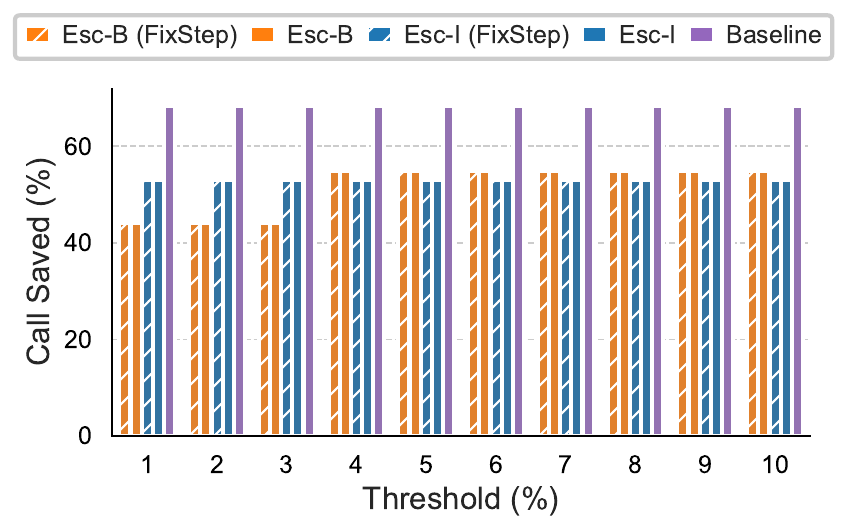}}
\subfigure[Learning Curve]{ \label{fig:twophase_covskip:job:k20:lc}
    \includegraphics[width=0.49\columnwidth]{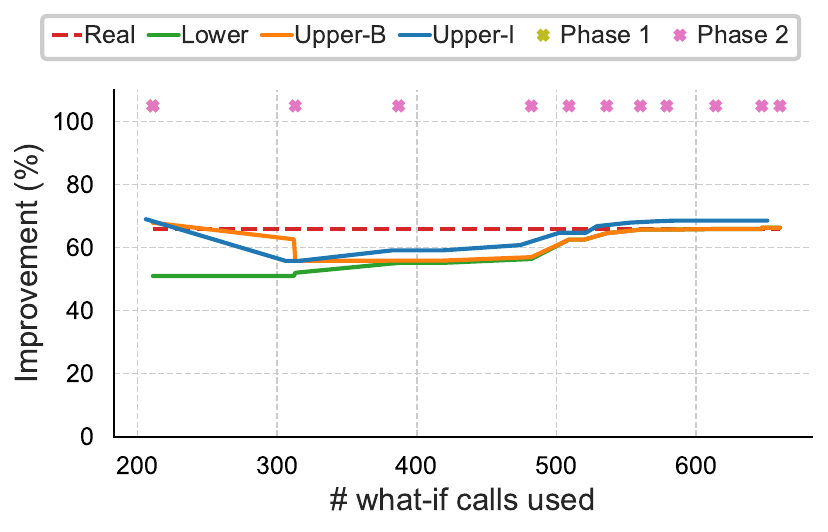}}
\vspace{-1.5em}
\caption{Two-phase greedy search (with Wii-Coverage), JOB, $K=20$, $B=20k$}
\label{fig:twophase_covskip:job:k20}
\vspace{-1em}
\end{figure*}

\begin{figure*}
\centering
\subfigure[Time Overhead]{ \label{fig:twophase_covskip:real-d:k20:overhead}
    \includegraphics[width=0.49\columnwidth]{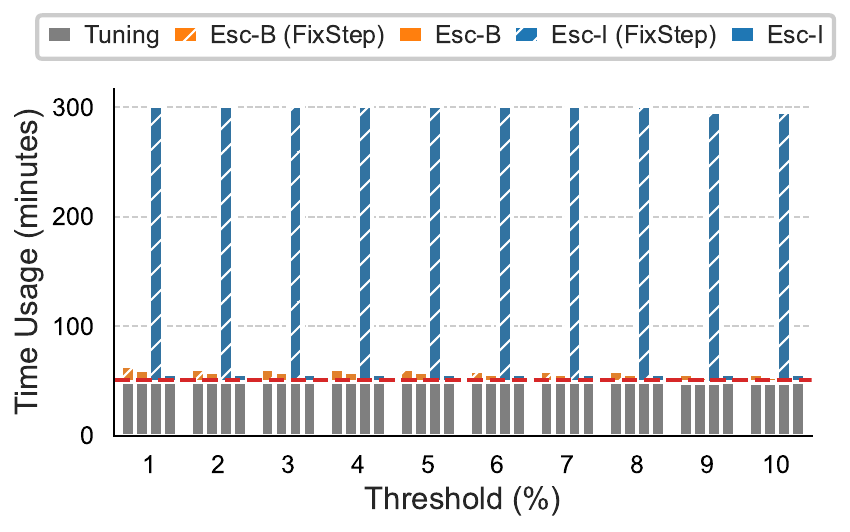}}
\subfigure[Improvement Loss]{ \label{fig:twophase_covskip:real-d:k20:impr-loss}
    \includegraphics[width=0.49\columnwidth]{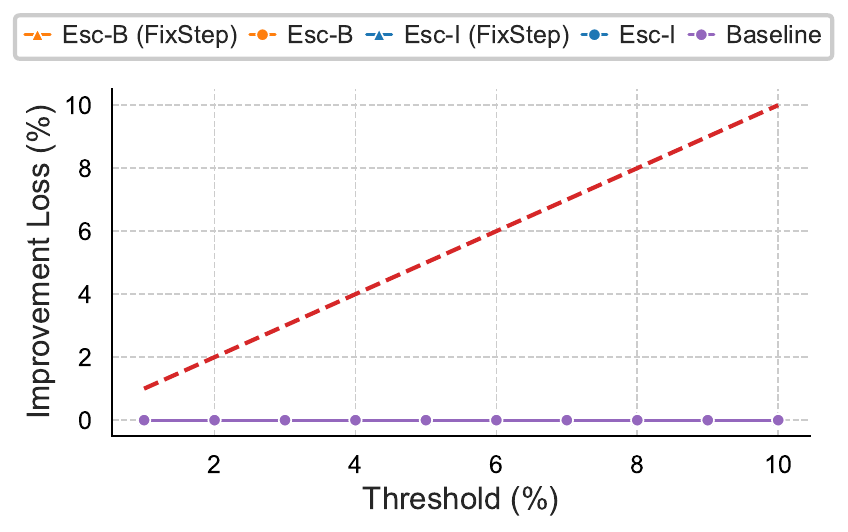}}
\subfigure[What-If Call Savings]{ \label{fig:twophase_covskip:real-d:k20:call-save}
    \includegraphics[width=0.49\columnwidth]{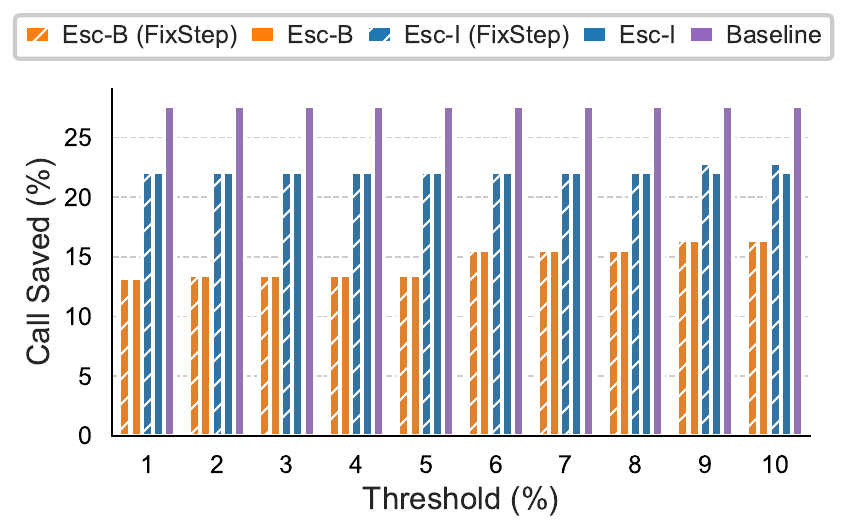}}
\subfigure[Learning Curve]{ \label{fig:twophase_covskip:real-d:k20:lc}
    \includegraphics[width=0.49\columnwidth]{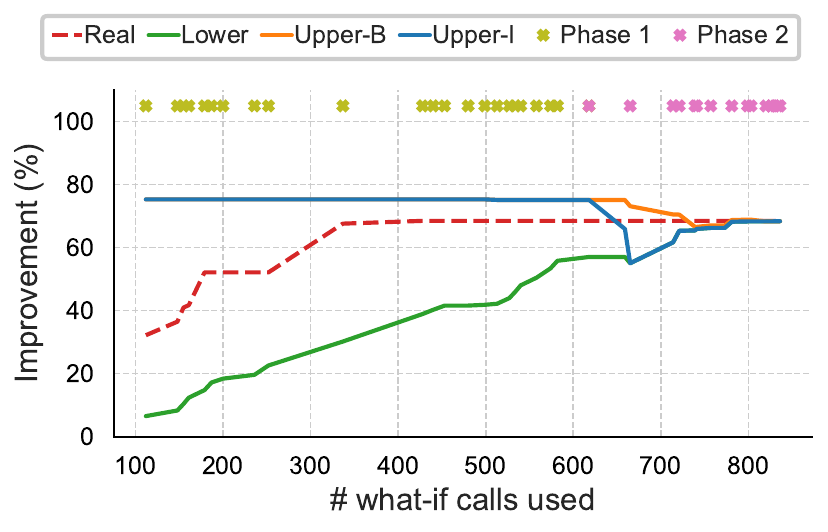}}
\vspace{-1.5em}
\caption{Two-phase greedy search (with Wii-Coverage), Real-D, $K=20$, $B=20k$}
\label{fig:twophase_covskip:real-d:k20}
\vspace{-1em}
\end{figure*}

\begin{figure*}
\centering
\subfigure[Time Overhead]{ \label{fig:mcts_covskip:tpch:k20:overhead}
    \includegraphics[width=0.49\columnwidth]{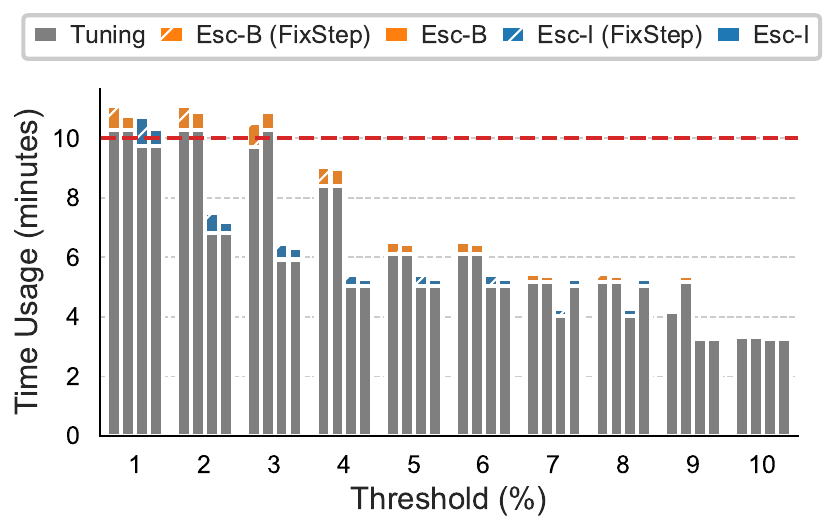}}
\subfigure[Improvement Loss]{ \label{fig:mcts_covskip:tpch:k20:impr-loss}
    \includegraphics[width=0.49\columnwidth]{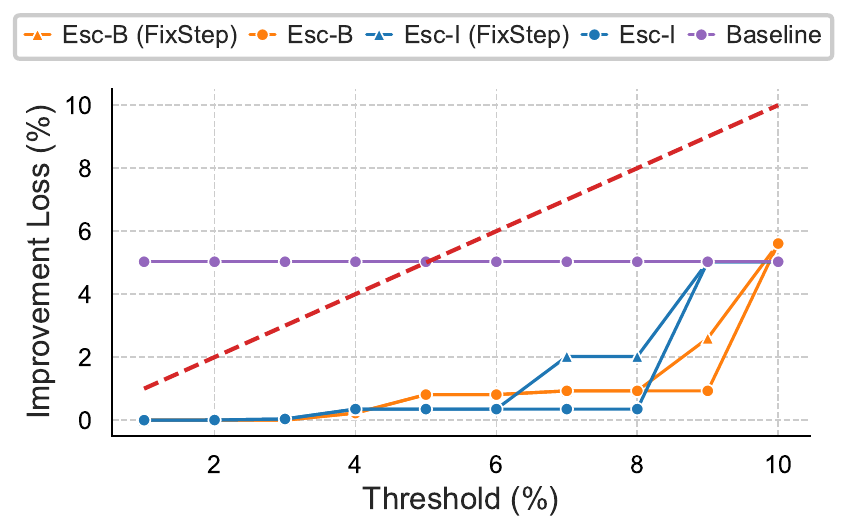}}
\subfigure[What-If Call Savings]{ \label{fig:mcts_covskip:tpch:k20:call-save}
    \includegraphics[width=0.49\columnwidth]{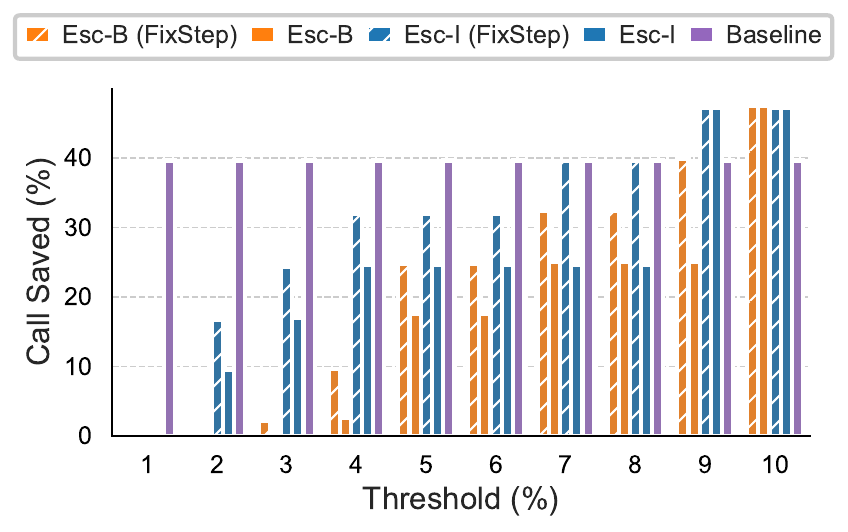}}
\subfigure[Learning Curve]{ \label{fig:mcts_covskip:tpch:k20:lc}
    \includegraphics[width=0.49\columnwidth]{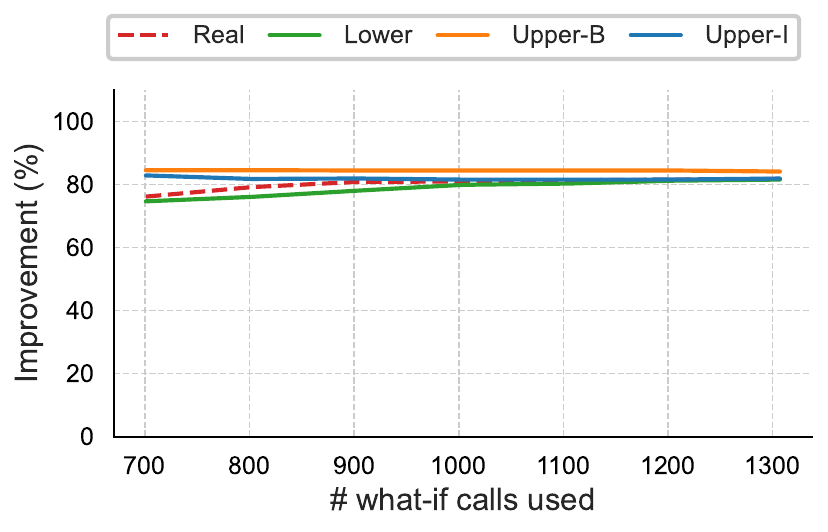}}
\vspace{-1.5em}
\caption{MCTS (with Wii-Coverage), TPC-H, $K=20$, $B=20k$}
\label{fig:mcts_covskip:tpch:k20}
\vspace{-1em}
\end{figure*}

\begin{figure*}
\centering
\subfigure[Time Overhead]{ \label{fig:mcts_covskip:tpcds:k20:overhead}
    \includegraphics[width=0.49\columnwidth]{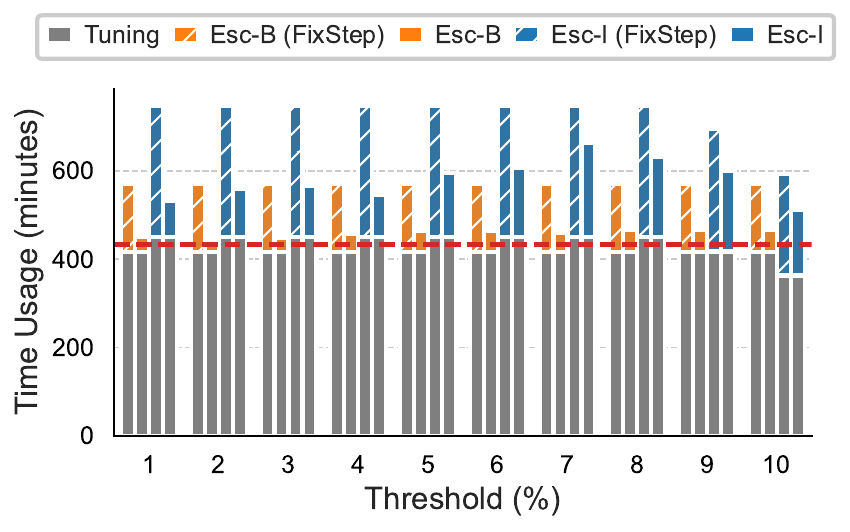}}
\subfigure[Improvement Loss]{ \label{fig:mcts_covskip:tpcds:k20:impr-loss}
    \includegraphics[width=0.49\columnwidth]{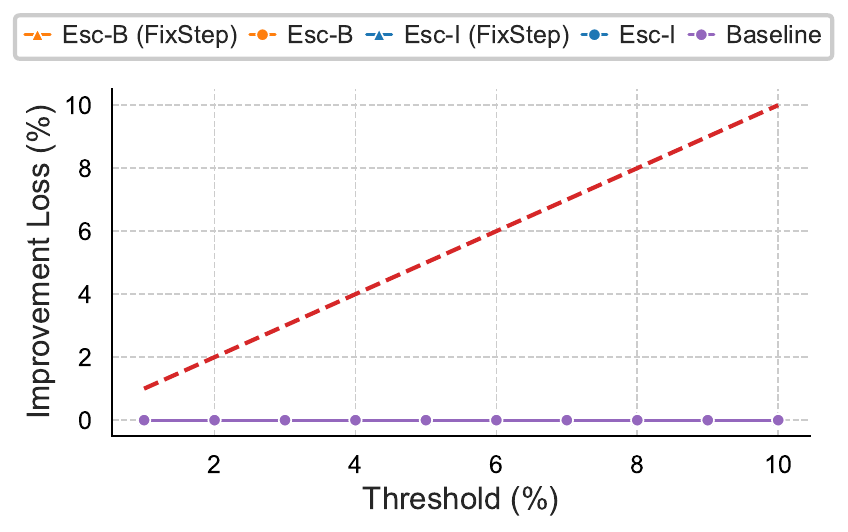}}
\subfigure[What-If Call Savings]{ \label{fig:mcts_covskip:tpcds:k20:call-save}
    \includegraphics[width=0.49\columnwidth]{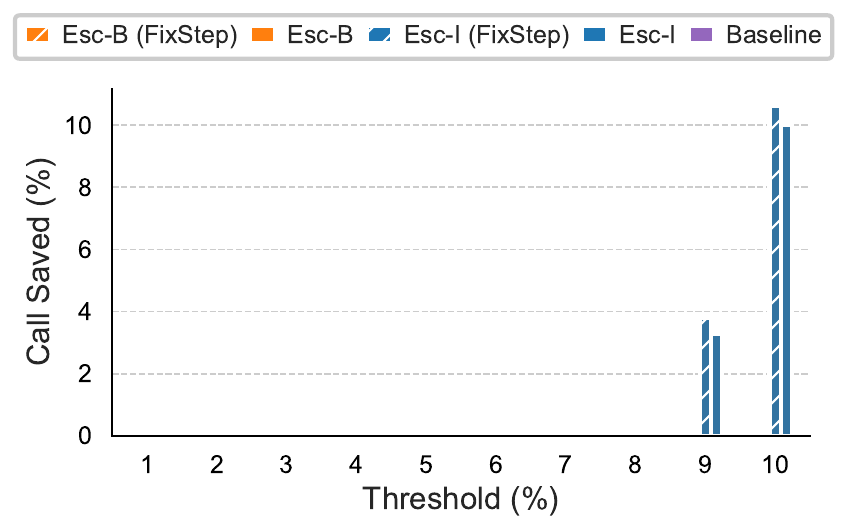}}
\subfigure[Learning Curve]{ \label{fig:mcts_covskip:tpcds:k20:lc}
    \includegraphics[width=0.49\columnwidth]{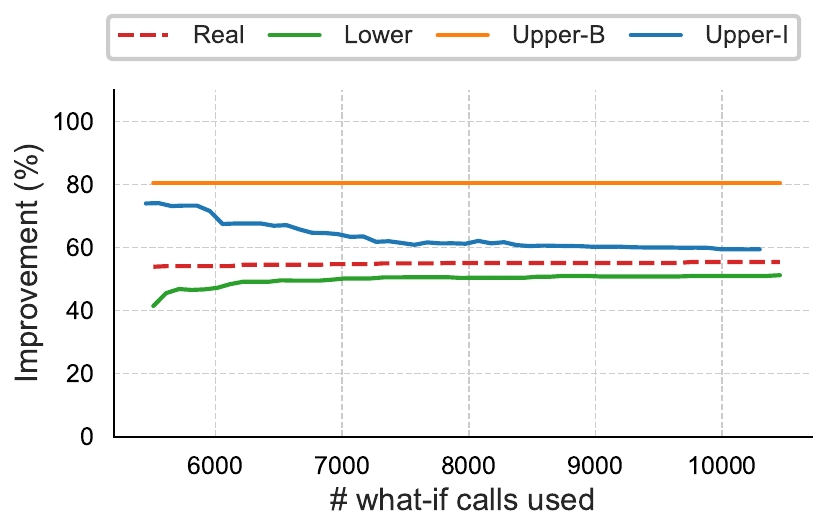}}
\vspace{-1.5em}
\caption{MCTS (with Wii-Coverage), TPC-DS, $K=20$, $B=20k$}
\label{fig:mcts_covskip:tpcds:k20}
\vspace{-1em}
\end{figure*}

\begin{figure*}
\centering
\subfigure[Time Overhead]{ \label{fig:mcts_covskip:job:k20:overhead}
    \includegraphics[width=0.49\columnwidth]{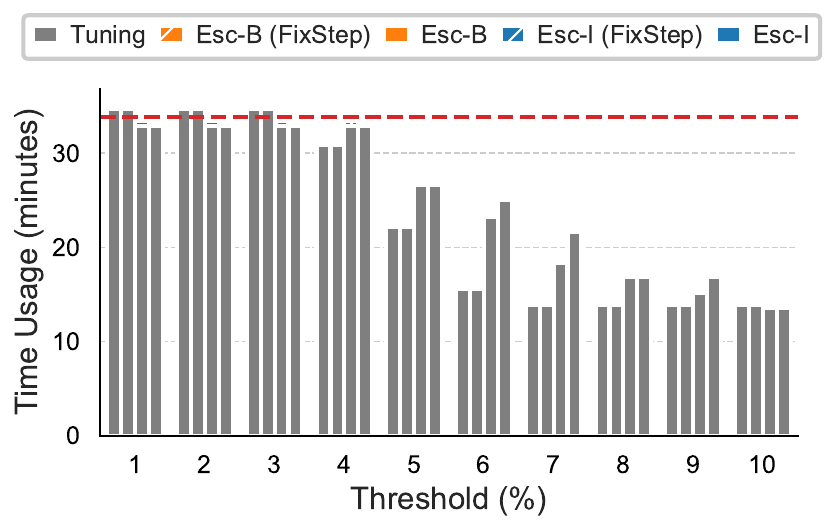}}
\subfigure[Improvement Loss]{ \label{fig:mcts_covskip:job:k20:impr-loss}
    \includegraphics[width=0.49\columnwidth]{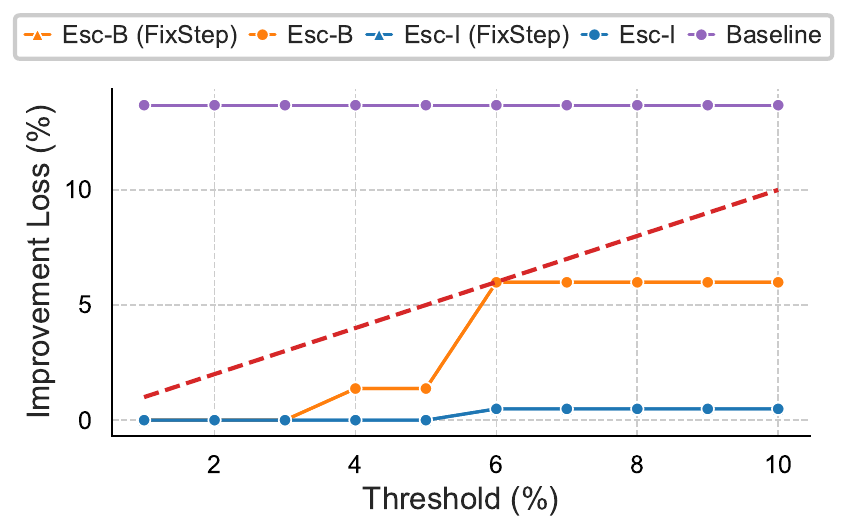}}
\subfigure[What-If Call Savings]{ \label{fig:mcts_covskip:job:k20:call-save}
    \includegraphics[width=0.49\columnwidth]{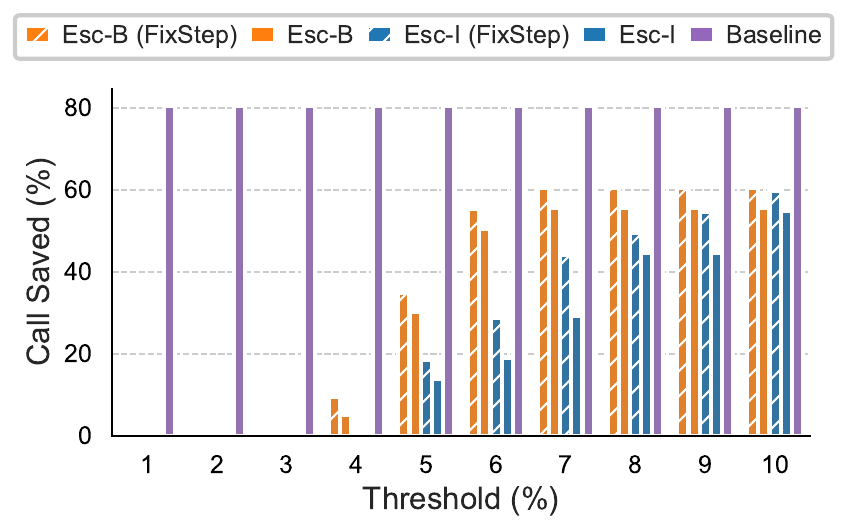}}
\subfigure[Learning Curve]{ \label{fig:mcts_covskip:job:k20:lc}
    \includegraphics[width=0.49\columnwidth]{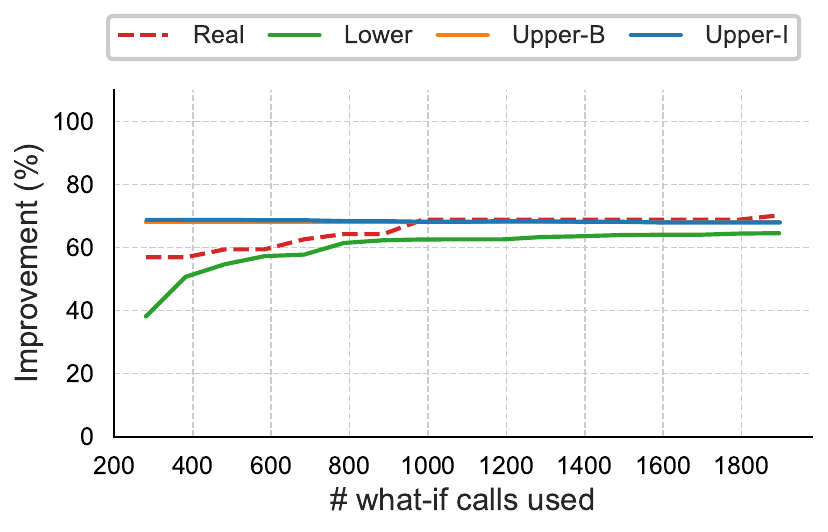}}
\vspace{-1.5em}
\caption{MCTS (with Wii-Coverage), JOB, $K=20$, $B=20k$}
\label{fig:mcts_covskip:job:k20}
\vspace{-1em}
\end{figure*}

\begin{figure*}
\centering
\subfigure[Time Overhead]{ \label{fig:mcts_covskip:real-m:k20:overhead}
    \includegraphics[width=0.49\columnwidth]{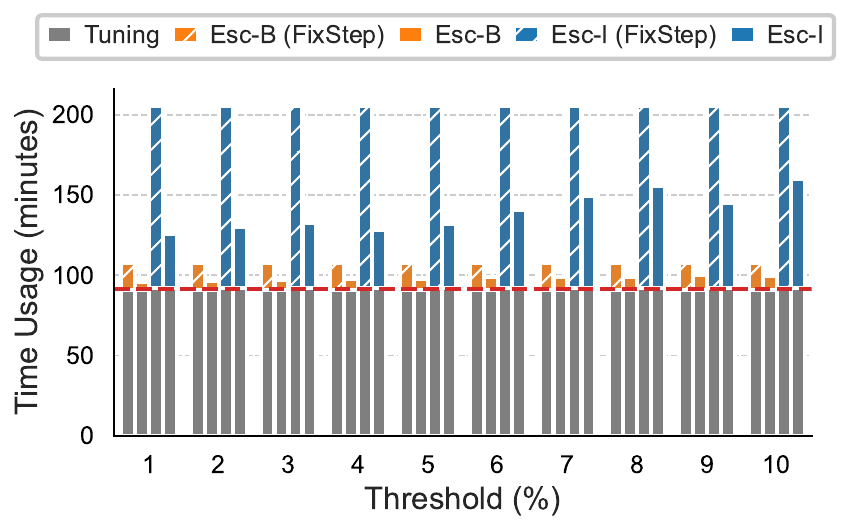}}
\subfigure[Improvement Loss]{ \label{fig:mcts_covskip:real-m:k20:impr-loss}
    \includegraphics[width=0.49\columnwidth]{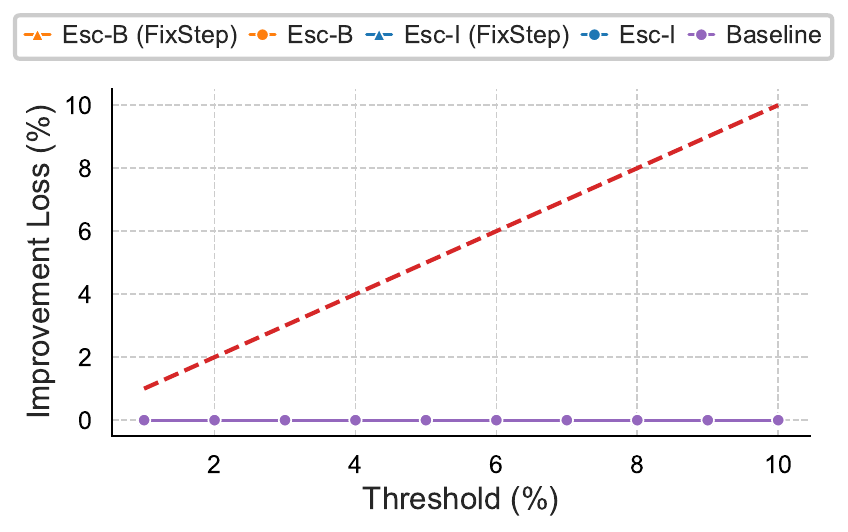}}
\subfigure[What-If Call Savings]{ \label{fig:mcts_covskip:real-m:k20:call-save}
    \includegraphics[width=0.49\columnwidth]{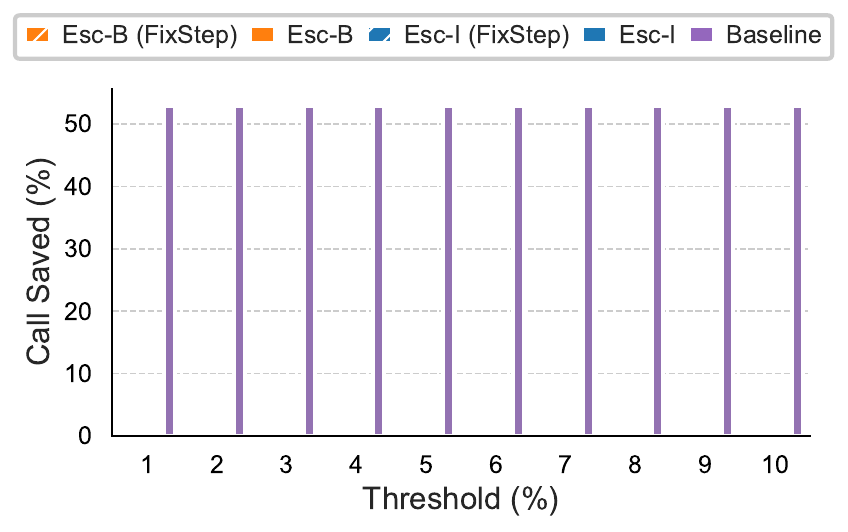}}
\subfigure[Learning Curve]{ \label{fig:mcts_covskip:real-m:k20:lc}
    \includegraphics[width=0.49\columnwidth]{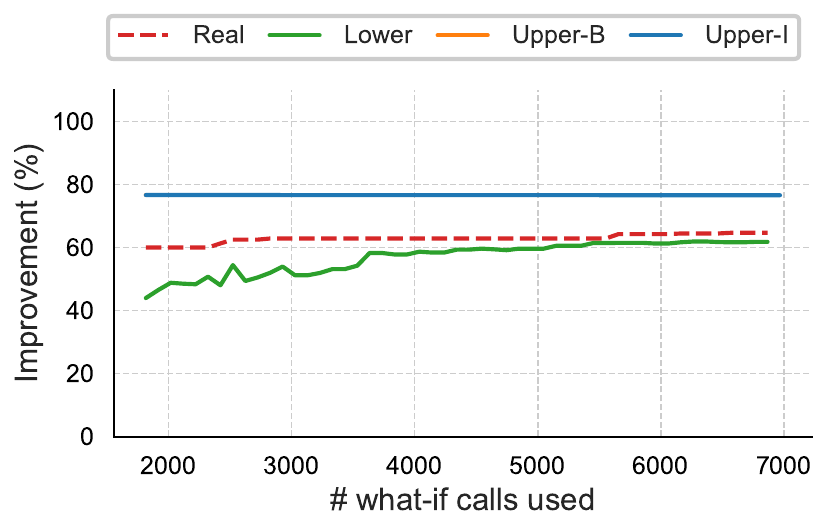}}
\vspace{-1.5em}
\caption{MCTS (with Wii-Coverage), Real-M, $K=20$, $B=20k$}
\label{fig:mcts_covskip:real-m:k20}
\vspace{-1em}
\end{figure*}

Figures~\ref{fig:twophase_covskip:tpch:k20} to~\ref{fig:twophase_covskip:real-d:k20} further present the results for the Wii-enhanced version of \emph{two-phase greedy search} when the coverage-based refinement is enabled, whereas Figures~\ref{fig:mcts_covskip:tpch:k20} to~\ref{fig:mcts_covskip:real-m:k20} present the corresponding results for \emph{MCTS}.
Again, we set $K=20$ in these evaluations.

For \emph{two-phase greedy search}, enabling the coverage-based refinement can sometimes help Wii further speed up the convergence of index tuning. For example, on \textbf{TPC-DS} it only needs 2,800 what-if calls to finish (ref. Figure~\ref{fig:twophase_covskip:tpcds:k20:lc}), compared to the 6,000 what-if calls required by Wii without the coverage-based refinement (ref. Figure~\ref{fig:twophase_skip:tpcds:k20:lc}).
This improved convergence, however, does not hinder the effectiveness of early-stopping.
As shown in Figure~\ref{fig:twophase_covskip:tpcds:k20:call-save}, we can further save 20\% what-if calls by using \textbf{\sysname-B} and 25\% what-if calls by using \textbf{\sysname-I} with no improvement loss (ref. Figure~\ref{fig:twophase_covskip:tpcds:k20:impr-loss}) and little extra time overhead (ref. Figure~\ref{fig:twophase_covskip:tpcds:k20:overhead}).

\begin{figure*}
\centering
\subfigure[Time Overhead]{ \label{fig:mcts_skip:real-m:k20:overhead}
    \includegraphics[width=0.49\columnwidth]{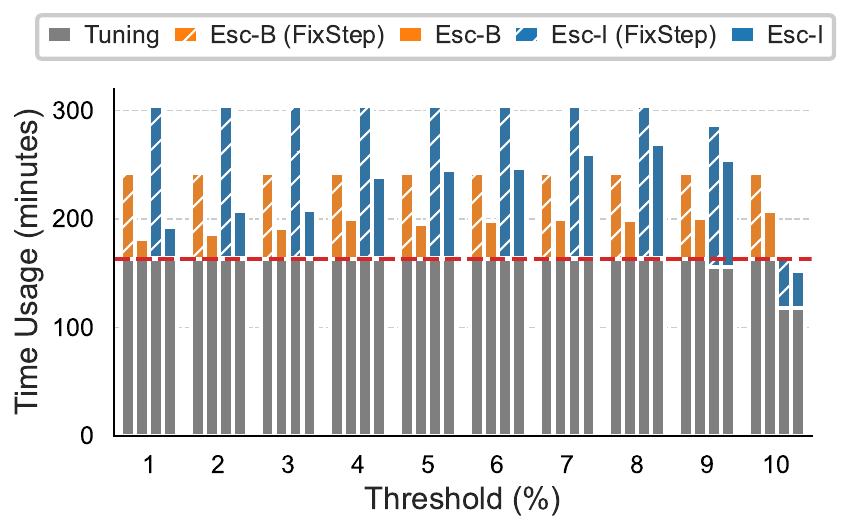}}
\subfigure[Improvement Loss]{ \label{fig:mcts_skip:real-m:k20:impr-loss}
    \includegraphics[width=0.49\columnwidth]{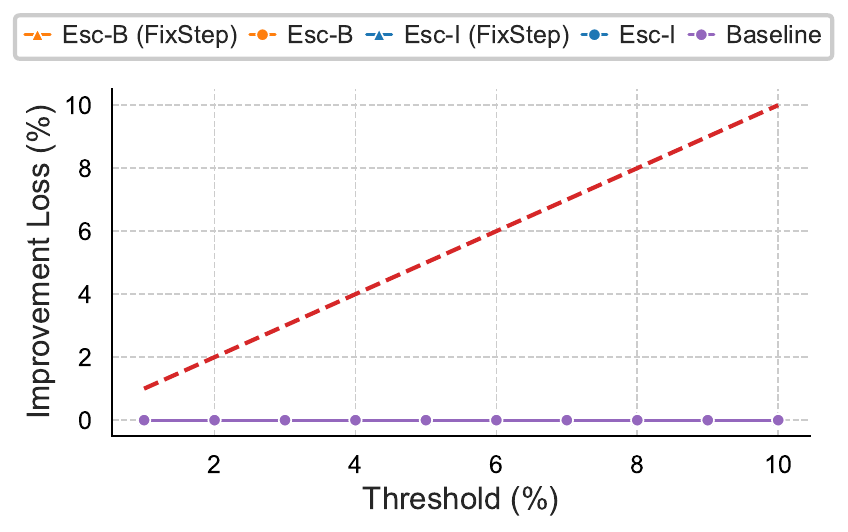}}
\subfigure[What-If Call Savings]{ \label{fig:mcts_skip:real-m:k20:call-save}
    \includegraphics[width=0.49\columnwidth]{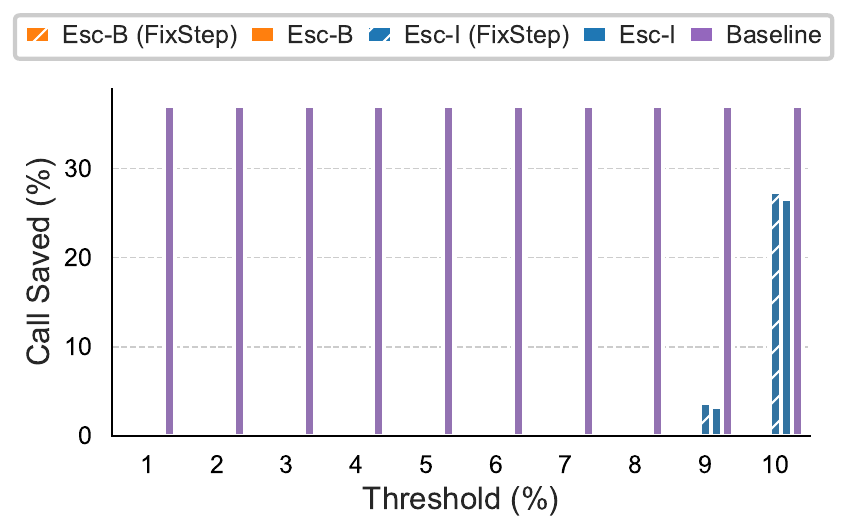}}
\subfigure[Learning Curve]{ \label{fig:mcts_skip:real-m:k20:lc}
    \includegraphics[width=0.49\columnwidth]{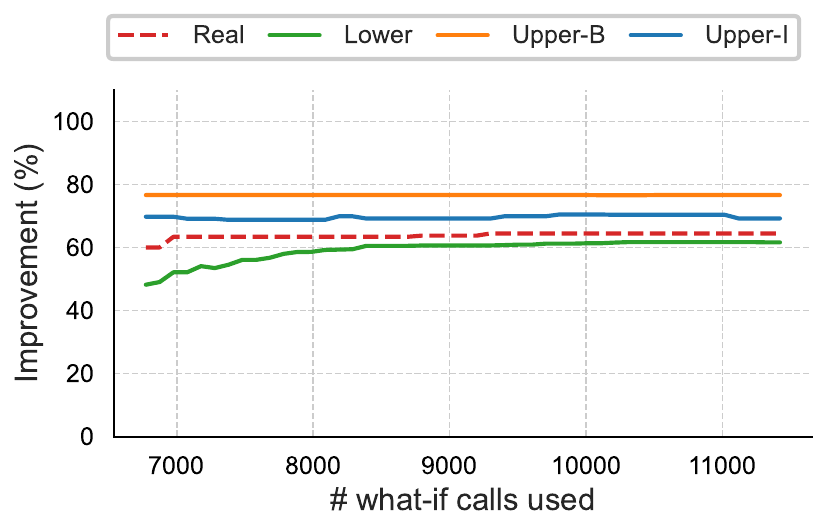}}
\vspace{-1.5em}
\caption{MCTS (with Wii), Real-M, $K=20$, $B=20k$}
\label{fig:mcts_skip:real-m:k20}
\vspace{-1em}
\end{figure*}

For \emph{MCTS}, on the other hand, enabling the coverage-based refinement is not so effective in terms of speeding up the convergence of index tuning, though we still observe some improvements on \textbf{Real-D} where the number of what-if calls required is reduced from 9,000 (ref. Figure~\ref{fig:mcts_skip:real-d:k20:lc}) to 6,000 (ref. Figure~\ref{fig:mcts_covskip:real-d:k20:lc}).
Moreover, using \textbf{\sysname-I} for early-stopping verification can further save 60\% to 80\% of what-if calls when varying the improvement-loss threshold $\epsilon$ from 7\% to 10\%, as Figure~\ref{fig:mcts_covskip:real-d:k20:call-save} presents.
One may also have noticed that \sysname seems not working for \textbf{Real-M}, as Figure~\ref{fig:mcts_covskip:real-m:k20:call-save} shows.
While this is true, we cannot conclude that this is a regressed case given that the coverage-based refinement has brought the number of what-if calls required from 11,500 (ref. Figure~\ref{fig:mcts_skip:real-m:k20:lc}) down to 7,000 (ref. Figure~\ref{fig:mcts_covskip:real-m:k20:lc}).
As a result, the maximum savings of 25\% what-if calls, as shown in Figure~\ref{fig:mcts_skip:real-m:k20:call-save}, still imply that around $11,500\times 75\%=8625$ what-if calls were allocated, which remains more than the 7,000 what-if calls with the coverage-based refinement enabled.

\vspace{-1em}
\subsection{Results with $K=10$}

\begin{figure*}
\centering
\subfigure[Time Overhead]{ \label{fig:twophase:tpch:k10:overhead}
    \includegraphics[width=0.49\columnwidth]{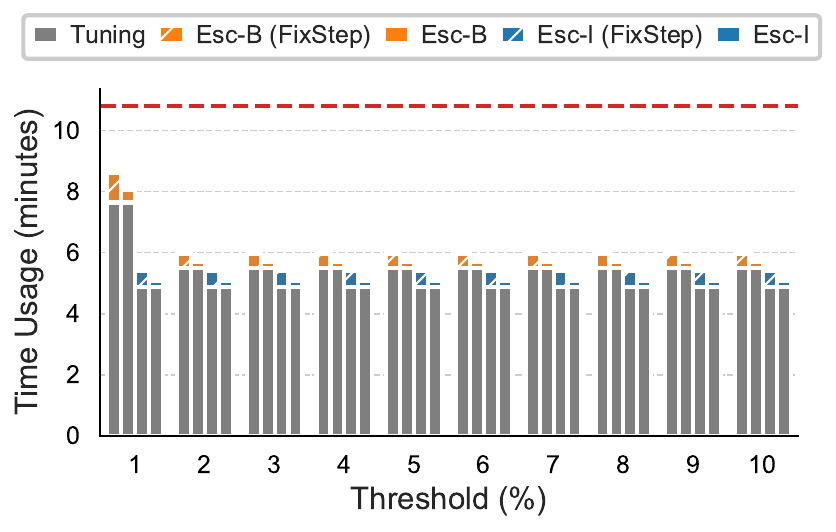}}
\subfigure[Improvement Loss]{ \label{fig:twophase:tpch:k10:impr-loss}
    \includegraphics[width=0.49\columnwidth]{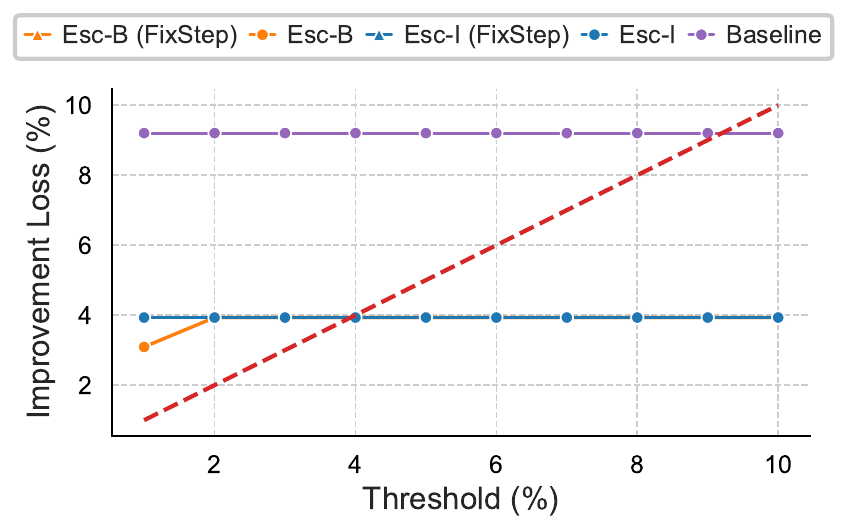}}
\subfigure[What-If Call Savings]{ \label{fig:twophase:tpch:k10:call-save}
    \includegraphics[width=0.49\columnwidth]{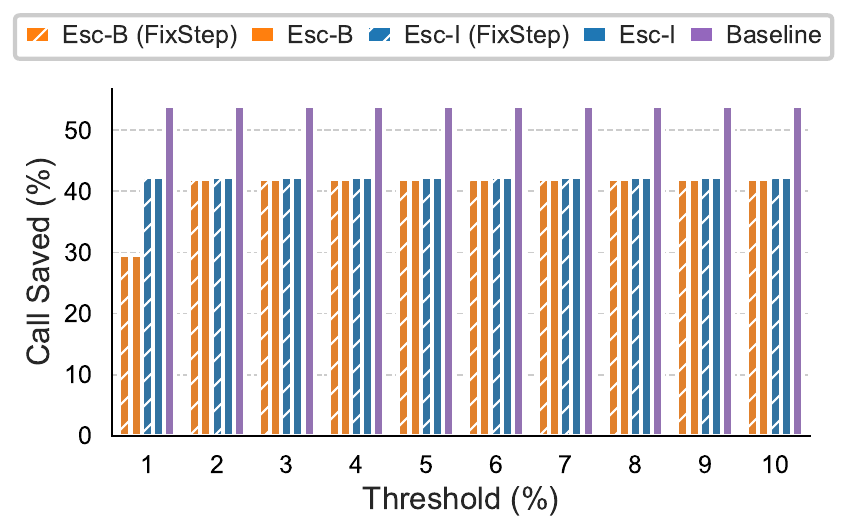}}
\subfigure[Learning Curve]{ \label{fig:twophase:tpch:k10:lc}
    \includegraphics[width=0.49\columnwidth]{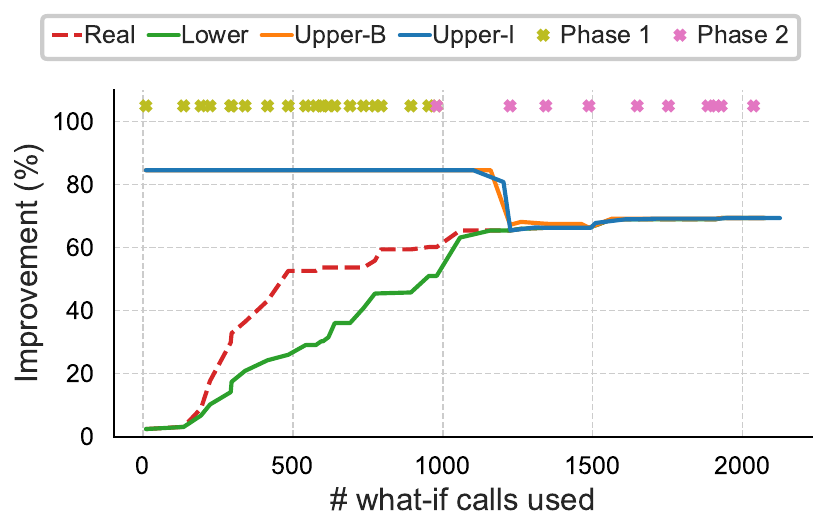}}
\vspace{-1.5em}
\caption{Two-phase greedy search, TPC-H, $K=10$, $B=20k$}
\label{fig:twophase:tpch:k10}
\vspace{-1em}
\end{figure*}

\begin{figure*}
\centering
\subfigure[Time Overhead]{ \label{fig:twophase_skip:tpch:k10:overhead}
    \includegraphics[width=0.49\columnwidth]{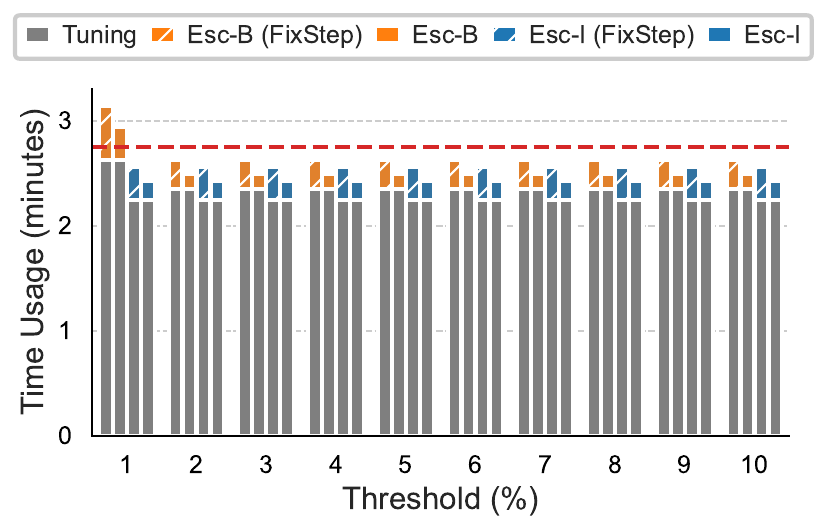}}
\subfigure[Improvement Loss]{ \label{fig:twophase_skip:tpch:k10:impr-loss}
    \includegraphics[width=0.49\columnwidth]{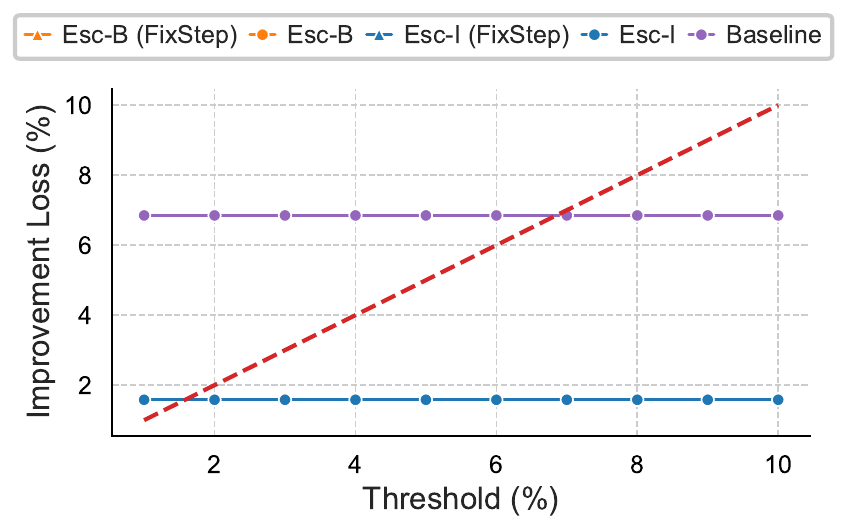}}
\subfigure[What-If Call Savings]{ \label{fig:twophase_skip:tpch:k10:call-save}
    \includegraphics[width=0.49\columnwidth]{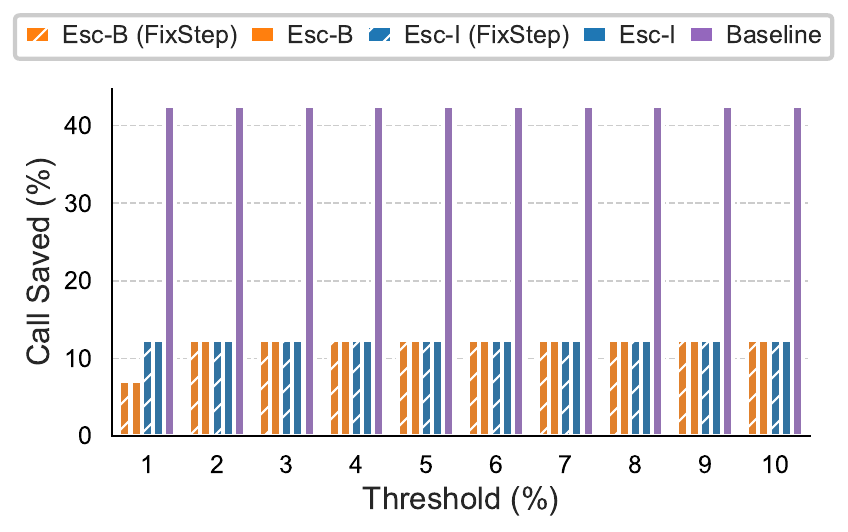}}
\subfigure[Learning Curve]{ \label{fig:twophase_skip:tpch:k10:lc}
    \includegraphics[width=0.49\columnwidth]{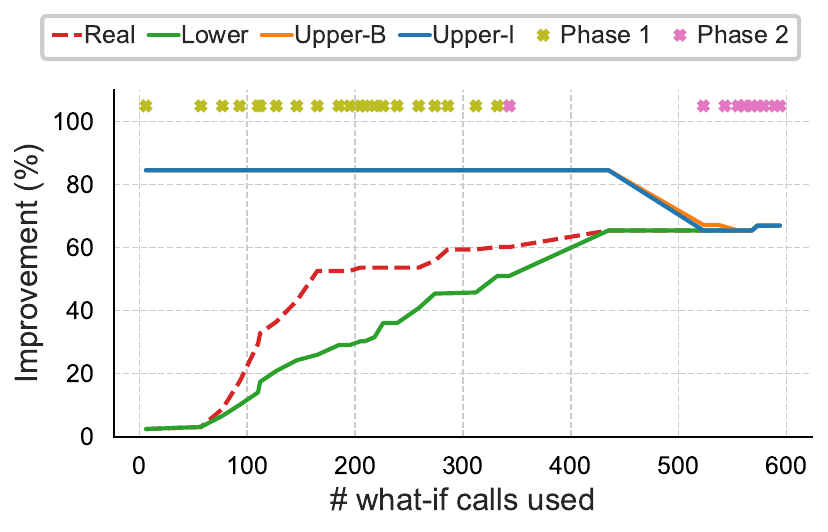}}
\vspace{-1.5em}
\caption{Two-phase greedy search (with Wii), TPC-H, $K=10$, $B=20k$}
\label{fig:twophase_skip:tpch:k10}
\vspace{-1em}
\end{figure*}

\begin{figure*}
\centering
\subfigure[Time Overhead]{ \label{fig:twophase_covskip:tpch:k10:overhead}
    \includegraphics[width=0.49\columnwidth]{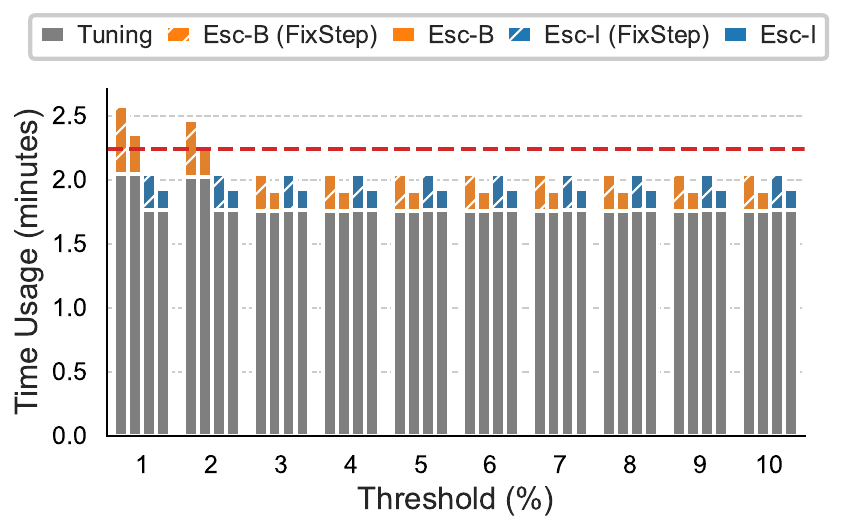}}
\subfigure[Improvement Loss]{ \label{fig:twophase_covskip:tpch:k10:impr-loss}
    \includegraphics[width=0.49\columnwidth]{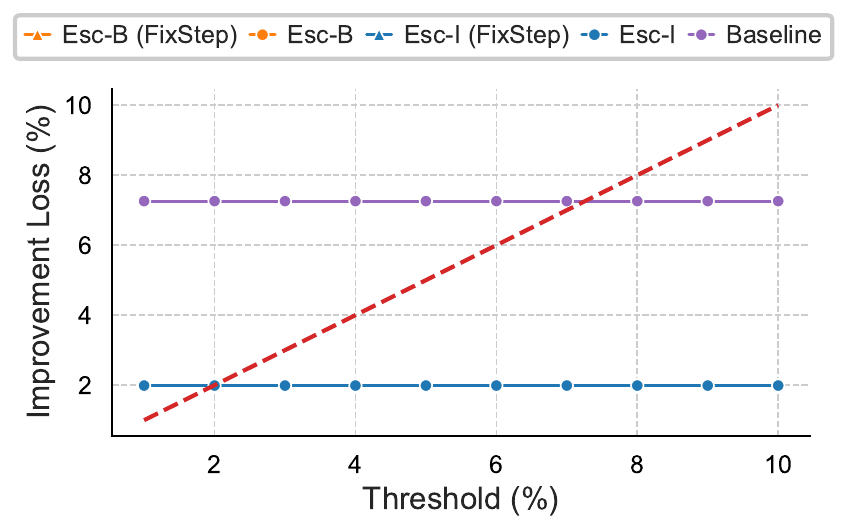}}
\subfigure[What-If Call Savings]{ \label{fig:twophase_covskip:tpch:k10:call-save}
    \includegraphics[width=0.49\columnwidth]{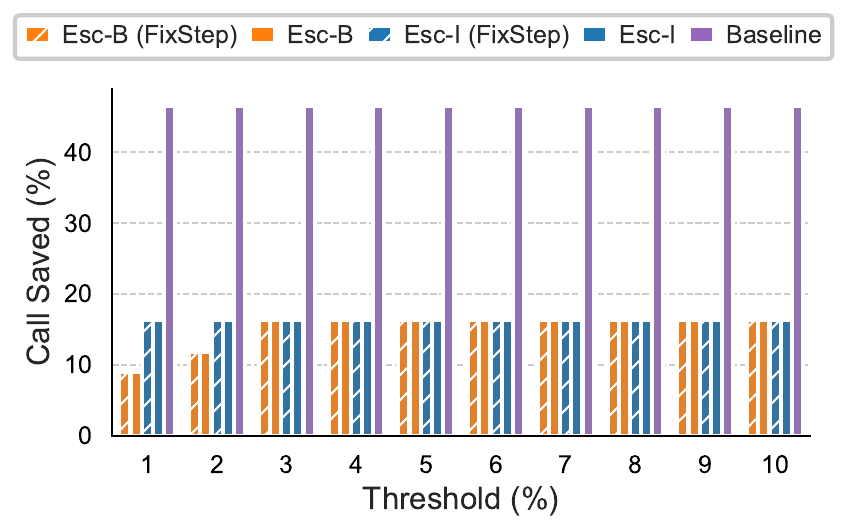}}
\subfigure[Learning Curve]{ \label{fig:twophase_covskip:tpch:k10:lc}
    \includegraphics[width=0.49\columnwidth]{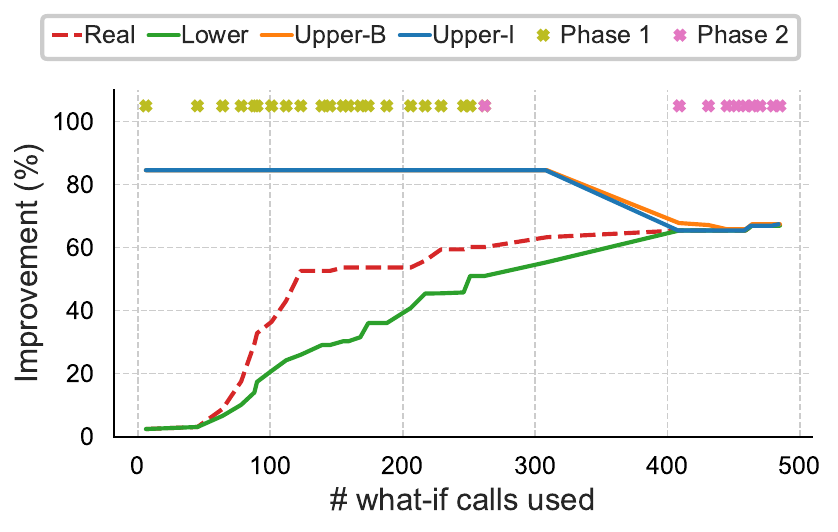}}
\vspace{-1.5em}
\caption{Two-phase greedy search (with Wii-Coverage), TPC-H, $K=10$, $B=20k$}
\label{fig:twophase_covskip:tpch:k10}
\vspace{-1em}
\end{figure*}

\begin{figure*}
\centering
\subfigure[Time Overhead]{ \label{fig:twophase:tpcds:k10:overhead}
    \includegraphics[width=0.49\columnwidth]{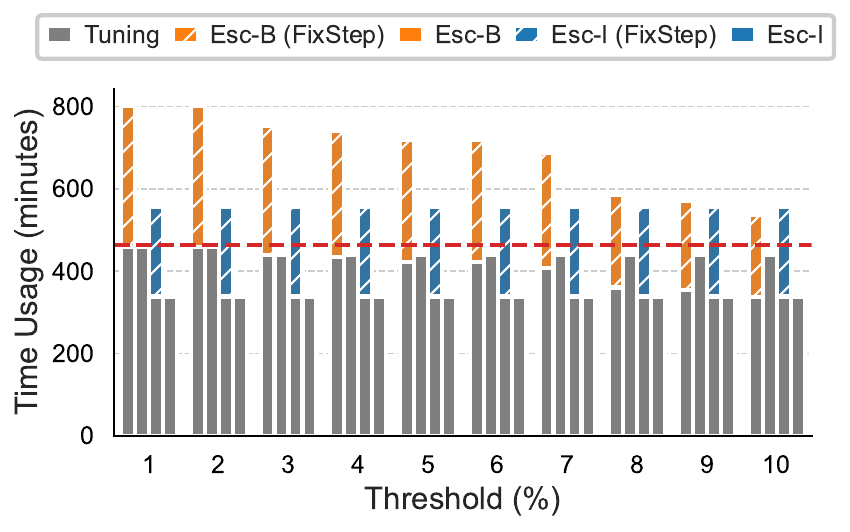}}
\subfigure[Improvement Loss]{ \label{fig:twophase:tpcds:k10:impr-loss}
    \includegraphics[width=0.49\columnwidth]{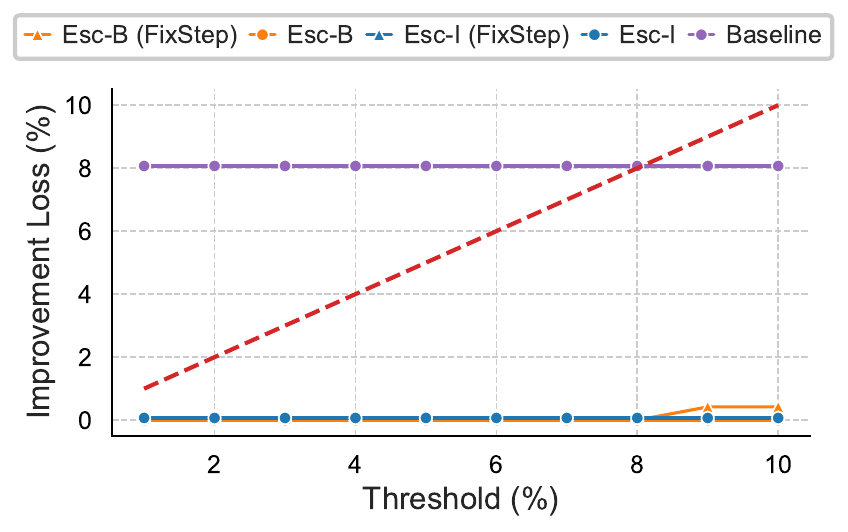}}
\subfigure[What-If Call Savings]{ \label{fig:twophase:tpcds:k10:call-save}
    \includegraphics[width=0.49\columnwidth]{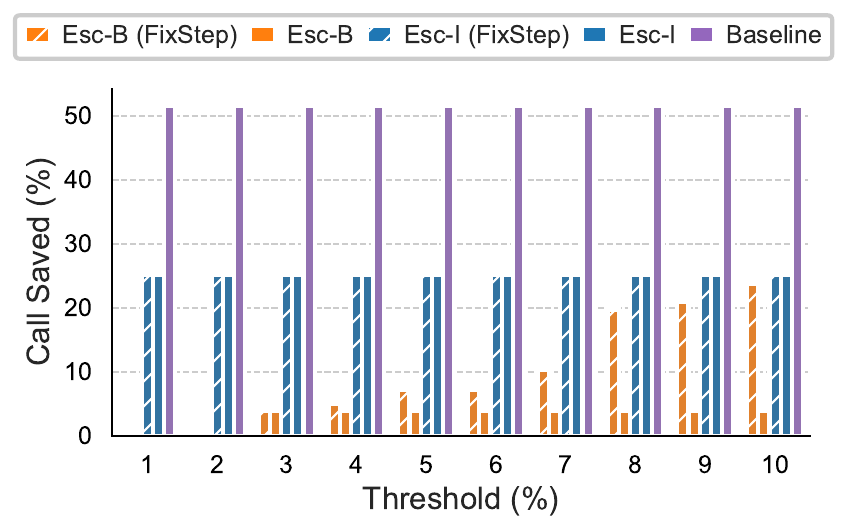}}
\subfigure[Learning Curve]{ \label{fig:twophase:tpcds:k10:lc}
    \includegraphics[width=0.49\columnwidth]{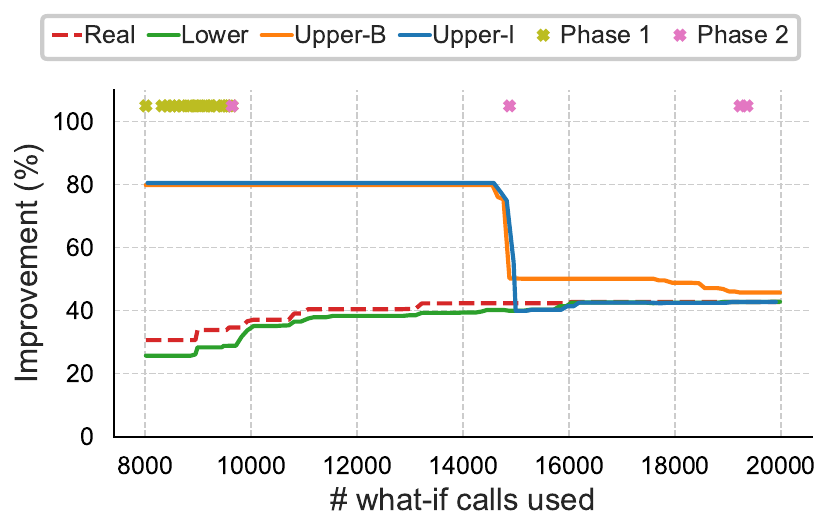}}
\vspace{-1.5em}
\caption{Two-phase greedy search, TPC-DS, $K=10$, $B=20k$}
\label{fig:twophase:tpcds:k10}
\vspace{-1em}
\end{figure*}

\begin{figure*}
\centering
\subfigure[Time Overhead]{ \label{fig:twophase_skip:tpcds:k10:overhead}
    \includegraphics[width=0.49\columnwidth]{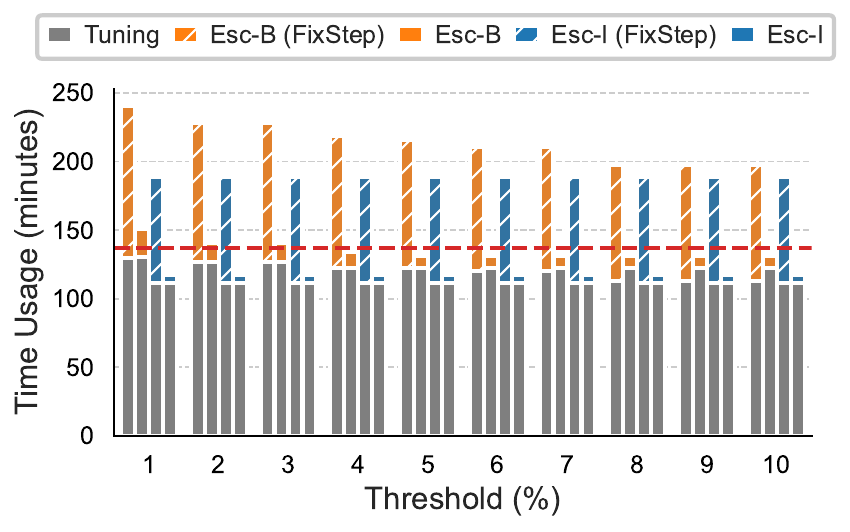}}
\subfigure[Improvement Loss]{ \label{fig:twophase_skip:tpcds:k10:impr-loss}
    \includegraphics[width=0.49\columnwidth]{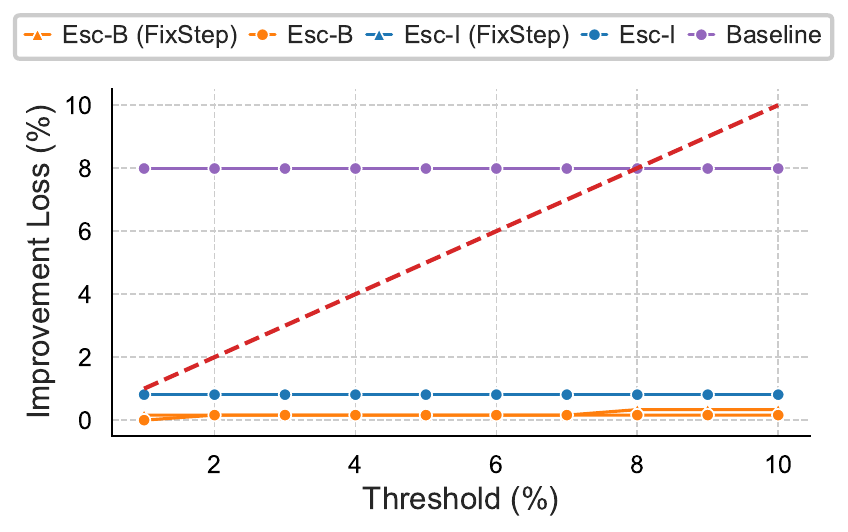}}
\subfigure[What-If Call Savings]{ \label{fig:twophase_skip:tpcds:k10:call-save}
    \includegraphics[width=0.49\columnwidth]{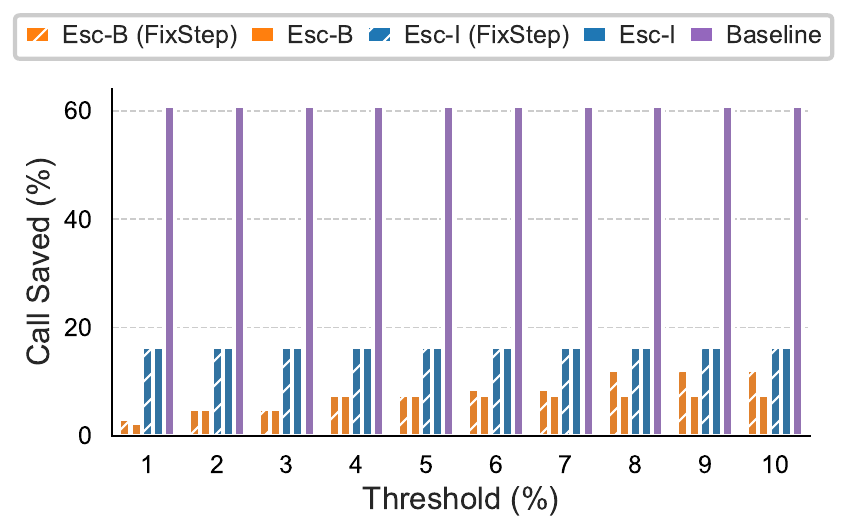}}
\subfigure[Learning Curve]{ \label{fig:twophase_skip:tpcds:k10:lc}
    \includegraphics[width=0.49\columnwidth]{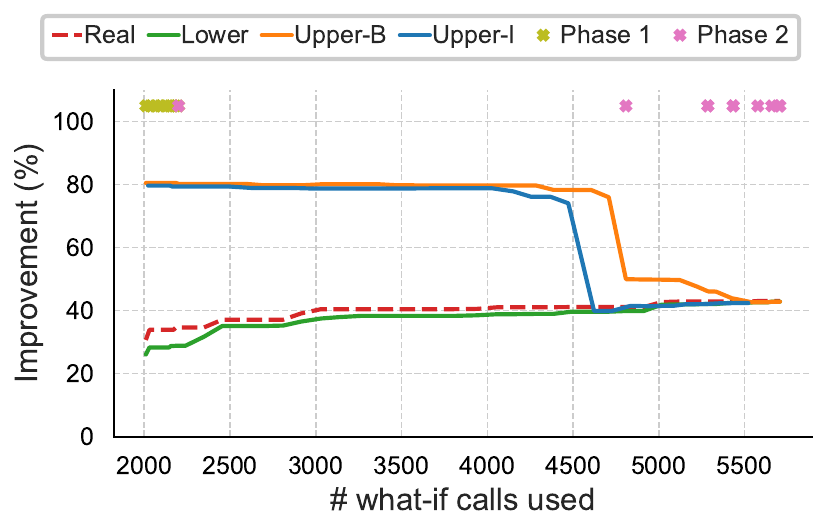}}
\vspace{-1.5em}
\caption{Two-phase greedy search (with Wii), TPC-DS, $K=10$, $B=20k$}
\label{fig:twophase_skip:tpcds:k10}
\vspace{-1em}
\end{figure*}

\begin{figure*}
\centering
\subfigure[Time Overhead]{ \label{fig:twophase_covskip:tpcds:k10:overhead}
    \includegraphics[width=0.49\columnwidth]{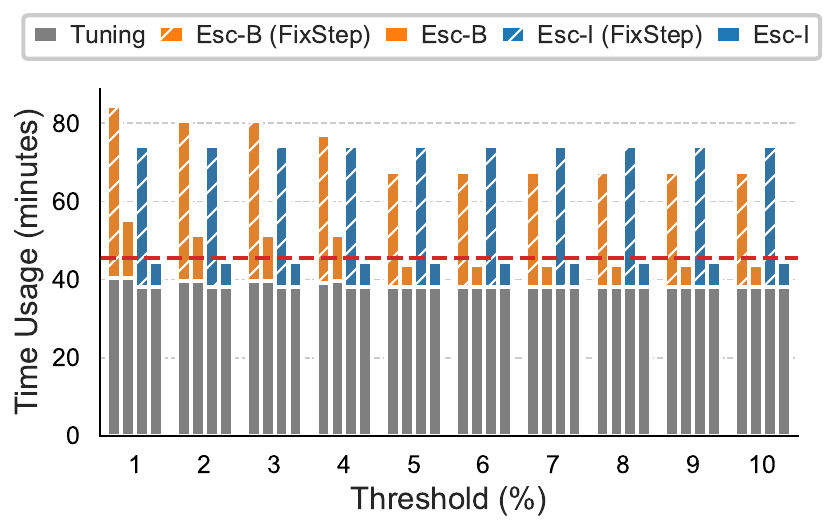}}
\subfigure[Improvement Loss]{ \label{fig:twophase_covskip:tpcds:k10:impr-loss}
    \includegraphics[width=0.49\columnwidth]{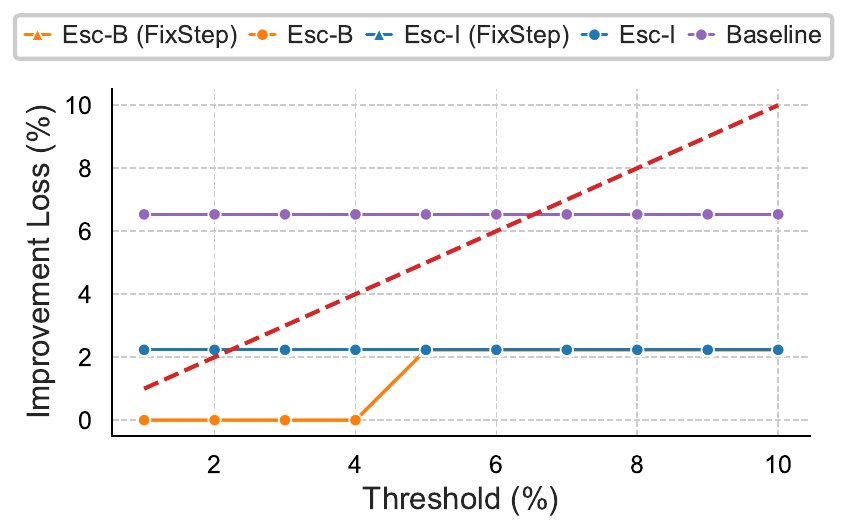}}
\subfigure[What-If Call Savings]{ \label{fig:twophase_covskip:tpcds:k10:call-save}
    \includegraphics[width=0.49\columnwidth]{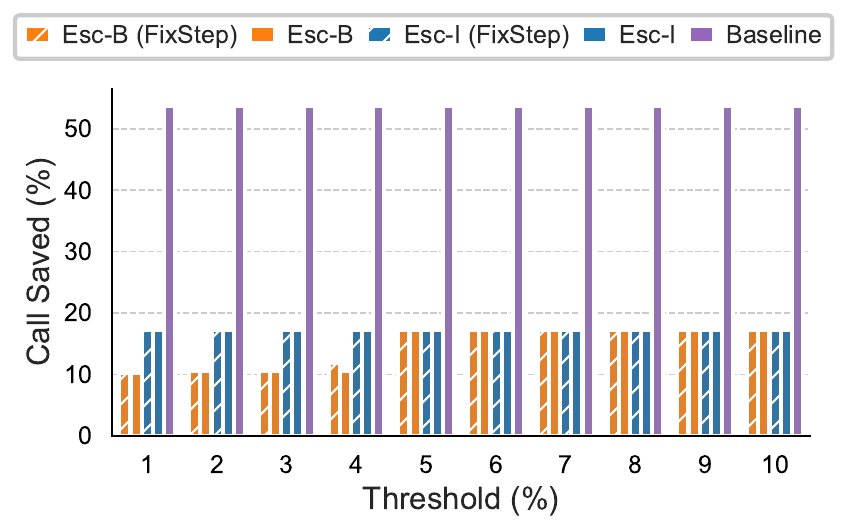}}
\subfigure[Learning Curve]{ \label{fig:twophase_covskip:tpcds:k10:lc}
    \includegraphics[width=0.49\columnwidth]{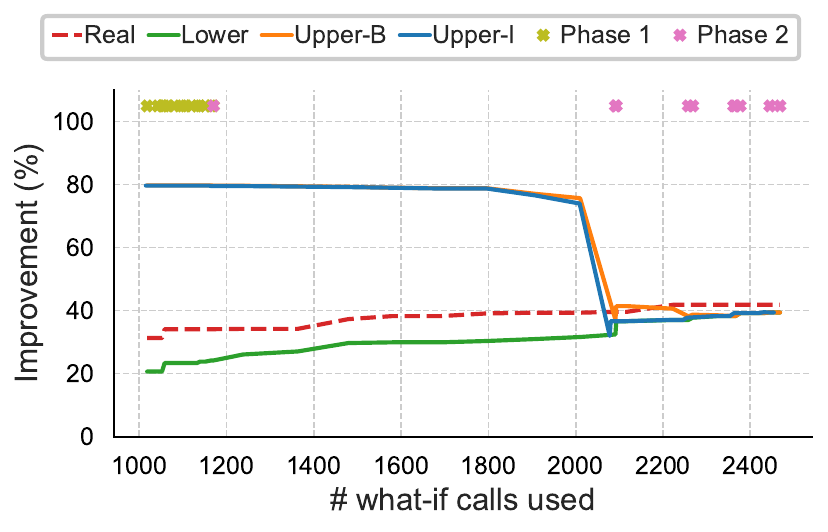}}
\vspace{-1.5em}
\caption{Two-phase greedy search (with Wii-Coverage), TPC-DS, $K=10$, $B=20k$}
\label{fig:twophase_covskip:tpcds:k10}
\vspace{-1em}
\end{figure*}

\begin{figure*}
\centering
\subfigure[Time Overhead]{ \label{fig:twophase:job:k10:overhead}
    \includegraphics[width=0.49\columnwidth]{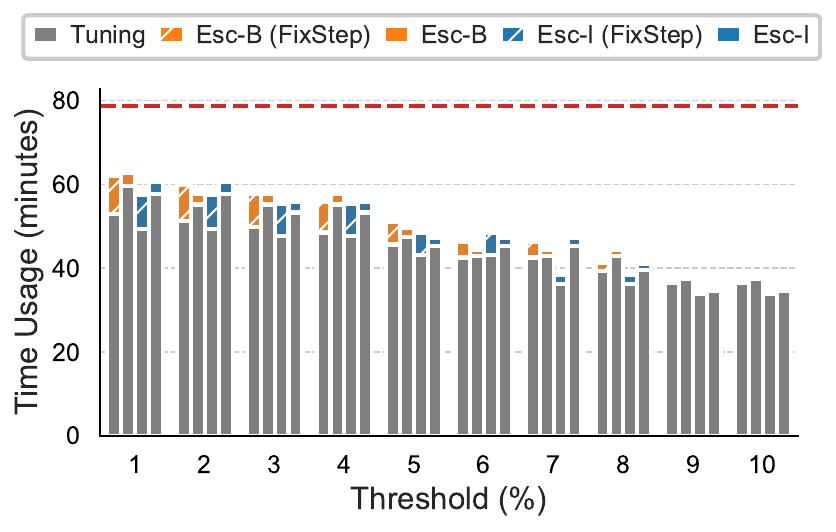}}
\subfigure[Improvement Loss]{ \label{fig:twophase:job:k10:impr-loss}
    \includegraphics[width=0.49\columnwidth]{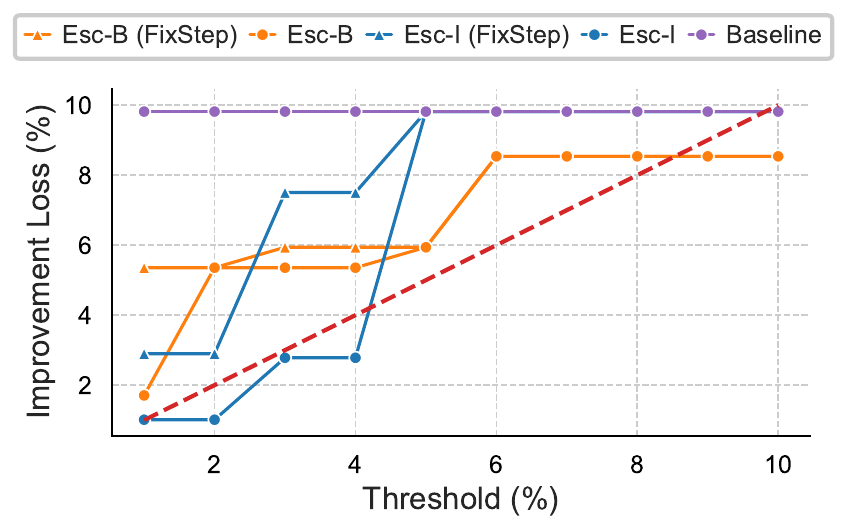}}
\subfigure[What-If Call Savings]{ \label{fig:twophase:job:k10:call-save}
    \includegraphics[width=0.49\columnwidth]{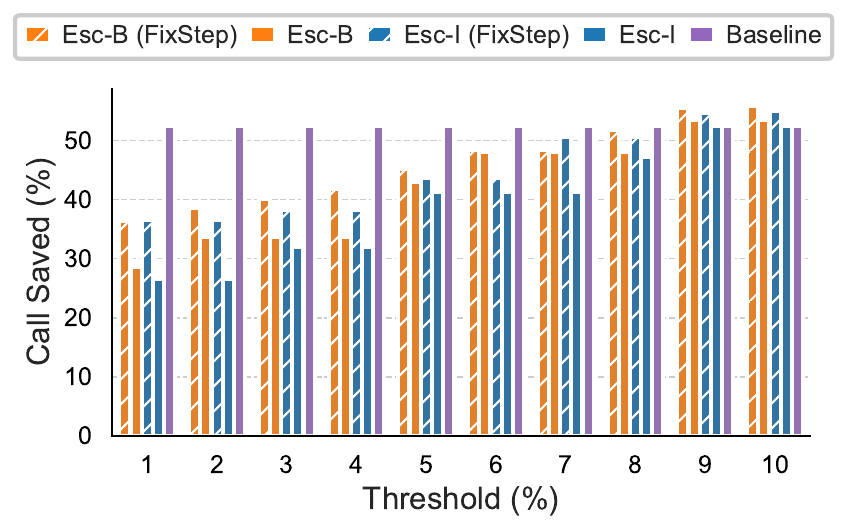}}
\subfigure[Learning Curve]{ \label{fig:twophase:job:k10:lc}
    \includegraphics[width=0.49\columnwidth]{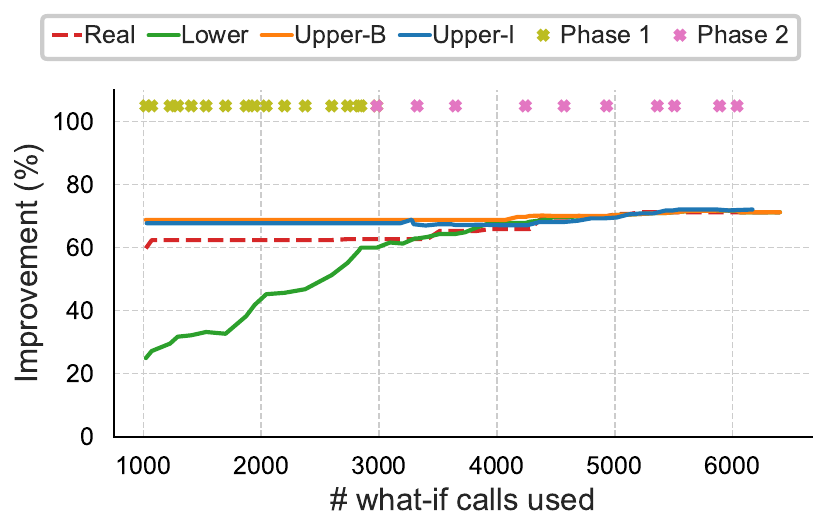}}
\vspace{-1.5em}
\caption{Two-phase greedy search, JOB, $K=10$, $B=20k$}
\label{fig:twophase:job:k10}
\vspace{-1em}
\end{figure*}

\begin{figure*}
\centering
\subfigure[Time Overhead]{ \label{fig:twophase_skip:job:k10:overhead}
    \includegraphics[width=0.49\columnwidth]{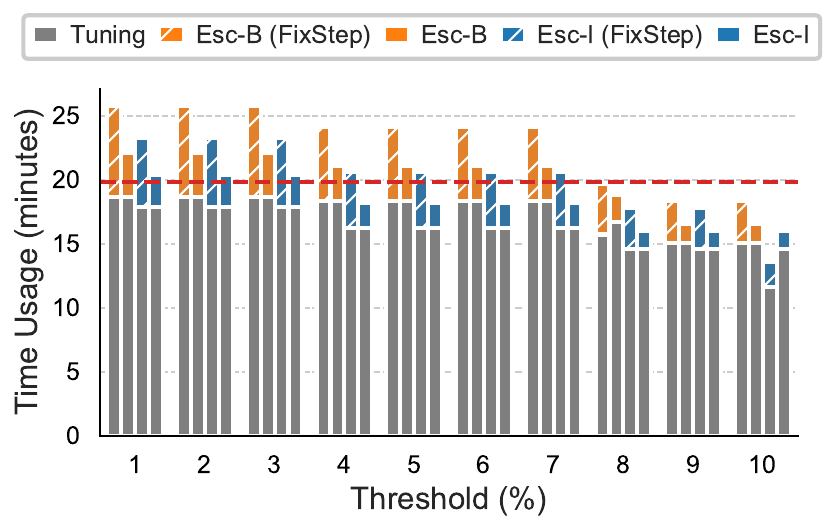}}
\subfigure[Improvement Loss]{ \label{fig:twophase_skip:job:k10:impr-loss}
    \includegraphics[width=0.49\columnwidth]{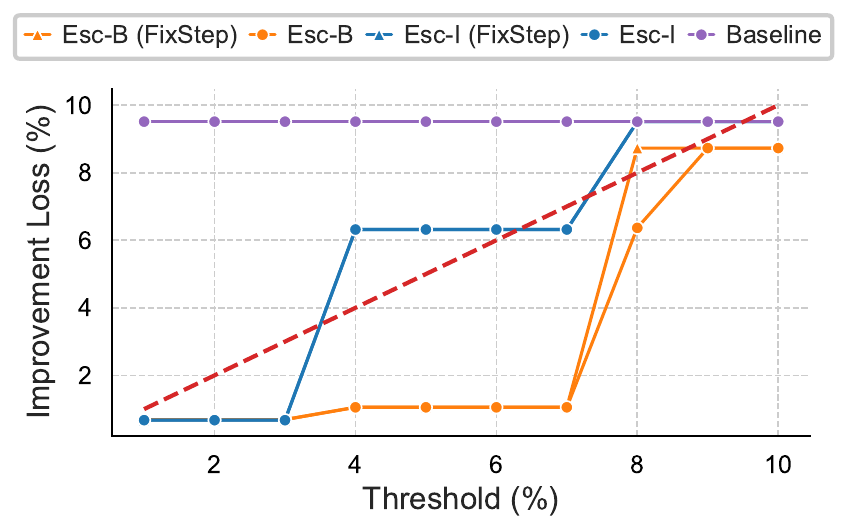}}
\subfigure[What-If Call Savings]{ \label{fig:twophase_skip:job:k10:call-save}
    \includegraphics[width=0.49\columnwidth]{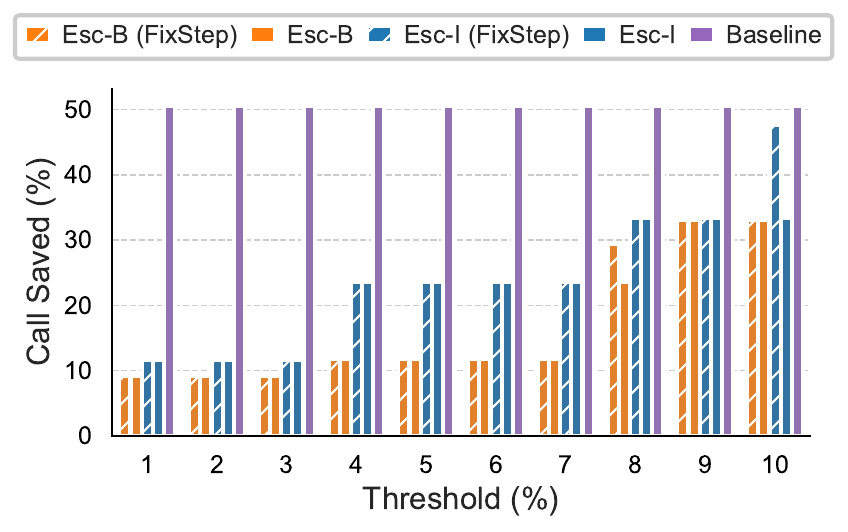}}
\subfigure[Learning Curve]{ \label{fig:twophase_skip:job:k10:lc}
    \includegraphics[width=0.49\columnwidth]{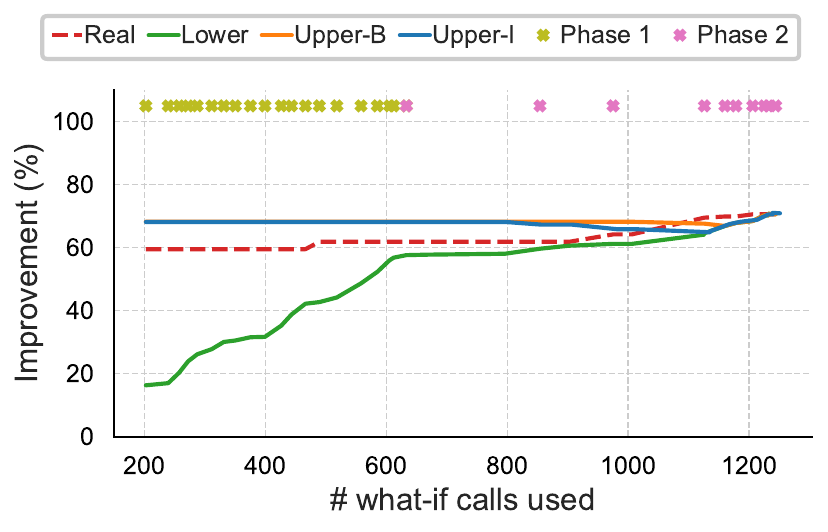}}
\vspace{-1.5em}
\caption{Two-phase greedy search (with Wii), JOB, $K=10$, $B=20k$}
\label{fig:twophase_skip:job:k10}
\vspace{-1em}
\end{figure*}

\begin{figure*}
\centering
\subfigure[Time Overhead]{ \label{fig:twophase_covskip:job:k10:overhead}
    \includegraphics[width=0.49\columnwidth]{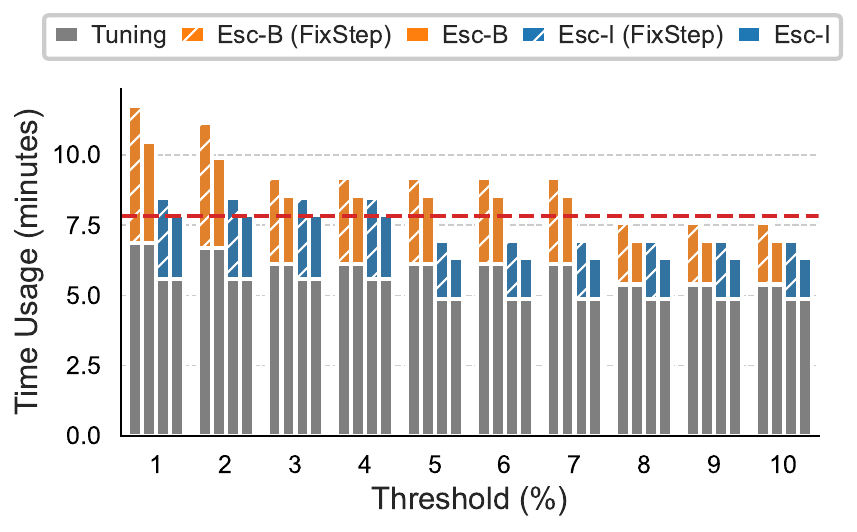}}
\subfigure[Improvement Loss]{ \label{fig:twophase_covskip:job:k10:impr-loss}
    \includegraphics[width=0.49\columnwidth]{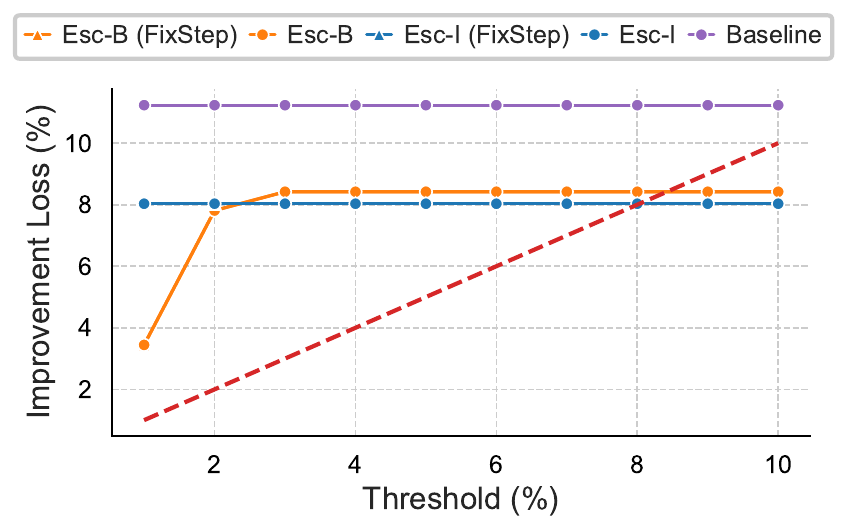}}
\subfigure[What-If Call Savings]{ \label{fig:twophase_covskip:job:k10:call-save}
    \includegraphics[width=0.49\columnwidth]{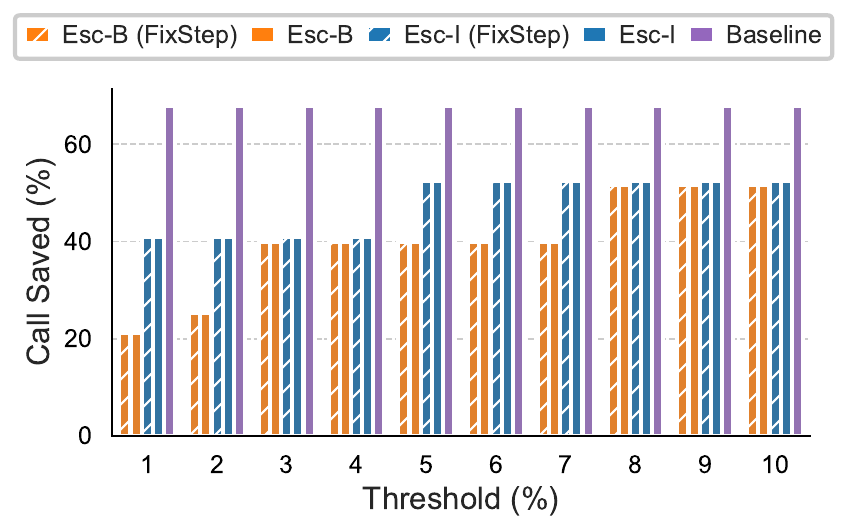}}
\subfigure[Learning Curve]{ \label{fig:twophase_covskip:job:k10:lc}
    \includegraphics[width=0.49\columnwidth]{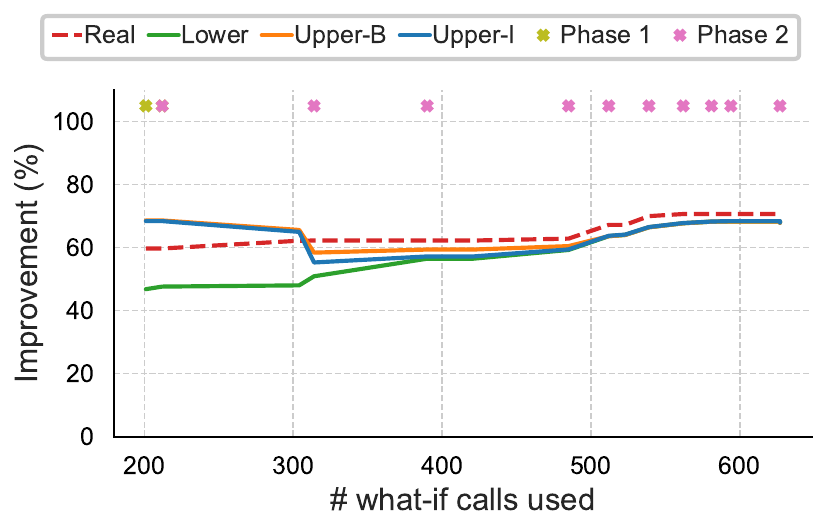}}
\vspace{-1.5em}
\caption{Two-phase greedy search (with Wii-Coverage), JOB, $K=10$, $B=20k$}
\label{fig:twophase_covskip:job:k10}
\vspace{-1em}
\end{figure*}

\begin{figure*}
\centering
\subfigure[Time Overhead]{ \label{fig:twophase:real-d:k10:overhead}
    \includegraphics[width=0.49\columnwidth]{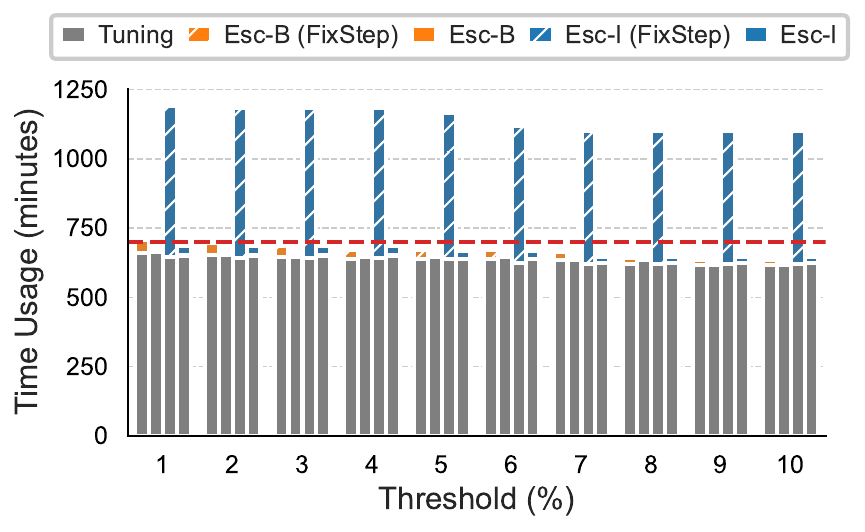}}
\subfigure[Improvement Loss]{ \label{fig:twophase:real-d:k10:impr-loss}
    \includegraphics[width=0.49\columnwidth]{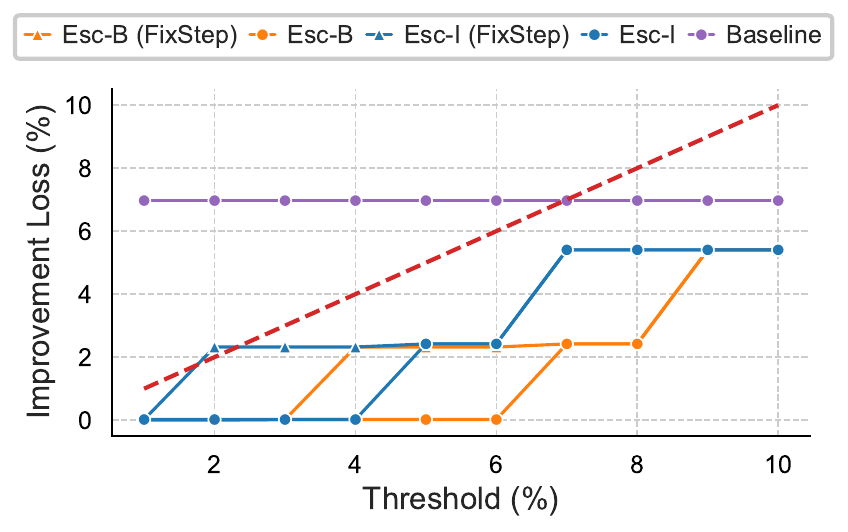}}
\subfigure[What-If Call Savings]{ \label{fig:twophase:real-d:k10:call-save}
    \includegraphics[width=0.49\columnwidth]{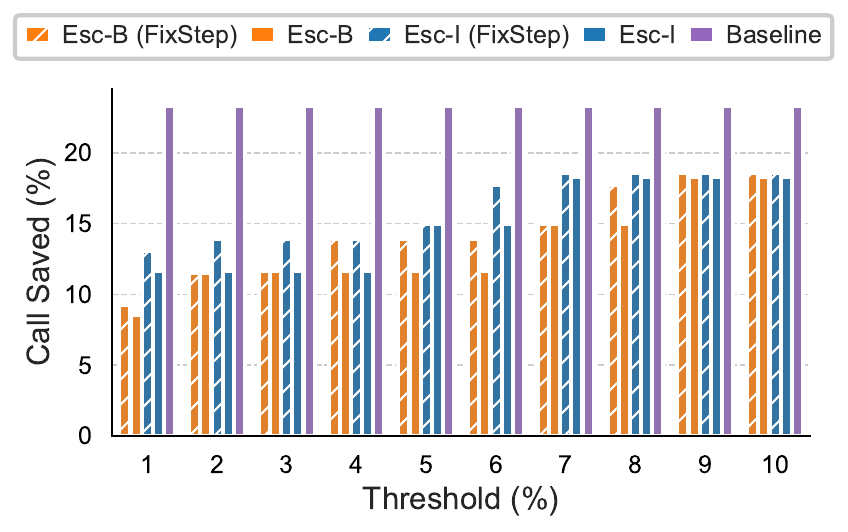}}
\subfigure[Learning Curve]{ \label{fig:twophase:real-d:k10:lc}
    \includegraphics[width=0.49\columnwidth]{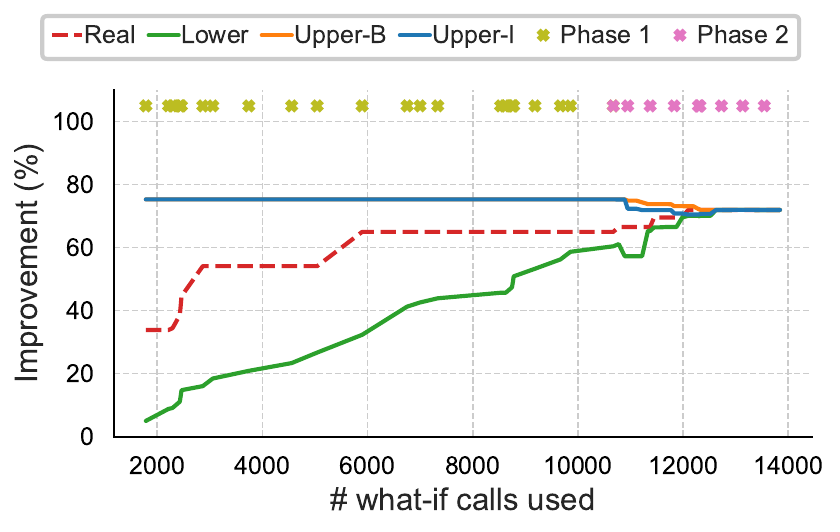}}
\vspace{-1.5em}
\caption{Two-phase greedy search, Real-D, $K=10$, $B=20k$}
\label{fig:twophase:real-d:k10}
\vspace{-1em}
\end{figure*}

\begin{figure*}
\centering
\subfigure[Time Overhead]{ \label{fig:twophase_skip:real-d:k10:overhead}
    \includegraphics[width=0.49\columnwidth]{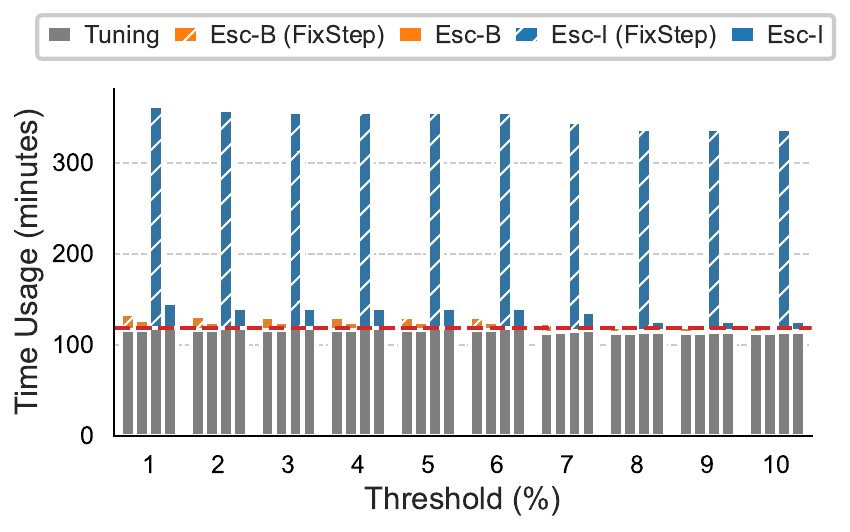}}
\subfigure[Improvement Loss]{ \label{fig:twophase_skip:real-d:k10:impr-loss}
    \includegraphics[width=0.49\columnwidth]{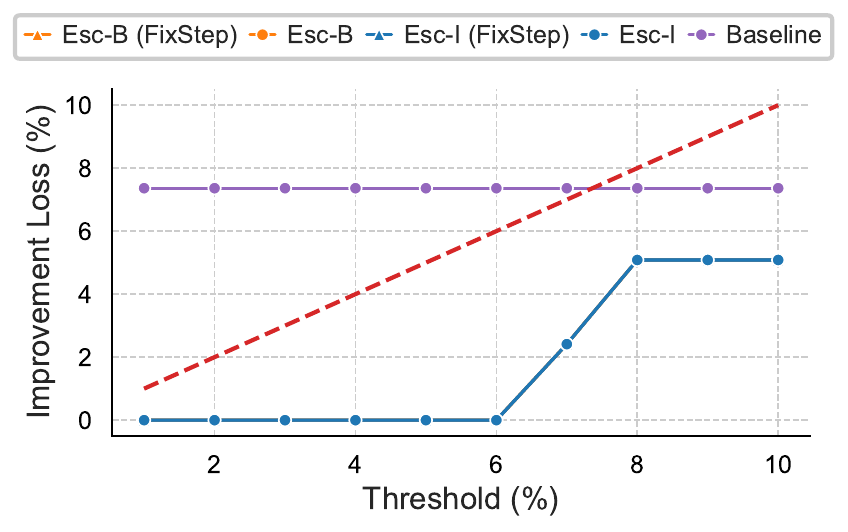}}
\subfigure[What-If Call Savings]{ \label{fig:twophase_skip:real-d:k10:call-save}
    \includegraphics[width=0.49\columnwidth]{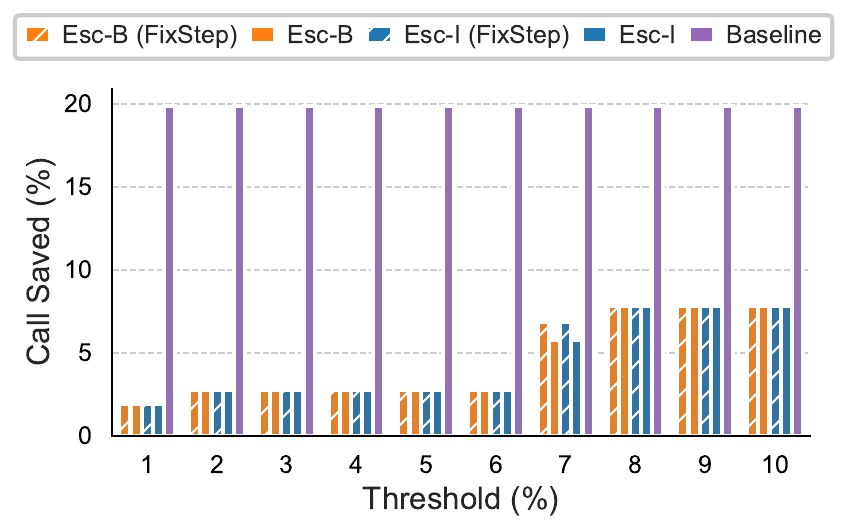}}
\subfigure[Learning Curve]{ \label{fig:twophase_skip:real-d:k10:lc}
    \includegraphics[width=0.49\columnwidth]{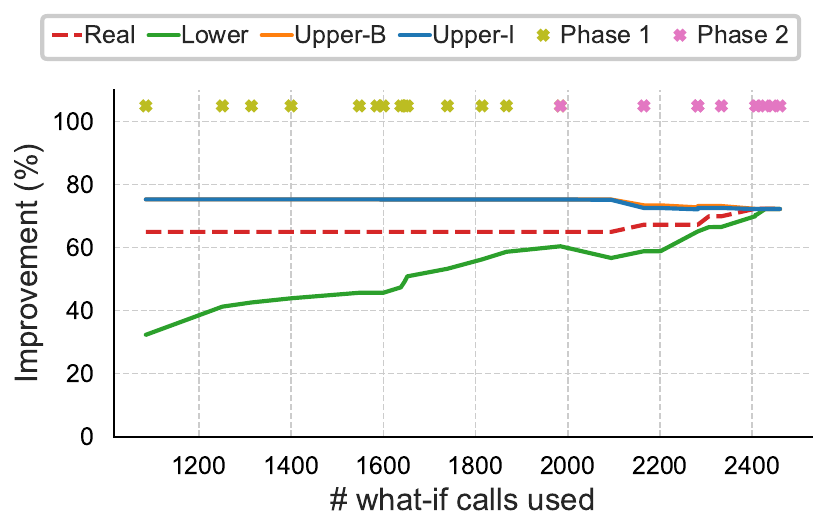}}
\vspace{-1.5em}
\caption{Two-phase greedy search (with Wii), Real-D, $K=10$, $B=20k$}
\label{fig:twophase_skip:real-d:k10}
\vspace{-1em}
\end{figure*}

\begin{figure*}
\centering
\subfigure[Time Overhead]{ \label{fig:twophase_covskip:real-d:k10:overhead}
    \includegraphics[width=0.49\columnwidth]{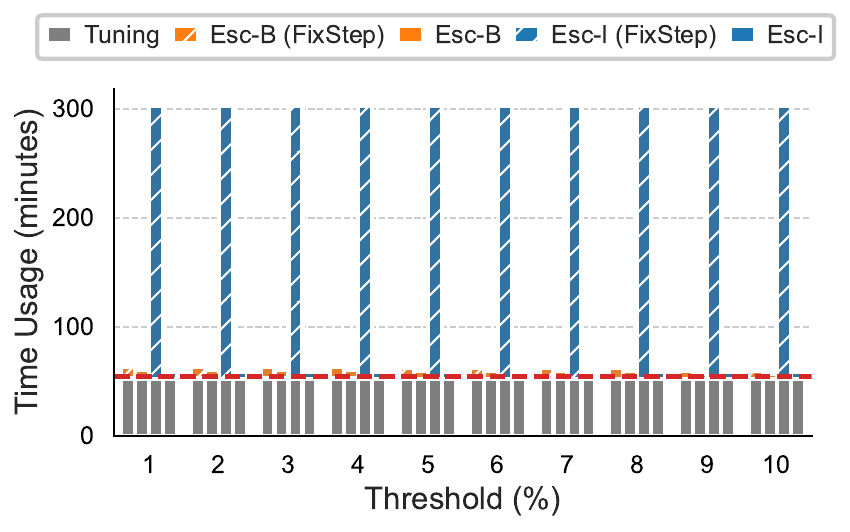}}
\subfigure[Improvement Loss]{ \label{fig:twophase_covskip:real-d:k10:impr-loss}
    \includegraphics[width=0.49\columnwidth]{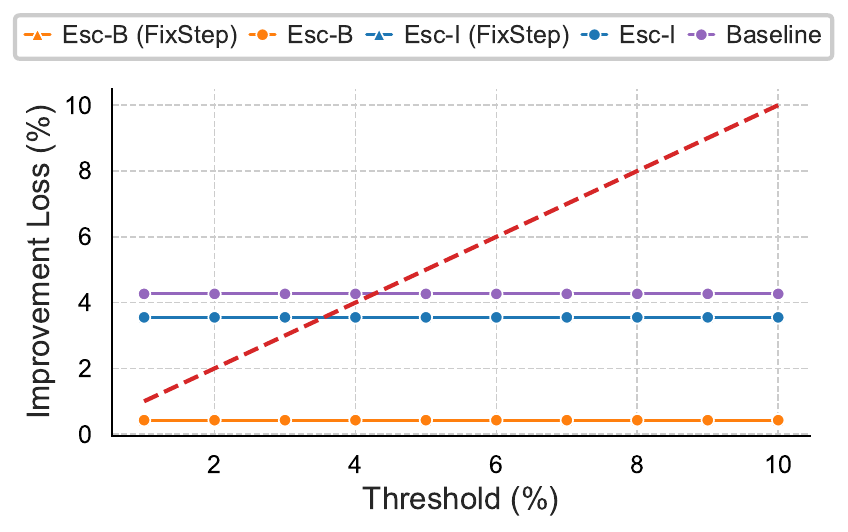}}
\subfigure[What-If Call Savings]{ \label{fig:twophase_covskip:real-d:k10:call-save}
    \includegraphics[width=0.49\columnwidth]{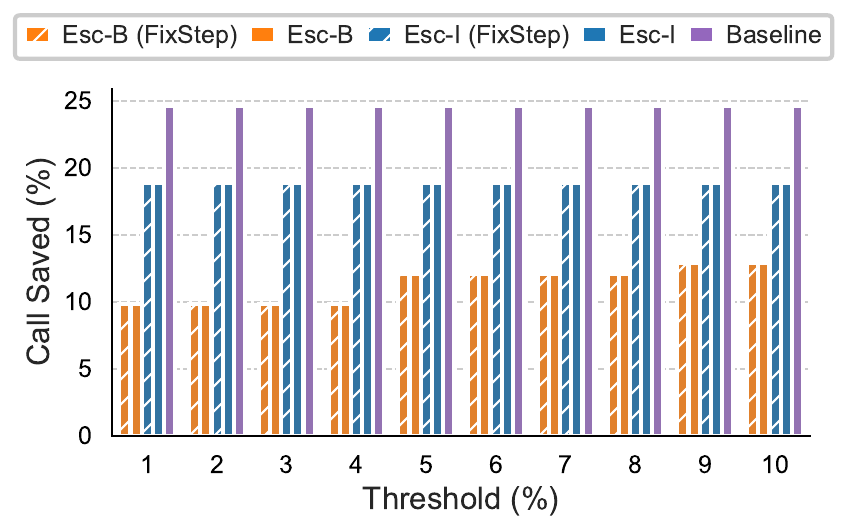}}
\subfigure[Learning Curve]{ \label{fig:twophase_covskip:real-d:k10:lc}
    \includegraphics[width=0.49\columnwidth]{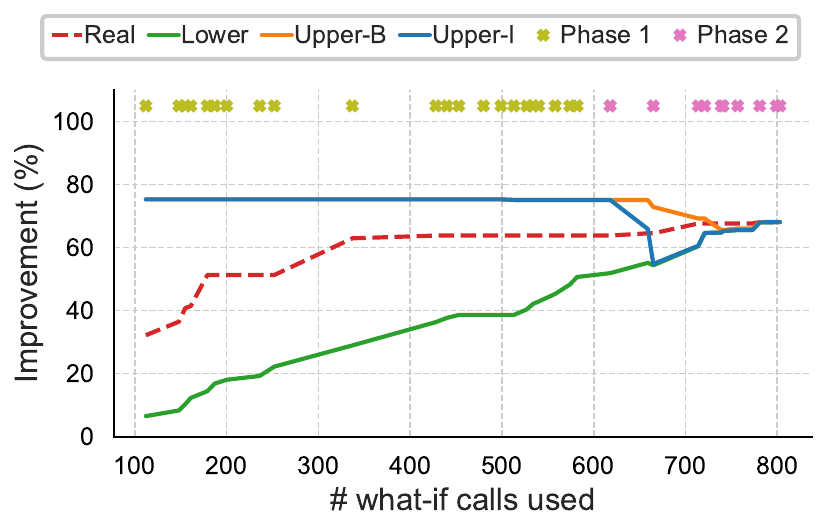}}
\vspace{-1.5em}
\caption{Two-phase greedy search (with Wii-Coverage), Real-D, $K=10$, $B=20k$}
\label{fig:twophase_covskip:real-d:k10}
\vspace{-1em}
\end{figure*}

\begin{figure*}
\centering
\subfigure[Time Overhead]{ \label{fig:twophase:real-m:k10:overhead}
    \includegraphics[width=0.49\columnwidth]{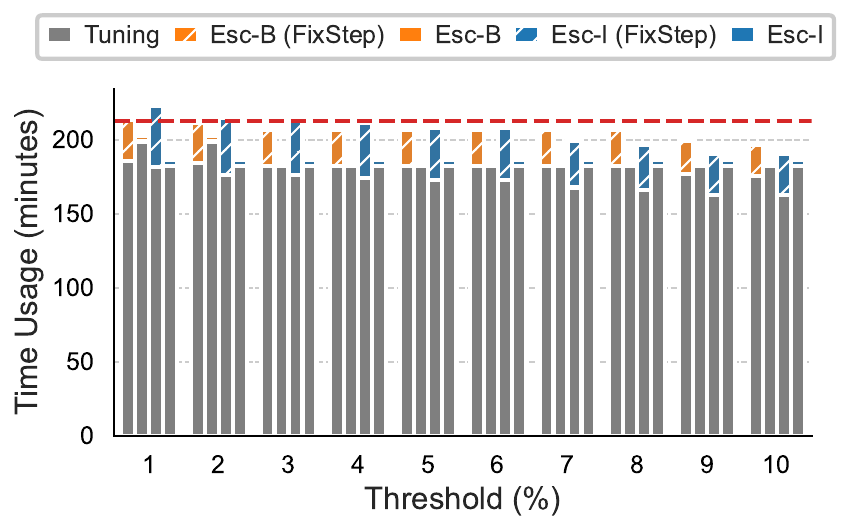}}
\subfigure[Improvement Loss]{ \label{fig:twophase:real-m:k10:impr-loss}
    \includegraphics[width=0.49\columnwidth]{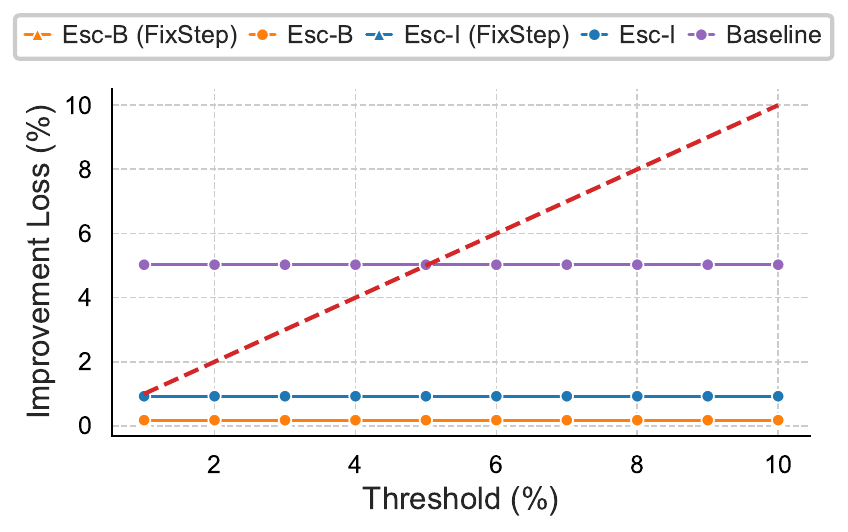}}
\subfigure[What-If Call Savings]{ \label{fig:twophase:real-m:k10:call-save}
    \includegraphics[width=0.49\columnwidth]{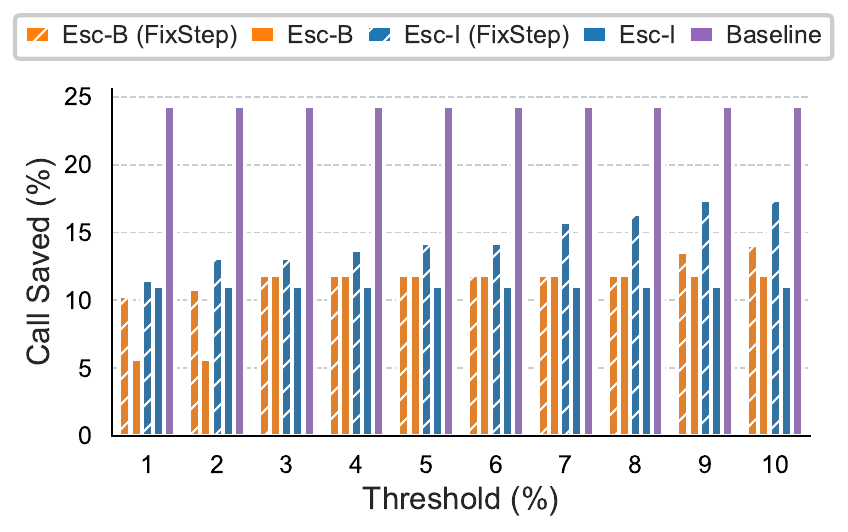}}
\subfigure[Learning Curve]{ \label{fig:twophase:real-m:k10:lc}
    \includegraphics[width=0.49\columnwidth]{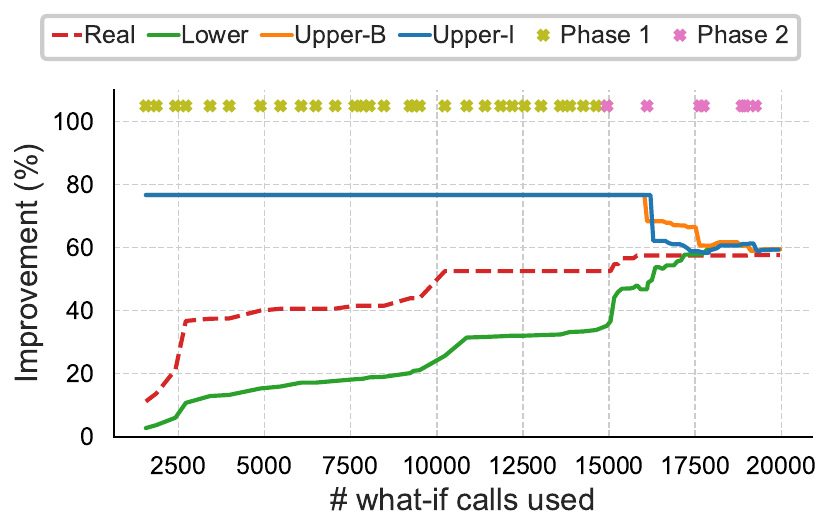}}
\vspace{-1.5em}
\caption{Two-phase greedy search, Real-M, $K=10$, $B=20k$}
\label{fig:twophase:real-m:k10}
\vspace{-1em}
\end{figure*}

\begin{figure*}
\centering
\subfigure[Time Overhead]{ \label{fig:twophase_skip:real-m:k10:overhead}
    \includegraphics[width=0.49\columnwidth]{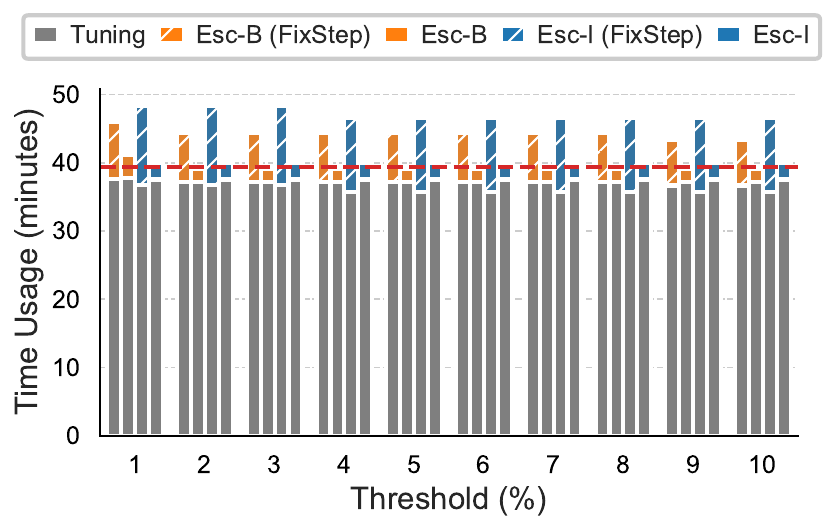}}
\subfigure[Improvement Loss]{ \label{fig:twophase_skip:real-m:k10:impr-loss}
    \includegraphics[width=0.49\columnwidth]{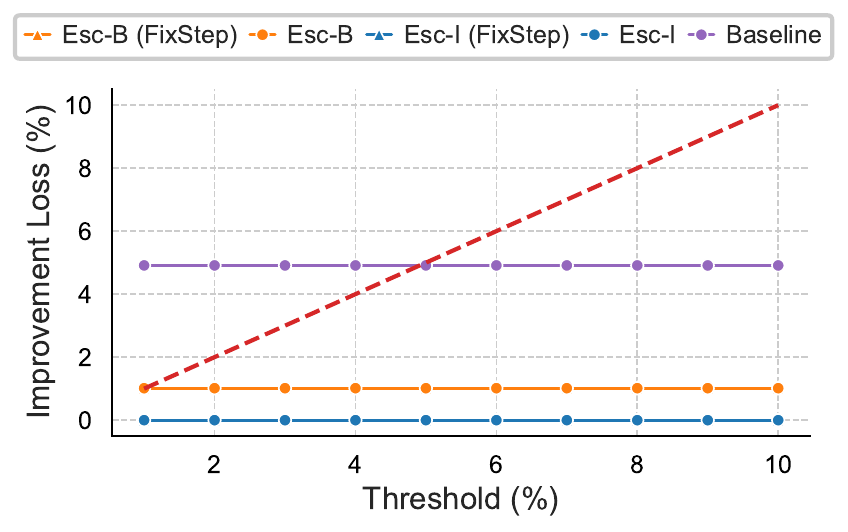}}
\subfigure[What-If Call Savings]{ \label{fig:twophase_skip:real-m:k10:call-save}
    \includegraphics[width=0.49\columnwidth]{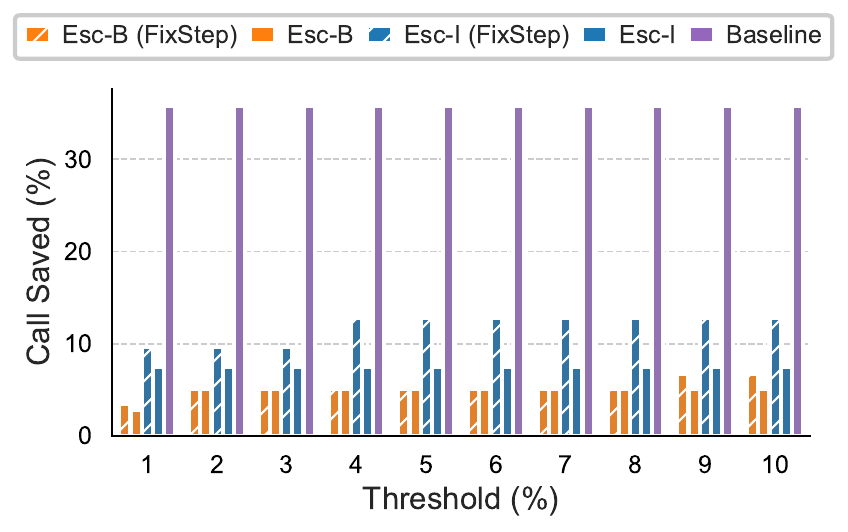}}
\subfigure[Learning Curve]{ \label{fig:twophase_skip:real-m:k10:lc}
    \includegraphics[width=0.49\columnwidth]{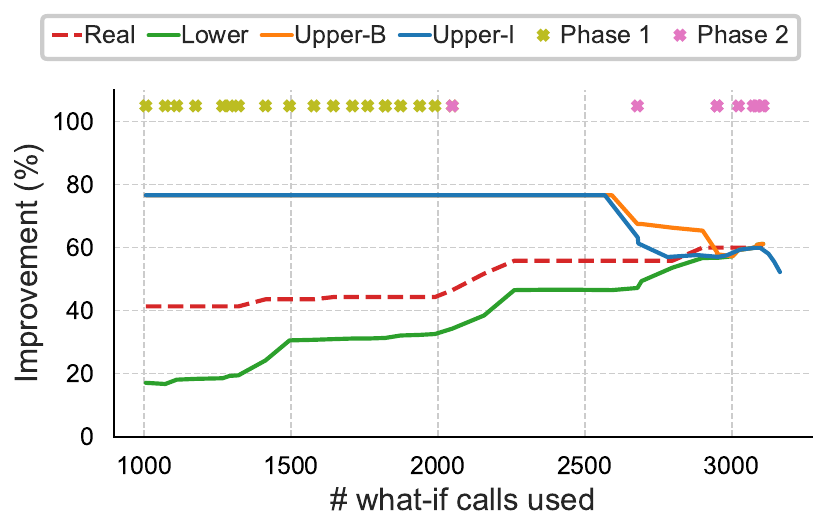}}
\vspace{-1.5em}
\caption{Two-phase greedy search (with Wii), Real-M, $K=10$, $B=20k$}
\label{fig:twophase_skip:real-m:k10}
\vspace{-1em}
\end{figure*}

\begin{figure*}
\centering
\subfigure[Time Overhead]{ \label{fig:twophase_covskip:real-m:k10:overhead}
    \includegraphics[width=0.49\columnwidth]{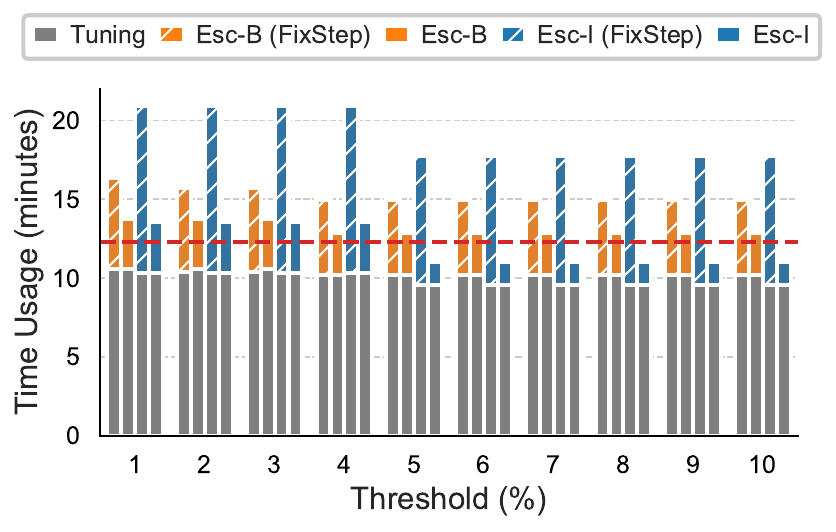}}
\subfigure[Improvement Loss]{ \label{fig:twophase_covskip:real-m:k10:impr-loss}
    \includegraphics[width=0.49\columnwidth]{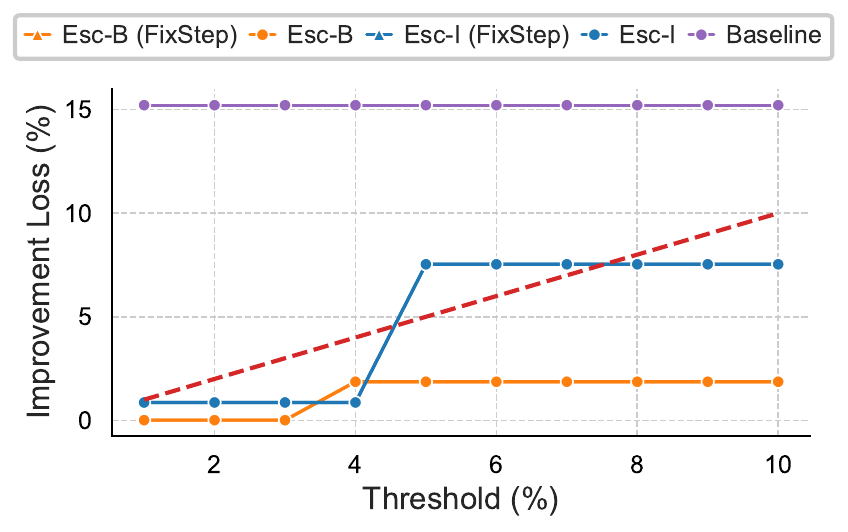}}
\subfigure[What-If Call Savings]{ \label{fig:twophase_covskip:real-m:k10:call-save}
    \includegraphics[width=0.49\columnwidth]{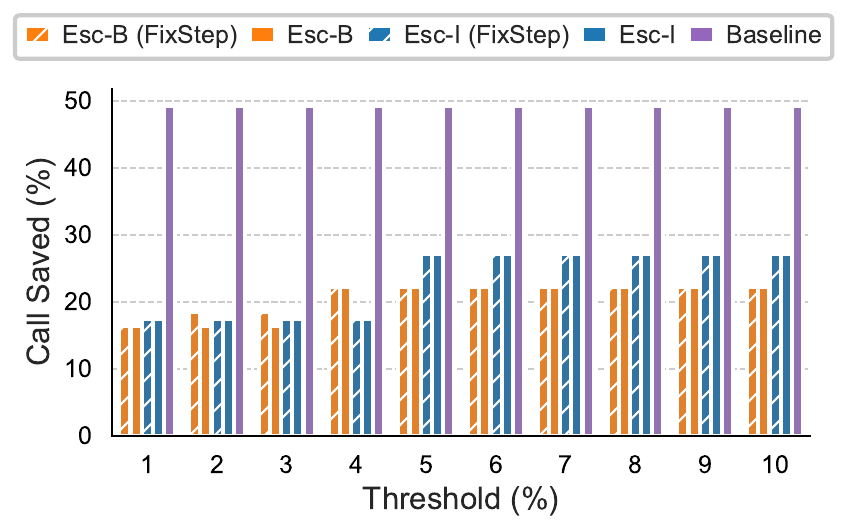}}
\subfigure[Learning Curve]{ \label{fig:twophase_covskip:real-m:k10:lc}
    \includegraphics[width=0.49\columnwidth]{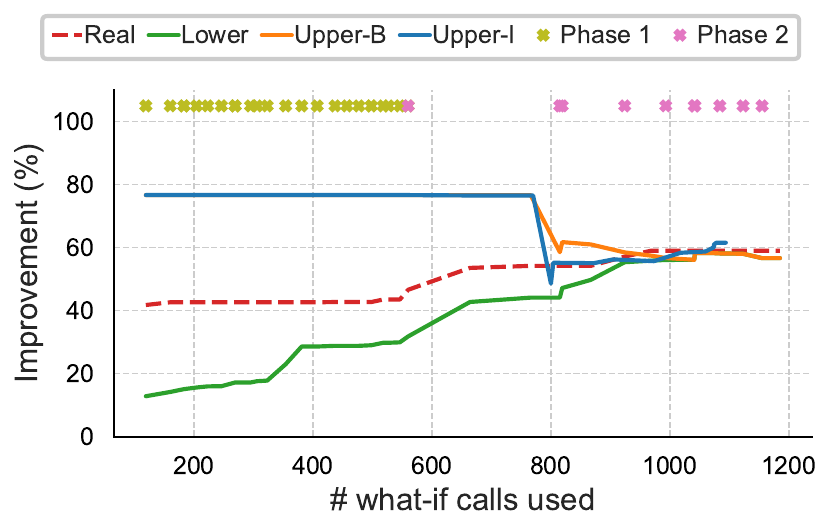}}
\vspace{-1.5em}
\caption{Two-phase greedy search (with Wii-Coverage), Real-M, $K=10$, $B=20k$}
\label{fig:twophase_covskip:real-m:k10}
\vspace{-1em}
\end{figure*}

\begin{figure*}
\centering
\subfigure[Time Overhead]{ \label{fig:mcts:tpch:k10:overhead}
    \includegraphics[width=0.49\columnwidth]{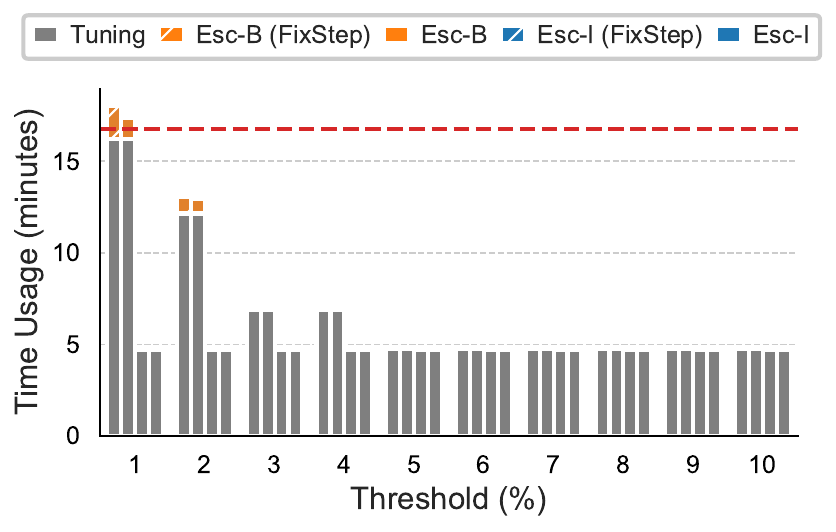}}
\subfigure[Improvement Loss]{ \label{fig:mcts:tpch:k10:impr-loss}
    \includegraphics[width=0.49\columnwidth]{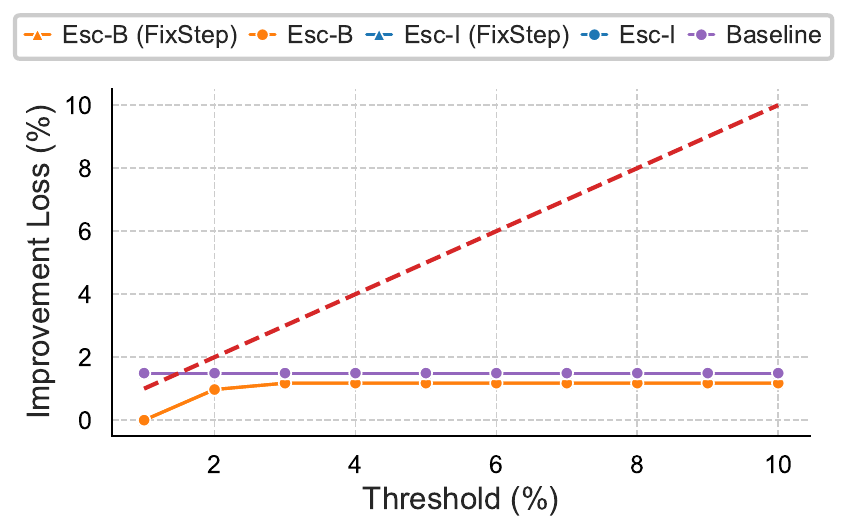}}
\subfigure[What-If Call Savings]{ \label{fig:mcts:tpch:k10:call-save}
    \includegraphics[width=0.49\columnwidth]{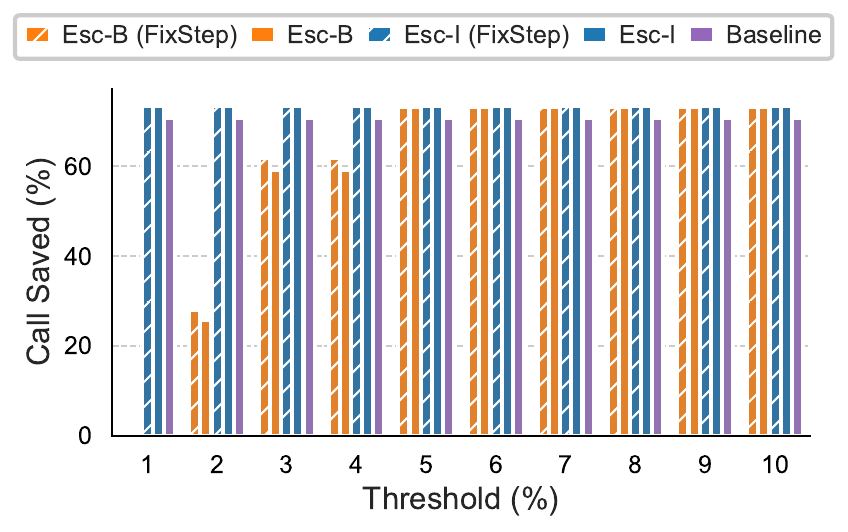}}
\subfigure[Learning Curve]{ \label{fig:mcts:tpch:k10:lc}
    \includegraphics[width=0.49\columnwidth]{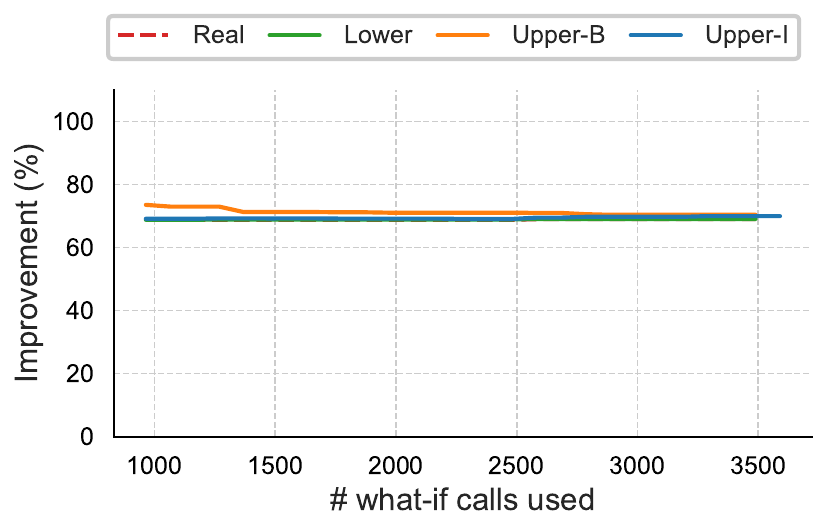}}
\vspace{-1.5em}
\caption{MCTS, TPC-H, $K=10$, $B=20k$}
\label{fig:mcts:tpch:k10}
\vspace{-1em}
\end{figure*}

\begin{figure*}
\centering
\subfigure[Time Overhead]{ \label{fig:mcts_skip:tpch:k10:overhead}
    \includegraphics[width=0.49\columnwidth]{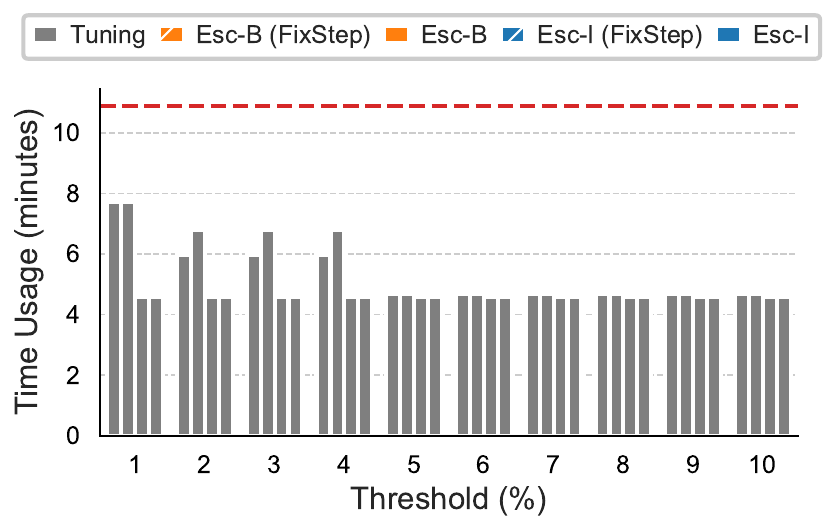}}
\subfigure[Improvement Loss]{ \label{fig:mcts_skip:tpch:k10:impr-loss}
    \includegraphics[width=0.49\columnwidth]{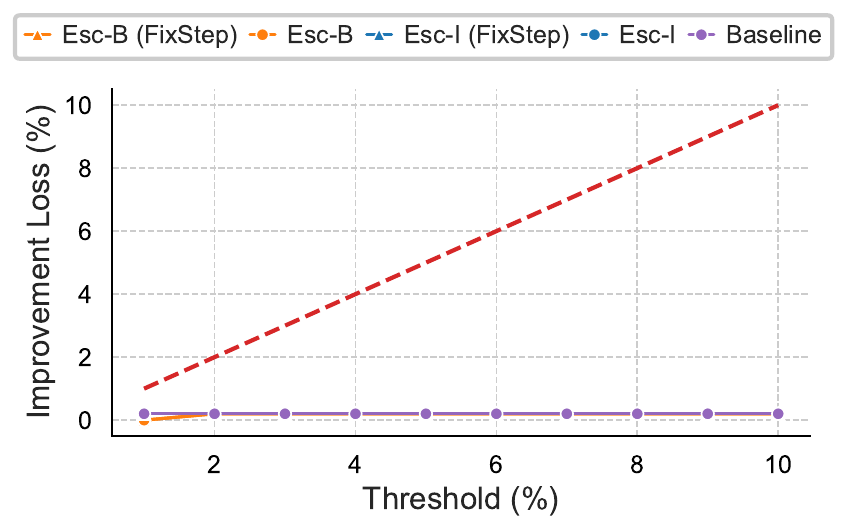}}
\subfigure[What-If Call Savings]{ \label{fig:mcts_skip:tpch:k10:call-save}
    \includegraphics[width=0.49\columnwidth]{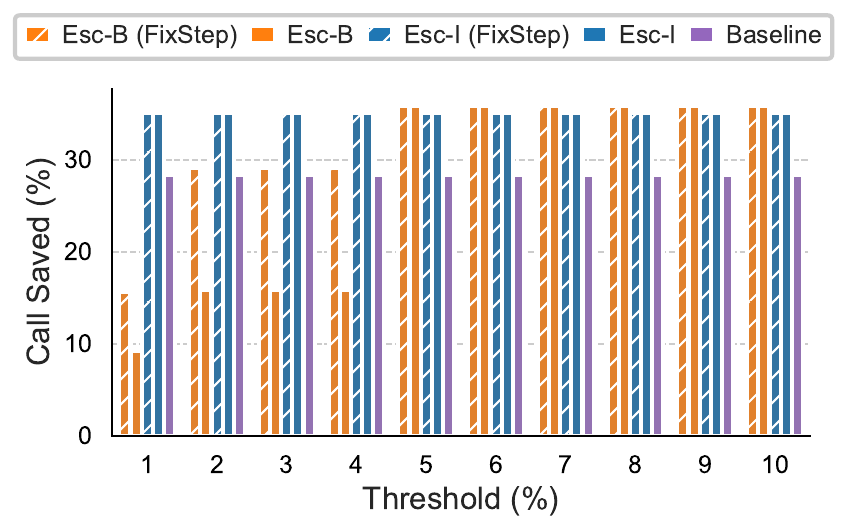}}
\subfigure[Learning Curve]{ \label{fig:mcts_skip:tpch:k10:lc}
    \includegraphics[width=0.49\columnwidth]{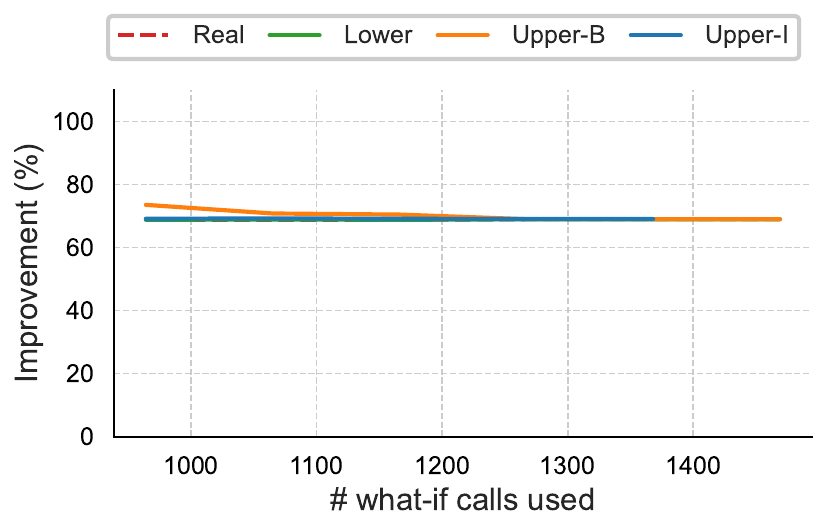}}
\vspace{-1.5em}
\caption{MCTS (with Wii), TPC-H, $K=10$, $B=20k$}
\label{fig:mcts_skip:tpch:k10}
\vspace{-1em}
\end{figure*}

\begin{figure*}
\centering
\subfigure[Time Overhead]{ \label{fig:mcts_covskip:tpch:k10:overhead}
    \includegraphics[width=0.49\columnwidth]{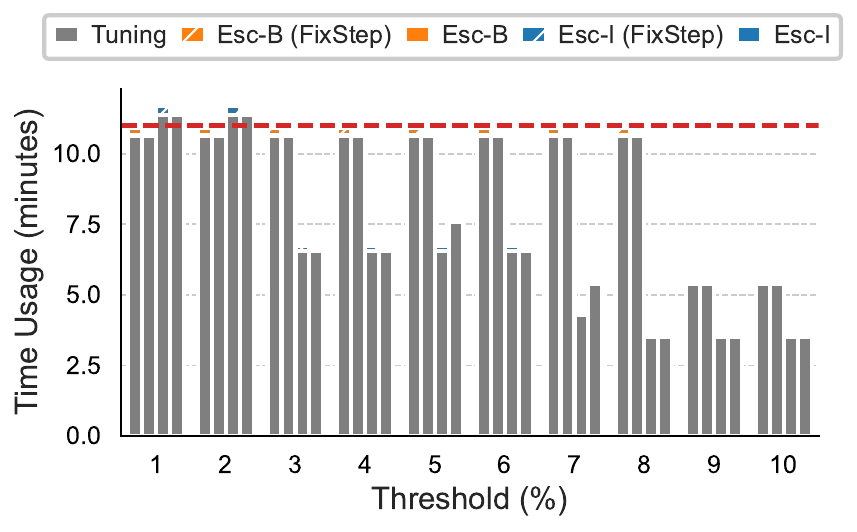}}
\subfigure[Improvement Loss]{ \label{fig:mcts_covskip:tpch:k10:impr-loss}
    \includegraphics[width=0.49\columnwidth]{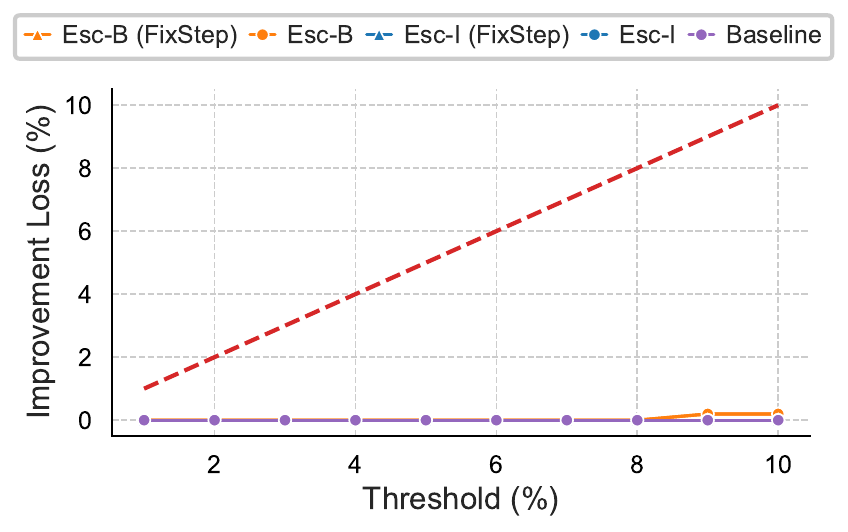}}
\subfigure[What-If Call Savings]{ \label{fig:mcts_covskip:tpch:k10:call-save}
    \includegraphics[width=0.49\columnwidth]{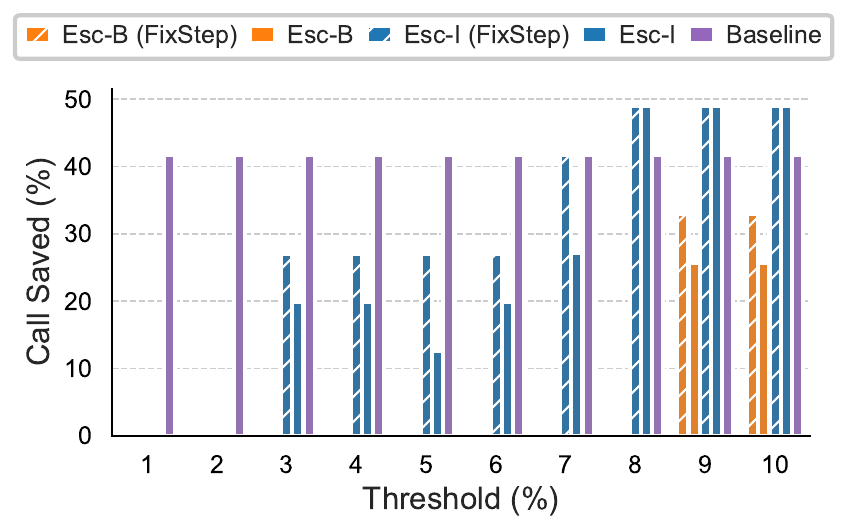}}
\subfigure[Learning Curve]{ \label{fig:mcts_covskip:tpch:k10:lc}
    \includegraphics[width=0.49\columnwidth]{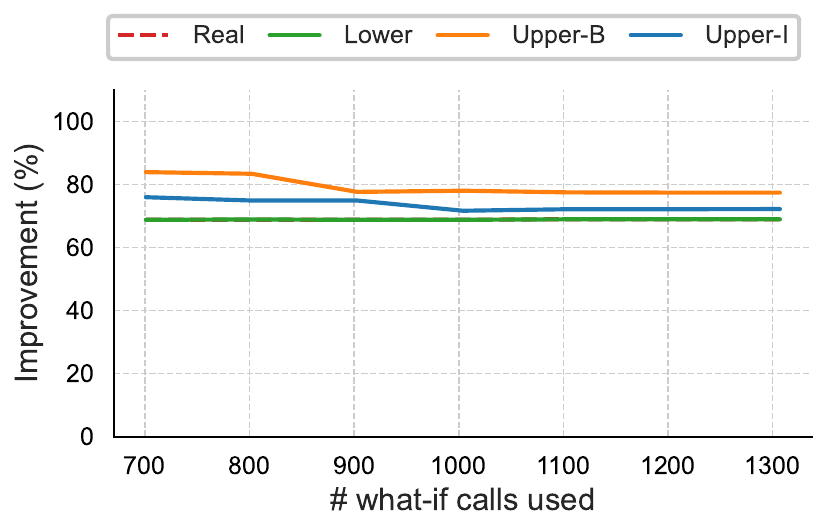}}
\vspace{-1.5em}
\caption{MCTS (with Wii-Coverage), TPC-H, $K=10$, $B=20k$}
\label{fig:mcts_covskip:tpch:k10}
\vspace{-1em}
\end{figure*}

\begin{figure*}
\centering
\subfigure[Time Overhead]{ \label{fig:mcts:tpcds:k10:overhead}
    \includegraphics[width=0.49\columnwidth]{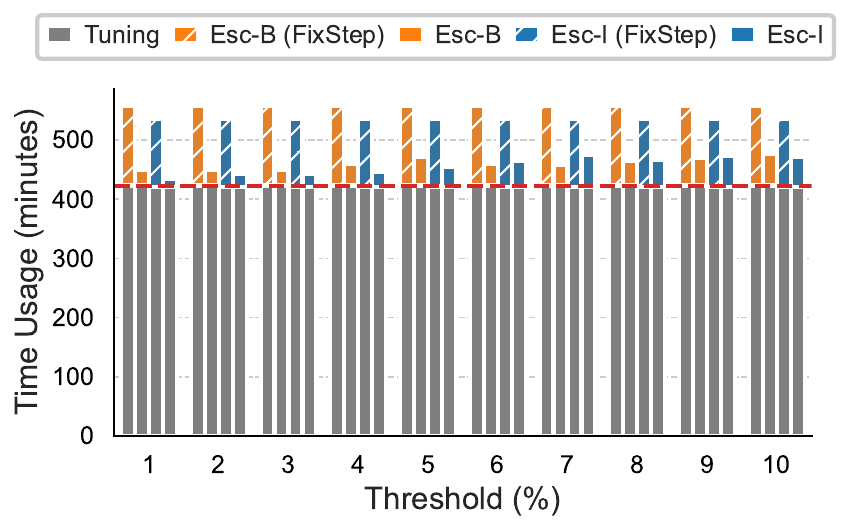}}
\subfigure[Improvement Loss]{ \label{fig:mcts:tpcds:k10:impr-loss}
    \includegraphics[width=0.49\columnwidth]{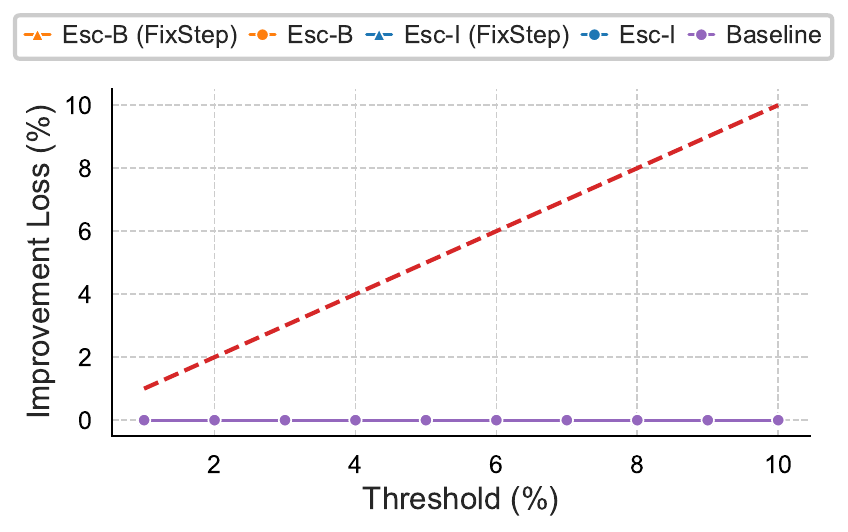}}
\subfigure[What-If Call Savings]{ \label{fig:mcts:tpcds:k10:call-save}
    \includegraphics[width=0.49\columnwidth]{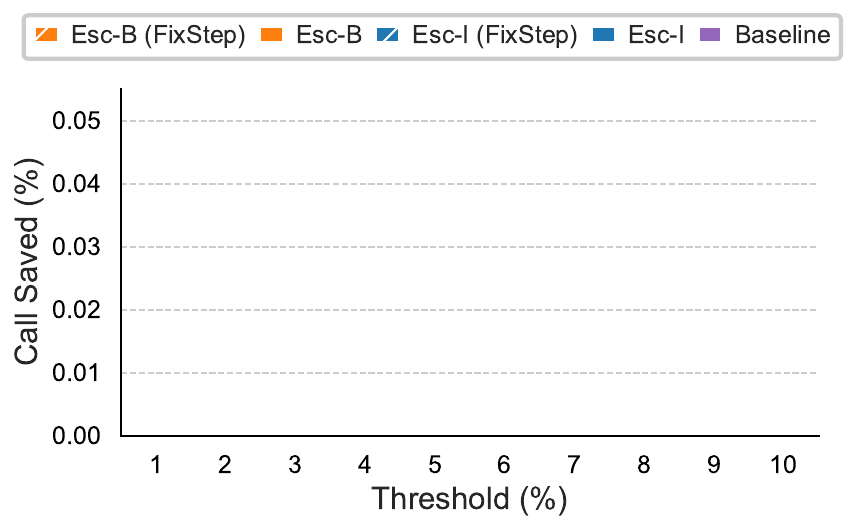}}
\subfigure[Learning Curve]{ \label{fig:mcts:tpcds:k10:lc}
    \includegraphics[width=0.49\columnwidth]{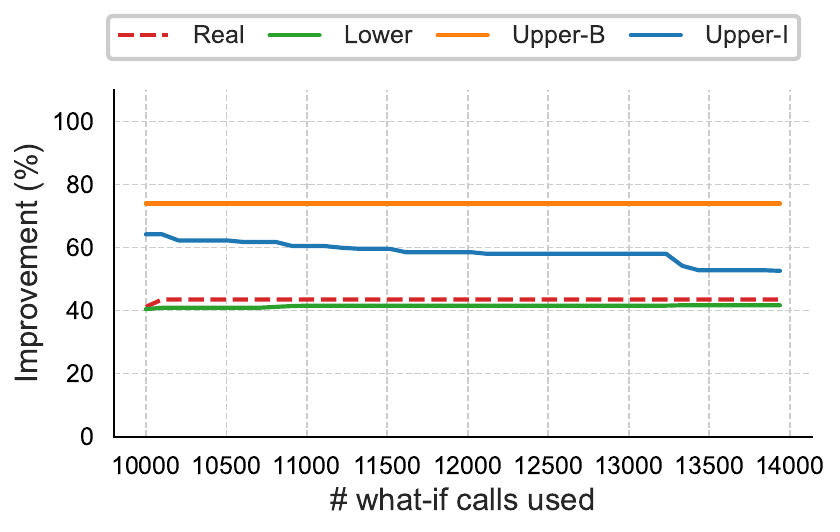}}
\vspace{-1.5em}
\caption{MCTS, TPC-DS, $K=10$, $B=20k$}
\label{fig:mcts:tpcds:k10}
\vspace{-1em}
\end{figure*}

\begin{figure*}
\centering
\subfigure[Time Overhead]{ \label{fig:mcts_skip:tpcds:k10:overhead}
    \includegraphics[width=0.49\columnwidth]{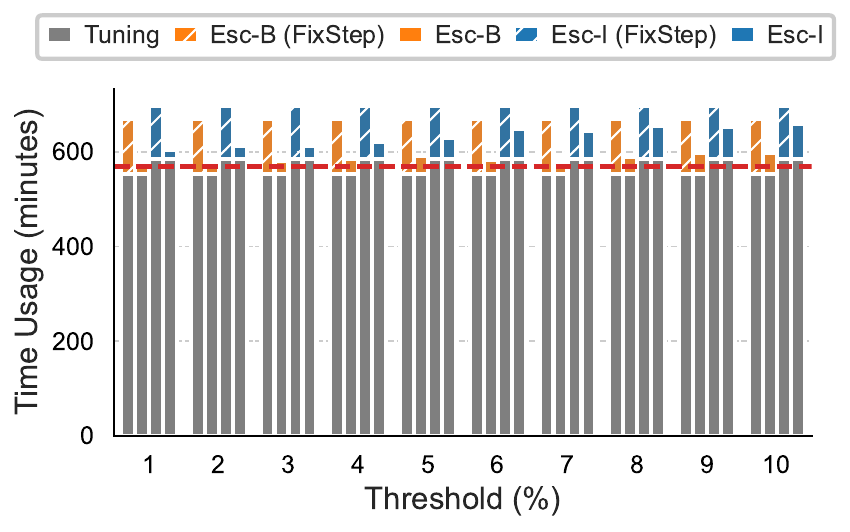}}
\subfigure[Improvement Loss]{ \label{fig:mcts_skip:tpcds:k10:impr-loss}
    \includegraphics[width=0.49\columnwidth]{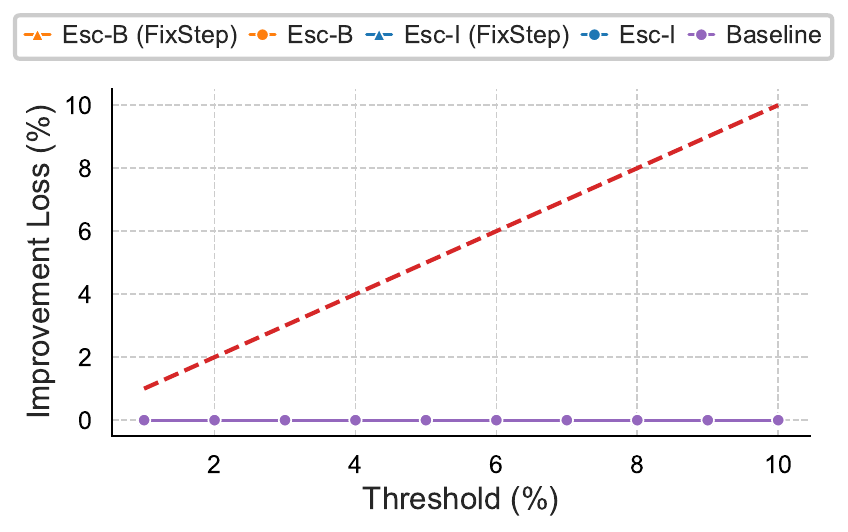}}
\subfigure[What-If Call Savings]{ \label{fig:mcts_skip:tpcds:k10:call-save}
    \includegraphics[width=0.49\columnwidth]{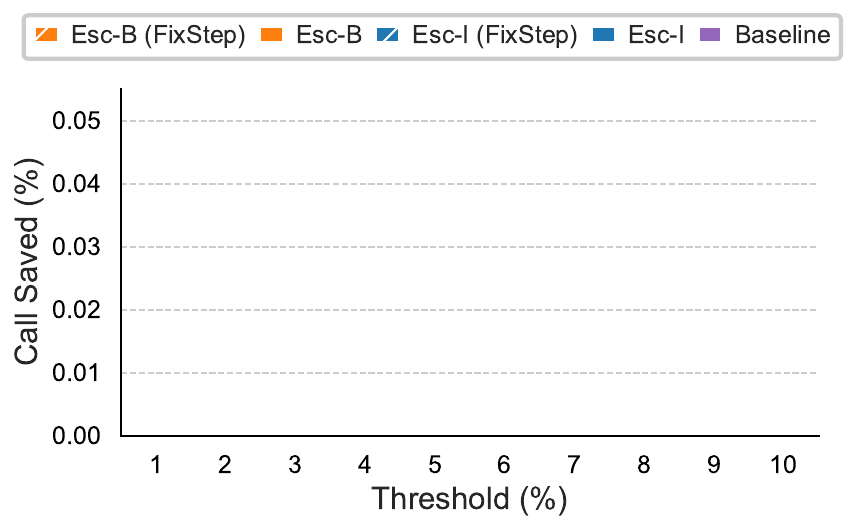}}
\subfigure[Learning Curve]{ \label{fig:mcts_skip:tpcds:k10:lc}
    \includegraphics[width=0.49\columnwidth]{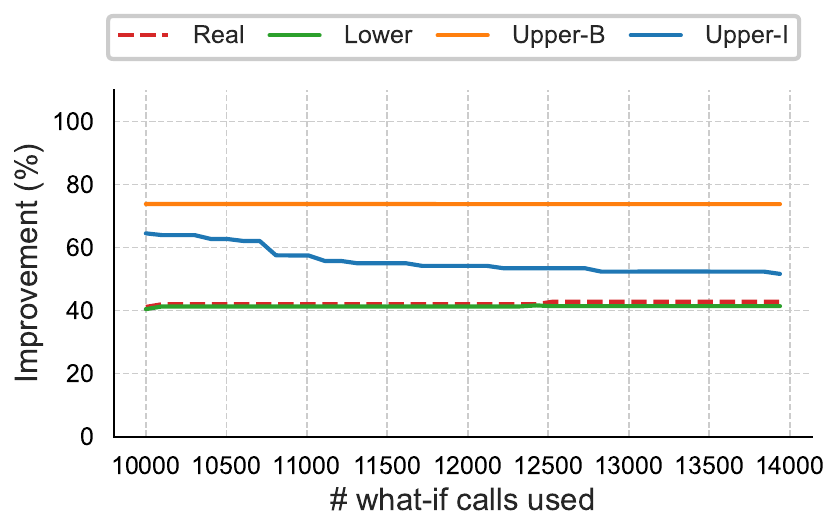}}
\vspace{-1.5em}
\caption{MCTS (with Wii), TPC-DS, $K=10$, $B=20k$}
\label{fig:mcts_skip:tpcds:k10}
\vspace{-1em}
\end{figure*}

\begin{figure*}
\centering
\subfigure[Time Overhead]{ \label{fig:mcts_covskip:tpcds:k10:overhead}
    \includegraphics[width=0.49\columnwidth]{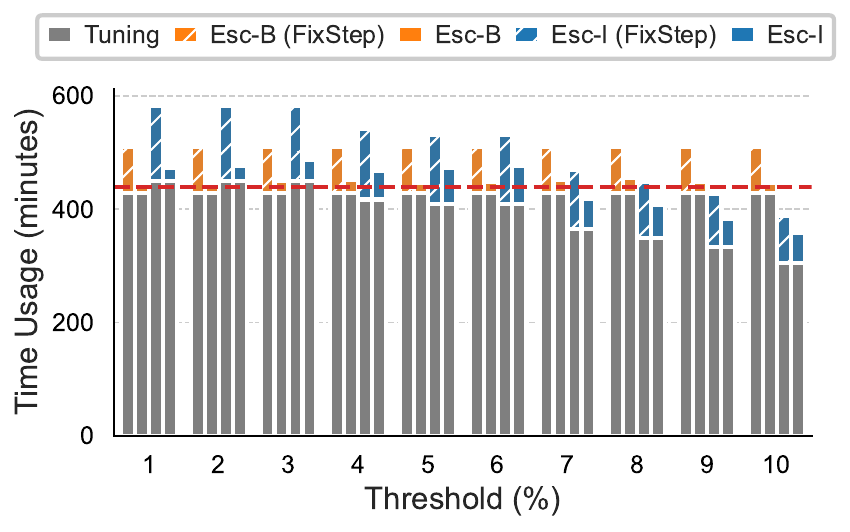}}
\subfigure[Improvement Loss]{ \label{fig:mcts_covskip:tpcds:k10:impr-loss}
    \includegraphics[width=0.49\columnwidth]{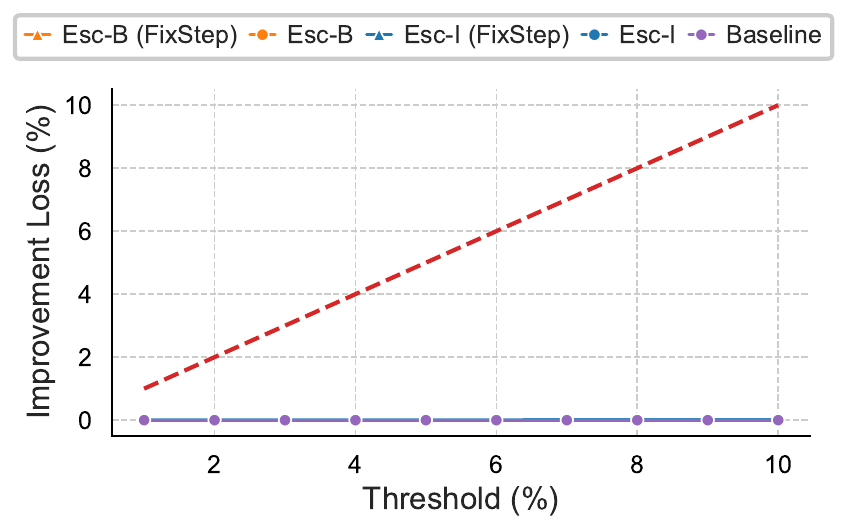}}
\subfigure[What-If Call Savings]{ \label{fig:mcts_covskip:tpcds:k10:call-save}
    \includegraphics[width=0.49\columnwidth]{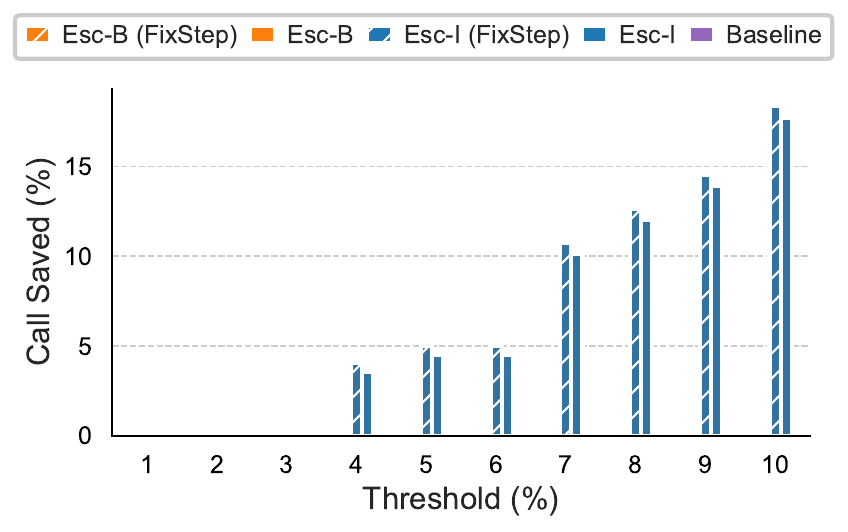}}
\subfigure[Learning Curve]{ \label{fig:mcts_covskip:tpcds:k10:lc}
    \includegraphics[width=0.49\columnwidth]{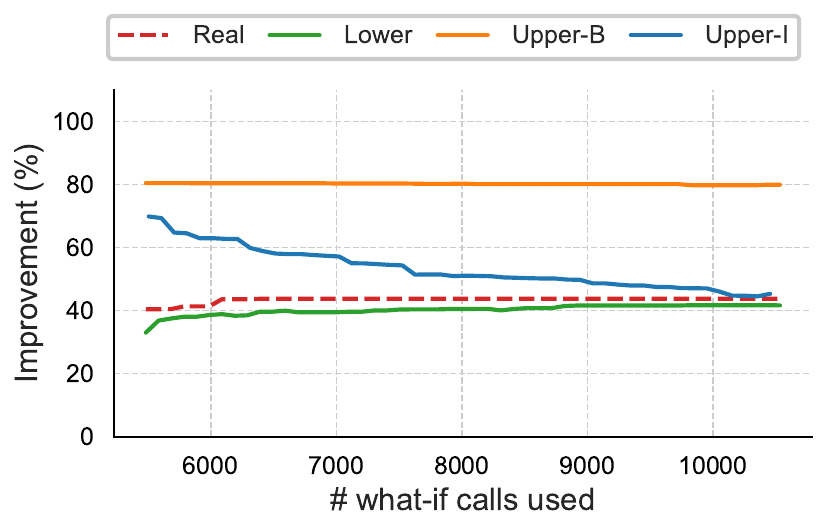}}
\vspace{-1.5em}
\caption{MCTS (with Wii-Coverage), TPC-DS, $K=10$, $B=20k$}
\label{fig:mcts_covskip:tpcds:k10}
\vspace{-1em}
\end{figure*}

\begin{figure*}
\centering
\subfigure[Time Overhead]{ \label{fig:mcts:job:k10:overhead}
    \includegraphics[width=0.49\columnwidth]{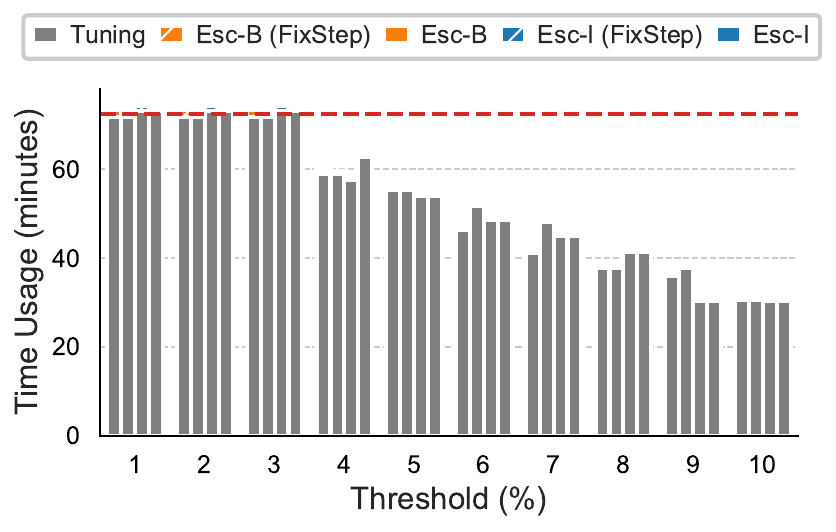}}
\subfigure[Improvement Loss]{ \label{fig:mcts:job:k10:impr-loss}
    \includegraphics[width=0.49\columnwidth]{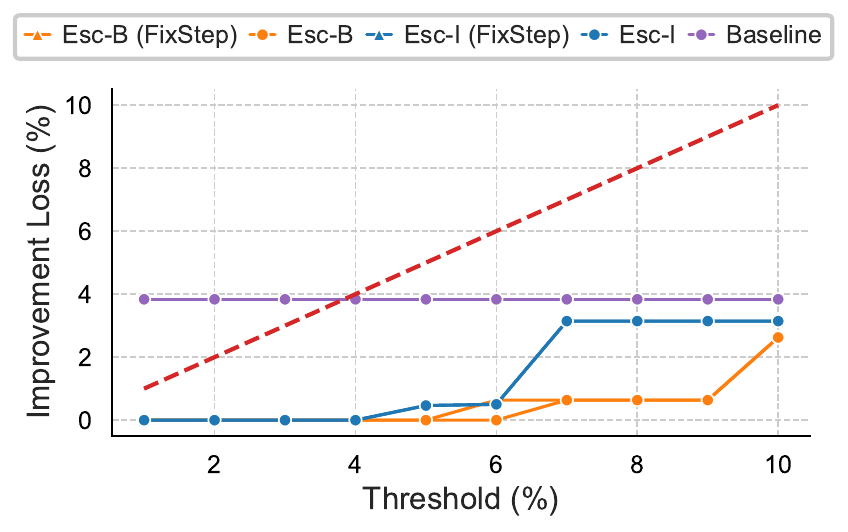}}
\subfigure[What-If Call Savings]{ \label{fig:mcts:job:k10:call-save}
    \includegraphics[width=0.49\columnwidth]{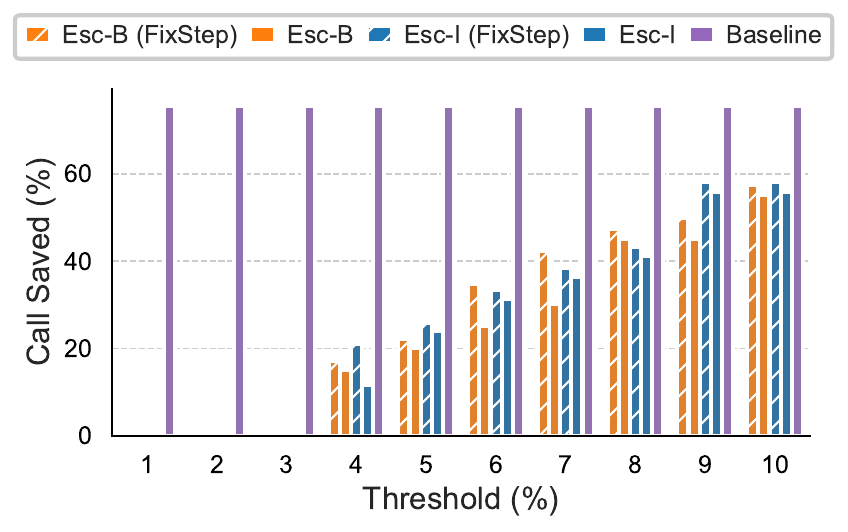}}
\subfigure[Learning Curve]{ \label{fig:mcts:job:k10:lc}
    \includegraphics[width=0.49\columnwidth]{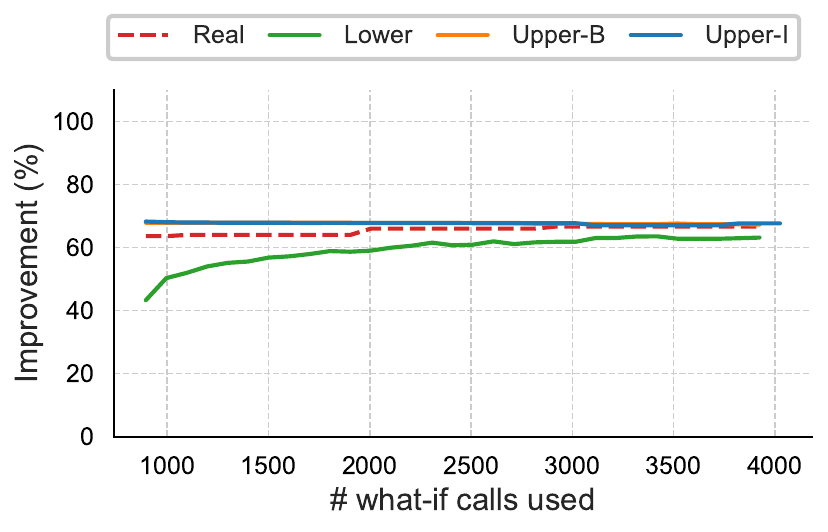}}
\vspace{-1.5em}
\caption{MCTS, JOB, $K=10$, $B=20k$}
\label{fig:mcts:job:k10}
\vspace{-1em}
\end{figure*}

\begin{figure*}
\centering
\subfigure[Time Overhead]{ \label{fig:mcts_skip:job:k10:overhead}
    \includegraphics[width=0.49\columnwidth]{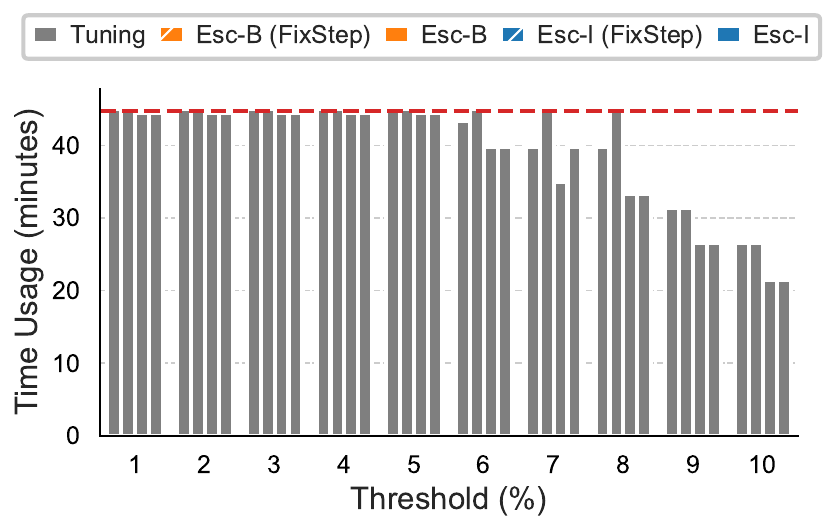}}
\subfigure[Improvement Loss]{ \label{fig:mcts_skip:job:k10:impr-loss}
    \includegraphics[width=0.49\columnwidth]{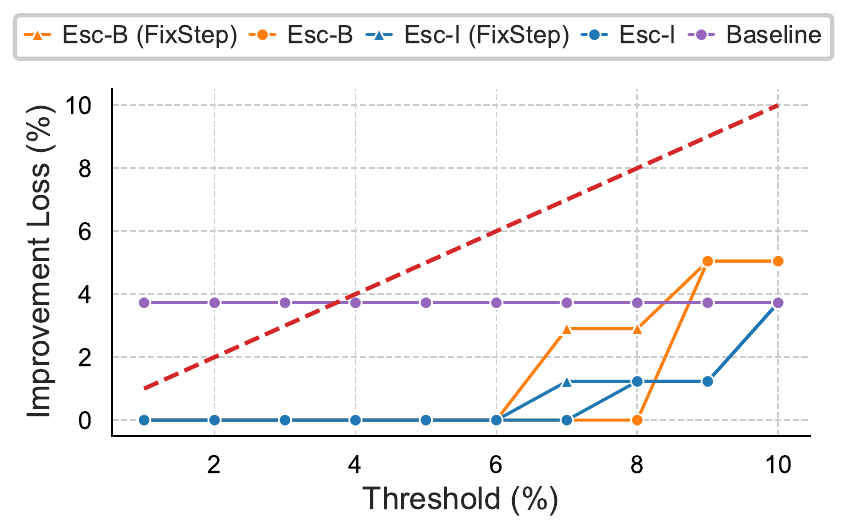}}
\subfigure[What-If Call Savings]{ \label{fig:mcts_skip:job:k10:call-save}
    \includegraphics[width=0.49\columnwidth]{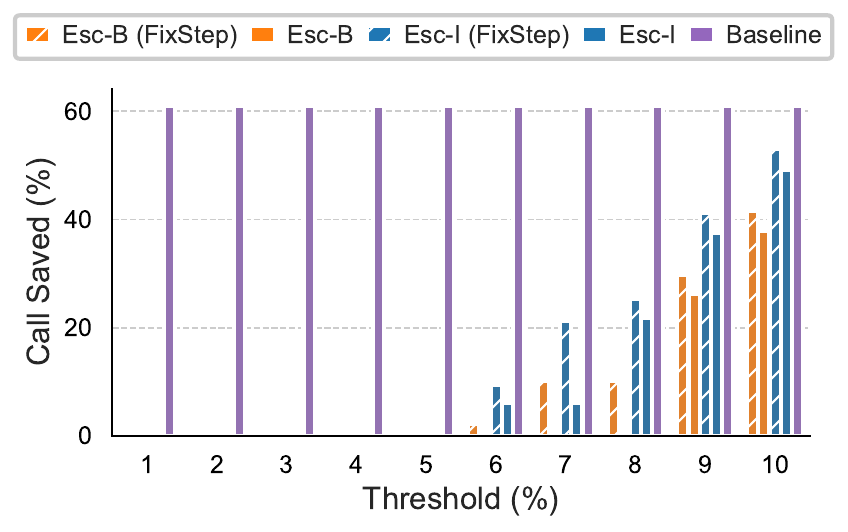}}
\subfigure[Learning Curve]{ \label{fig:mcts_skip:job:k10:lc}
    \includegraphics[width=0.49\columnwidth]{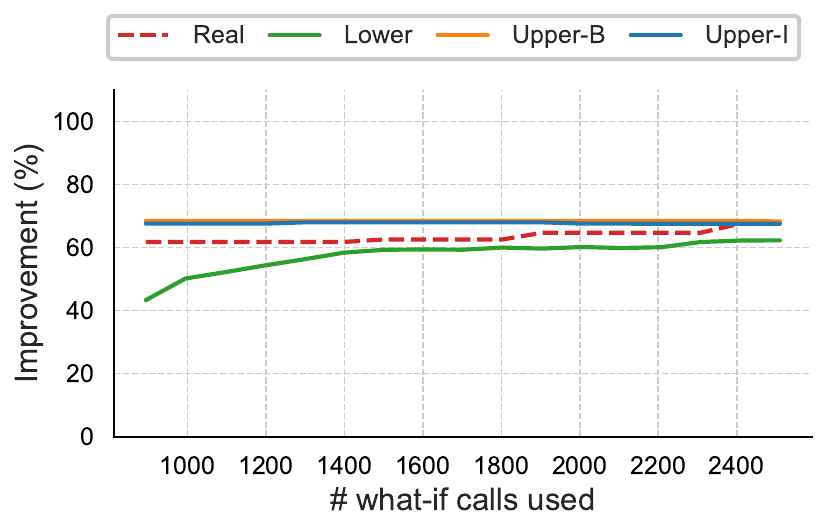}}
\vspace{-1.5em}
\caption{MCTS (with Wii), JOB, $K=10$, $B=20k$}
\label{fig:mcts_skip:job:k10}
\vspace{-1em}
\end{figure*}

\begin{figure*}
\centering
\subfigure[Time Overhead]{ \label{fig:mcts_covskip:job:k10:overhead}
    \includegraphics[width=0.49\columnwidth]{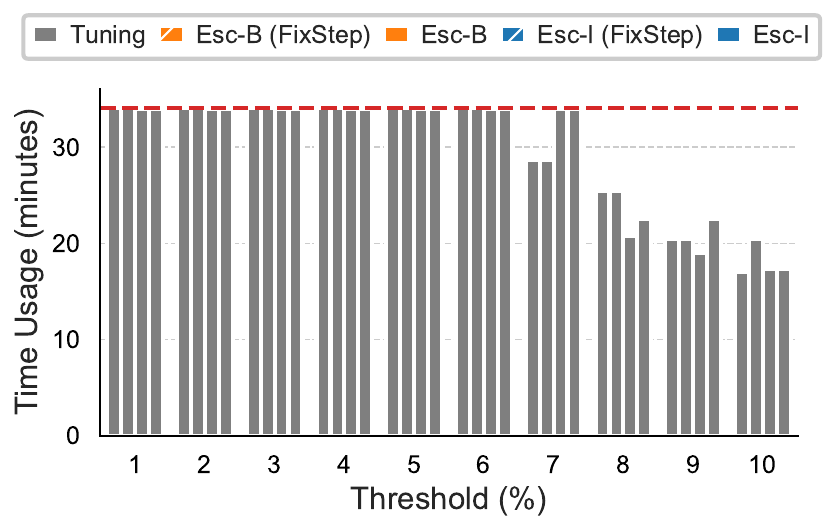}}
\subfigure[Improvement Loss]{ \label{fig:mcts_covskip:job:k10:impr-loss}
    \includegraphics[width=0.49\columnwidth]{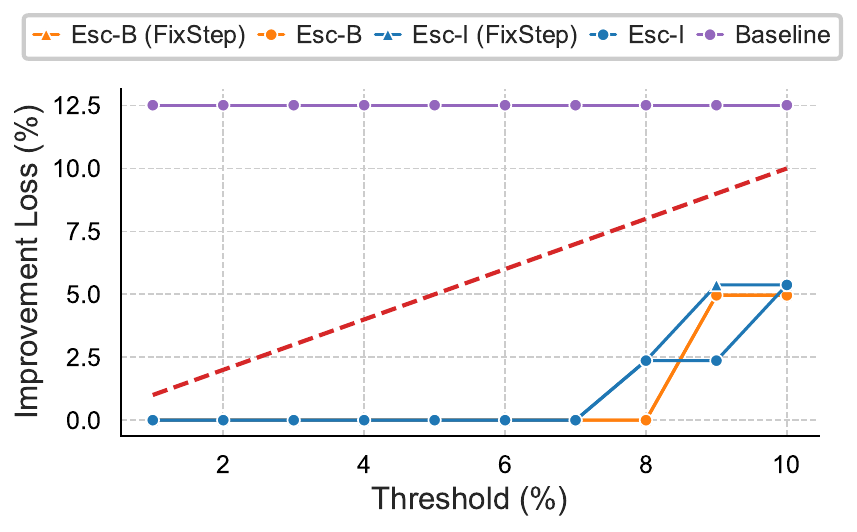}}
\subfigure[What-If Call Savings]{ \label{fig:mcts_covskip:job:k10:call-save}
    \includegraphics[width=0.49\columnwidth]{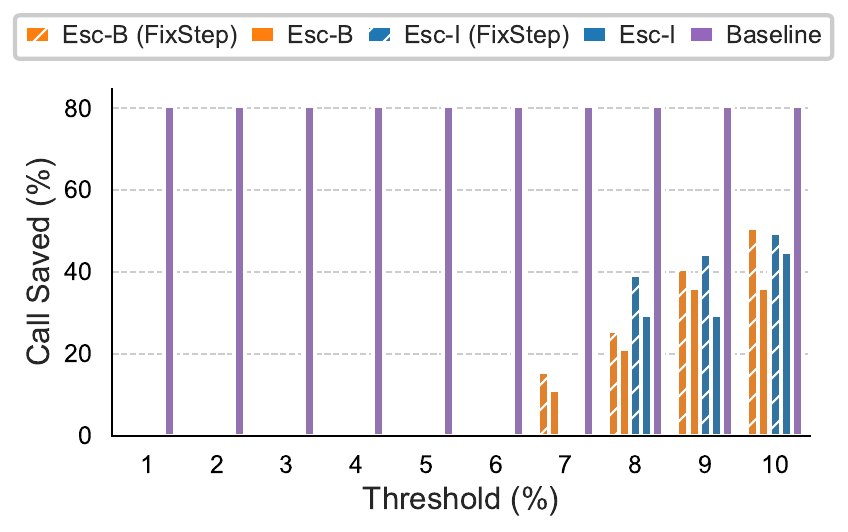}}
\subfigure[Learning Curve]{ \label{fig:mcts_covskip:job:k10:lc}
    \includegraphics[width=0.49\columnwidth]{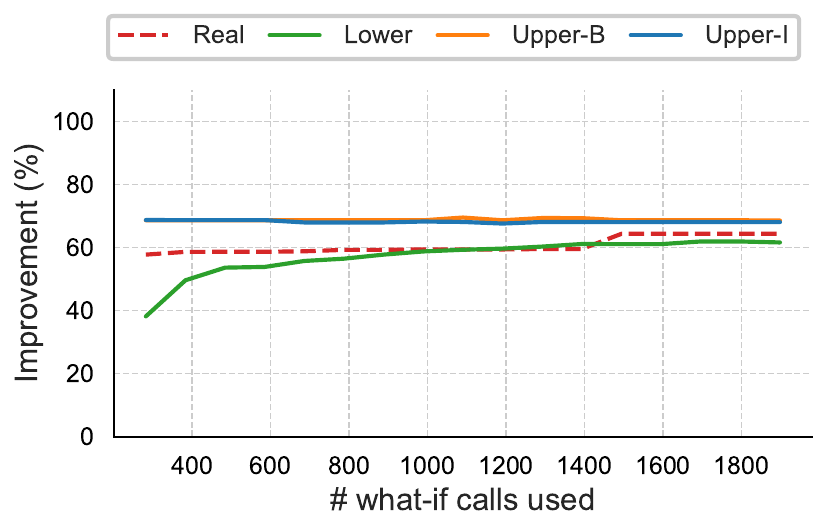}}
\vspace{-1.5em}
\caption{MCTS (with Wii-Coverage), JOB, $K=10$, $B=20k$}
\label{fig:mcts_covskip:job:k10}
\vspace{-1em}
\end{figure*}

\begin{figure*}
\centering
\subfigure[Time Overhead]{ \label{fig:mcts:real-d:k10:overhead}
    \includegraphics[width=0.49\columnwidth]{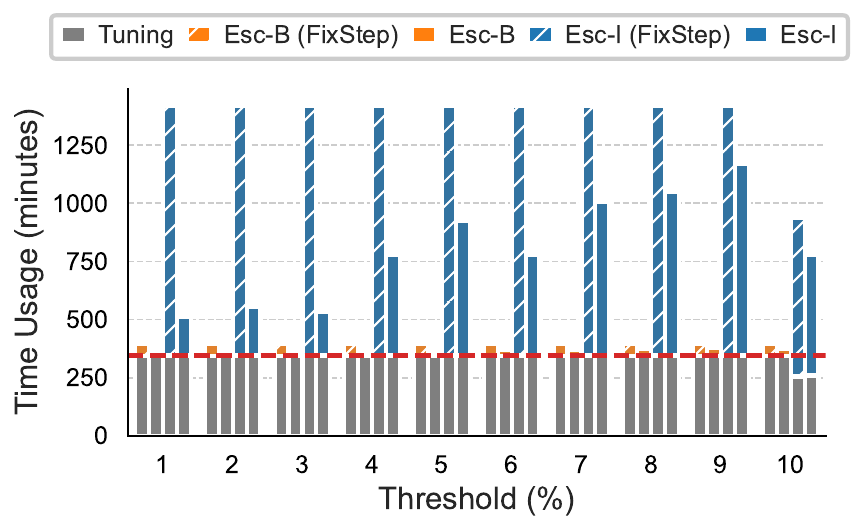}}
\subfigure[Improvement Loss]{ \label{fig:mcts:real-d:k10:impr-loss}
    \includegraphics[width=0.49\columnwidth]{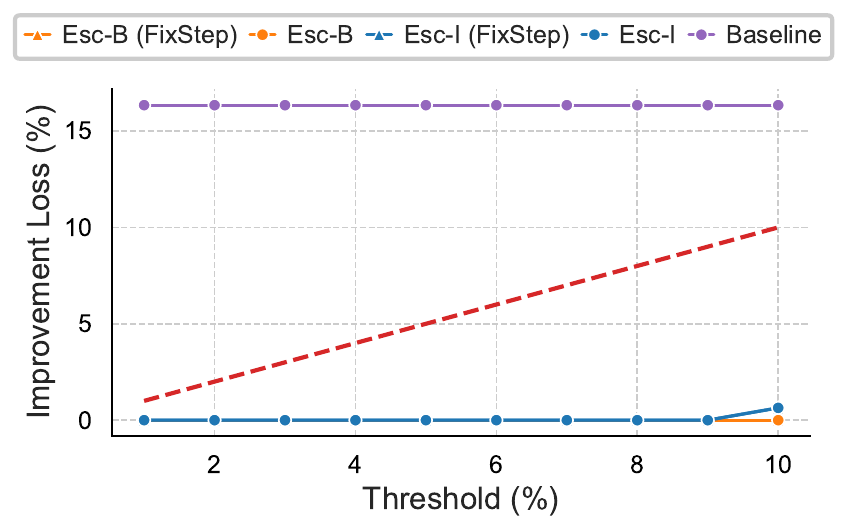}}
\subfigure[What-If Call Savings]{ \label{fig:mcts:real-d:k10:call-save}
    \includegraphics[width=0.49\columnwidth]{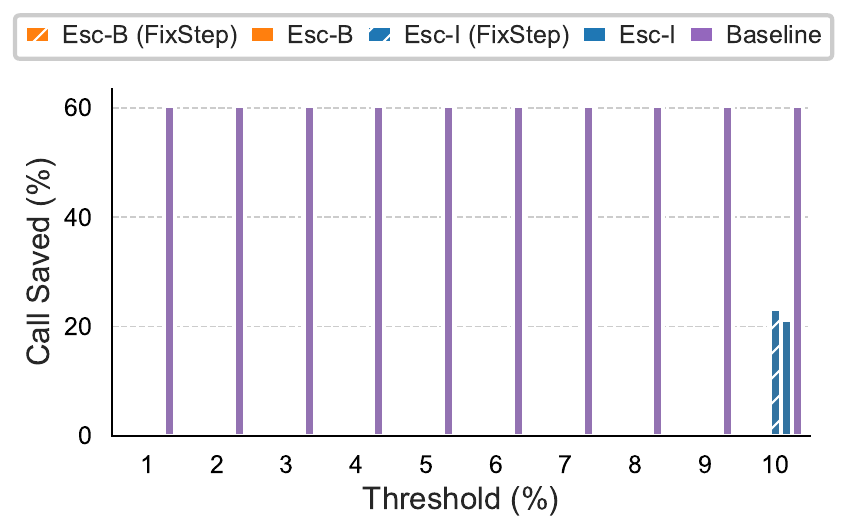}}
\subfigure[Learning Curve]{ \label{fig:mcts:real-d:k10:lc}
    \includegraphics[width=0.49\columnwidth]{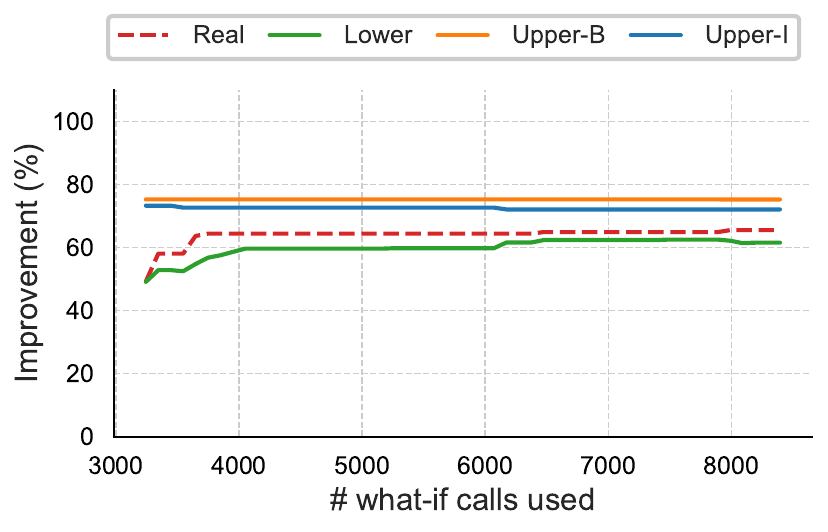}}
\vspace{-1.5em}
\caption{MCTS, Real-D, $K=10$, $B=20k$}
\label{fig:mcts:real-d:k10}
\vspace{-1em}
\end{figure*}

\begin{figure*}
\centering
\subfigure[Time Overhead]{ \label{fig:mcts_skip:real-d:k10:overhead}
    \includegraphics[width=0.49\columnwidth]{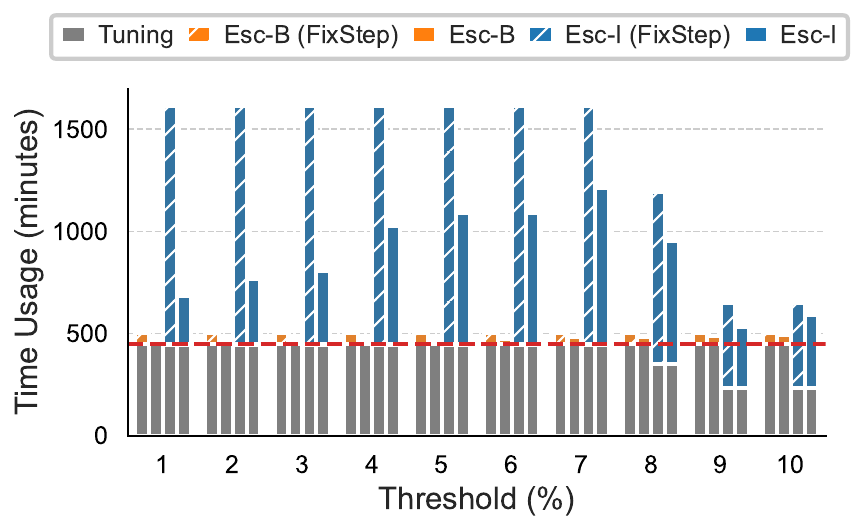}}
\subfigure[Improvement Loss]{ \label{fig:mcts_skip:real-d:k10:impr-loss}
    \includegraphics[width=0.49\columnwidth]{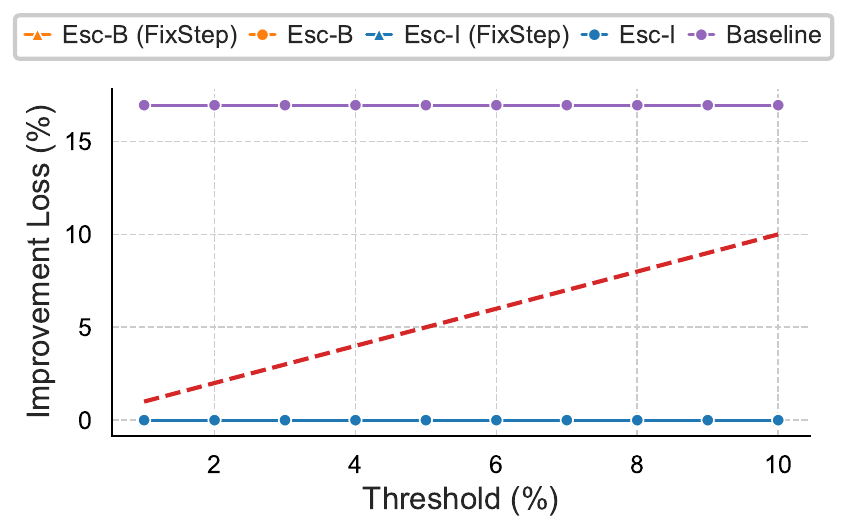}}
\subfigure[What-If Call Savings]{ \label{fig:mcts_skip:real-d:k10:call-save}
    \includegraphics[width=0.49\columnwidth]{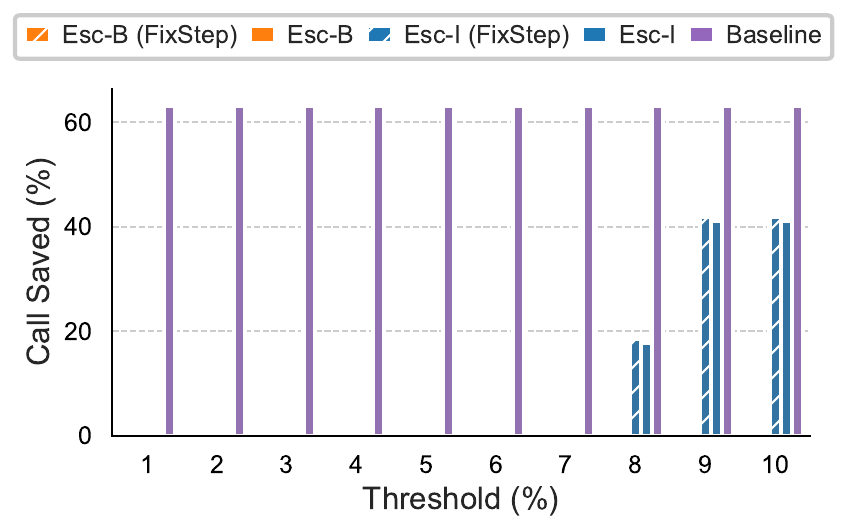}}
\subfigure[Learning Curve]{ \label{fig:mcts_skip:real-d:k10:lc}
    \includegraphics[width=0.49\columnwidth]{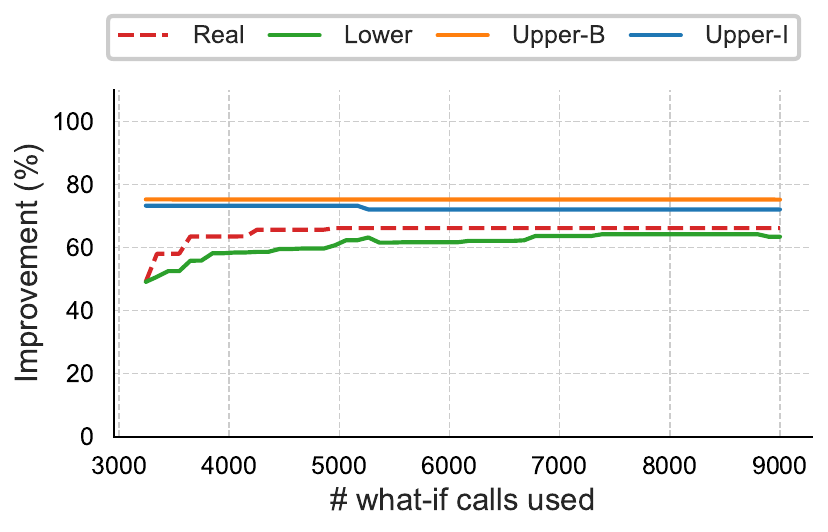}}
\vspace{-1.5em}
\caption{MCTS (with Wii), Real-D, $K=10$, $B=20k$}
\label{fig:mcts_skip:real-d:k10}
\vspace{-1em}
\end{figure*}

\begin{figure*}
\centering
\subfigure[Time Overhead]{ \label{fig:mcts_covskip:real-d:k10:overhead}
    \includegraphics[width=0.49\columnwidth]{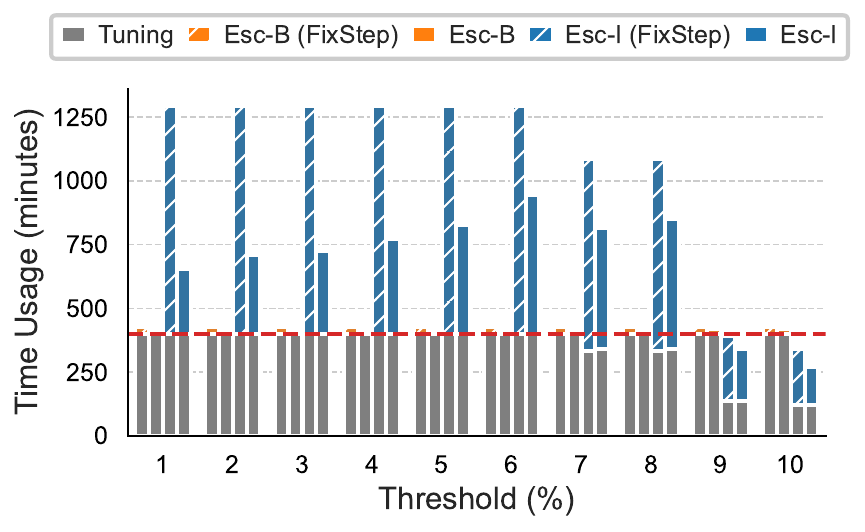}}
\subfigure[Improvement Loss]{ \label{fig:mcts_covskip:real-d:k10:impr-loss}
    \includegraphics[width=0.49\columnwidth]{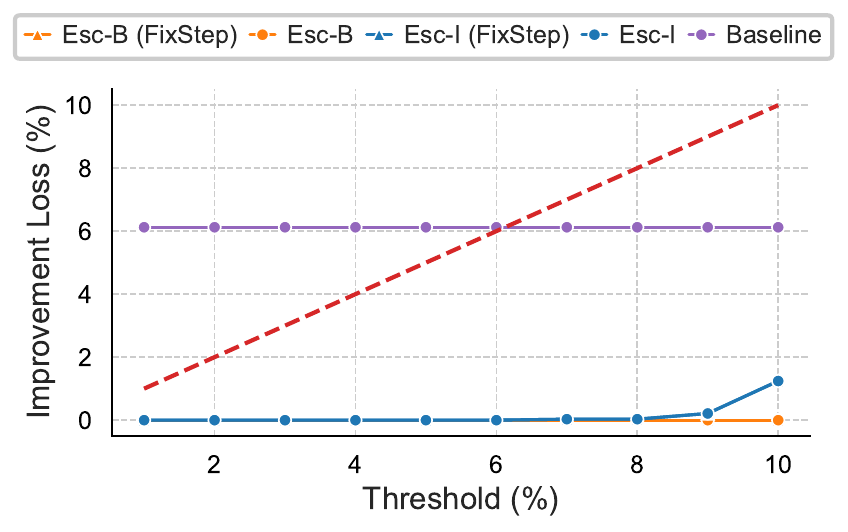}}
\subfigure[What-If Call Savings]{ \label{fig:mcts_covskip:real-d:k10:call-save}
    \includegraphics[width=0.49\columnwidth]{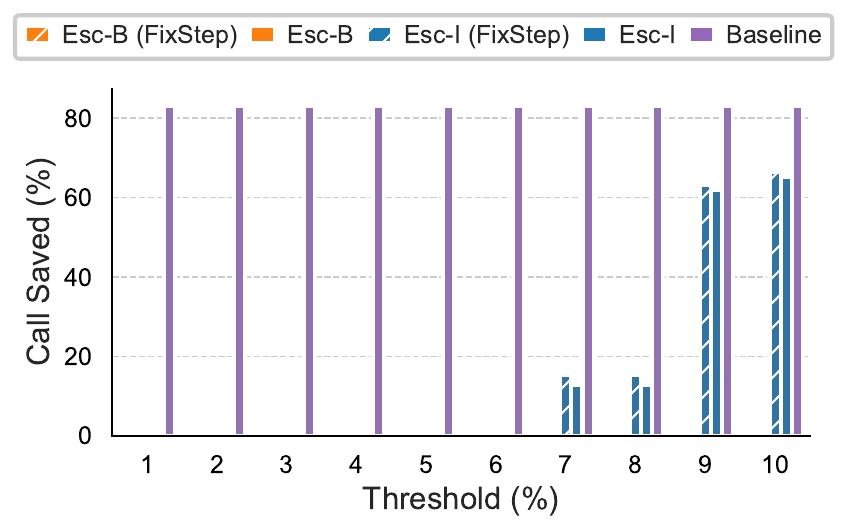}}
\subfigure[Learning Curve]{ \label{fig:mcts_covskip:real-d:k10:lc}
    \includegraphics[width=0.49\columnwidth]{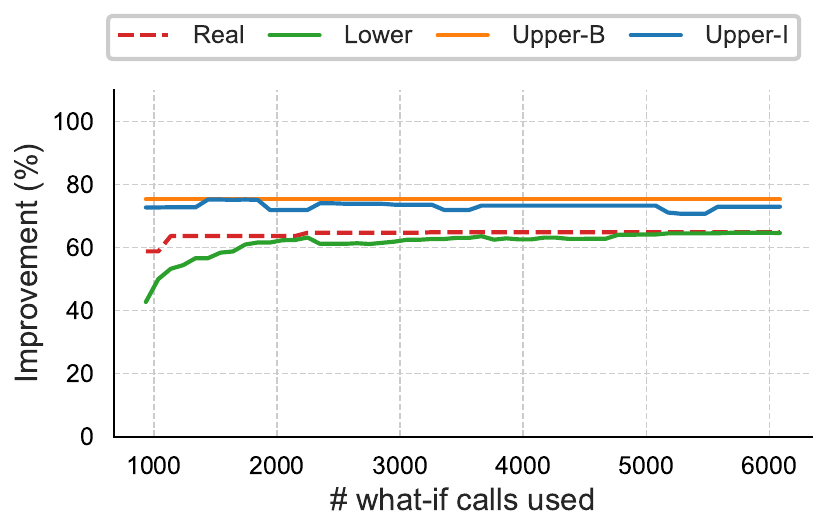}}
\vspace{-1.5em}
\caption{MCTS (with Wii-Coverage), Real-D, $K=10$, $B=20k$}
\label{fig:mcts_covskip:real-d:k10}
\vspace{-1em}
\end{figure*}

\begin{figure*}
\centering
\subfigure[Time Overhead]{ \label{fig:mcts:real-m:k10:overhead}
    \includegraphics[width=0.49\columnwidth]{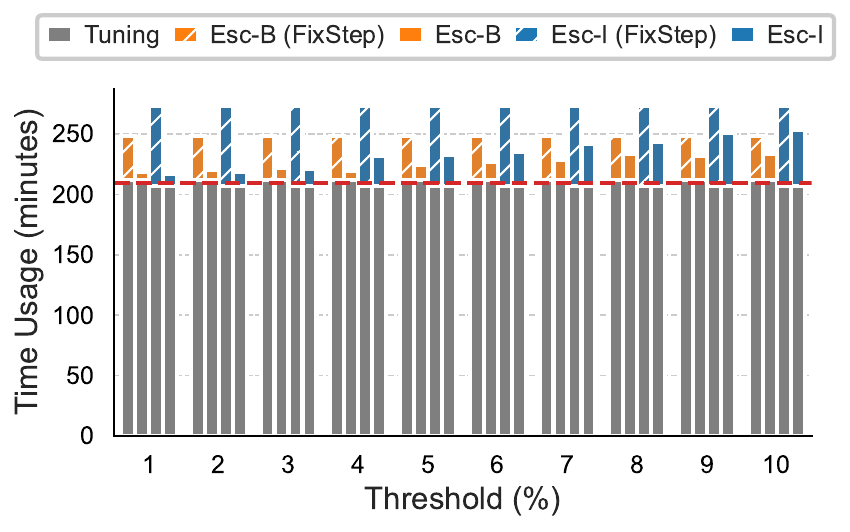}}
\subfigure[Improvement Loss]{ \label{fig:mcts:real-m:k10:impr-loss}
    \includegraphics[width=0.49\columnwidth]{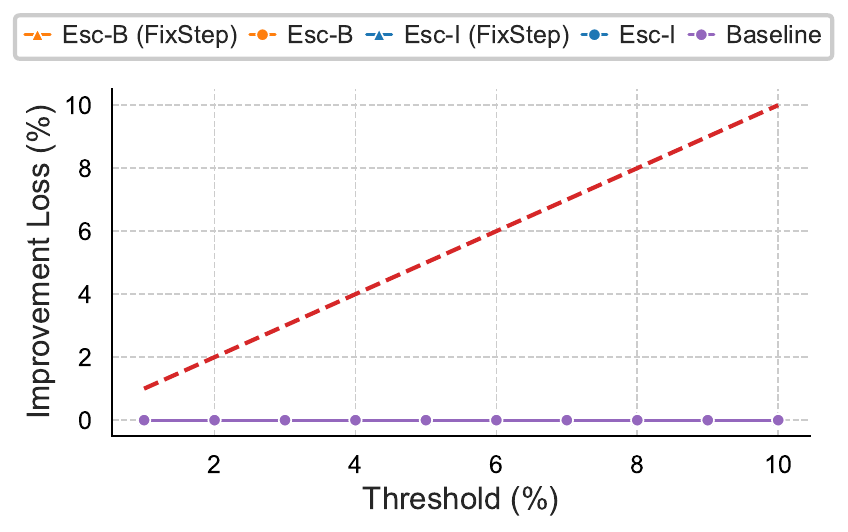}}
\subfigure[What-If Call Savings]{ \label{fig:mcts:real-m:k10:call-save}
    \includegraphics[width=0.49\columnwidth]{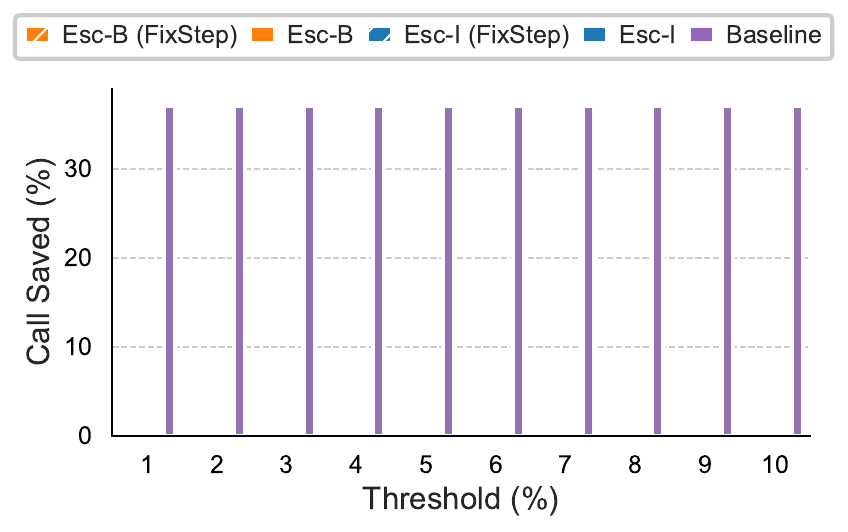}}
\subfigure[Learning Curve]{ \label{fig:mcts:real-m:k10:lc}
    \includegraphics[width=0.49\columnwidth]{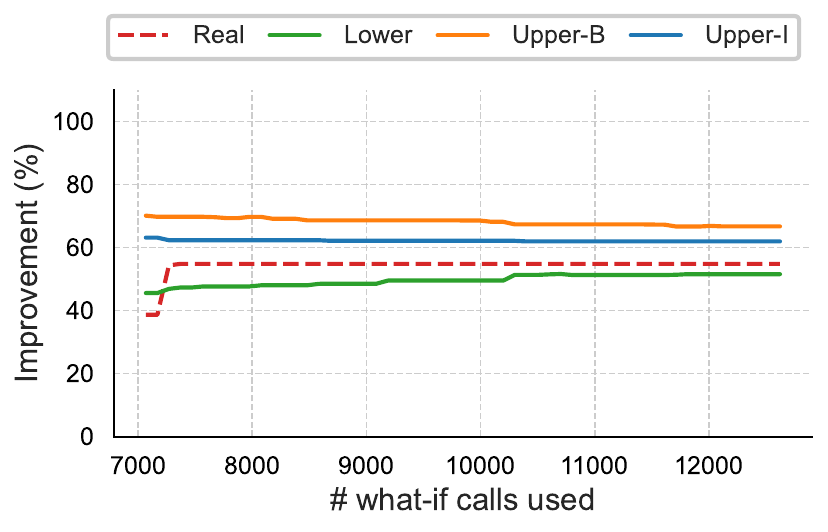}}
\vspace{-1.5em}
\caption{MCTS, Real-M, $K=10$, $B=20k$}
\label{fig:mcts:real-m:k10}
\vspace{-1em}
\end{figure*}

\begin{figure*}
\centering
\subfigure[Time Overhead]{ \label{fig:mcts_skip:real-m:k10:overhead}
    \includegraphics[width=0.49\columnwidth]{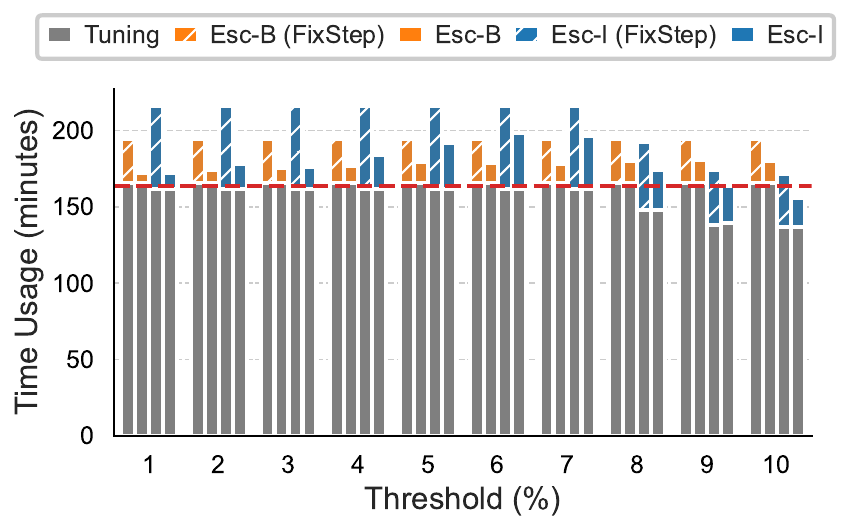}}
\subfigure[Improvement Loss]{ \label{fig:mcts_skip:real-m:k10:impr-loss}
    \includegraphics[width=0.49\columnwidth]{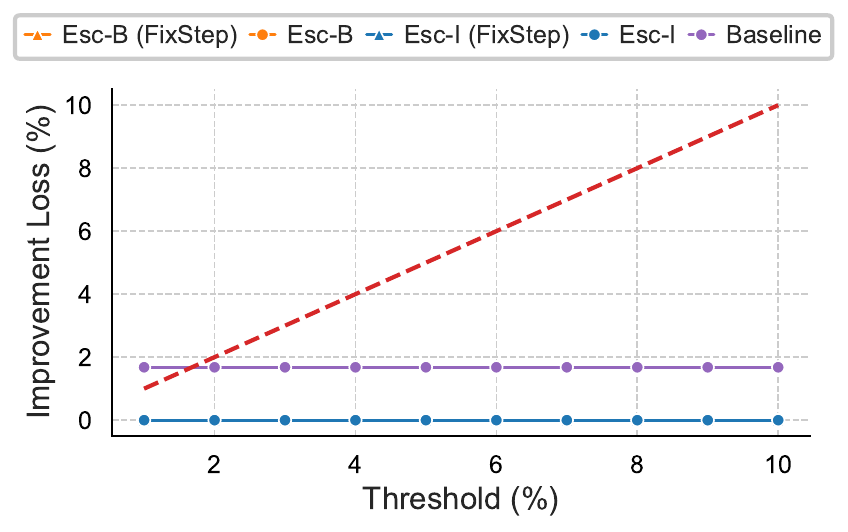}}
\subfigure[What-If Call Savings]{ \label{fig:mcts_skip:real-m:k10:call-save}
    \includegraphics[width=0.49\columnwidth]{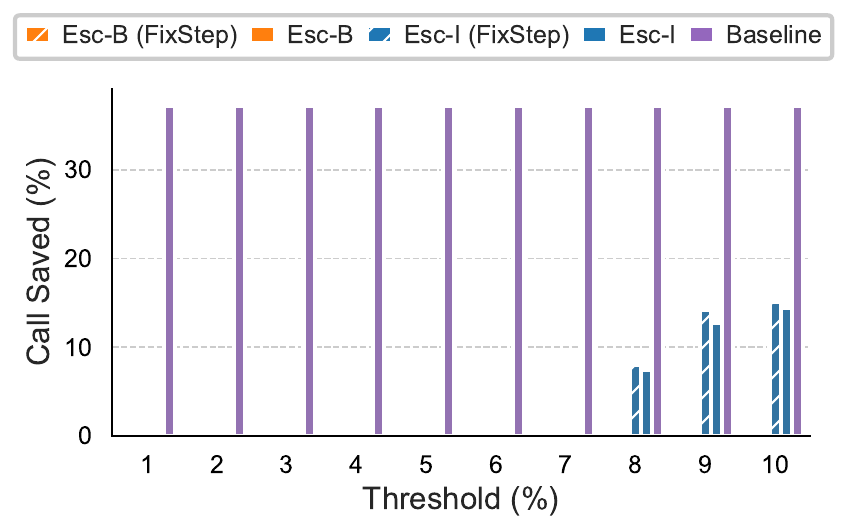}}
\subfigure[Learning Curve]{ \label{fig:mcts_skip:real-m:k10:lc}
    \includegraphics[width=0.49\columnwidth]{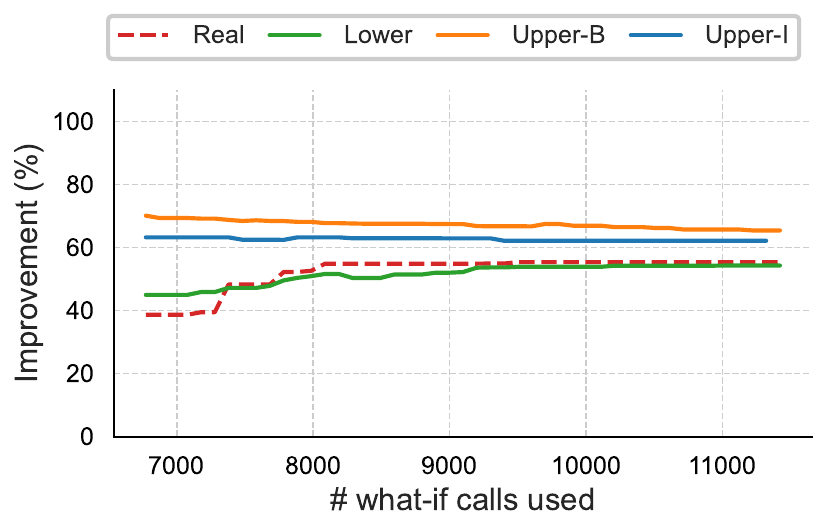}}
\vspace{-1.5em}
\caption{MCTS (with Wii), Real-M, $K=10$, $B=20k$}
\label{fig:mcts_skip:real-m:k10}
\vspace{-1em}
\end{figure*}

\begin{figure*}
\centering
\subfigure[Time Overhead]{ \label{fig:mcts_covskip:real-m:k10:overhead}
    \includegraphics[width=0.49\columnwidth]{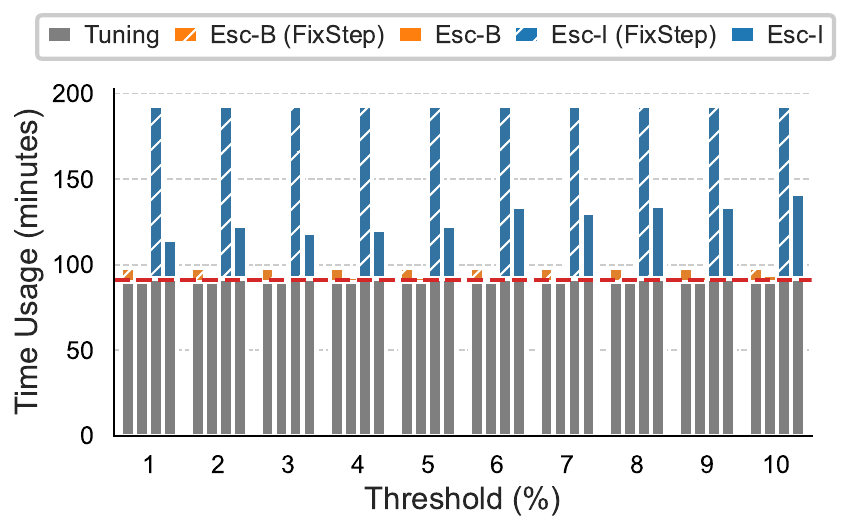}}
\subfigure[Improvement Loss]{ \label{fig:mcts_covskip:real-m:k10:impr-loss}
    \includegraphics[width=0.49\columnwidth]{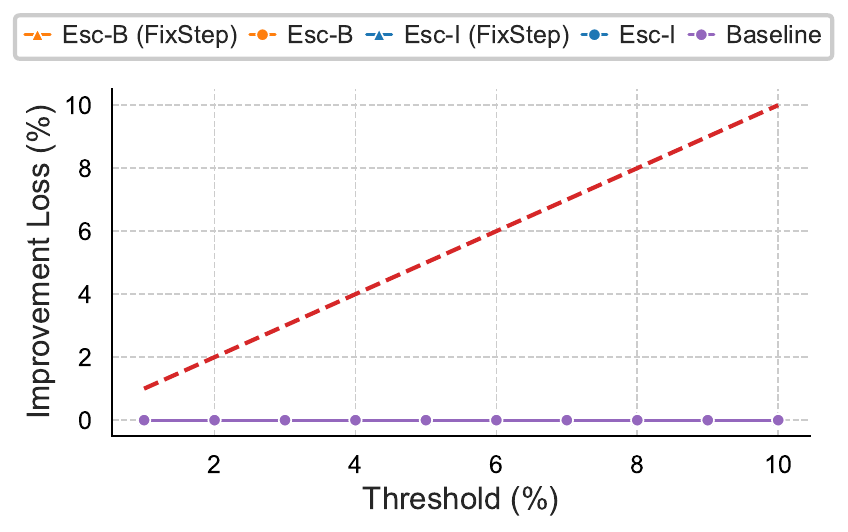}}
\subfigure[What-If Call Savings]{ \label{fig:mcts_covskip:real-m:k10:call-save}
    \includegraphics[width=0.49\columnwidth]{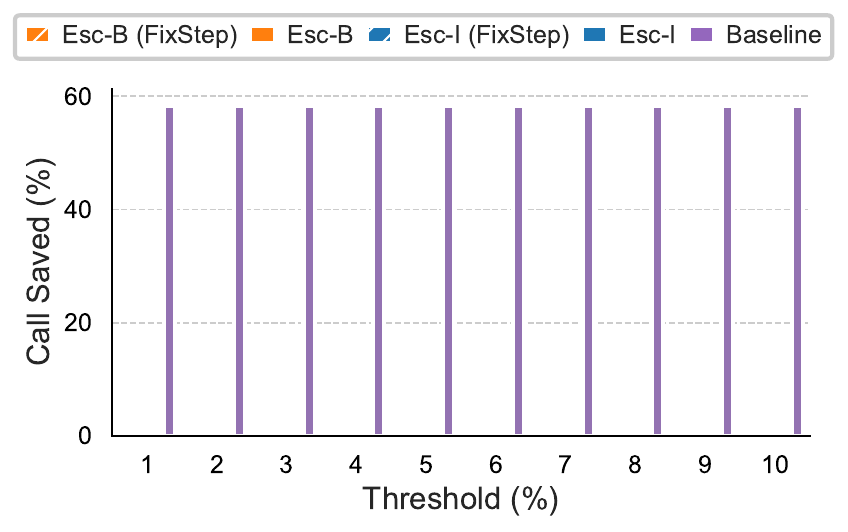}}
\subfigure[Learning Curve]{ \label{fig:mcts_covskip:real-m:k10:lc}
    \includegraphics[width=0.49\columnwidth]{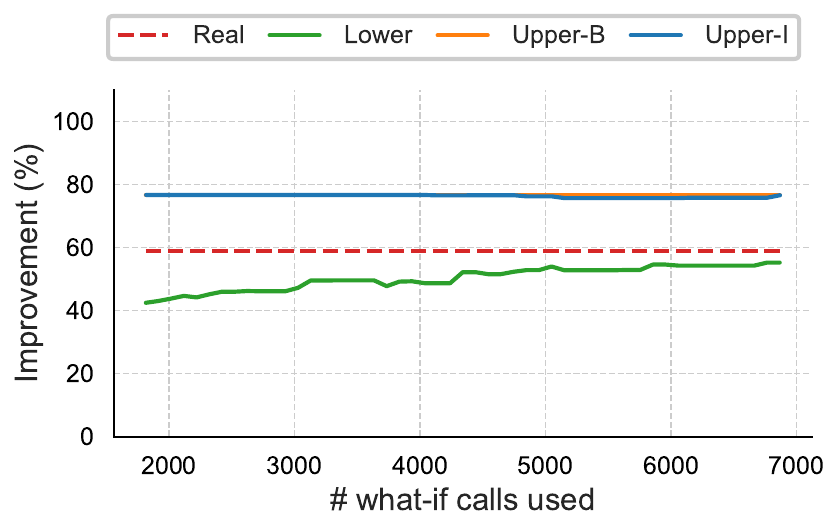}}
\vspace{-1.5em}
\caption{MCTS (with Wii-Coverage), Real-M, $K=10$, $B=20k$}
\label{fig:mcts_covskip:real-m:k10}
\vspace{-1em}
\end{figure*}

Figures~\ref{fig:twophase:tpch:k10} to~\ref{fig:mcts_covskip:real-m:k10} present evaluation results when setting $K=10$.
Overall, the observations are similar to those when setting $K=20$.


\vspace{-1em}
\subsection{More Discussion and Analysis}


\subsubsection{Index Interaction}

Continuing with Example~\ref{example:interaction}, suppose that we are running the simulated greedy search in Procedure~\ref{proc:simulation-greedy} without knowing the true what-if call cost $c(q_1, \{z_1,z_2\})$.
Assuming that $\{z_1,z_2\}\subseteq C^*_B$, when computing $L(q_1, C_B^*)$ in Equation~\ref{eq:c_prim_q_C_t_u_K} we need to subtract the MCI upper bounds of both $z_1$ and $z_2$, i.e., 
$u(q_1,z_1)+u(q_1, z_2) =\Delta(q_1, \{z_1\}) + \Delta(q_1, \{z_2\})\approx 2\cdot \Delta(q_1, \{z_1,z_2\}).$
Hence, the lower bound $L(q_1, C_B^*)$ so derived can be loose in the presence of (strong) index interactions.

Note that index interaction is \emph{query-dependent}. 
To see this, consider the same $z_1$ and $z_2$ in Example~\ref{example:interaction} but a different SQL query $q_2$ in Figure~\ref{fig:index-interaction-example}. 
Since $z_1$ can hardly be used for evaluating $q_2$, we have $\Delta(q_2, \{z_1\})\approx 0$ (see~\cite{full-version} for details).
As a result, in the presence of both $z_1$ and $z_2$, the query optimizer will pick $z_2$ over $z_1$; hence, we have $\Delta(q_2, \{z_1, z_2\})=\Delta(q_2, \{z_2\})\approx \Delta(q_2, \{z_1\})+\Delta(q_2, \{z_2\})$.
Therefore, $z_1$ and $z_2$ \emph{do not interact} in the case of $q_2$.
More discussion on this can be found in Appendix~\ref{appendix:index-interaction:query-dependency}.

These examples suggest that, for a given query $q$, we can use the relationship between $\Delta(q, \{z_1, z_2\})$ and $\Delta(q, \{z_1\})+\Delta(q, \{z_2\})$ to quantify index interaction.
On one hand, if $\Delta(q, \{z_1, z_2\})=\Delta(q, \{z_1\})+\Delta(q, \{z_2\})$, then there is no index interaction.
On the other hand, the discrepancy between these two quantities indicates the degree of index interaction.

\vspace{-0.5em}
\paragraph*{Discussion}
Consider two extreme cases of $\mathcal{I}(z, C | q)=0$: 
\begin{itemize}[leftmargin=*]
    \item If $\mathcal{I}(z, C | q)=0$, it implies $\Delta(q, C\cup \{z\})=\Delta_U(q, C\cup \{z\})$.
        As a result, we have $\Delta(q, C\cup \{z\})=\Delta(q, C)+\Delta(q, \{z\})$.
        Or equivalently, the MCI 
        $\delta(q, z, C)=c(q, C)-c(q, C\cup\{z\})=\Delta(q, C\cup \{z\})-\Delta(q, C)=\Delta(q, \{z\}).$
        This suggests that the extra improvement by including $z$ on top of $C$ achieves its maximum possible, since $\Delta(q, \{z\})$ is an MCI upper-bound of $\delta(q, z, C)$ based on Equation~\ref{eq:upper-bound:mci:call-level}.
        Therefore, it indicates that there is almost \emph{no interaction} between $z$ and the indexes contained by $C$.
    \item If $\mathcal{I}(z, C | q)=1$, it implies $\Delta(q, C\cup \{z\})=\Delta_L(q, C\cup \{z\})$, namely, $\Delta(q, C\cup \{z\})=\max\{\Delta(q, C), \Delta(q, \{z\})\}$. Assume that $\Delta(q, C)\geq \Delta(q, \{z\})$. It follows that $\Delta(q, C\cup \{z\})=\Delta(q, C)$. This means that including $z$ into $C$ does not bring in any extra improvement, which indicates that there is \emph{strong interaction} between $z$ and the indexes contained by $C$. On the other hand, if $\Delta(q, C)<\Delta(q, \{z\})$, then $\Delta(q, C\cup \{z\})=\Delta(q, \{z\})$. However, this cannot happen in the simulated greedy search unless $C=\emptyset$.
\end{itemize}
In summary, we can use the index-configuration interaction $\mathcal{I}(z, C | q)$ as an indicator of the extra cost improvement of the next index $z$ to be selected w.r.t. the configuration $C$ that has been selected in the simulated greedy search.
Weak interaction (with small value of $\mathcal{I}(z, C | q)$) implies large extra cost improvement by incorporating $z$ into $C$, whereas strong interaction (with large value of $\mathcal{I}(z, C | q)$) implies small extra cost improvement.



\vspace{-0.5em}
\paragraph*{Motivation of Threshold-based Refinement}

To demonstrate the motivation behind the threshold-based refinement of the lower bound based on index interaction, we conduct the following empirical study.
For each workload detailed in Section~\ref{sec:eval:settings:workloads}, we pick the top 10 costly queries and for each query we evaluate the what-if costs of all \emph{singleton} configurations.
We then select the top 50 indexes w.r.t. their cost improvement (CI) in decreasing order and evaluate the what-if costs of all $50\times 49=2,450$ configurations that contain a pair of the top-50 indexes.
Finally, we compute the pairwise index interactions $\mathcal{I}(z_1, z_2 | q)$ and the pairwise index similarity $\mathcal{S}(z_1, z_2 | q)$ as well as their correlation.

\begin{figure*}
\centering
\subfigure{\includegraphics[width=0.19\textwidth]{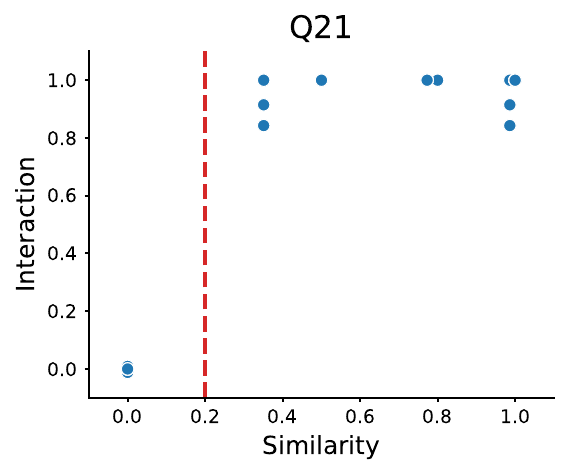}}
\subfigure{\includegraphics[width=0.19\textwidth]{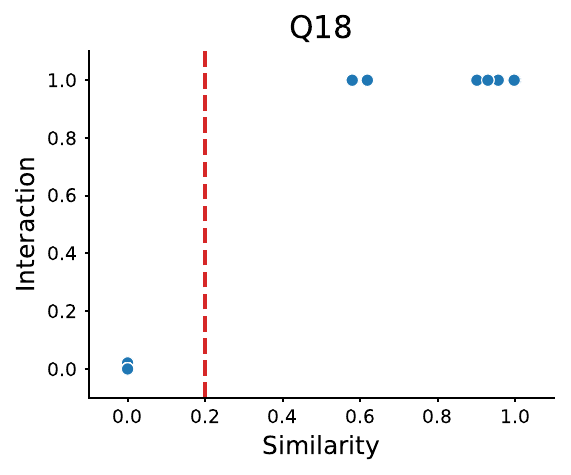}}
\subfigure{\includegraphics[width=0.19\textwidth]{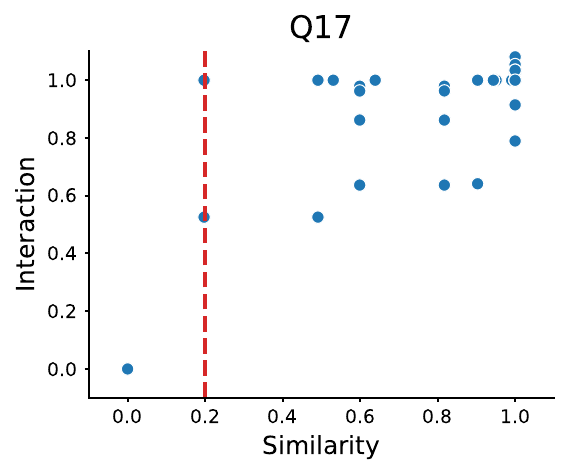}}
\subfigure{\includegraphics[width=0.19\textwidth]{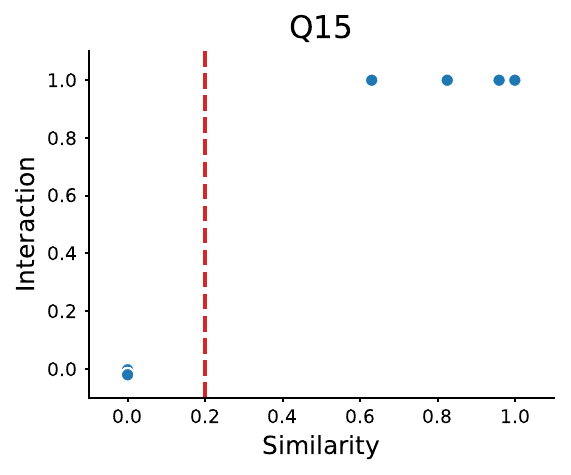}}
\subfigure{\includegraphics[width=0.19\textwidth]{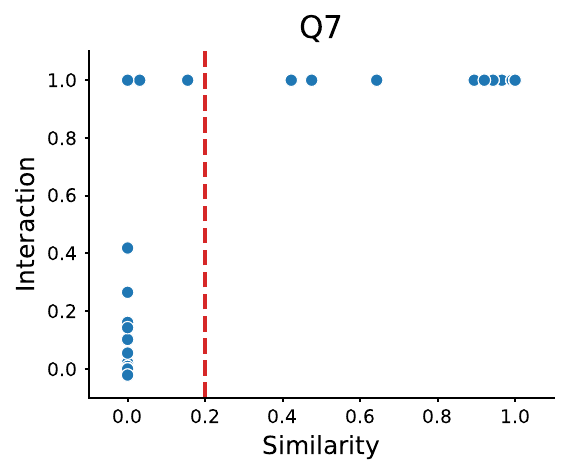}}
\subfigure{\includegraphics[width=0.19\textwidth]{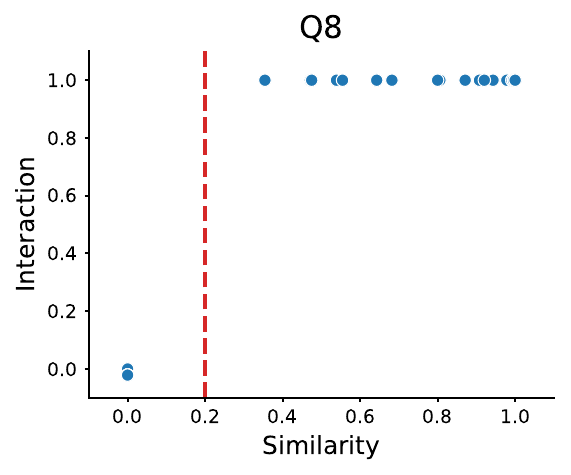}}
\subfigure{\includegraphics[width=0.19\textwidth]{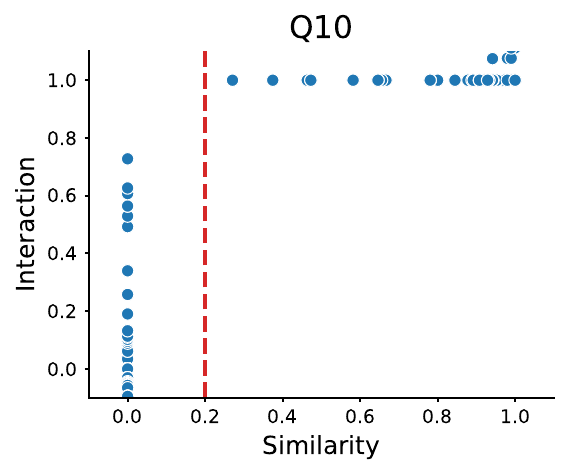}}
\subfigure{\includegraphics[width=0.19\textwidth]{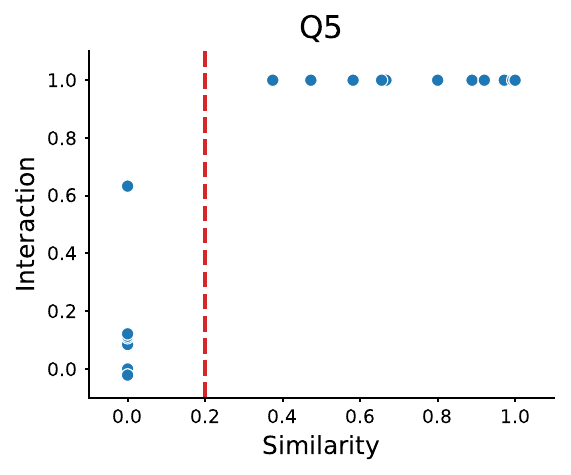}}
\subfigure{\includegraphics[width=0.19\textwidth]{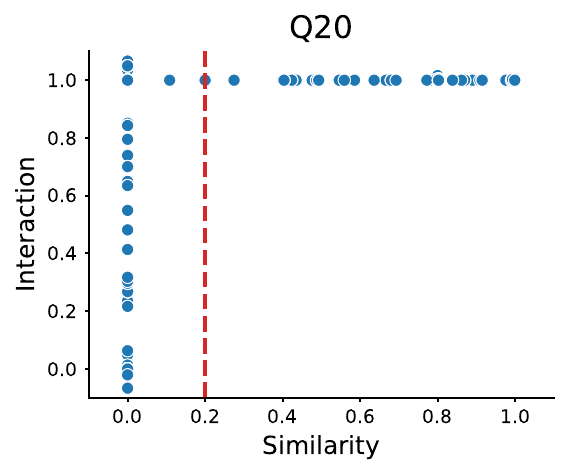}}
\subfigure{\includegraphics[width=0.19\textwidth]{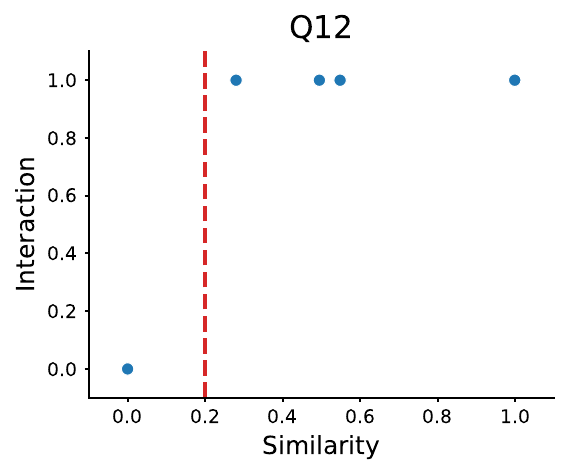}}
\vspace{-1.5em}
\caption{Relationship between pairwise index interaction and pairwise index similarity (TPC-H).}
\vspace{-1.5em}
\label{fig:interaction-tpch}
\end{figure*}

Figure~\ref{fig:interaction-tpch} presents the correlation results for the top-10 queries from the \textbf{TPC-H} workload, where the $x$-axis represents the pairwise index similarity and the $y$-axis represents the pairwise index interaction.
We observe that there is no strong correlation between the pairwise index similarity and index interaction.
Indeed, for most of the queries, there is a sudden jump on the pairwise index interaction when the pairwise index similarity increases.
That is, when the pairwise index similarity exceeds a certain threshold (e.g., the dashed line in each plot of Figure~\ref{fig:interaction-tpch} that represents an index similarity of 0.2), the pairwise index interaction will increase to a high value (e.g., close to 1).




\vspace{-0.5em}
\paragraph*{Impact of Index-Interaction Threshold}

\begin{figure*}
\centering
\subfigure{\includegraphics[width=0.19\textwidth]{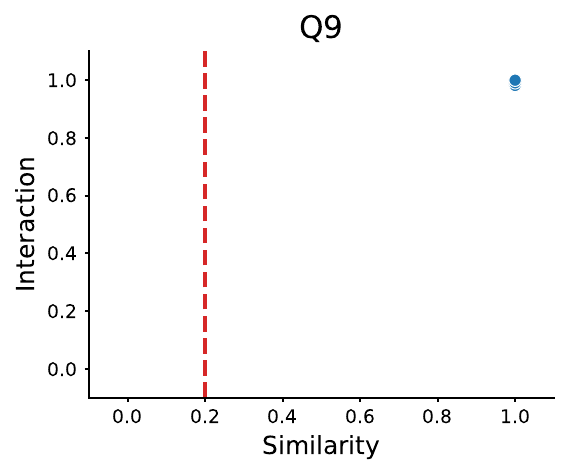}}
\subfigure{\includegraphics[width=0.19\textwidth]{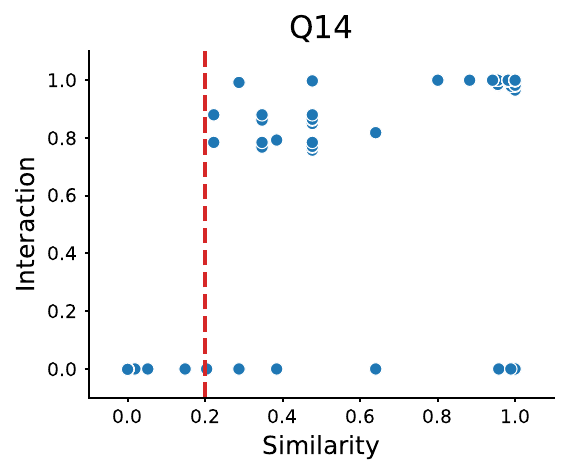}}
\subfigure{\includegraphics[width=0.19\textwidth]{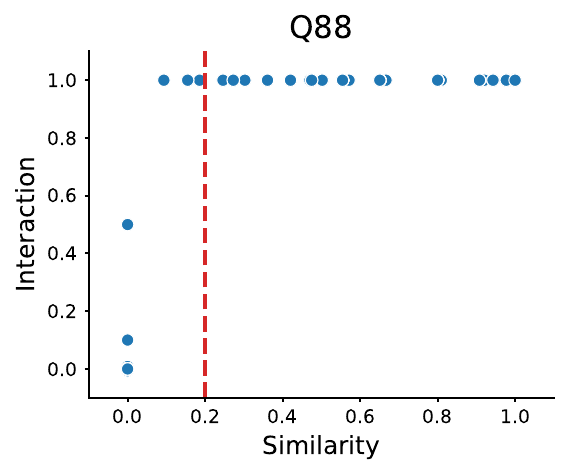}}
\subfigure{\includegraphics[width=0.19\textwidth]{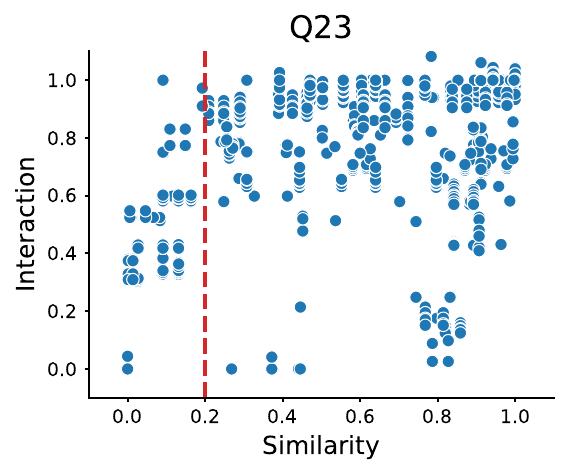}}
\subfigure{\includegraphics[width=0.19\textwidth]{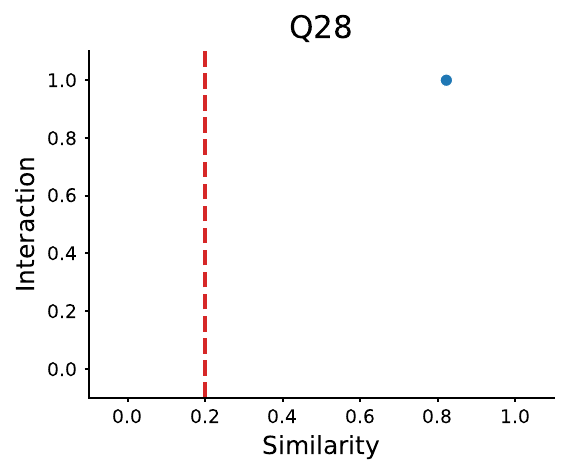}}
\subfigure{\includegraphics[width=0.19\textwidth]{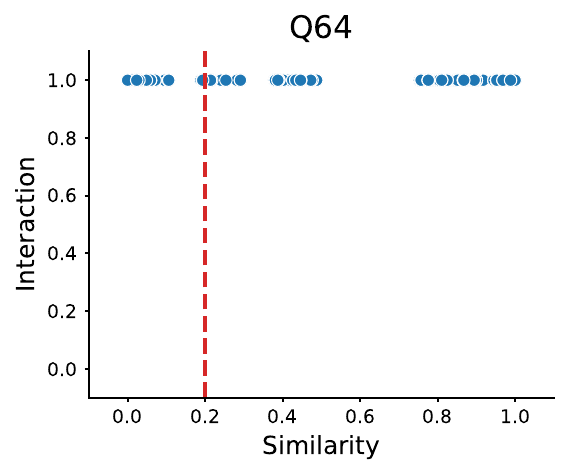}}
\subfigure{\includegraphics[width=0.19\textwidth]{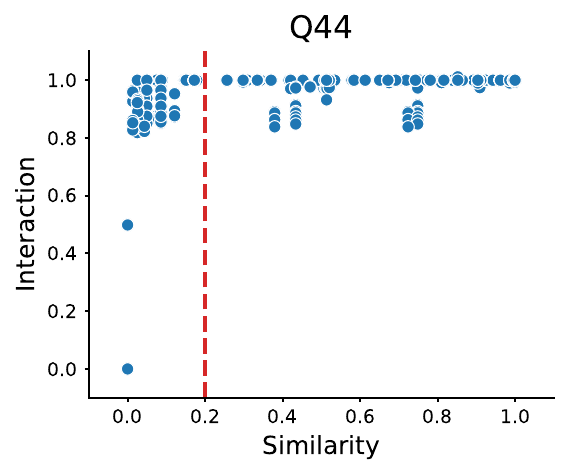}}
\subfigure{\includegraphics[width=0.19\textwidth]{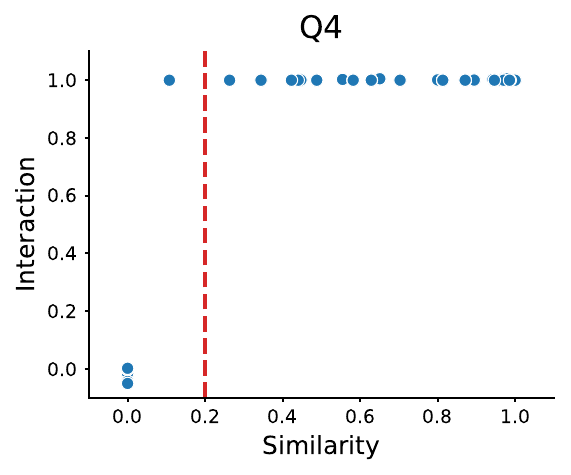}}
\subfigure{\includegraphics[width=0.19\textwidth]{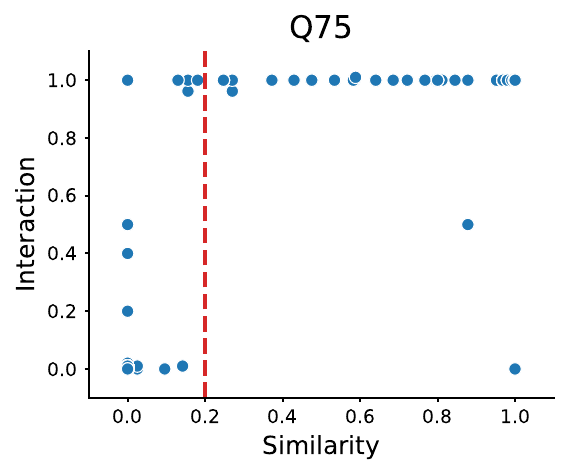}}
\subfigure{\includegraphics[width=0.19\textwidth]{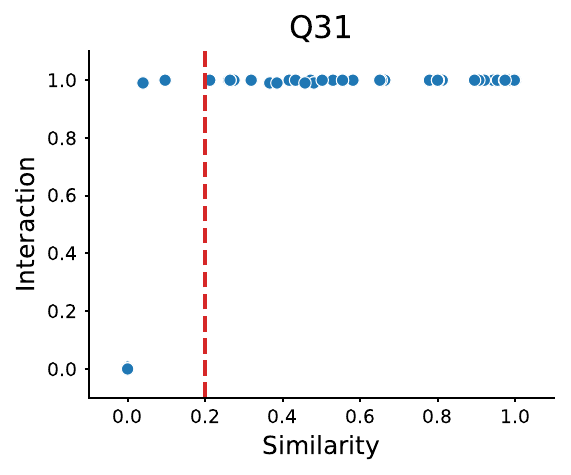}}
\vspace{-1em}
\caption{Relationship between pairwise index interaction and pairwise index similarity (TPC-DS).}
\label{fig:interaction-tpcds}
\end{figure*}

\begin{figure*}
\centering
\subfigure{\includegraphics[width=0.19\textwidth]{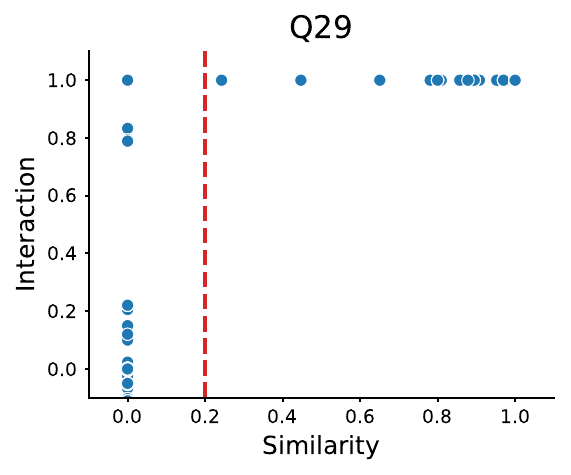}}
\subfigure{\includegraphics[width=0.19\textwidth]{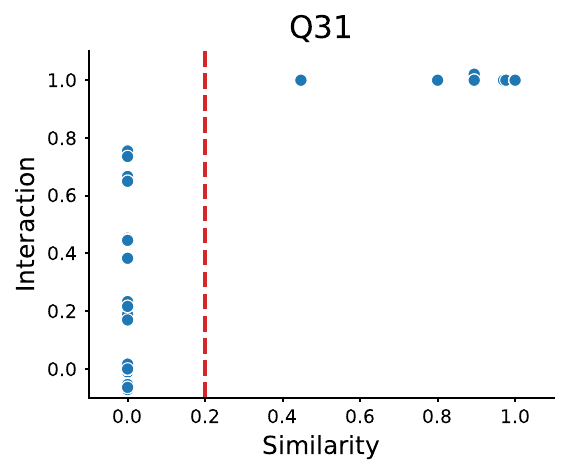}}
\subfigure{\includegraphics[width=0.19\textwidth]{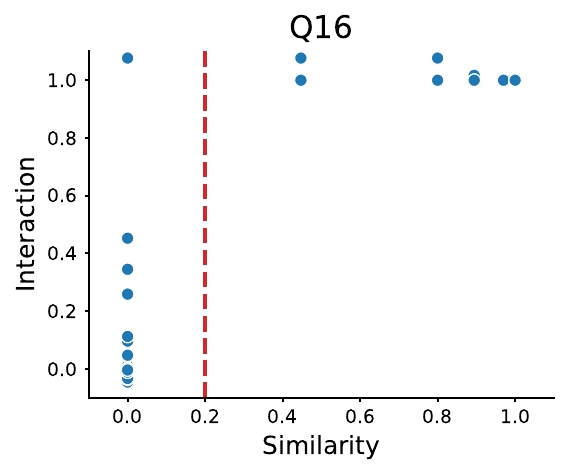}}
\subfigure{\includegraphics[width=0.19\textwidth]{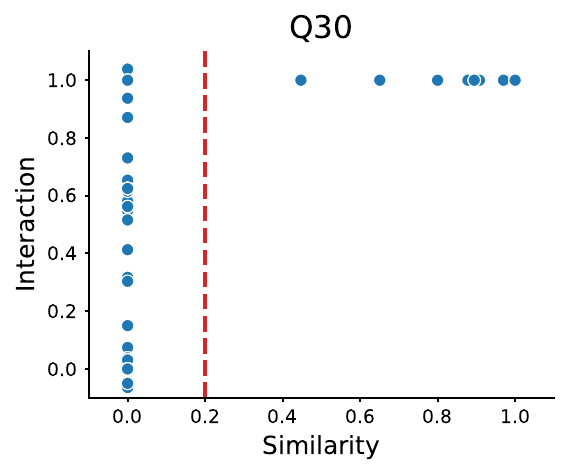}}
\subfigure{\includegraphics[width=0.19\textwidth]{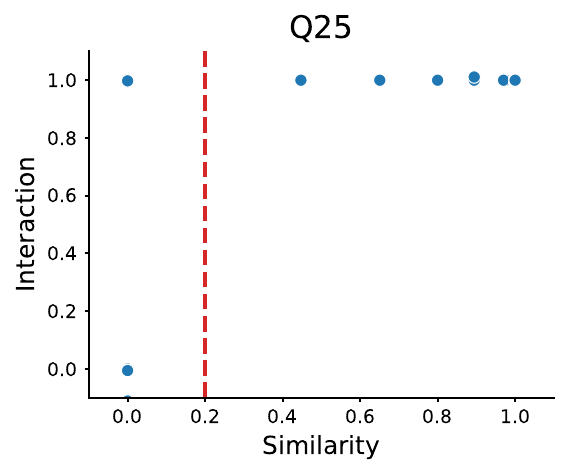}}
\subfigure{\includegraphics[width=0.19\textwidth]{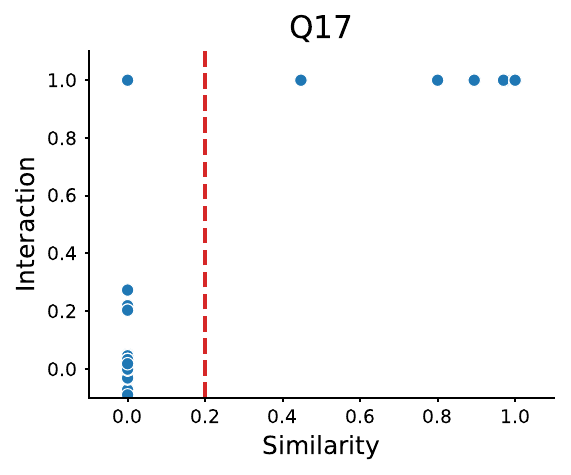}}
\subfigure{\includegraphics[width=0.19\textwidth]{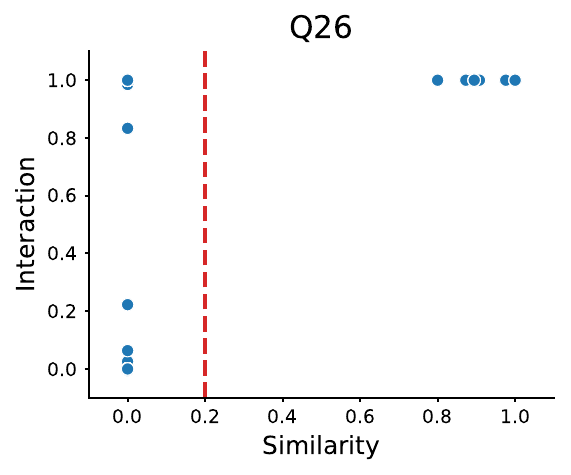}}
\subfigure{\includegraphics[width=0.19\textwidth]{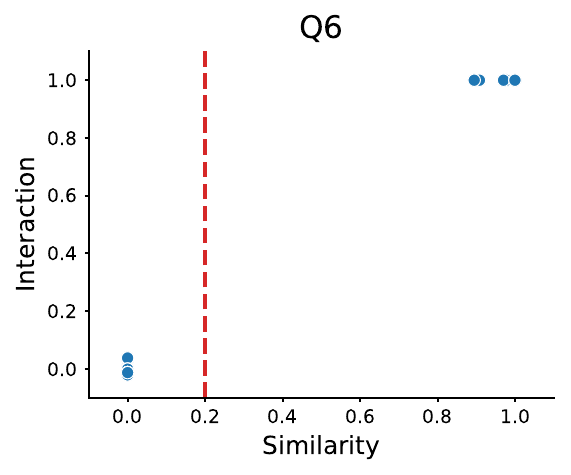}}
\subfigure{\includegraphics[width=0.19\textwidth]{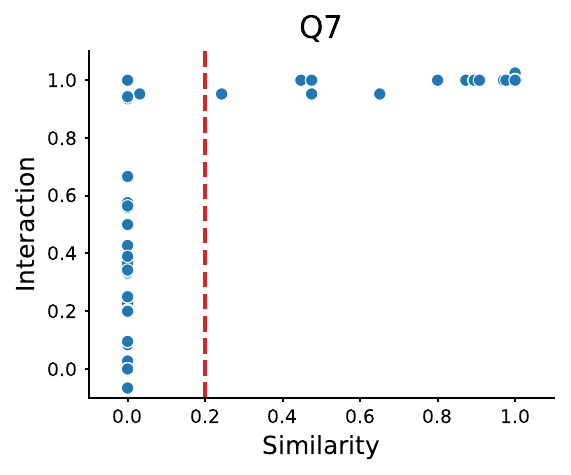}}
\subfigure{\includegraphics[width=0.19\textwidth]{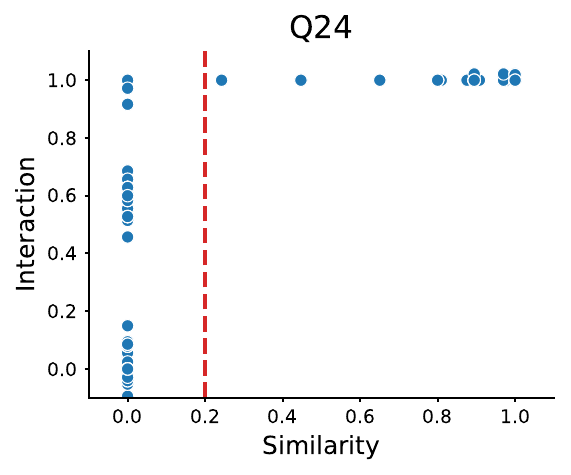}}
\vspace{-1em}
\caption{Relationship between pairwise index interaction and pairwise index similarity (JOB).}
\label{fig:interaction-job}
\end{figure*}

\begin{figure*}
\centering
\subfigure{\includegraphics[width=0.19\textwidth]{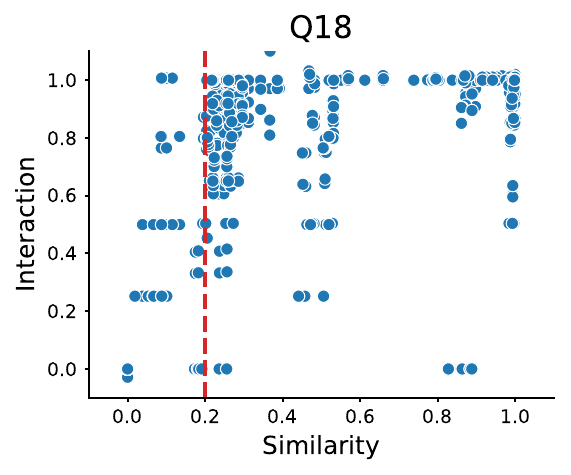}}
\subfigure{\includegraphics[width=0.19\textwidth]{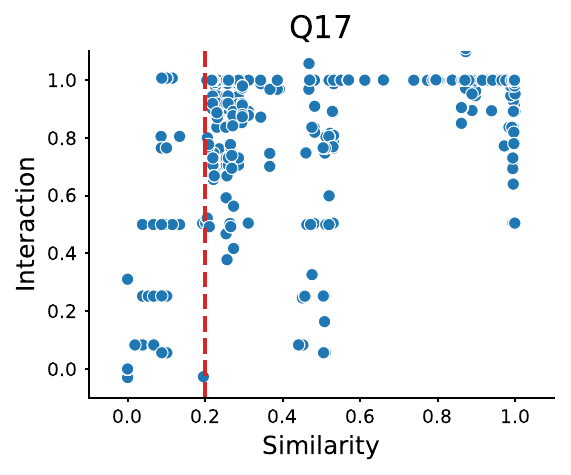}}
\subfigure{\includegraphics[width=0.19\textwidth]{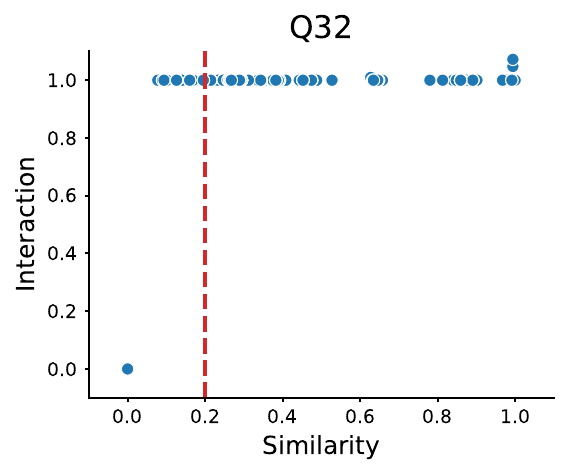}}
\subfigure{\includegraphics[width=0.19\textwidth]{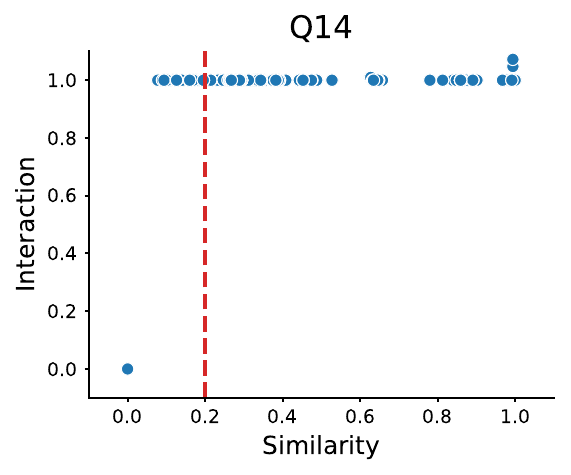}}
\subfigure{\includegraphics[width=0.19\textwidth]{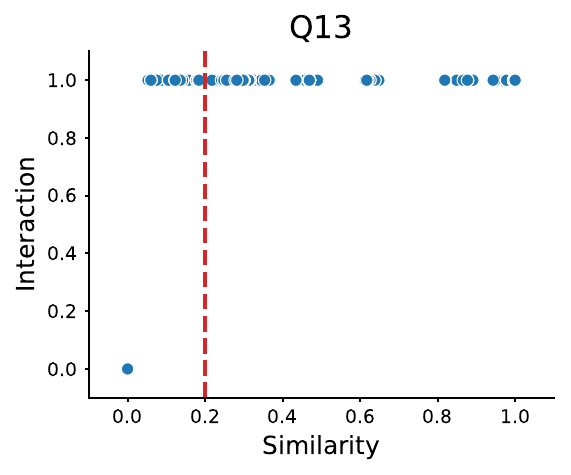}}
\subfigure{\includegraphics[width=0.19\textwidth]{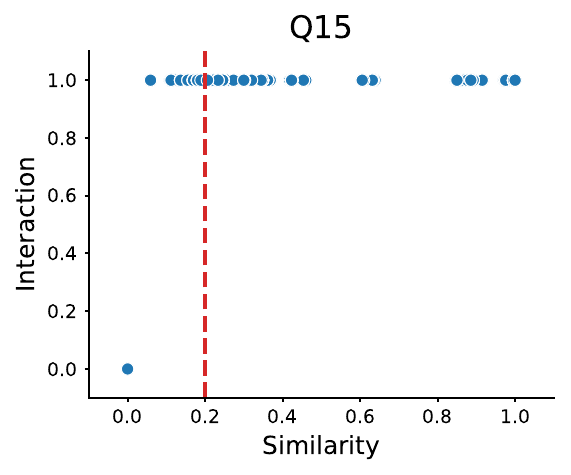}}
\subfigure{\includegraphics[width=0.19\textwidth]{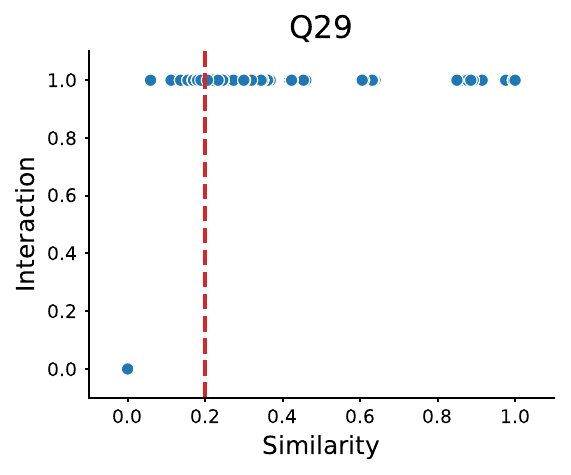}}
\subfigure{\includegraphics[width=0.19\textwidth]{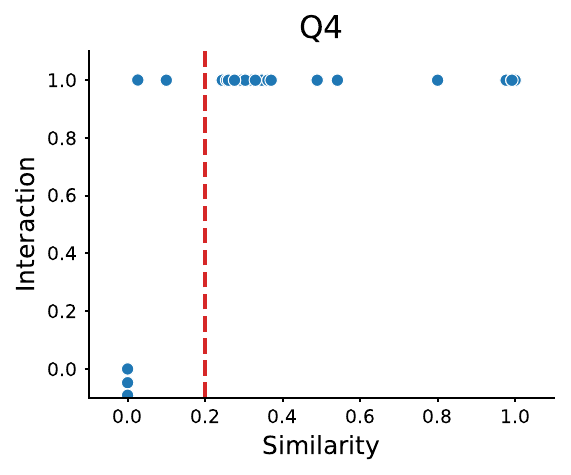}}
\subfigure{\includegraphics[width=0.19\textwidth]{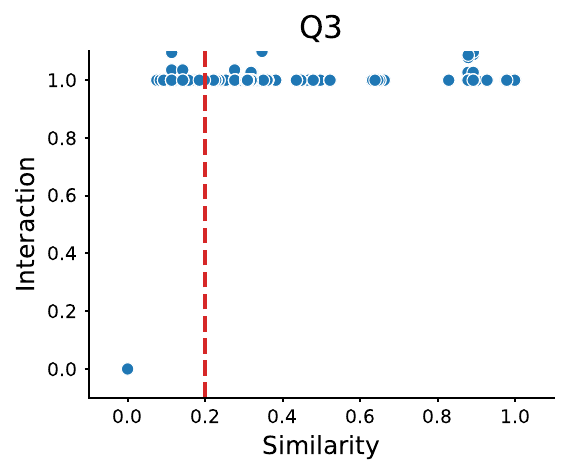}}
\subfigure{\includegraphics[width=0.19\textwidth]{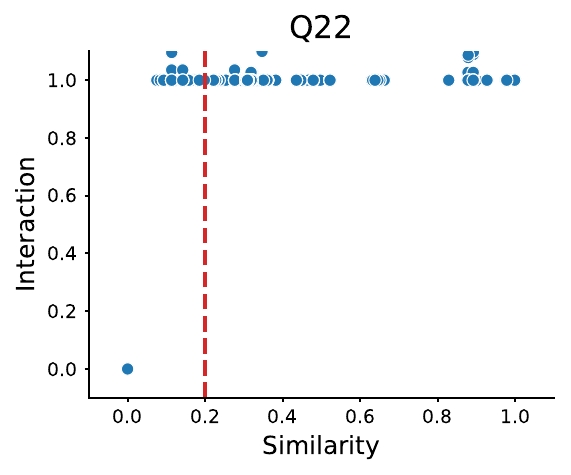}}
\vspace{-1em}
\caption{Relationship between pairwise index interaction and pairwise index similarity (Real-D).}
\label{fig:interaction-real-d}
\end{figure*}

\begin{figure*}
\centering
\subfigure{\includegraphics[width=0.19\textwidth]{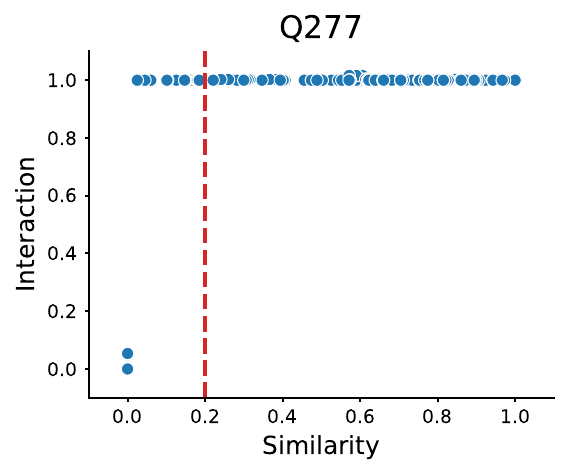}}
\subfigure{\includegraphics[width=0.19\textwidth]{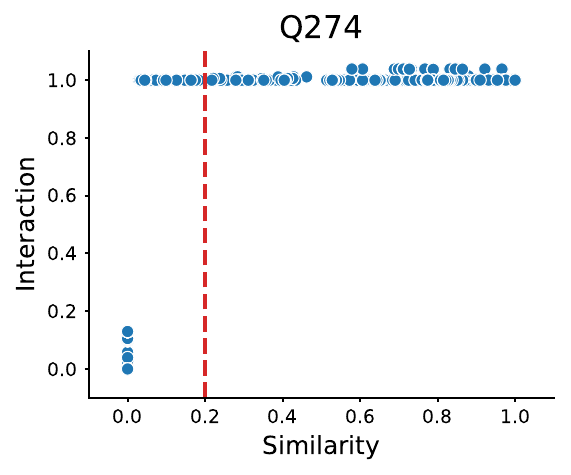}}
\subfigure{\includegraphics[width=0.19\textwidth]{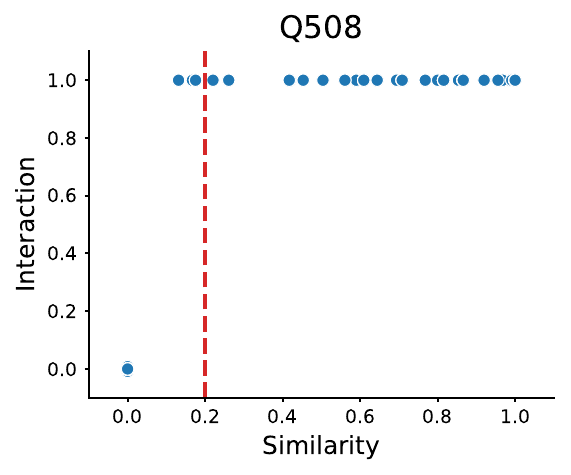}}
\subfigure{\includegraphics[width=0.19\textwidth]{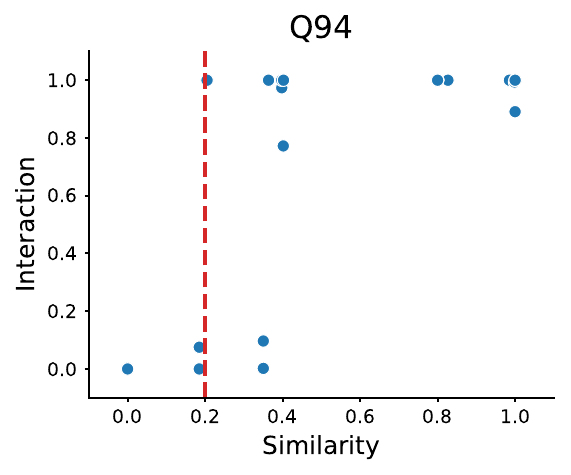}}
\subfigure{\includegraphics[width=0.19\textwidth]{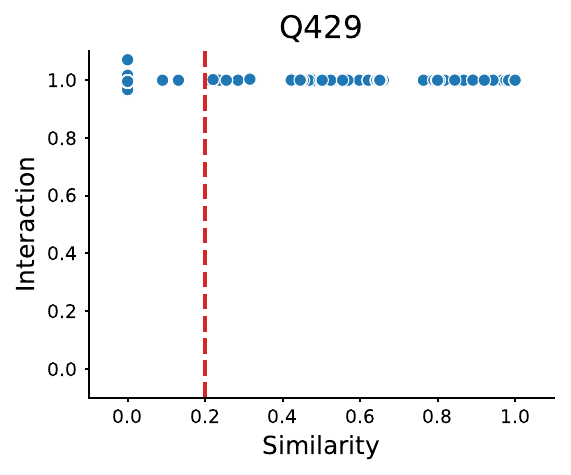}}
\subfigure{\includegraphics[width=0.19\textwidth]{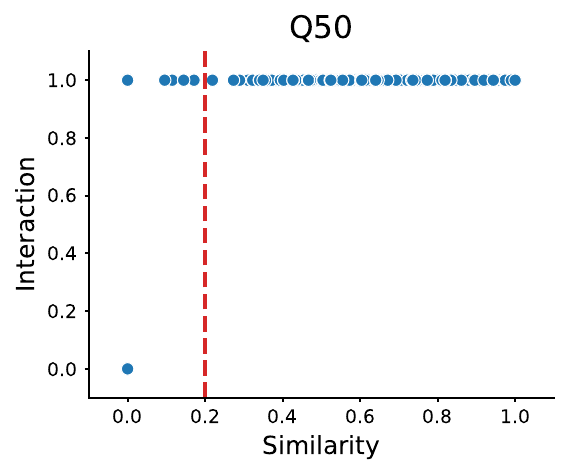}}
\subfigure{\includegraphics[width=0.19\textwidth]{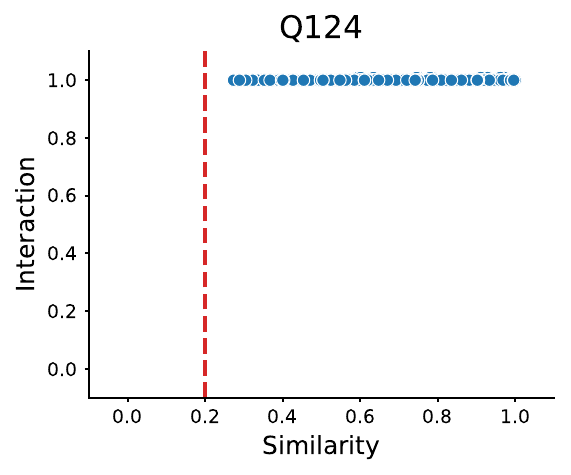}}
\subfigure{\includegraphics[width=0.19\textwidth]{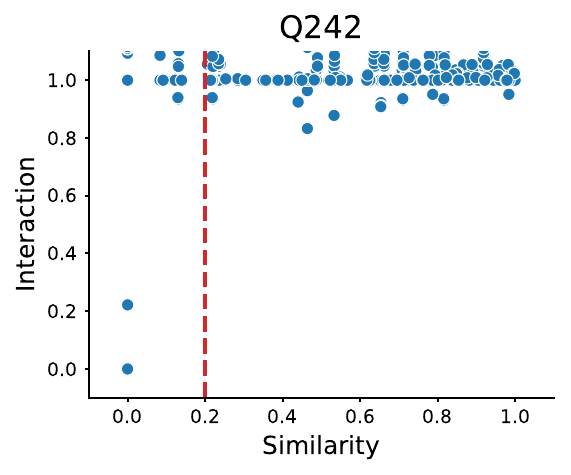}}
\subfigure{\includegraphics[width=0.19\textwidth]{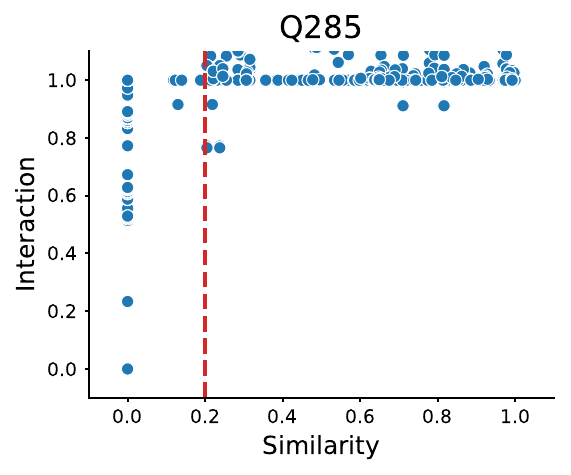}}
\subfigure{\includegraphics[width=0.19\textwidth]{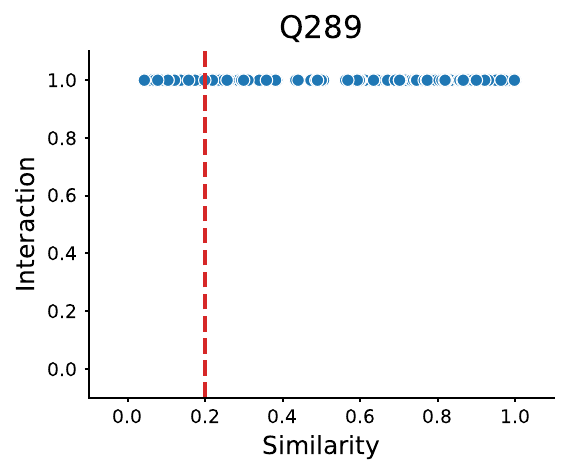}}
\vspace{-1em}
\caption{Relationship between pairwise index interaction and pairwise index similarity (Real-M).}
\label{fig:interaction-real-m}
\end{figure*}

The threshold $\tau$ that controls the degree of index interaction has an impact on the upper bound $\eta_U(W, C_B^*)$ for triggering early-stopping.
Our initial thought was that $\tau$ may be workload-dependent, namely, each workload needs its own customized threshold.
However, in our evaluation, we find that using a relatively small non-zero value, e.g., $\tau=0.2$, works consistently well across the workloads tested.
To understand this behavior better, Figures~\ref{fig:interaction-tpcds} to~\ref{fig:interaction-real-m} present evaluation results of index interaction on the other workloads \textbf{TPC-DS}, \textbf{JOB}, \textbf{Real-D}, and \textbf{Real-M}.
From the plots, we can see that for most queries, when the index similarity is larger than $0.2$, the index interaction is close to $1$ already. 
Indeed, there are some exceptional queries where the index interaction has no correlation with the index similarity. 
However, based on our experimental results, such violations have little impact on the efficacy of early-stopping.

\vspace{-0.5em}
\subsubsection{Impact of Step Size in Early-Stopping Verification}

The step size $s$ used by the generic ESVS proposed in Section~\ref{sec:integration:early-stopping:check} deals with the trade-off between (1) the extra time overhead of invoking ESV and (2) the savings on what-if calls when early-stopping is triggered.
Clearly, if we invoke ESV more frequently by using a smaller step size, we will incur higher extra time overhead but also result in larger savings on what-if calls.

\begin{figure*}
\centering
\subfigure[Time Overhead]{ \label{fig:mcts_covskip:real-d:k20:overhead}
    \includegraphics[width=0.49\columnwidth]{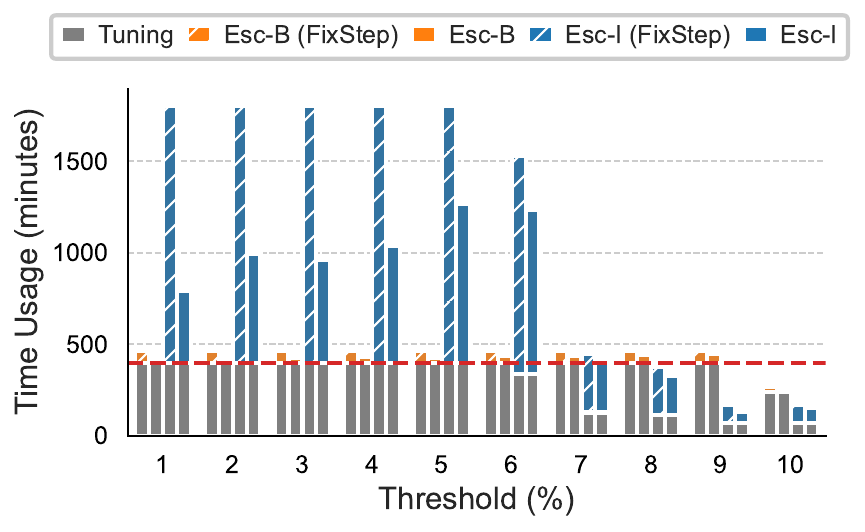}}
\subfigure[Improvement Loss]{ \label{fig:mcts_covskip:real-d:k20:impr-loss}
    \includegraphics[width=0.49\columnwidth]{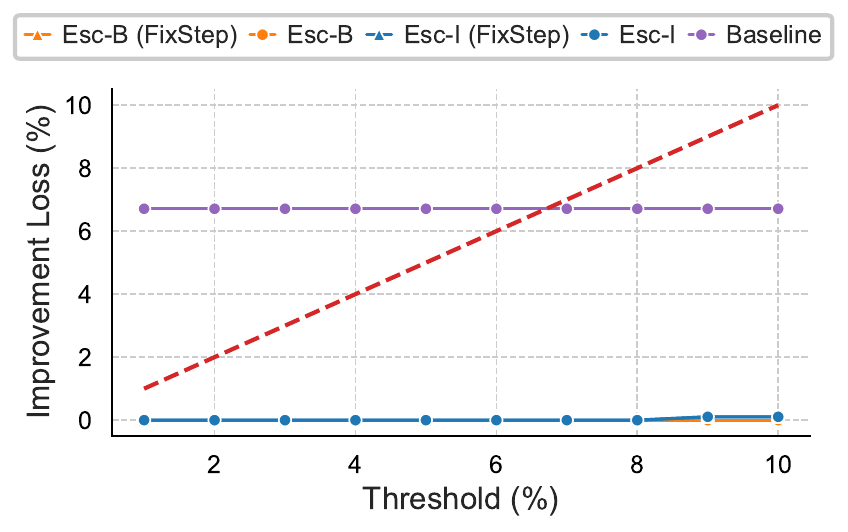}}
\subfigure[What-If Call Savings]{ \label{fig:mcts_covskip:real-d:k20:call-save}
    \includegraphics[width=0.49\columnwidth]{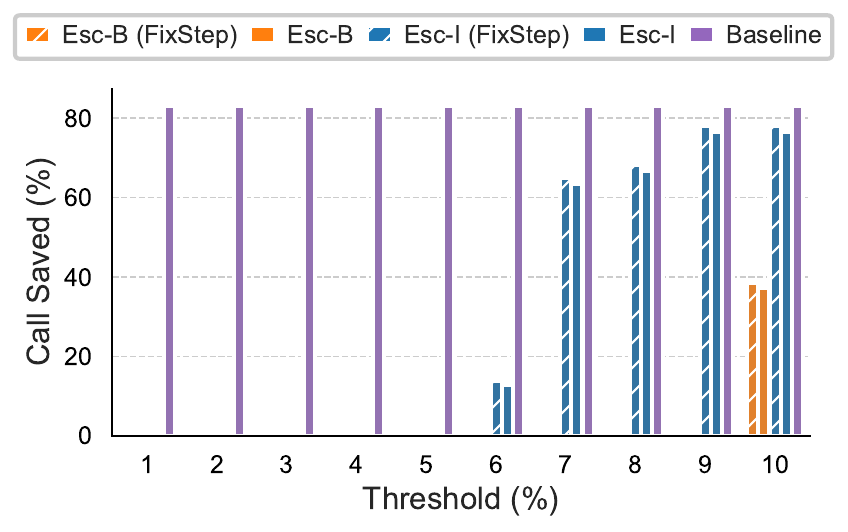}}
\subfigure[Learning Curve]{ \label{fig:mcts_covskip:real-d:k20:lc}
    \includegraphics[width=0.49\columnwidth]{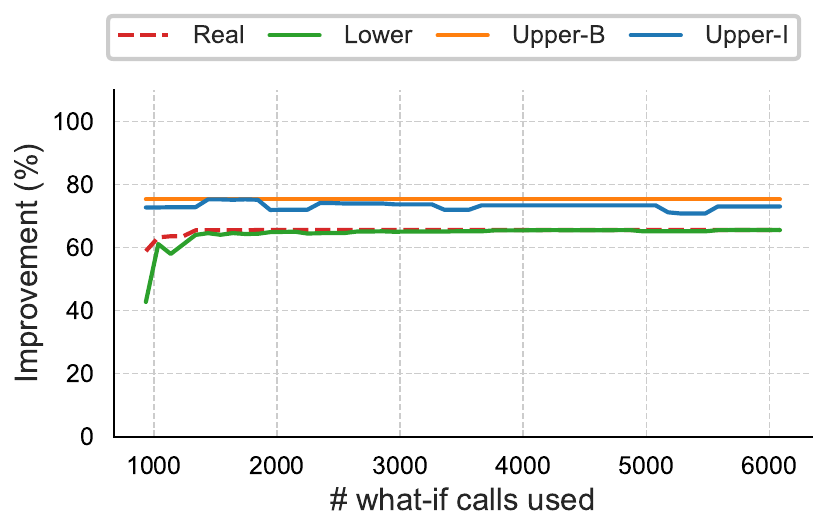}}
\vspace{-1.5em}
\caption{MCTS (with Wii-Coverage), Real-D, $K=20$, $B=20k$}
\label{fig:mcts_covskip:real-d:k20}
\vspace{-1em}
\end{figure*}
\begin{figure*}
\centering
\subfigure[Time Overhead]{ \label{fig:mcts:real-d:k20:overhead:s500}
    \includegraphics[width=0.49\columnwidth]{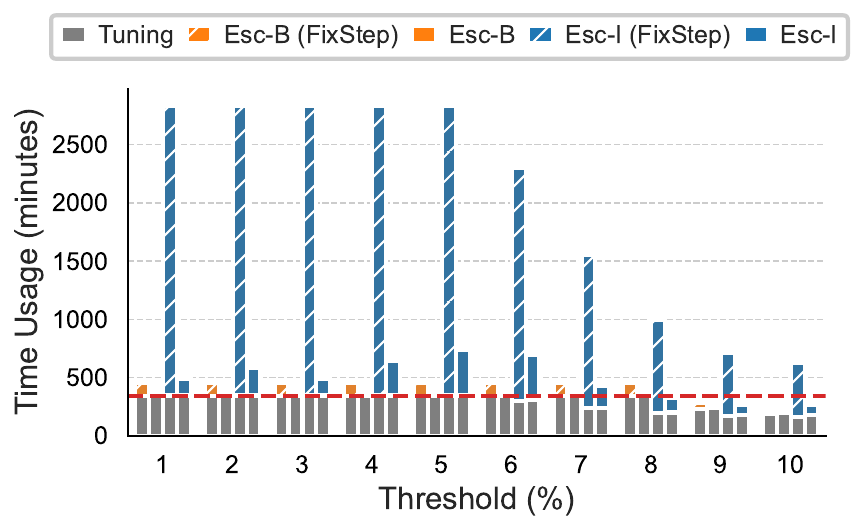}}
\subfigure[Improvement Loss]{ \label{fig:mcts:real-d:k20:impr-loss:s500}
    \includegraphics[width=0.49\columnwidth]{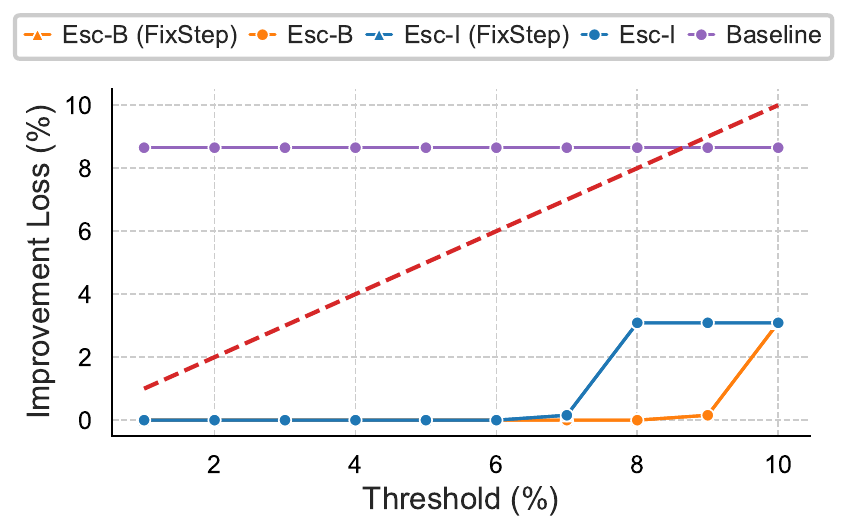}}
\subfigure[What-If Call Savings]{ \label{fig:mcts:real-d:k20:call-save:s500}
    \includegraphics[width=0.49\columnwidth]{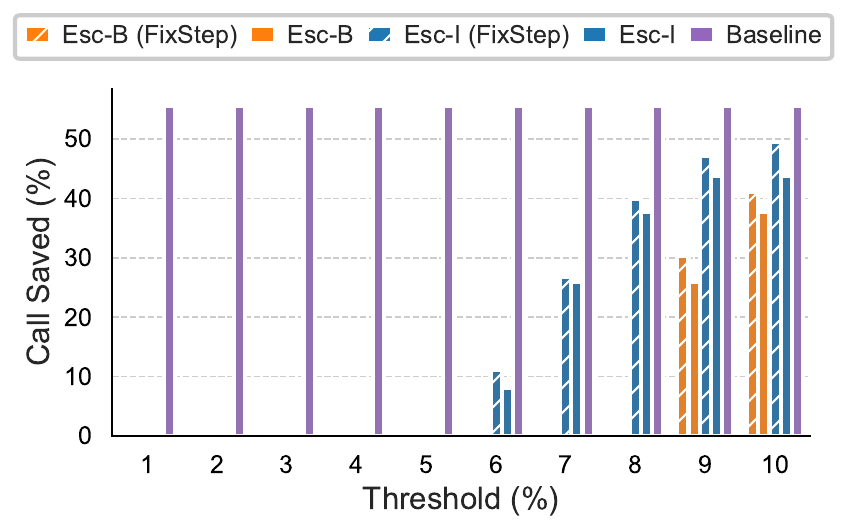}}
\subfigure[Learning Curve]{ \label{fig:mcts:real-d:k20:lc:s500}
    \includegraphics[width=0.49\columnwidth]{figs/new/skip-disable/real-d/mcts_real-d_K20_B20000_lc-eps-converted-to.pdf}}
\vspace{-1.5em}
\caption{MCTS, Real-D, $K=20$, $B=20k$, step size $s=500$}
\label{fig:mcts:real-d:k20:s500}
\vspace{-1em}
\end{figure*}
\begin{figure*}
\centering
\subfigure[Time Overhead]{ \label{fig:mcts:real-m:k20:overhead:s500}
    \includegraphics[width=0.49\columnwidth]{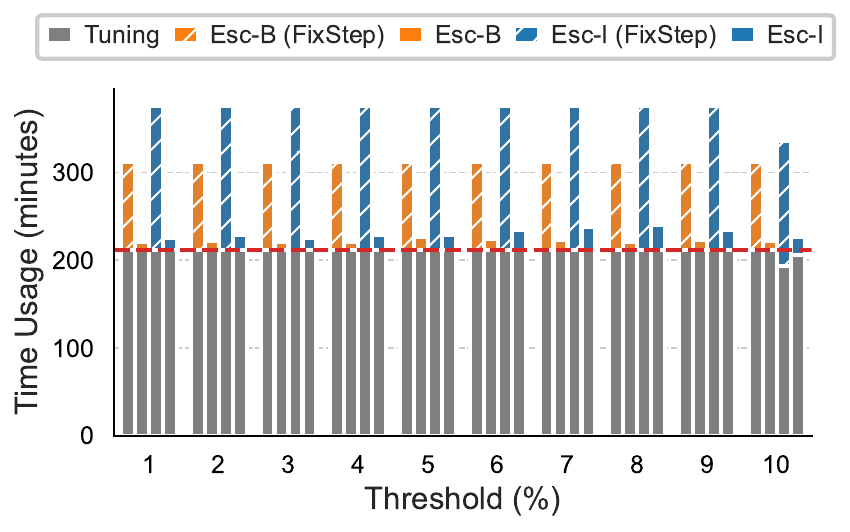}}
\subfigure[Improvement Loss]{ \label{fig:mcts:real-m:k20:impr-loss:s500}
    \includegraphics[width=0.49\columnwidth]{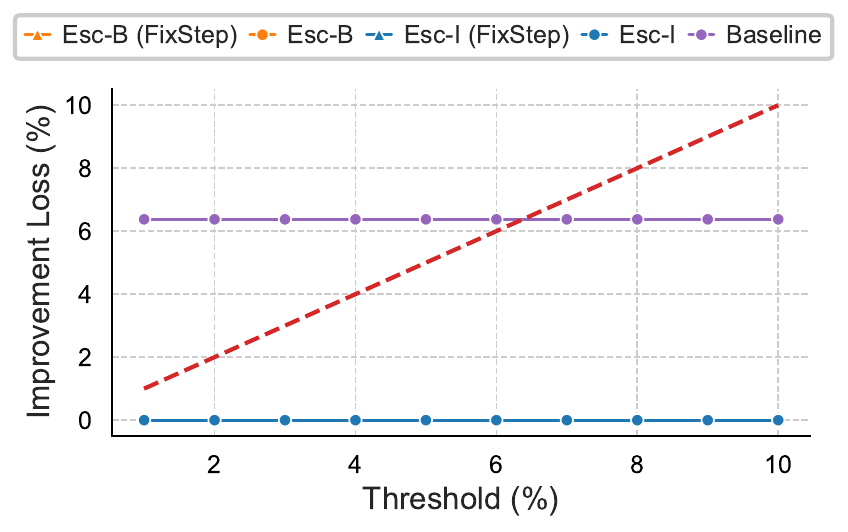}}
\subfigure[What-If Call Savings]{ \label{fig:mcts:real-m:k20:call-save:s500}
    \includegraphics[width=0.49\columnwidth]{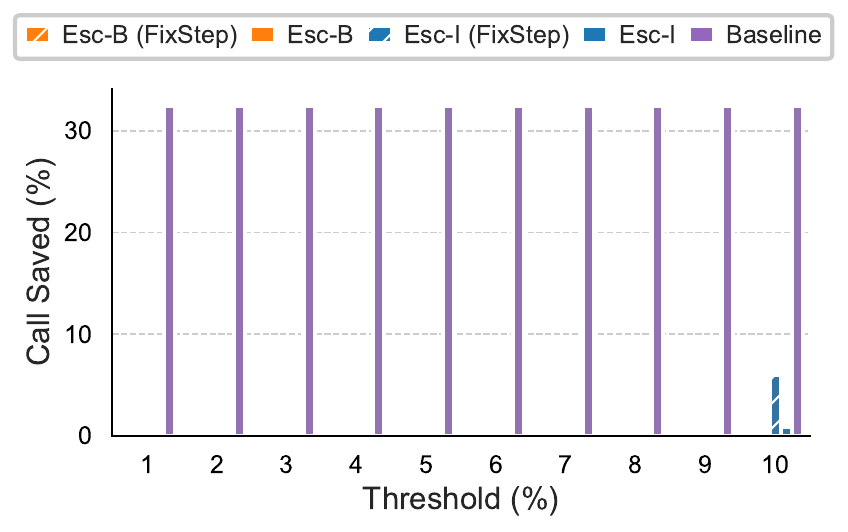}}
\subfigure[Learning Curve]{ \label{fig:mcts:real-m:k20:lc}
    \includegraphics[width=0.49\columnwidth]{figs/new/skip-disable/real-m/mcts_real-m-small_K20_B20000_lc-eps-converted-to.pdf}}
\vspace{-1.5em}
\caption{MCTS, Real-M, $K=20$, $B=20k$, step size $s=500$}
\label{fig:mcts:real-m:k20:s500}
\vspace{-1em}
\end{figure*}
\begin{figure*}
\centering
\subfigure[Time Overhead]{ \label{fig:mcts_skip:real-d:k20:overhead:s500}
    \includegraphics[width=0.49\columnwidth]{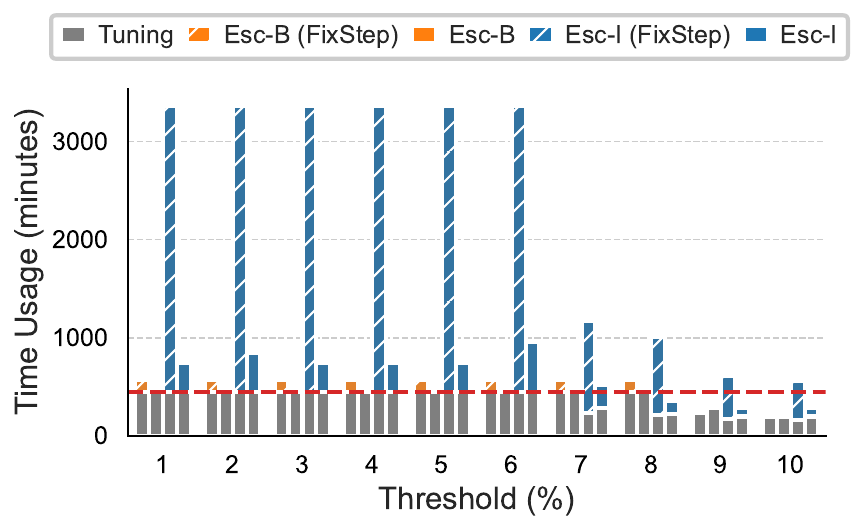}}
\subfigure[Improvement Loss]{ \label{fig:mcts_skip:real-d:k20:impr-loss:s500}
    \includegraphics[width=0.49\columnwidth]{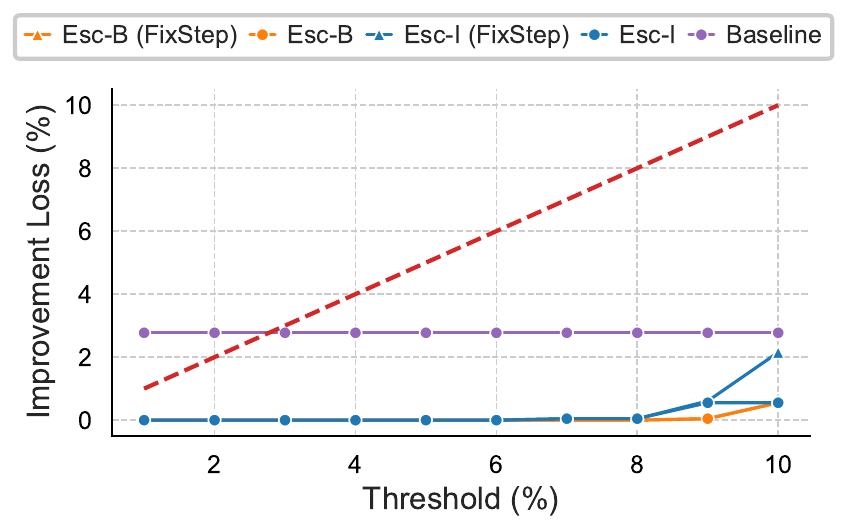}}
\subfigure[What-If Call Savings]{ \label{fig:mcts_skip:real-d:k20:call-save:s500}
    \includegraphics[width=0.49\columnwidth]{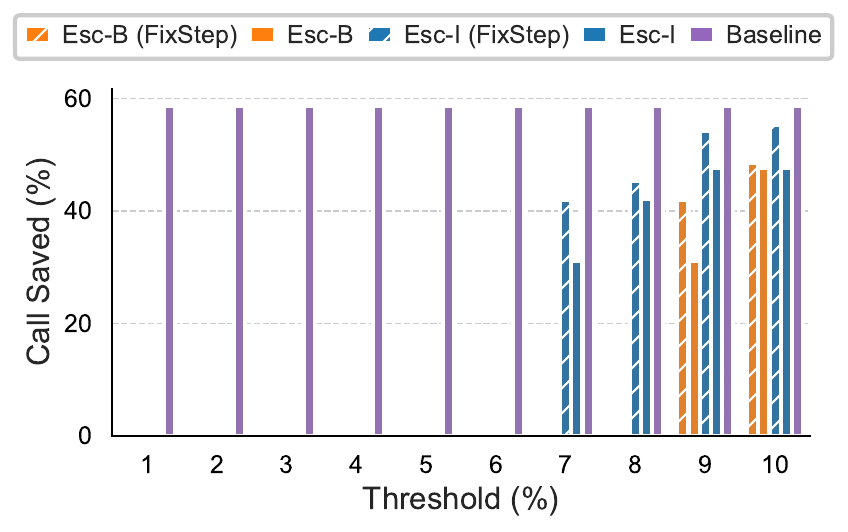}}
\subfigure[Learning Curve]{ \label{fig:mcts_skip:real-d:k20:lc}
    \includegraphics[width=0.49\columnwidth]{figs/new/skip-bound/real-d/mcts-skip_real-d_K20_B20000_lc-eps-converted-to.pdf}}
\vspace{-1.5em}
\caption{MCTS (with Wii), Real-D, $K=20$, $B=20k$, step size $s=500$}
\label{fig:mcts_skip:real-d:k20:s500}
\vspace{-1em}
\end{figure*}
\begin{figure*}
\centering
\subfigure[Time Overhead]{ \label{fig:mcts_skip:real-m:k20:overhead:s500}
    \includegraphics[width=0.49\columnwidth]{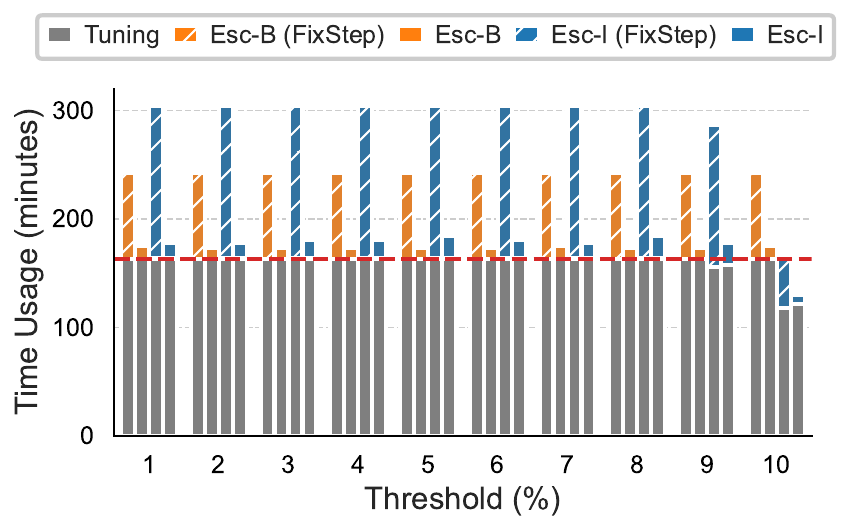}}
\subfigure[Improvement Loss]{ \label{fig:mcts_skip:real-m:k20:impr-loss:s500}
    \includegraphics[width=0.49\columnwidth]{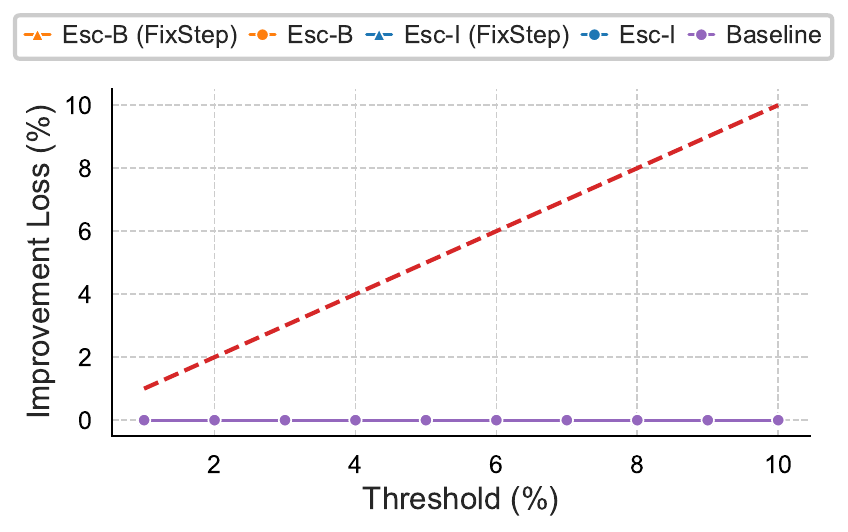}}
\subfigure[What-If Call Savings]{ \label{fig:mcts_skip:real-m:k20:call-save:s500}
    \includegraphics[width=0.49\columnwidth]{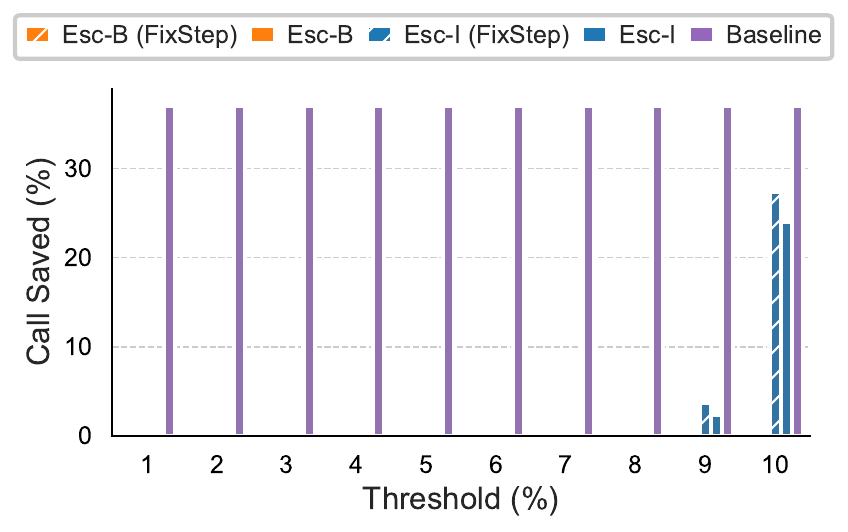}}
\subfigure[Learning Curve]{ \label{fig:mcts_skip:real-m:k20:lc}
    \includegraphics[width=0.49\columnwidth]{figs/new/skip-bound/real-m/mcts-skip_real-m-small_K20_B20000_lc-eps-converted-to.pdf}}
\vspace{-1.5em}
\caption{MCTS (with Wii), Real-M, $K=20$, $B=20k$, step size $s=500$}
\label{fig:mcts_skip:real-m:k20:s500}
\vspace{-1em}
\end{figure*}
\begin{figure*}
\centering
\subfigure[Time Overhead]{ \label{fig:mcts_covskip:real-d:k20:overhead:s500}
    \includegraphics[width=0.49\columnwidth]{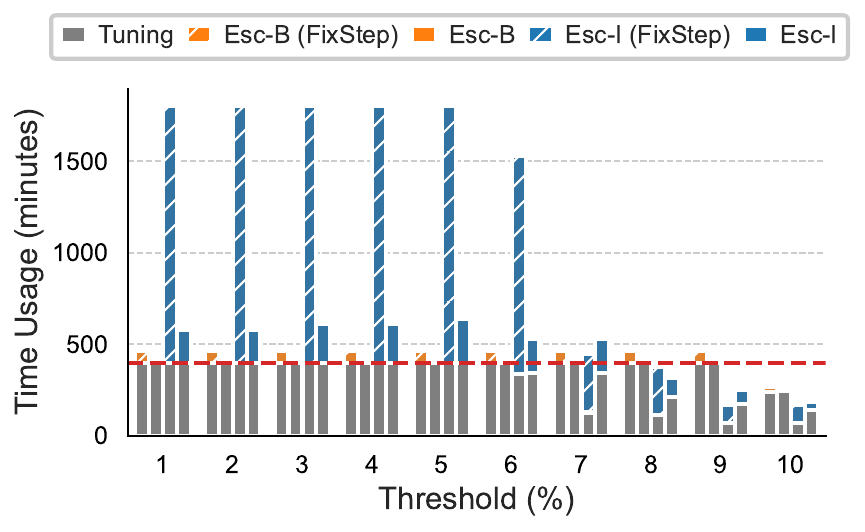}}
\subfigure[Improvement Loss]{ \label{fig:mcts_covskip:real-d:k20:impr-loss:s500}
    \includegraphics[width=0.49\columnwidth]{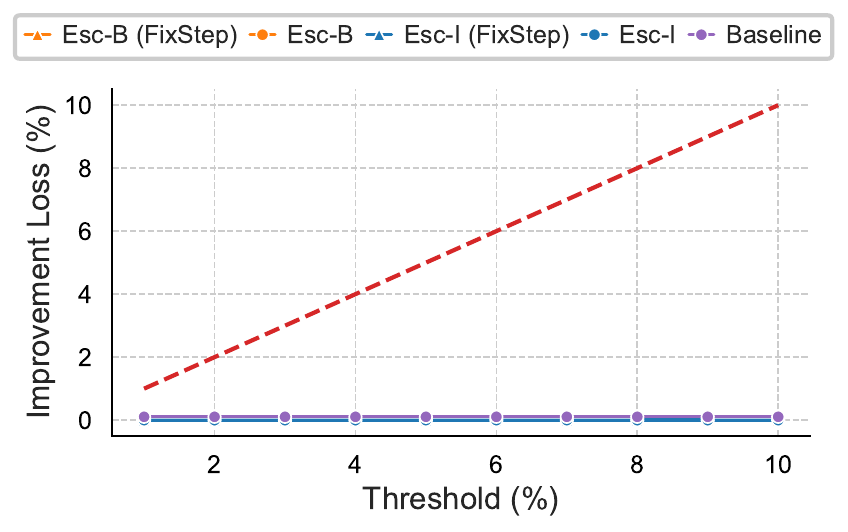}}
\subfigure[What-If Call Savings]{ \label{fig:mcts_covskip:real-d:k20:call-save:s500}
    \includegraphics[width=0.49\columnwidth]{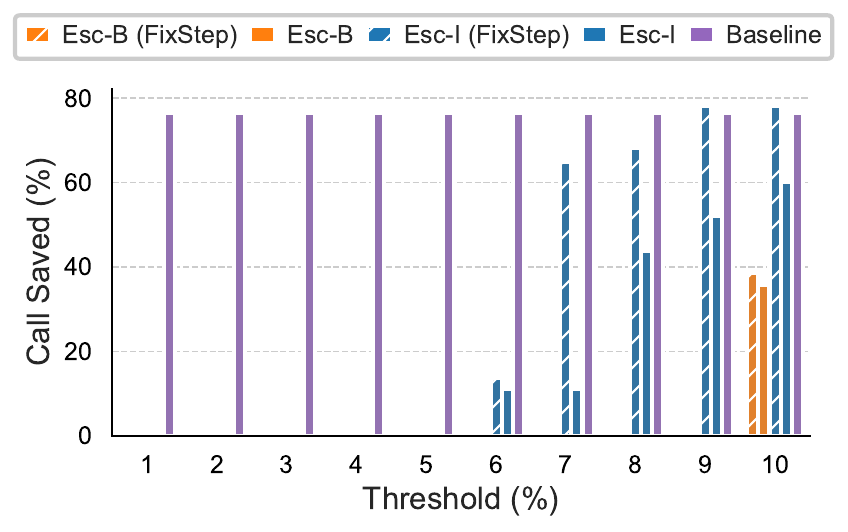}}
\subfigure[Learning Curve]{ \label{fig:mcts_covskip:real-d:k20:lc}
    \includegraphics[width=0.49\columnwidth]{figs/new/skip-coverage/real-d/mcts-covskip_real-d_K20_B20000_lc-eps-converted-to.pdf}}
\vspace{-1.5em}
\caption{MCTS (with Wii-Coverage), Real-D, $K=20$, $B=20k$, step size $s=500$}
\label{fig:mcts_covskip:real-d:k20:s500}
\vspace{-1em}
\end{figure*}
\begin{figure*}
\centering
\subfigure[Time Overhead]{ \label{fig:mcts_covskip:real-m:k20:overhead:s500}
    \includegraphics[width=0.49\columnwidth]{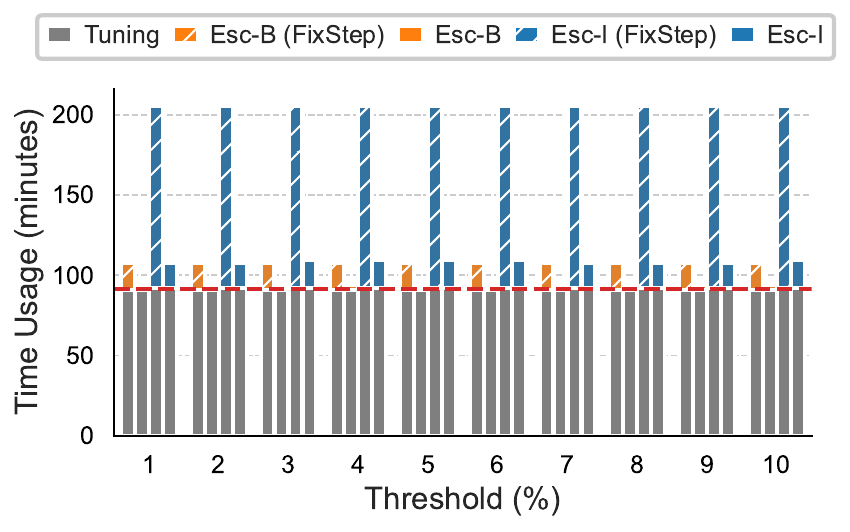}}
\subfigure[Improvement Loss]{ \label{fig:mcts_covskip:real-m:k20:impr-loss:s500}
    \includegraphics[width=0.49\columnwidth]{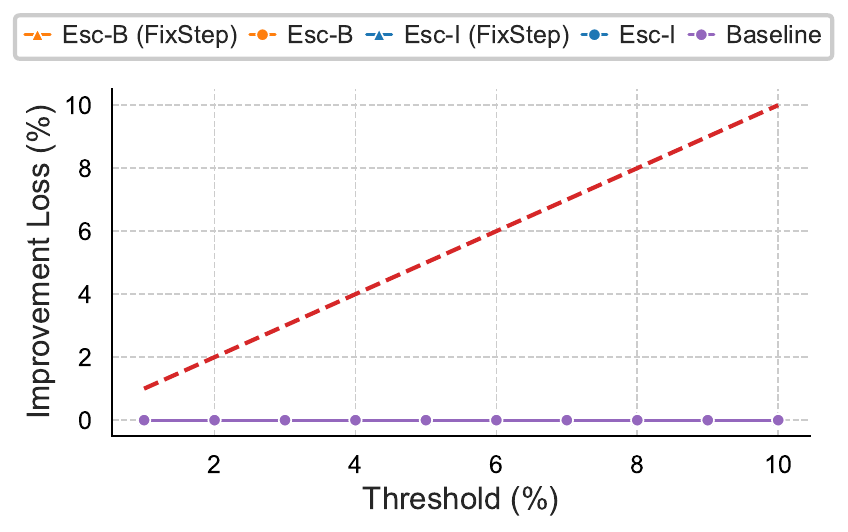}}
\subfigure[What-If Call Savings]{ \label{fig:mcts_covskip:real-m:k20:call-save:s500}
    \includegraphics[width=0.49\columnwidth]{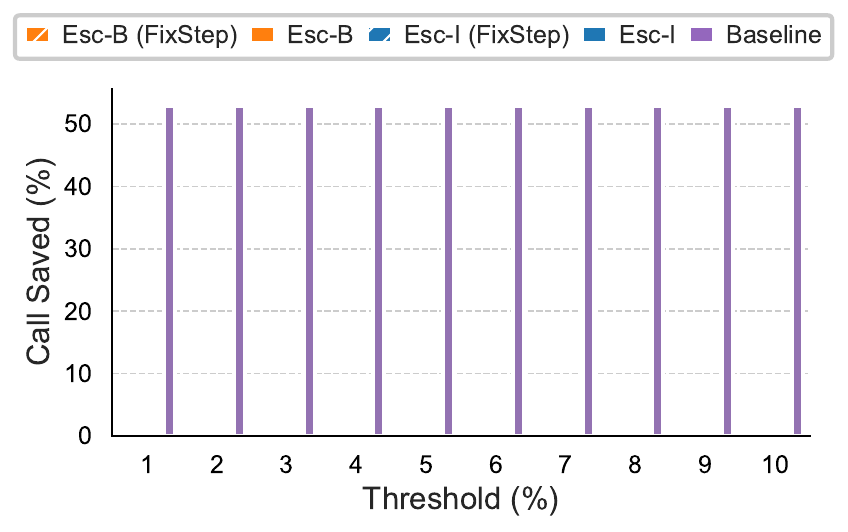}}
\subfigure[Learning Curve]{ \label{fig:mcts_covskip:real-m:k20:lc}
    \includegraphics[width=0.49\columnwidth]{figs/new/skip-coverage/real-m/mcts-covskip_real-m-small_K20_B20000_lc-eps-converted-to.pdf}}
\vspace{-1.5em}
\caption{MCTS (with Wii-Coverage), Real-M, $K=20$, $B=20k$, step size $s=500$}
\label{fig:mcts_covskip:real-m:k20:s500}
\vspace{-1em}
\end{figure*}

To demonstrate the impact of the step size $s$, Figures~\ref{fig:mcts:real-d:k20:s500} to~\ref{fig:mcts_covskip:real-m:k20:s500} present evaluation results for running \emph{MCTS} on top of the two real workloads \textbf{Real-D} and \textbf{Real-M}, by setting $s=500$ instead of $s=100$ as used in the previous evaluation.
We chose these two workloads because of the relatively higher extra time overhead introduced by early-stopping, using either \textbf{\sysname-B} or \textbf{\sysname-I}.
We can see the aforementioned trade-off by comparing the results of $s=500$ with those of $s=100$.
For example, comparing Figure~\ref{fig:mcts:real-m:k20:overhead:s500} with Figure~\ref{fig:mcts:real-m:k20:Time Usage}, we observe that the extra time overhead of using either \textbf{\sysname-B} or \textbf{\sysname-I} is significantly lowered on \textbf{Real-M} when setting $s=500$; however, on the other hand, the (percentage) savings on the what-if calls are also reduced, if we compare Figure~\ref{fig:mcts:real-m:k20:call-save:s500} with Figure~\ref{fig:mcts:real-m:k20:call-save}.
As another example, comparing Figure~\ref{fig:mcts_covskip:real-d:k20:overhead:s500} with Figure~\ref{fig:mcts_covskip:real-d:k20:overhead}, we can see that the extra time overhead of using \textbf{\sysname-I} is significantly lowered on \textbf{Real-D} when enabling coverage-based Wii-enhancement and setting $s=500$; however, the savings on the what-if calls can suffer from considerable drop (e.g., from 80\% to 60\% with $\epsilon=10\%$, or from 60\% to 10\% with $\epsilon=7\%$) by comparing Figure~\ref{fig:mcts_covskip:real-d:k20:call-save:s500} with Figure~\ref{fig:mcts_covskip:real-d:k20:call-save}.

\section{Proofs}

\subsection{Proof of Theorem~\ref{theorem:query-level-lower-bound}}

\paragraph*{Proof of the Lower Bound}
We have two observations for $u^{(i)}(q, z)$.
\begin{property}\label{property:monotone}
$u^{(i)}(q, z)\geq u^{(j)}(q,z)$ for $i\leq j$, due to the submodularity assumption (i.e., Assumption~\ref{assumption:submodular}).
\end{property}\label{property:fixed}
Note that the only interesting case here is when $u^{(i)}(q, z)$ and $u^{(j)}(q,z)$ are different, where we can apply the submodularity assumption to prove the ``monotonicity'' of $u^{(i)}(q, z)$.
\begin{property}\label{property:no-update}
$u^{(j)}(q,z_i)=u^{(i)}(q, z_i)$ for all $j\geq i$, since $u(q, z_i)$ will not be updated after $z_i$ is selected by greedy search.
\end{property}




Our next goal is to show
\begin{equation}\label{eq:u:monotone:greedy}
\sum\nolimits_{i=1}^k u^{(i)}(q,z_i)\leq \sum\nolimits_{j=1}^K u^{(j)}(q, z'_j).
\end{equation}
Without loss of generality, assume that the two sequences $\{z_1, ..., z_k\}$ and $\{z'_1, ..., z'_K\}$ start to diverge at some greedy step $s$. That is, $z_i=z'_i$ for $i<s$. We have the following observation:
\begin{property}\label{property:prefix}
If we order all the remaining indexes $z$ based on $u^{(s)}(q, z)$ in decreasing order, then $u^{(s)}(q, z'_s)$, ..., $u^{(s)}(q, z'_k)$ is the prefix in this ordering.
\end{property}
Since we will not have more what-if calls and therefore updates after the greedy step $s$ (otherwise the two sequences will not diverge at the greedy step $s$ since they follow the same budget allocation strategy), by Property~\ref{property:no-update} we have 
$$\sum\nolimits_{j=s}^K u^{(j)}(q, z'_j)=\sum\nolimits_{j=s}^K u^{(s)}(q, z'_j).$$
On the other hand, by Property~\ref{property:prefix} we must have $u^{(s)}(q, z_s)\leq u^{(s)}(q, z'_s)$, ..., $u^{(s)}(q, z_k)\leq u^{(s)}(q, z'_k)$.
As a result,
$$\sum\nolimits_{i=s}^k u^{(s)}(q, z_i)\leq \sum\nolimits_{j=s}^k u^{(s)}(q, z'_j)\leq \sum\nolimits_{j=s}^K u^{(s)}(q, z'_j).$$
By Property~\ref{property:monotone}, we further have
$$\sum\nolimits_{i=s}^k u^{(i)}(q, z_i)\leq \sum\nolimits_{i=s}^k u^{(s)}(q, z_i).$$
Therefore, we have proved
$$\sum\nolimits_{i=s}^k u^{(i)}(q, z_i)\leq\sum\nolimits_{j=s}^K u^{(s)}(q, z'_j)$$
and therefore Equation~\ref{eq:u:monotone:greedy} as well.

As a result, by Equations~\ref{eq:query-level:lower-bound:uncomputable}, \ref{eq:c_prim_q_C_t_u_K}, and~\ref{eq:u:monotone:greedy}, it follows that
\begin{eqnarray*}
L(q, C_B^*) &=& c(q,\emptyset)-\sum\nolimits_{j=1}^K u^{(j)}(q, z'_j)\\
&\leq & c(q, \emptyset)-\sum\nolimits_{i=1}^k u^{(i)}(q,z_i)\leq c(q, C_B^*).
\end{eqnarray*}

\subsection{Proof of Theorem~\ref{theorem:lower-bound-workload-level}}

\begin{proof}
There are three cases that we need to consider: (1) both $t$ and $B$ are in Phase 1; (2) $t$ is in Phase 1 and $B$ is in Phase 2; and (3) both $t$ and $B$ are in Phase 2.

For case (1) and (2) we have $u^{(j)}(q, z'_j)$ be always the same as Equation~\ref{eq:upper-bound:mci:call-level}, by the update step (b) of Procedure~\ref{proc:mci-update:two-phase},
which is the largest possible value of $u(q, z'_j)$.
We use $u^{\max}(q, z)$ to denote this largest possible value based on Equation~\ref{eq:upper-bound:mci:call-level}.
As before, without loss of generality, let $C_B^*=\{z_1, ..., z_K\}$. Let $C_i$ be the configuration selected in greedy step $i$ when $B$ what-if calls are allocated.
We have $C_i=C_{i-1}\cup\{z_i\}$ and
\begin{equation}\label{eq:c_B_star_two_phase}
c(W, \emptyset) - c(W, C_B^*)=\sum\nolimits_{i=1}^K \Big(c(W, C_{i-1})-c(W, C_i)\Big).
\end{equation}
For any query $q\in W$, we have $c(q, C_{i-1})-c(q, C_i)\leq u^{(i)}(q, z_i)$.
Therefore, $c(W, C_{i-1})-c(W, C_i)\leq\sum_{q\in W}u^{(i)}(q, z_i)$, and thus
\begin{eqnarray*}
c(W, C_B^*)&=& c(W, \emptyset) - \sum\nolimits_{i=1}^K \Big(c(W, C_{i-1})-c(W, C_i)\Big)\\
&\geq & c(W, \emptyset) -\sum\nolimits_{i=1}^K\sum\nolimits_{q\in W}u^{(i)}(q, z_i)\\
&=& c(W, \emptyset) -\sum\nolimits_{q\in W}\sum\nolimits_{i=1}^K u^{(i)}(q, z_i)\\
&\geq & c(W, \emptyset) -\sum\nolimits_{q\in W}\sum\nolimits_{i=1}^K u^{\max}(q, z_i)\\
&\geq & c(W, \emptyset) -\sum\nolimits_{q\in W}\sum\nolimits_{j=1}^K u^{\max}(q, z'_j).
\end{eqnarray*}
The last step is based on the fact that the simulated greedy search in Procedure~\ref{proc:simulation-greedy} will return the $K$ indexes $z$ with the largest $u^{\max}(q, z)$.
Since $u^{(j)}(q, z'_j)=u^{\max}(q, z'_j)$, we conclude that
\begin{eqnarray*}
c(W, C_B^*)&\geq & c(W, \emptyset) -\sum\nolimits_{q\in W}\sum\nolimits_{j=1}^K u^{(j)}(q, z'_j)\\
&=& c(W, \emptyset) -\sum\nolimits_{j=1}^K \sum\nolimits_{q\in W} u^{(j)}(q, z'_j)\\
&=& c(W, \emptyset) -\sum\nolimits_{j=1}^K u^{(j)}(W, z'_j)\\
&=& L(W, C_B^*).
\end{eqnarray*}

For case (3), we are in the same situation as regular greedy search, by the update step (c) in Procedure~\ref{proc:mci-update:two-phase}. 
Therefore, we can conclude that $L(W, C_B^*)\leq c(W, C_B^*)$ (see Procedure~\ref{proc:maintain-mci:greedy-search} and Theorem 1 of~\cite{Wii} for more details).
\end{proof}

\subsection{Proof of Lemma~\ref{lemma:concave}}

\begin{proof}
If $f$ is strictly concave and twice-differentiable, then $f''(b)<0$.
Consider the Taylor expansion of the function $f(x)$ at any particular point $b$. We have
$$f(x)=f(b)+f'(b)(x-b)+\frac{f''(\xi)}{2}(x-b)^2,$$
where $\xi$ is some number between $x$ and $b$.
Since $f''(\xi)<0$, it follows that
$$f(x)<f(b)+f'(b)(x-b).$$
In particular, this holds for $x=0$. As a result,
$$f(0)<f(b)-f'(b)\cdot b.$$
Since $f(0)=I_0=0$ by default, it follows that
$f'(b)<\frac{f(b)}{b}$, which completes the proof of the lemma.
\end{proof}

\subsection{Proof of Theorem~\ref{theorem:concave}}

\begin{proof}
By the definition of improvement rate (IR), we have 
$$r_j=\frac{I_j-I_0}{B_j-B_0}=\frac{I_j}{B_j}=\frac{f(B_j)}{B_j}.$$
By the definition of latest improvement rate (LIR), we have
$$l_j = \frac{I_j - I_{j-1}}{B_j - B_{j-1}}=\frac{f(B_j)-f(B_{j-1})}{B_j - B_{j-1}}.$$
Applying the Lagrange mean-value theorem, there exists some $\theta\in(B_{j-1}, B_j)$ such that
$$f(B_j)-f(B_{j-1})=f'(\theta)(B_j-B_{j-1}).$$
As a result, $l_j=f'(\theta).$
Since $f''(b)<0$ given that $f$ is strictly concave, $f'(b)$ is strictly decreasing.
It then follows that
$f'(\theta)<f'(B_j)$, since $\theta\in(B_{j-1}, B_j)$. By Lemma~\ref{lemma:concave}, $f'(B_j)<\frac{f(B_j)}{B_j}$. As a result, we have
$$l_j=f'(\theta)<f'(B_j)<\frac{f(B_j)}{B_j}=r_j,$$
which completes the proof of the theorem.
\end{proof}

\vspace{-0.5em}
\section{More Technical Details}

\subsection{More on Greedy Search}

\begin{algorithm}[t]
\small
  \SetAlgoLined
  \KwIn{$W$, the workload; $\mathcal{I}$, the candidate indexes; $K$, the number of indexes allowed.}
  \KwOut{$L(W, C_B^*)$, the lower bound of the what-if cost $c(W, C_B^*)$.}
  \SetAlgoLined
  $C^u_t\leftarrow\emptyset$, $S\leftarrow 0$\;
  \While{$\mathcal{I}\neq\emptyset$ and $|C^u_t|<K$} {
    $C^{\max}\leftarrow C^u_t$, $u^{\max}\leftarrow 0$\;
    \ForEach{index $z\in\mathcal{I}$} {
        $C_z\leftarrow C^u_t\cup\{z\}$, $u(W,z)\leftarrow\sum_{q\in W} u(q, z)$\;
        \uIf{$u(W, z)>u^{\max}$}{
            $C^{\max}\leftarrow C_z$, $u^{\max}\leftarrow u(W, z)$\;
        }
    }
    \uIf{$u^{\max}>0$} {
        $C^u_t\leftarrow C^{\max}$, $S\leftarrow S + u^{\max}$, $\mathcal{I}\leftarrow\mathcal{I}-C^u_t$\;
    }\uElse {
        \textbf{break}\;
    }
  }
  $L(W, C_B^*)\leftarrow c(W, \emptyset)-S$\;
  \Return{$L(W, C_B^*)$\;}
  \caption{Simulated greedy search for $L(W, C_B^*)$.}
\label{alg:simulated-greedy-search:lower-bound}
\end{algorithm}

The simulated greedy search outlined in Procedure~\ref{proc:simulation-greedy} can be used for computing both the workload-level lower bound $L(W, C_B^*)$ and upper bound $U(W, C_t^*)$.
Algorithms~\ref{alg:simulated-greedy-search:lower-bound} and~\ref{alg:simulated-greedy-search:upper-bound} illustrate the details.

\begin{algorithm}[t]
\small
  \SetAlgoLined
  \KwIn{$W$, the workload; $\mathcal{I}$, the candidate indexes; $C_t$, the best configuration at time $t$; $K$, the number of indexes allowed.}
  \KwOut{$U(W, C_t^*)$, the upper bound of the what-if cost $c(W, C_t^*)$.}
  \SetAlgoLined
    $C_t^*\leftarrow C_t$, $\mathcal{I}\leftarrow\mathcal{I}-C_t$, $\cost^{\min}\leftarrow d(W, C_t)$\;
  \While{$\mathcal{I}\neq\emptyset$ and $|C_t^*|<K$} {
    $C\leftarrow C_t^*$, $\cost\leftarrow\cost^{\min}$\;
    \ForEach{index $z\in\mathcal{I}$} {
        $C_z\leftarrow C_t^*\cup\{z\}$, $d(W, C_z)\leftarrow\sum_{q\in W} d(q, C_z)$\;
        \uIf{$d(W, C_z)<\cost$}{
            $C\leftarrow C_z$, $\cost\leftarrow d(W, C_z)$\;
        }
    }
    \uIf{$\cost < \cost^{\min}$} {
        $C_t^*\leftarrow C$, $\cost^{\min}\leftarrow\cost$, $\mathcal{I}\leftarrow\mathcal{I}-C_t^*$\;   
    }\uElse{
        \textbf{break}\;
    }
  }
  $U(W, C_t^*)\leftarrow d(W, C_t^*)$\;
  \Return{$U(W, C_t^*)$\;}
  \caption{Simulated greedy search for $U(W, C_t^*)$.}
\label{alg:simulated-greedy-search:upper-bound}
\end{algorithm}

\vspace{-0.5em}
\subsection{More On Index Interaction}

\subsubsection{Query Dependency}\label{appendix:index-interaction:query-dependency}
Index interaction is \emph{query-dependent}. That is, the same indexes may interact on one query but not on another.
To see this, consider the same $z_1$ and $z_2$ in Example~\ref{example:interaction} but a different SQL query $q_2$ in Figure~\ref{fig:index-interaction-example}, which involves a conjunctive range predicate on three columns $a$, $b$, and $c$. Although one can use $z_1$ for a partial evaluation of $q_2$ by first getting row ID's of the tuples that satisfy the predicate $b>10$ via $z_1$ and then scanning the table $R$ to fetch the rows w.r.t. to the row ID's, the execution cost of this query evaluation plan is likely to be higher than a simple table scan over $R$ unless the predicate $b>10$ is very selective.
As a result, we can assume $\Delta(q_2, \{z_1\})\approx 0$.
On the other hand, one can easily use $z_2$ to evaluate $q_2$ without referencing the table $R$.
In the presence of both $z_1$ and $z_2$, the query optimizer will then pick $z_2$ over $z_1$; hence, we have $\Delta(q_2, \{z_1, z_2\})=\Delta(q_2, \{z_2\})\approx \Delta(q_2, \{z_1\})+\Delta(q_2, \{z_2\})$.
Therefore, $z_1$ and $z_2$ \emph{do not interact} in the case of $q_2$.

\subsubsection{Optimization for MCTS}
In our experimental evaluation (see Section~\ref{sec:eval}), we found that the \emph{conditional benefit} $\mu^{(j)}(q, z'_j)$ defined by Equation~\ref{eq:conditional-benefit} can significantly improve the lower bound for \emph{two-phase greedy search}. 
However, for \emph{MCTS}, the lower bound often barely changes even with the refined $\mu^{(j)}(q, z'_j)$, due to the presence of many query-index pairs with unknown what-if costs.
Recall that, for such a query-index pair $(q, z)$, we have to use $u(q, z)=\Delta(q, \Omega_q)$, which is perhaps too conservative.
To alleviate this dilemma, we further refine $\mu^{(j)}(q, z'_j)$ for index $z'_j$ with interaction below the threshold (i.e., $\mathcal{S}(z'_j, C_{j-1} | q) \leq \tau$) as
$$
\mu^{(j)}(q, z'_j) = 
\begin{cases}
u^{(j)}(q, z'_j), & \text{if } \Delta(q, \{z'_j\}) \text{ is known}; \\
avg_{z \in \mathcal{K}(q, z'_j)}\Delta(q, \{z\}), & \text{otherwise}.
\end{cases}$$
Here, $\mathcal{K}(q,z'_j)=\{z | z \in I \land \mathcal{S}(z, z'_j|q) > \tau \land \Delta(q, \{z\}) \text{ is known} \}$. That is, for a query-index pair $(q, z)$ with unknown what-if cost, we initialize its MCI upper-bound by averaging the MCI upper-bounds of indexes with known what-if costs that are similar to $z$ w.r.t. $q$.

\vspace{-0.5em}
\subsection{More on Generic Verification Scheme}

\subsubsection{Significance of Concavity}
We measure the \emph{significance} of the potential concavity of the tuning curve.
Specifically, we \emph{project} the percentage improvement at $B_{j+1}$ using the improvement rates $l_j$ and $r_j$ and compare it with $I_{j+1}$ to decide whether we want to invoke verification at the time point $j+1$.
By Equation~\ref{eq:proj-impr}, the projected percentage improvements are
$$p_{j+1}^r=p_j^r(B_{j+1})=I_j+r_j(B_{j+1}-B_j)=I_j\cdot\frac{B_{j+1}}{B_j}$$
if we use the improvement rate $r_j$ and 
$$p_{j+1}^l=p_j^l(B_{j+1})=I_j+l_j(B_{j+1}-B_j)=I_j+\frac{I_j-I_{j-1}}{B_j-B_{j-1}}\cdot(B_{j+1}-B_j)$$
if we use the latest improvement rate $l_j$.
For the fixed-step verification scheme, we have 
$B_{j+1}-B_j=B_j-B_{j-1}=s.$
As a result, it follows that
$p_{j+1}^l=I_j + (I_j-I_{j-1})=2I_j - I_{j-1}.$
We now define the projected \emph{improvement gap} between $p_{j+1}^r$ and $p_{j+1}^l$ as
$$\Delta_{j+1}=p_{j+1}^r-p_{j+1}^l=I_j\cdot\Big(\frac{B_{j+1}}{B_j}-2\Big)+I_{j-1}.$$
Clearly, $\Delta_{j+1}>0$ since $l_j<r_j$. Moreover, the larger $\Delta_{j+1}$ is, the more significant the corresponding \emph{concavity} is. Therefore, intuitively we should have a higher probability of invoking verification.

\begin{algorithm}[t]
\small
  \SetAlgoLined
  \KwIn{$f$, the index tuning curve; $\{B_i\}_{i=1}^n$, a fixed-step ESVS with step size $s$; 
  $B_{j+1}$, the step to decide whether to invoke ESV ($j\geq 0$); $\sigma$, the threshold for significance of concavity.}
  \SetAlgoLined
  \uIf{$l_j\geq r_j$} {
        \textbf{return}; // Do NOT invoke verification at $B_{j+1}$.
    }\uElse{
        Compute $p_{j+1}^l$, $p_{j+1}^r$, and observe $I_{j+1}\leftarrow f(B_{j+1})$\;
        \uIf{$p_{j+1}^l<p_{j+1}^r<I_{j+1}$} {
            \textbf{return}; // Do NOT invoke verification at $B_{j+1}$.
        }\uElse{
            $\Delta_{j+1}\leftarrow p_{j+1}^r-p_{j+1}^l$, 
            $\delta_{j+1}\leftarrow p_{j+1}^r - I_{j+1}$,
            $\sigma_{j+1}\leftarrow\frac{\delta_{j+1}}{\Delta_{j+1}}$\;
            \uIf{$I_{j+1}<p_{j+1}^l$, or $p_{j+1}^l< I_{j+1} < p_{j+1}^r$ but $\sigma_{j+1} \geq \sigma$}{
                Invoke ESV, and obtain $L_j(W, C_t^*)$ and $U_j(W, C_B^*)$\;
                \uIf{ESV returns \emph{true}} {
                    Terminate index tuning\;
                }\Else{
                    // Refine improvements for step $j+2$ (and later).\\
                    $r_{j+1}\leftarrow\min\{\frac{I_{j+1}}{B_{j+1}}, \frac{U_{j+1}-I_{j+1}}{s}\}$\;
                    $l_{j+1}\leftarrow\min\{\frac{I_{j+1}-I_j}{s}, \frac{U_{j+1}-I_{j+1}}{s}\}$\;
                }
            }\uElse{
                \textbf{return}; // Do NOT invoke verification at $B_{j+1}$.
            }
        }
    }
  \caption{A generic early-stopping verification scheme.}
\label{alg:generic-esvs}
\end{algorithm}

\vspace{-0.5em}
\subsubsection{The Generic ESVS}
Algorithm~\ref{alg:generic-esvs} presents the details of the generic ESVS at $B_{j+1}$ ($j\geq 0$) without the probabilistic mechanism for invoking ESV, which will be further detailed below.

\subsubsection{A Probabilistic Mechanism for Invoking ESV}

One problem of Algorithm~\ref{alg:generic-esvs} is that, if the observed improvement is flat (i.e., $l_i=0$) but the lower and upper bounds are not converging yet, then it may result in unnecessary ESV invocations. 
We therefore need to further consider the convergence of the bounds.

Let $L_j(W, C_t^*)$ and $U_j(W, C_B^*)$ be the lower and upper bounds returned (from line 10 of Algorithm~\ref{alg:generic-esvs}). 
We define 
$G_j(W, C_t^*, C_B^*)=U_j(W, C_B^*) - L_j(W, C_t^*)$ as the gap between the lower/upper bounds and further define 
$\rho_j(W, C_t^*, C_B^*)=\frac{G_j(W, C_t^*, C_B^*)}{\epsilon}$ as the \emph{relative gap} w.r.t. the threshold $\epsilon$ of improvement loss. Intuitively, the smaller $\rho_j$ is, the more likely that we can stop index tuning the next time when we check for early-stopping. Clearly, $\rho_j>1$. As a result, $0<\frac{1}{\rho_j}<1$ and we can use $\lambda_j=\frac{1}{\rho_j}$ to measure the probability of early-stopping after time point $j$.
A higher $\lambda_j$ implies a lower $\rho_j$ and thus higher chance of early-stopping (as $G_j$ is closer to $\epsilon$). 

To apply this mechanism, at line 10 of Algorithm~\ref{alg:generic-esvs}, instead of always invoking ESV, we invoke it with probability $\lambda_j$.


\clearpage



\clearpage

\end{document}